\documentclass[12pt]{report}
\PassOptionsToPackage{breaklinks}{hyperref}
\usepackage{amsmath}
\usepackage{amssymb}
\usepackage{cite}
\newcommand{\ds}{\displaystyle}
\newcommand{\vp}{\varphi}
\newcommand{\tr}{\mbox{tr}(}
\begin{document}

\begin{titlepage}
\begin{flushleft}
DESY 03-079 \\
 HU-EP-03/25 \\
quant-ph/0307069
 \end{flushleft} \vspace{0.5cm}
\begin{center}
{\Large Quantization of the Optical Phase Space \vspace{0.5cm} \\
 ${\cal S}^2 =\{\vp \bmod{2\pi}, I >0\}$
 \vspace{0.5cm}\\
in Terms of the Group $SO^{\uparrow}(1,2)$}
 \vspace{1.0cm}
\\  {\large H.A.\
Kastrup\footnote{E-mail: Hans.Kastrup@desy.de}} \vspace{0.4cm}
\\ {DESY, Theory Group \\ Notkestr.\ 85, D-22603 Hamburg
\\Germany} \end{center}\vspace{0.4cm}

 \begin{center}{\bf Abstract}\end{center}
The problem of quantizing properly the canonical pair ``angle and action 
 variables '', $\vp$ and $I$, is almost as old as quantum mechanics itself
 and since decades
an intensively debated but still unresolved issue in quantum optics.
 The present paper proposes
a new approach to the problem, namely quantization in terms of the group
$SO(1,2)$: The crucial point is that the phase space $ {\cal S}^2 =
\{\vp \bmod{2\pi},\,I> 0\}$ has the global structure $S^1 \times \mathbb{R}^+$
(a simple cone) and cannot be quantized in the conventional manner. As
 the group $SO(1,2)$ acts transitively, effectively and Hamilton-like on that
space its irreducible unitary representations of the positive discrete series
provide the appropriate quantum theoretical framework. The phase space $ {\cal
S}^2$ has the conic structure  of an orbifold $\mathbb{R}^2/Z_2$. That
 structure is closely  related to a $Z_2$ gauge symmetry which
 corresponds to the center of a 2-fold covering of $SO(1,2)$, the
 symplectic group
$Sp(2,\mathbb{R})$. The basic variables on the phase space are the functions
$h_0 =I\,,\,h_1 = I\,\cos\vp$ and $h_2 =-I\,\sin\vp$ the Poisson brackets of
which obey the Lie algebra $\mathfrak{so}(1,2)$. In the quantum theory they
are represented by the self-adjoint Lie algebra generators $K_0,K_1$ and $K_2$ 
of a unitary representation, where $K_0$ has the spectrum
 $\{k+n,\,n=0,1,\ldots; k>0\}$. A crucial prediction is that the classical
Pythagorean relation $h_1^2+h_2^2 =h_0^2$ can be violated
 in the quantum theory.
For each representation one can define three different types of coherent
states the complex phases of which may be ``measured'' by means of
 the operators $K_1$ and $K_2$ alone without introducing any new phase
 operators! The $SO(1,2)$ structure of optical squeezing and
 interference properties 
as well as that of the harmonic oscillator are analyzed in detail. 
The additional
coherent states can be used for the introduction of  (Husimi type) ``Q''
distributions and (Sudarshan-Glauber type) ``P'' representations of the 
density operator. The three operators $K_0, K_1$ and $K_2$ are fundamental
in the sense that one can construct composite position and momentum operators
out of them! The new framework poses quite a number of fascinating
experimental and theoretical challenges. 
\end{titlepage}

\tableofcontents

\chapter{Introduction and overview}
The problem how to quantize  modulus and  phase of a wave as
some kind of canonically conjugate action and angle variables and
 relate them to
genuine self-adjoint operators in Hilbert spaces is a very old one
and appears - according to the still ongoing controversial
discussions in the field of quantum optics - not yet settled.

 (See,
e.g.\ the reviews
\cite{rev1,pau1,rev2,rev2a,rev2b,rev3,rev4,rev5,rev6,rev8} and the
textbooks \cite{texb1,texb2,texb3,texb4,texb5,texb6,texb7,texb8})

A solution of that theoretical problem becomes more and more
urgent, however, because the fascinating experiments in quantum optics become
increasingly more refined and allow to differentiate between
different theoretical schemes.

The present paper addresses the problem from a new point of view,
namely quantization of classical phase spaces in terms of groups
and their irreducible unitary representations \cite{is1,gui}. This
approach to the quantization of classical systems provides a
genuine extension of the conventional quantization procedure which
is applicable to phase spaces of the type $\mathbb{R}^{2n}$ only. 

The more general approach used in this paper allows for a quantization
of phase spaces which have a {\em global} topological structure which is
different from that of  $\mathbb{R}^{2n}$ and which makes predictions
which can be tested experimentally.

\section{The problem}
Let me illustrate the essential origin of the difficulties with the
conventional quantization procedure applied to the phase-modulus pair by
 the harmonic oscillator with the Hamiltonian (the mass $m$ and the
frequency $\omega$ are scaled to $1$)
\begin{equation}
  \label{eq:1}
H(q,p)=\frac{1}{2}\,p^2+\frac{1}{2}\,q^2~,(q,p) \in
 \mathbb{R}^2\,,~ \{q,p\}_{q,p}=1\,,
\end{equation}
where
\begin{equation}
  \label{eq:2}
 \{f_1,f_2\}_{q,p} \equiv
\partial_{q}f_1\,\partial_p f_2-\partial_p f_1\,
\partial_{q}f_2
\end{equation}
denotes the Poisson bracket for the phase
space functions $f_1(q,p)$ and $f_2(q,p)$. It is intimately
connected to the symplectic form (see Appendix A.1)
\begin{equation}
  \label{eq:740}
   \omega = dq \wedge dp\,.
\end{equation}

  The transformation to
  action and  angle variables $I >0$ and $\vp \in (-\pi\, \pi]$,
 \begin{eqnarray}
 \label{eq:1501}q(\vp,I)&=&  \sqrt{2\,I}\,\cos\vp\,, \\
 \label{eq:1003}p(\vp,I)&=& -\sqrt{2\, I}\,\sin\vp\,, \end{eqnarray}
 is locally canonical because 
\begin{equation}
  \label{eq:3}
 \frac{\partial(q,p)}{\partial(\varphi,I)}=1\,,~\mbox{ or equivalently }\,:~
     dq \wedge dp =  d\vp \wedge dI\,, 
\end{equation}
for its
  functional determinant. 

 The transformation \ref{eq:1501} - \ref{eq:1003} yields
\begin{equation}
  \label{eq:4}
 H = I\,,~~ \{\vp,I\}_{\vp,I}=1\,,
\end{equation}
where
\begin{equation}
  \label{eq:5}
 (\vp,I) \in { \cal{S}}^2_{\vp,I} =
  \{(\varphi \in \mathbb{R} \bmod{2\pi},I\,>\,0)\}\,,
\end{equation}
and where now
  the Poisson bracket applies to functions $h_j(\vp,I)\,,\,j=1,2$:
\begin{equation}
  \label{eq:6}
 \{h_1,h_2\}_{\vp,I} \equiv
\partial_{\vp}h_1\,\partial_I h_2-\partial_I h_1\,
\partial_{\vp}h_2\,.
\end{equation}
This new
  phase space ${ \cal{S}}^2_{\vp,I}$ no longer has the global topology
  $\mathbb{R}^2$ we started from, \ref{eq:1},  but is homeomorphic to
\begin{equation}
  \label{eq:7}
 S^1 \times
 \mathbb{R}^+\,,~~\mathbb{R}^+ =\{ r \in \mathbb{R}\,, r>0
 \}\,,
\end{equation}
where $S^1$ denotes the unit circle. This is the
 topology of a simple cone with the vertex deleted or that of $\mathbb{R}^2
 -(0,0)$, i.e. the plane without the origin! 

The reason for deleting the origin or to demand $I >0$ is the following:

At first sight that deletion does not to appear necessary because the
 functional
determinant \ref{eq:3} is regular at $I=0$. However, the transformation
formulae \ref{eq:1501} and \ref{eq:1003} contain the factor $\sqrt{I}$ which
is {\em not analytic} at $I=0$! 

Another way of looking at the problem is to introduce polar coordinates
$\vp$ and $\rho = \sqrt{I}$. Now we get the functional determinant
\begin{equation}
  \label{eq:741}
 \frac{\partial(q,p)}{\partial(\varphi,\rho)}= 2\, \rho\,,  
\end{equation}
which shows that the transformation is not regular for $\rho=0$!

 Thus, the symplectic space \ref{eq:5} confronts us with a non-trivial
  topology which prevents a ``naive'' quantization
 approach even though we {\em locally} still have the equality \ref{eq:3}
 of the symplectic forms!

We know from the Aharonov-Bohm effect \cite{ahar} that a ``punctured plane''
can  yield very interesting physical effects and that one should take the
hole seriously, being it as small as it may! The crucial point is that
one cannot contract loops around it to a point.

We shall see below that the picture of a simple cone with its vertex point
deleted will be more adequate than the picture of a punctured plane. 

  Before pointing out explicitly the well-known
 difficulties with quantizing the canonical pair $(\vp,I)$ naively,
 let me write down the complex ``amplitudes''
\begin{eqnarray}
 \label{eq:1502}a&=&\frac{1}{\sqrt{2}}(q+i\,p) =
 I^{1/2}e^{-i\,\vp}\,,\\
\label{eq:1004}\bar{a}
&=&\frac{1}{\sqrt{2}}(q-i\,p)
= I^{1/2}e^{i\,\vp}\,, \\ \label{eq:1005}I & =& \bar{a}\,a\,, \end{eqnarray}
 which become annihilation and creation operators for the quantized
system. 

(Here and in the following $\bar{a}$ denotes the complex conjugate
of $a$!)

 The conventional recipe for quantizing a 2-dimensional
classical phase space - like that of the harmonic oscillator -
consists in replacing the  pair $(q,p)$ of canonical variables by
self-adjoint operators $(Q,P)$ in an appropriate Hilbert space,
where the classical Poisson bracket \ref{eq:1} is replaced by the
quantum mechanical commutator\footnote{I assume $\hbar = 1$  throughout
the whole paper}:
\begin{equation}
  \label{eq:8}
  (q,p) \to  (Q, P)\,,~~ \{q,p\}=1 \to [Q, P] = i\,.
\end{equation}
This commutator implies that {\em self-adjoint} $Q$ and $P$ both have
the full real line $\mathbb{R}$ as their spectrum, reflecting the
fact that we started from a classical phase space with the global
structure $\mathbb{R}^2$.

 We shall see more details of this below
when we discuss the Weyl-Heisen\-berg group generated by the operators 
\ref{eq:8}.

Instead of  the classical Hamiltonian \ref{eq:1} we now get the Hamilton
 operator
\begin{eqnarray}  \label{eq:1503}\hat{H} &=&
\frac{1}{2}\,P^2+\frac{1}{2}\,Q^2
\\ \label{eq:1006}&=&   \hat{N}+\frac{1}{2}\,,~~
\hat{N}=\hat{a}^+\hat{a}\,,
\\  \label{eq:1007}\hat{a} &=&
\frac{1}{\sqrt{2}}(Q+i\,P)\,,
\\  \label{eq:1008}\hat{a}^+ &=&
\frac{1}{\sqrt{2}}(Q-i\,P)\,,
\\ \label{eq:1009}&&  [\hat{a},\hat{a}^+]=1\,. \end{eqnarray} $\hat{H}$ has the
well-known normalizable eigenstates $|n \rangle\,,n=0,1,\ldots$,
and the  spectrum $E_n= n+1/2\,$. 

 We now
come to the crucial point: 

 In view of the canonical character of
the transformations \ref{eq:1501} and \ref{eq:1003} and the  Poisson bracket
\ref{eq:4} one might be tempted to quantize $\vp$ and $I$ by the
replacements \cite{dir1,hei}
\begin{equation}
  \label{eq:9}
 \vp \to \hat{\vp}\,,~~ I \to
\hat{N}\,,~~ \{\vp,I\}=1 \to [\hat{\vp},\hat{N}]= i\,.
\end{equation}
That
commutator, however, implies an immediate contradiction when
one writes down its number state matrix elements:
\begin{equation}
  \label{eq:10}
\langle n_2|[\hat{\vp},\hat{N}]|n_1\rangle = (n_1-n_2)\langle
n_2|\hat{\vp}|n_1 \rangle = i\, \delta_{n_2n_1}\,.
\end{equation}
For
$n_2=n_1$ we get $0=i\;$! 
\section{Some history}
 That the commutator \ref{eq:9} does not
make sense was noticed very early - even before Dirac
proposed it - by London \cite{lon1} who essentially used the
argument just presented in the framework of the then just invented
matrix mechanics. 

London had seen two earlier papers of Dirac \cite{dir2} in which
he also dealt with the problem of quantizing angle and action variables of
classical mechanics and in which Dirac  suggested operator versions of the  
complex amplitudes \ref{eq:1502} and \ref{eq:1004}, without yet postulating
the commutator \ref{eq:9}. 

London took up the issue of quantizing the canonical pair $w$ and $J$ 
of angle and action variables and started by observing that the quantized 
quantity
$J\,w-w\,J$ cannot be a diagonal matrix in the framework of matrix mechanics
if $J$ is diagonal. This is exactly the argument from the end of the last
section above. 

Then London goes on, introduces an operator
\begin{equation}
  \label{eq:742}
  E(i\, n\,w) = \sum_{\nu =0}^{\infty} \frac{(i\,n\,w)^{\nu}}{\nu!} =E^n(iw)\,
,~n=1,2,\ldots\,,
\end{equation}
and discusses  properties of that formal 
 power series, e.g.
\begin{equation}
  \label{eq:744}
  E^{-1}(iw)J\,E(iw) -J = \boldsymbol{1}\,.
\end{equation}
 
For the harmonic oscillator he writes down the operator transformation
\begin{equation}
  \label{eq:743}
P=\frac{1}{\sqrt{2}}(\sqrt{J}\,E+E^{-1}\sqrt{J})\,,~
Q =\frac{1}{i\sqrt{2}}(\sqrt{J}\,E-E^{-1}\sqrt{J})\,,\,E=E(iw)\,,
\end{equation}
which he calls canonical because it yields the 
commutator \ref{eq:8}. 

(Like London I have used the same letters $w$ and $J$ for the classical
and the quantized quantities.)

In his second paper on the subject \cite{lon2} London used wave functions,
 Hilbert spaces and unitary transformations between Hilbert spaces:
 He transformed, e.g.\ the wave functions $\exp[i\,n\,(w= \vp)]$ of the
 harmonic oscillator (described in  the Hilbert space \ref{eq:15})
into the usual Hermite functions (see Ch.\ 4 of the present paper for details).
He  realized that there
was no such operator like $w= \hat{\vp}$ (I change notations now) but
 that operators like
\begin{equation}
  \label{eq:11}
E_-= \widehat{e^{-i\,\vp}}\,,~E_+ =\widehat{e^{i\,\vp}}~~(= E(i\,w))
\end{equation}
can
make sense according to the formal power series \ref{eq:742} and that
 they have the properties 
\begin{equation}
  \label{eq:12}
E_-|n\rangle = |n-1\rangle\,,~E_+|n\rangle = |n+1\rangle\,,~
E_-\hat{N}E_+=\hat{N}+1~.
\end{equation}
London also made the correct
mathematical observation  that $\hat{\vp}$ cannot be a
self-adjoint (multiplication) operator in a Hilbert space with a
scalar product
\begin{equation}
  \label{eq:15}
 (\psi_2,\psi_1)=
\frac{1}{2\,\pi}\int_0^{2\,\pi}
d\vp\,\bar{\psi}_2(\vp)\psi_1(\vp)~,
\end{equation}
because $\vp$ is not
periodic. (As to a modern discussion of that problem see Ref.\
\cite{reed1}). 

 The mathematical inconsistency \ref{eq:10} of the
commutator \ref{eq:9} was rediscovered -- without the knowledge of
London's previous work -- in the sixties \cite{lou,suss} as well as
the possible usefulness  of the operators $E_-$
and $E_+$ from \ref{eq:11} and their properties \ref{eq:12}. 

If one  inverts Dirac's  formal operator polar decompositions
\begin{equation}
  \label{eq:745}
  \hat{a} =E_-\,\sqrt{\hat{N}}\,,~~~\hat{a}^+ =\sqrt{\hat{N}}\, E_+\,,
\end{equation}
naively, one gets the representations
\begin{equation}
  \label{eq:746}
  E_- = \hat{a}\, \hat{N}^{-1/2}\,,~~~E_+=\hat{N}^{-1/2}\,\hat{a}^+\,.
\end{equation}

 As to the introduction of these operators the almost only new
 element which was added 
in the sixties of the last century, 
compared
to that what was known to London (and Dirac),  was the
observation that the operator $E_-$ from Eq.\ \ref{eq:746} is
not defined when applied to the ground state $|n=0\rangle$. This
deficiency was cured in the following way \cite{suss, carr2}:

 If
$f(\hat{N})$ is an ``appropriate'' function of the number operator
$\hat{N}$ then -- as a consequence of the Eqs.\ \ref{eq:1006}
 and \ref{eq:1009} --  the
relations
\begin{equation}
  \label{eq:16}
 \hat{a}\,f(\hat{N})=f(\hat{N}+1)\,\hat{a}~,
~f(\hat{N})\,\hat{a}^+ = \hat{a}^+\,f(\hat{N}+1)
\end{equation}
hold.

``Appropriate'' here means that both sides of the Eqs.\ \ref{eq:16} are
well-defined operators in the Hilbert space they act in. Though
the operator $E_-$ from Eq.\ \ref{eq:746} is not appropriate in this
sense, the relations \ref{eq:16} were used in order to make $E_-$ - and
$E_+$ - well-defined:
\begin{equation}
  \label{eq:17}
 E_- =
(\hat{N}+1)^{-1/2}\hat{a}~,~ E_+ =\hat{a}^+(\hat{N}+1)^{-1/2}~,
\end{equation}
where $E_-|n=0\rangle = 0$ is now obvious.

(Already Dirac used the relations \ref{eq:16} in special cases
\cite{dir1}!)

 The operators \ref{eq:17}
still have the properties \ref{eq:12} from which it follows that
\begin{equation}
  \label{eq:18}
E_-E_+=1\,,~ E_+E_-= 1-\hat{P}_0\,,
\end{equation}
where $\hat{P_0}$ is the
projection operator onto the ground state $|n=0\rangle$. 

The
second of the relations \ref{eq:18} shows that the operator $E_-$ is not
unitary, but merely isometric. 

In mathematics operators with the
properties \ref{eq:12} are called ``shift operators'' and they have been
studied extensively \cite{shiop}. 

 Susskind and Glogower
\cite{suss} defined operators for $cosine$ and $sine$ as
\begin{eqnarray} \label{eq:1504}\hat{C} \equiv \widehat{\cos\vp}&=&
\frac{1}{2}(E_+ +E_-)~, \\ \label{eq:1010}\hat{S} \equiv \widehat{\sin\vp} &=&
\frac{1}{2i}(E_+ -E_-)~\end{eqnarray} and discussed several of
their properties. 

 The development of those years was nicely
summarized in a thorough review by Carruthers and Nieto
\cite{rev1}. For a personal account by Nieto of those times and their problems 
 see Ref.\ \cite{niet0}.
\section{Central elements of the new approach}

If one tries to analyze and trace back all the apparent and real problems
and difficulties which have been discussed over the decades since
the early days of quantum mechanics in connection with the
quantization of moduli and phases, one always ends up with the
non-trivial global structure of the symplectic (phase) space \ref{eq:5}.

Quantizing this symplectic space requires a new approach and group
theoretical quantization \cite{is1,gui} provides such an approach!

I shall first outline how this approach works in the case of the conventional
quantization scheme and then discuss the appropriate generalizations for
the phase space \ref{eq:5}.
\subsection{Group theoretical background of the conventional quantization
procedure}
\subsubsection{The classical phase space}
Let me briefly recall essential group-theoretical properties
of the 2-dimensional phase space
\begin{equation}
  \label{eq:13}
  {\cal S}^2_{q,p} =\{(q,p) \in \mathbb{R}^2 \}\,,
\end{equation}
  associated with the system \ref{eq:1} or similar ones which have that 
phase space:

 Because of the Poisson bracket \ref{eq:2} 
 the phase space \ref{eq:13} does not only form a
 vector space, but also has a
3-dimensional (nilpotent) Lie algebra $\tilde{\mathfrak{t}}^{2+1}_{WH} $
 associated with
it.

 The index ``WH'' stands for ``Weyl'' and ``Heisenberg'', because it has
become customary to speak of the corresponding group as the ``Weyl-Heisenberg
group'' (or ``Weyl group'' or ``Heisenberg'' group). To name that group
 after Weyl
is certainly justified, but it might be debatable to single out the name
of Heisenberg in view of the equally important  contributions of Born, Jordan
and Dirac to the commutator structure of the quantized canonical variables!

The letter $t$ in $\tilde{\mathfrak{t}}^{2+1}_{WH}$ stands for
 ``translations'' (see below) and the ``tilde'' for ``central extension''
(see below, too).

The Lie algebra is 3-dimensional because the Poisson (Lie) bracket
of $q$ and $p$ is a fixed real number - compared to the {\em variables}
 $q$ and $p$ - , 
\begin{equation}
  \label{eq:14}
  \{q,p\}_{q,p} =1\,,~~\{q,1\}_{q,p} =0\,,~~\{p,1\}_{q,p}=0\,,
\end{equation}
i.e.\ the Lie algebra is
generated by $q$, $p$ and the real number $1$.

 The commutator of 2
general Lie algebra elements
\begin{equation}
  \label{eq:19}
 l_j = a_j\,q + b_j\,p
+r_j~,~j=1,2,
\end{equation}
is given by
\begin{equation}
  \label{eq:20}
 \{l_1,l_2\}_{q,p} = a_1\,b_2 -
a_2\,b_1~,
\end{equation}
i.e.\ a real number.

 As
\begin{equation}
  \label{eq:21}
 \{l,\{l_1,l_2\}_{q,p}\}_{q,p} =0
\end{equation}
for an arbitrary $l \in \tilde{\mathfrak{t}}^{2+1}_{WH} $, the Lie
algebra is nilpotent, see, e.g.\ Ref.\ \cite{hel}.

 The center of the
Lie algebra is generated by the number $1$. 

If we consider $q,\,p$ and $1$ as basis of the Lie algebra, we may
characterize a general element \ref{eq:19} by a triple of real numbers
\begin{equation}
  \label{eq:747}
  (a,b,r)\,.
\end{equation}
 According to Eq.\ \ref{eq:20} we then have the following general  Lie
algebra  commutator structure
\begin{equation}
  \label{eq:22}
[(a_1,b_1,r_1),(a_2,b_2,r_2)] = (0,0,a_1\,b_2 - a_2\,b_1)~.
\end{equation} for two of the column vectors \ref{eq:747}.

If we merely look at the first 2 components of the elements $(a,b,r)$ on both
sides of the Eq.\ \ref{eq:22} and ignore for a moment the 3rd component, 
then we have a 
  Lie algebra  of the 2-dimensional abelian group
 of translations which acts on functions on $\mathbb{S}^2_{q,p} \cong 
\mathbb{R}^2$ as follows:

 For
any smooth function $f(q,p)$ we have
\begin{equation}
  \label{eq:23}
\{a\,q+b\,p+r,f(q,p)\}_{q,p}=a\,\partial_p f(q,p)-b\,\partial_q f(q,p)~.
\end{equation}

Thus, the Lie algebra generator $q$ generates (infinitesimal)
translations in momentum space and $p$ generates (infinitesimal) translations
in coordinate space!

 Actually, however,  that Lie algebra of the
2-dimensional translation group is {\em not} a subalgebra of
 $\tilde{\mathfrak{t}}^{2+1}_{WH} $, because the Poisson
commutator of the two translation generators  $q$ and $p$ does not vanish,
but gives the 3rd generator, see \ref{eq:14}.

The Lie algebra $\tilde{\mathfrak{t}}^{2+1}_{WH} $ generates a group, the
so-called ``Weyl-Heisenberg'' group $\tilde{T}^{2+1}_{WH}$. Its group law is
obtained by exponentiating the Lie algebra elements $l_1$ and
$l_2$ and multiplying the result:
\begin{equation}
  \label{eq:24}
 e^{l_1}\circ e^{l_2} =
e^{l_1+l_2 + \{l_1,l_2\}_{q,p}/2}~.
\end{equation}
Here the Baker-Campbell-Hausdorff
formula \cite{mes}
\begin{equation}
  \label{eq:25}
 e^A \cdot e^B = e^{A+B+[A,B]/2}
\end{equation}
for the product of the exponentials of two
operators $A$ and $B$ has been used, in the special case that their commutator
$[A,B]$  commutes with both $A$ and $B$. 

The relation \ref{eq:24} for the exponentials shows that one
 can characterize
a group element by the  tripel $(a,b,r)$ which describes the
 exponentiated Lie algebra
element. 

From Eq.\ \ref{eq:24} one reads off the following group law:
\begin{equation}
  \label{eq:26}
(a_1,b_1,r_1)\circ (a_2,b_2,r_2) =
(a_1+a_2,b_1+b_2,r_1+r_2+(a_1\,b_2-a_2\,b_1)/2)~.
\end{equation}
The group $\tilde{T}^{2+1}_{WH}$
consists of 2-dimensional translations enlarged by a 1-di\-men\-sional
``central extension'' $\mathbb{R}$ of the additive group of real numbers! 
(As to central extensions of groups in physics see, e.g.\ the Refs.\ 
\cite{ish3})

 The action of the
2-dimensional translations
\begin{equation}
  \label{eq:748}
  q \to q -b\,,~~~p \to p+a\,,~~~a,b \in \mathbb{R}\,,
\end{equation}
 generated by the group $\tilde{T}^{2+1}_{WH} $ on the phase space
\ref{eq:13} has the following properties: 
\begin{enumerate}
\item It is {\em symplectic\,}: $d(q-b)\wedge d(p+a)=dq \wedge dp $.
\item  It is is {\em transitive\,}: Any  two points $(q_1,p_1)$ and
$(q_2,p_2)$ can be transformed into each other by an element of
$\tilde{T}^{2+1}_{WH}(a,b): (b=q_1-q_2,a=p_2-p_1)$. Here the point
 $(q=0,p=0)$ is
in no way special\,!
\item
 It is {\em effective\,}: If
$(a,b)\cdot (q,p)= (q-b,p+a)= (q,p)~ \forall (q,p)$, then $(a,b)= (0,0)$. 
\item  The abstract Lie algebra $\tilde{\mathfrak{t}}^{2+1}_{HW}$ 
defined by the commutators  \begin{equation}
    \label{eq:750}
    [A_1,A_2] = A_3\,,~~[A_1,A_3] = 0\,,~~[A_2,A_3] =0\,,
  \end{equation}
of the ``canonical'' group $\tilde{T}^{2+1}_{WH}$ is isomorphic to the
Poisson algebra of 3 {\em globally} defined functions $f_{A_1}(q,p),\,f_{A_2}$
and $f_{A_3}$ on \ref{eq:13}, namely
\begin{equation}
  \label{eq:751}
  \{f_{A_1},f_{A_2}\}_{q,p} = f_{A_3}\,,~~ \{f_{A_1},f_{A_3}\}_{q,p}
=0\,,~~ \{f_{A_2},f_{A_3}\}_{q,p}=0\,.
\end{equation}
The required 3 functions are obviously (see Eqs.\ \ref{eq:14})
\begin{equation}
  \label{eq:752}
  f_{A_1}(q,p)=q\,,~~f_{A_2}(q,p)=p\,,~~f_{A_3}(q,p)=1\,.
\end{equation}
\end{enumerate}
These properties are essential for a generalization of the following
 quantization procedure to other phase spaces like \ref{eq:5}!
\subsubsection{Group theoretical quantization of the phase space
 ${\cal S}^2_{q,p}\protect \cong \mathbb{R}^2$}
Quantization of the phase space \ref{eq:13} now
consists in determining the irreducible unitary representations of
the Weyl-Heisenberg group \ref{eq:26}\,! 

The important point for this
approach is that in  such a representation the unitary operators
$U(a)$ and $V(b)$ which implement the translations \ref{eq:748} 
are generated by {\em self-adjoint} operators
$Q$ and $P$:
\begin{equation}
  \label{eq:27}
 U(a)=e^{-i\,a\,Q}~,~ V(b)=e^{-i\,b\,P}~.
\end{equation}
Because of  \ref{eq:8} and \ref{eq:25} we have \begin{eqnarray}
V(b)\,U(a)&=&
e^{i\,a\,b/2}\,e^{-i\,(a\,Q + b\,P)}\,, \\
\label{eq:1011}U(a)\,V(b)&=& e^{-i\,a\,b/2}\,e^{-i\,(a\,Q + b\,P)}\,,
\end{eqnarray} and therefore Weyl's integrated {\em group} commutator
relation \cite{wey} \begin{eqnarray}
\label{eq:1505}U(a_1)\,U(a_2)&=&U(a_2)\,U(a_1)~, \\ \label{eq:1012}V(b_1)\,V(b_2) &=&
V(b_2)\,V(b_1)~,\\ \label{eq:1013}U^{-1}(a)\,V(b)\,U(a)\,V^{-1}(b)&=&
e^{i\,a\,b}\,, \end{eqnarray} instead of the {\em  Lie algebra} relation
\ref{eq:8} as introduced by Born, Heisenberg and Jordan and by Dirac as
well.

 Weyl's approach has the mathematical advantage that one is
dealing with {\em bounded} operators in Hilbert space and the famous
Stone - von Neumann  theorem \cite{sto} asserts that all
irreducible unitary representations of the operators $U(a)$ and
$V(b)$ with the properties \ref{eq:1505}-\ref{eq:1013} are
 unitarily equivalent to
the Schr\"odinger representation in the $L^2$-space of square-integrable
 functions on the real line $\mathbb{R}$.

 Notice, however, that the
fundamental ``observables'' of the system are the {\em Lie algebra}
 basis elements
$q$, $p$ and $1$ on the classical level and the corresponding operators
$Q$, $P$ and $\boldsymbol{1}$ on the quantum level,  the Lie algebra
 being that of the
group $\tilde{T}^{2+1}_{WH} $! 

The group law \ref{eq:26} is implemented again by applying the relation
\ref{eq:25} to the operator product 
\begin{equation}
  \label{eq:749}
  e^{-i\,(a_1\,Q+ b_1\,P +r_1)}\cdot
e^{-i\,(a_2\,Q+ b_2\,P +r_2)}\,.
\end{equation}
  
At first sight the
group-theoretical approach to the usual quantization of the phase
space \ref{eq:13} may appear complicated and even far-fetched. The reason
is that the quantizing ``canonical'' Weyl-Heisenberg group \ref{eq:26}
 is unusual
in the sense that it has the structure of a {\em central extension}
(\cite{ish3}) of an abelian translation group
which acts on that phase space \ref{eq:13}. The resulting
group is nilpotent and as such somewhat ``singular''.

Essential is, however, that the basic underlying ``canonical'' transformation
group on  \ref{eq:13} is the 2-dimensional  translation
 group of that space, with the properties listed above.
\subsection{The group $SO^{\uparrow}(1,2)$ as the canonical group of the
phase space ${\cal S}^2_{\vp,I}$} 
Contrary to the seemingly somewhat complicated Weyl-Heisenberg
 group \ref{eq:26} of the phase space \ref{eq:13}, the corresponding
``canonical'' group of the phase space \ref{eq:5} is much simpler, namely
the group
\begin{equation}
  \label{eq:753}
 SO^{\uparrow}(1,2)\,, 
\end{equation}
 the ``orthochronous proper'' Lorentz group
in $2+1$ (space-time) dimensions (see Eq. \ref{eq:603} for the precise
 definition). The group is also ``simple'' in the
mathematical sense \cite{hela}.

 The role of the symplectic group $Sp(2,\mathbb{R})$ as the canonical group
of the phase space $\mathbb{R}^2 -(0,0)$ was first discussed in the context
of a $U(1)$-gauge model by Loll \cite{lo} and more recently --
without the knowledge of Loll's paper -- in connection with the
quantization of Schwarzschild black holes \cite{bo,ka2}. 

After finishing the paper \cite{bo} I realized that the quantization
 formalism employed there also sheds new light
on the old -- still mainly unsolved -- problem of how to describe phase and
modulus in terms of self-adjoint operators in a suitable Hilbert
space associated with a corresponding physical system 
 \cite{ka1,ka3,ka4}. 

The general idea of a group theoretical quantization of a given phase
 space is outlined
 in Appendix A.1. Appendices A.2 and A.3 describe the application
of that general approach to the symplectic space \ref{eq:5} in terms of the
group \ref{eq:753} in detail. Mathematical properties of that group, its
double covering groups $SU(1,1) \cong SL(2,\mathbb{R}) = Sp(2,\mathbb{R})$
and the universal covering group of all of them are discussed in Appendix B.

In the following I briefly sketch and summarize the essential results in
order to illustrate the power and the richness of the theory and to indicate
the points for possible experimental tests of the framework.

The action of the transformation group \ref{eq:753} on the phase space
\ref{eq:5} should have all the general properties listed after Eq.\
\ref{eq:748}. We shall see that it indeed does have them.  
Most of the proofs can be found in Appendix A. 

In order to describe the action of $SO^{\uparrow}(1,2)$
the following parametrization of the space \ref{eq:5}  is convenient:

 A point $s \in {\cal S}^2_{\vp,I}$  can be represented
 by the matrix
\begin{equation}
  \label{eq:28}
   \underline{s}= \left(
\begin{array}{cc} I &  I\,e^{\ds -i\varphi} \\  I\,e^{\ds i\varphi}
 & I \end{array} \right) \equiv
\left(
\begin{array}{cc} h_0 &   h_1+i\,h_2 \\  h_1-i\, h_2
 & h_0 \end{array} \right)~.
\end{equation}

As 
\begin{equation}
  \label{eq:31}
 \det(\underline{s}) = h_0^2-h_1^2-h_2^2 =
I^2-(I\,\cos\vp)^2-(-I\,\sin\vp)^2 = 0
\end{equation}
and  $h_0 > 0$ the three (dependent)
coordinates
\begin{equation}
  \label{eq:754}
  h_0 = I\,,~~h_1 = I\,\cos\vp\,,~~h_2 = -I\,\sin\vp\,,
\end{equation}
  parametrize a 2-dimensional ``forward
light cone'' with the vertex deleted. It is well known that the group
$SO^{\uparrow}(1,2)$ acts transitively on that cone.

As in the case of the Lorentz or rotation group
 the transformation of $\underline{s}$ with respect to
 $SO^{\uparrow}(1,2)$ is best implemented in terms of the 2-fold
 covering group
\begin{equation}
  \label{eq:29}
 SU(1,1) =\left\{\tilde{g}= \left( \begin{array}{ll} \alpha & \beta \\
 \bar{\beta} & \bar{\alpha}
\end{array} \right)\,,~~ |\alpha|^2-|\beta|^2=1~ \right\}\,,
\end{equation}
namely
\begin{equation}
  \label{eq:30}
 \underline{s} \to \hat{\underline{s}} =
\tilde{g}\cdot \underline{s}\cdot \tilde{g}^+\,,
\end{equation}
where
$\tilde{g}^+$ denotes the hermitian conjugate of $\tilde{g}$. 

 The center
 \begin{equation}
   \label{eq:784}
   Z_2\,:~~\{E_2,-E_2\}\,,~~E_2 = \begin{pmatrix}1&0\\0&1 \end{pmatrix}\,,
 \end{equation}
 of $SU(1,1)$ leaves all  points $\underline{s}$
invariant, i.e.\ the group $SU(1,1)$ acts only almost effectively
on the phase space \ref{eq:5}, but $SO^{\uparrow}(1,2) \cong SU(1,1)/Z_2$
 itself acts effectively, i.e.\  only the identity element of 
$SO^{\uparrow}(1,2)$ leaves all points \ref{eq:28} fixed.

 More important, the transformations \ref{eq:30} leave the symplectic form 
 \begin{equation}
   \label{eq:755}
   \omega_{\vp,I} =d\vp \wedge dI
 \end{equation}
 invariant (Appendix A.2):
\begin{equation}
  \label{eq:32}
d\hat{\vp}\wedge d\hat{I} = d\vp \wedge dI~.
\end{equation}
The Lie algebra
$\mathfrak{so}(1,2)$ may be spanned by  three
generators $A_j\,,\,j=0,1,2,$ with the commutators
\begin{equation}
  \label{eq:33}
 [A_0,A_1]= -A_2\,,~~[A_0,A_2]=A_1\,,~~
[A_1,A_2]=A_0 \,.
\end{equation}
Here $A_0$ generates rotations and $A_1$  and $A_2$ generate ``Lorentz
boosts''! (See  Eqs.\ \ref{eq:757}-\ref{eq:765} of Appendix A.2.) 

 Each of those 1-dimensional
subgroups  generates a global vector field $\tilde{A}_j$ on the
phase space \ref{eq:5}. These have the form (Appendix A.2) \begin{eqnarray}
\label{eq:1018}\tilde{A}_0 &=& \partial_{\vp}\,, \\
\label{eq:1506}\tilde{A}_1
&=&\cos\vp\,\partial_{\vp}  +I\,\sin\vp\,\partial_I\,, \\
\label{eq:1017}\tilde{A}_2  &=& \sin\vp\,\partial_{\vp}-I\,\cos\vp\,
\partial_I\,.
  \end{eqnarray}
 The vector
fields generate the same Lie algebra \ref{eq:33} as the $A_j$ themselves.

 A crucial point now is that those vector fields are global
{\em Hamiltonian} ones, i.e.\ they have the form
\begin{equation}
  \label{eq:34}
 -X_h=
\partial_{\vp}h(\vp,I)\,\partial_I - \partial_I
h(\vp,I)\,\partial_{\vp}~,
\end{equation}
where $h(\vp,I)$ is a smooth
function on the phase (symplectic) space \ref{eq:5}. 

The most essential result of all this (details are in Appendix
A.2) is that the generating
functions for the  Hamiltonian vector fields \ref{eq:1018}-\ref{eq:1017}
 are just
the three coordinate functions $h_j$ introduced  in Eq.\ \ref{eq:28}:
\begin{equation}
  \label{eq:35}
  h_0(\vp,I) = I\,,~~ h_1(\vp,I)=I
\,\cos \vp\,,~~ h_2(\vp,I)=-I\,\sin \vp\,.
\end{equation}

These functions 
obey the Lie algebra $\mathfrak{so}(1,2)$ in terms of
the Poisson brackets \ref{eq:6}, too:
\begin{equation}
  \label{eq:36}
 \{h_0,h_1\}_{\vp,I}=
-h_2\,,~~\{h_0,h_2\}_{\vp,I}=h_1\,,~~ \{h_1,h_2\}_{\vp,I}=h_0\, ~.
\end{equation}
As in the case of
the Weyl-Heisenberg group above where the basic observables on the 
phase space $ {\cal S}^2_{q,p}$ are given by the generators
 $q$ and $p$ (and $1$) of
the Lie algebra $\mathfrak{t}^{2+1}_{WH}$,
 now the three functions $h_j(\vp,I)$ from \ref{eq:35} are to be chosen as
the {\em basic} classical observables on the phase space \ref{eq:5}\,!
  They obviously suffice to expand any ``decent'' function  $f(\vp
 \bmod{2\pi}, I)$ on ${\cal S}^2_{\vp,I}$!

It is important to understand the following point: The resulting
Hamiltonian functions \ref{eq:35} are solely determined by the vector fields
\ref{eq:1018} - \ref{eq:1017} the form of which is a consequence of
 the action of the associated 
1-parameter subgroups on the phase space \ref{eq:5} (see Eqs.\ \ref{RI_{12}}-
\ref{Avps} and \ref{BI_{12}}-\ref{Bspc} of Appendix A), with the invariance
property \ref{eq:32}. Their form \ref{eq:35} 
is {\em not} a consequence of the very
convenient but not cogent parametrization \ref{eq:28} of the conic space
\ref{eq:5}.
 
We have to conclude that the canonical group $SO^{\uparrow}(1,2)$
 and its action on the symplectic
space \ref{eq:5} determine the basic ``observables'' $h_j$ on that space all
by itself!
   
 This systematic result justifies an early
 suggestion by Louisell \cite{lou} that one should use $\cos\vp$
 and $\sin \vp$ instead of $\vp$ itself when trying to quantize the
 latter. 

 At this point  one might ask:

Why then not use the functions 
\begin{equation}
  \label{eq:767}
 \tilde{h}_1 = \cos\vp\,,~
\tilde{h}_2 = \sin\vp\,,~ \tilde{h}_3 = I\,, 
\end{equation} as basic observables?

  These functions, however,
generate the Lie algebra of the {\em Euclidean group  $E(2)$ in
the plane\,}:
\begin{equation}
  \label{eq:768}
 \{\tilde{h}_3,\tilde{h}_1\}_{\vp,I}=
\tilde{h}_2,~~\{\tilde{h}_3,\tilde{h}_2\}_{\vp,I}= -\tilde{h}_1,~~
\{\tilde{h}_1,\tilde{h}_2\}_{\vp,I}=0\,,  
\end{equation}
  where  $\cos\vp$ and $\sin\vp$ commute now!
 
The latter property might be welcome, but the group $E(2)$ is
not suitable at all for our purpose:

 First, the group consists of
rotations $O(2)$ and two translations on the plane which do {\em not}
avoid the origin!

Second, the spectrum of the self-adjoint operator corresponding to the
modulus
$I$ in any irreducible unitary representation of $E(2)$
corresponds to the integers  $\mathbb{Z}$ \cite{is2,suga}, not to the
positive numbers $\mathbb{N}$, and therefore the quantized $I$
would have the wrong spectrum! 

 The deeper reason is that the  group
$E(2)$ is the quantizing group of $S^1\times  \mathbb{R}$
\cite{is2}, not of $S^1 \times \mathbb{R}^+$!

 The situation is
quite different for  the group $SO^{\uparrow}(1,2)$, where the
positive discrete series of irreducible unitary representations
provides a positive definite operator for the quantized action variable
$I$\,! 
\section{The relationship between the action of the group 
$SO^{\uparrow}(1,2)$ on ${\cal S}^2_{\vp,I}$ and the action of its
covering group
$Sp(2,\mathbb{R})$ on ${\cal S}^2_{q,p}\,$; \\ the $Z_2$ gauge symmetry}
We have just seen how the group $SO^{\uparrow}(1,2)$ acts on the phase
space \ref{eq:5}. On the other hand,  its double covering group, the symplectic
group $Sp(2,\mathbb{R})$, see \ref{eq:375}, acts on the space \ref{eq:13} as
follows:
\begin{equation}
  \label{eq:769}
  \begin{pmatrix}q \\p\end{pmatrix} \to g_1\cdot \begin{pmatrix}q
 \\p\end{pmatrix}\,,~ g_1 \in Sp(2,\mathbb{R})\,.
\end{equation} The transformations \ref{eq:769} leave the symplectic
 form \ref{eq:740} invariant, transform the point $(q=0,p=0)$ into itself and
act transitively on the compliment
\begin{equation}
  \label{eq:770}
  {\cal S}^2_{q,p;0} = {\cal S}^2_{q,p} -\{(0,0)\}\,.
\end{equation}

In order to see the difference between the simultaneous actions of the
symplectic group $Sp(2,\mathbb{R})$ on the spaces \ref{eq:770} and \ref{eq:5}
 (via $SO^{\uparrow}(1,2) = Sp(2,\mathbb{R})/Z_2$) let us look at the
following two examples:

The groups \ref{ka1} and \ref{a1} act on the space ${\cal S}^2_{q,p}$ as
\begin{eqnarray}
  \label{eq:771}
  R_1:~~~ q &\to& \cos(\theta/2)\,q +\sin(\theta/2)\,p\,, \\
p &\to&  -\sin(\theta/2)\,q + \cos(\theta/2)\,p\,,~\theta \in (-2\pi,2\pi]\,,
\label{eq:772} \\
A_1:~~~ q & \to & e^{t/2}\,q\,, \label{eq:773} \\
p &\to& e^{-t/2}\,p\,,~~t \in \mathbb{R}\,. \label{eq:774}
\end{eqnarray}
The transformations leave the symplectic form \ref{eq:740} invariant.

The groups induce simultaneous transformations on ${\cal S}^2_{\vp,I}$:
This space is parametrized by the coordinates \ref{eq:754} which transform
as a 3-vector under $SO^{\uparrow}(1,2)$ (see the formulae
 \ref{eq:757}-\ref{eq:762}):
 \begin{eqnarray}
   \label{eq:775}
 R_1:~~~h_0 &\to& h_0\,, \\
h_1 &\to& \cos\theta\, h_1-\sin\theta \,h_2\,, \label{eq:776}  \\
h_2 &\to& \sin\theta\,h_1 + \cos\theta\,h_2\,, \label{eq:777} \\
A_1: ~~~ h_0 &\to& \cosh t\,h_0 + \sinh t\, h_2\,, \label{eq:778}\\
h_1 &\to& h_1\,,  \label{eq:779} \\
h_2 & \to& \sinh t\,h_0 + \cosh t\,h_2\,. \label{eq:780}
 \end{eqnarray}
Here the transformations leave $h_0^2-h_1^2-h_2^2$ invariant!

The crucial point now is the following:
 
If $\theta =2\pi$ for the $R_1$ transformations, then the pair $(q,p)$ changes
sign, but for the triple $(h_0,h_1,h_2)$ we have the identity transformation!
This is due to the fact that $Sp(2,\mathbb{R})$ is a double covering of
$SO^{\uparrow}(1,2)$ and that the kernel of the homomorphism
\begin{equation}
  \label{eq:781}
  Sp(2,\mathbb{R}) \to SO^{\uparrow}(1,2)
\end{equation} is the center \ref{eq:784}: As to its transformation
properties with regards to the group $SO^{\uparrow}(1,2)$ the pair $(q,p)$
transforms as a ``spinor'', namely as a vector with respect to the double
covering $Sp(2,\mathbb{R} \cong SU(1,1)$, whereas
 the $h_j$ transform as a vector with
respect to $SO^{\uparrow}(1,2)$. The relationships here parallel completely
those for the well-known rotation group $SO(3)= SU(2)/Z_2$ and its
 spinor group $SU(2)\,$!

We come here to an essential point of the whole paper:

{\em The center \ref{eq:784} of the group $Sp(2, \mathbb{R})$ acts on the space
$\{(q,p) \in \mathbb{R}^2\}$ as the identity,
too, if we identify the points $(q,p)$ and $(-q,-p)$\,}! 

That is so say, if we pass from the space \ref{eq:13} to the quotient space
\begin{equation}
  \label{eq:783}
  \check{{\cal S}}^2_{q,p} =\{ (-q,-p) \equiv (q,p) \in \mathbb{R}^2\} \cong
\mathbb{R}^2/Z_2\,.
\end{equation}
The resulting space \ref{eq:783} is a simple cone with its tip (vertex) 
at the origin!

This can be seen as follows: Consider the $(q,p)$-plane: Rotate the lower
half of that plane around the $q$-axis till it lies on the upper half plane
such that the negative part of the  p-axis coincides with the positive one.
Afterwards rotate the left half of the upper half plane around the positive
$p$-axis till the negative part of the $q$-axis lies on the positive one.
Then glue these two $q$-half-axis together. The result is the cone just
mentioned.

We now recognize the essential point of the difference between the phase spaces
\ref{eq:5} and \ref{eq:13}: The phase space \ref{eq:5} is globally 
equivalent (homeomorphic and even diffeomorphic) to the quotient space
\ref{eq:783} if the point $(0,0)\,$,  the ``tip'' or vertex of the
 cone, is deleted!

We started from the local equality \ref{eq:3} and see now the global
difference between the spaces \ref{eq:13} and \ref{eq:5}. Notice, however,
that the local symplectic form \ref{eq:740} is invariant under the action
of the center \ref{eq:784}, too.

Quotient spaces of the type \ref{eq:783} where points of a given space
are identified by means  of a discontinuous transformation group are
called ``orbifolds'' \cite{orbi}.

The equivalence of the spaces \ref{eq:5} and \ref{eq:783} has far-reaching
consequences, because the $Z_2$ group \ref{eq:784} acts as kind of gauge group
on the phase space \ref{eq:13}: 

With regard to the phase space \ref{eq:5} only those functions of
$(q,p)$ are ``observables'' which are invariant under the $Z_2$ transformations
\ref{eq:784}, i.e.\ only even powers of $q$ and (or) $p$. Thus, the 
{\em original} $q$ or $p$
themselves are no observables in this sense!

There is a surprise, however: We may define ``composite'' canonical
coordinates
\begin{eqnarray}
  \label{eq:792}
  \tilde{q}(\vp,I)& =& \sqrt{2}\,\frac{h_1(\vp,I)}{\sqrt{h_0(\vp,I)}}
 =\sqrt{2\,I}\,\cos\vp\,, \\
\tilde{p}(\vp,I)&=& \sqrt{2}\, \frac{h_2(\vp,I)}{\sqrt{h_0(\vp,I)}} =
-\sqrt{2\,I}\,\sin\vp\,, \label{eq:796}
\end{eqnarray}
on the symplectic space \ref{eq:5}!

They obey the usual relation
\begin{equation}
  \label{eq:793}
 \{\tilde{q},\tilde{p}\}_{\vp,I} =1\,, 
\end{equation} but are $Z_2$ {\em  gauge invariant} functions now: 

According to their definition and according to the Eqs.\ \ref{eq:775}-
\ref{eq:777} they transform as
\begin{eqnarray}
  \label{eq:794}
  \tilde{q} & \to& \cos\theta \,\tilde{q} - \sin\theta\, \tilde{p}\,, \\
\tilde{p} &\to & \sin\theta\,\tilde{q} + \cos\theta \, \tilde{p}\,.
\label{eq:795}
\end{eqnarray}
One realizes the important difference to the transformation formulae
\ref{eq:771} and \ref{eq:772}: 

 The crucial point is that the coordinates
\ref{eq:792} and \ref{eq:796} are functions on ${\cal S}^2_{\vp,I} \cong
{\cal S}^2_{q,p;0}/Z_2$, whereas the original $q$ and $p$ are functions
on ${\cal S}^2_{q,p}$!

Notice that $\tilde{q}$ and $\tilde{p}$ have the property $(\tilde{q},
\tilde{p}) \neq (0,0)$.

The quantized version of the composite coordinates \ref{eq:792}
 and \ref{eq:796} will be discussed in Ch.\ 2.

More about the consequences of the gauge group \ref{eq:784} can be found
 in Secs. 6.3, 8.1.3 and Appendix A.3.

Gauge symmetries of the $Z_2$ type appearing here have been
 discussed by Prokhorov and Shabanov \cite{prok1,sha1}. 
\section{Overview}
In the following I briefly sketch the main topics and results of the
following chapters. The bulk of the references will be given within
the chapters and the appendices. Due to the length of the paper I could (did)
not always avoid using the same symbol (letter) for different things, but
their meaning will be clear from the context!
\subsection{Chapter 2}
 The quantized versions of the basic classical
observables \ref{eq:35} are the corresponding self-adjoint generators
\begin{equation}
  \label{eq:37}K_0=\hat{I}=\hat{h}_0\,,~~
 K_1=\widehat{I\,\cos\vp} =\hat{h}_1\,,~~
K_2=-\widehat{I\,\sin\vp} =\hat{h}_2\,,
\end{equation}
of the unitary 1-dimensional subgroups in the positive discrete
series of the unitary irreducible representations of
$SO^{\uparrow}(1,2)$ or one of its covering groups. The essential advantage
of this procedure is: Given a unitary representation of a {\em group}, then
the {\em Lie algebra generators} of its 1-dimensional subgroups
 are {\em self-adjoint\,}!

The $K_j$ obey
the relations
\begin{equation}
  \label{eq:38}
 [K_0,K_1]= i K_2,~~
[K_0,K_2]=-iK_1,~~[K_1,K_2]=-iK_0 ~~.
\end{equation}
A salient feature of the
positive series representations is that $K_0$, the generator of
the $O(2)$ subgroup, has the spectrum
\begin{equation}
  \label{eq:39}
 \mbox{spec}(K_0)= \{k+n,
k>0, n=0,1,\ldots\}~,
\end{equation}
where the number $k
>0$ characterizes the representation (like the number $j$ in the
case of $SU(2)$!) Its possible values depend on the group: For the
group $SO^{\uparrow}(1,2)$ $k$ can take the positive integer values
$k= 1,2,\ldots$ and for the, e.g.\ 2-fold covering $SU(1,1) \cong
SL(2, \mathbb{R}) = Sp(2, \mathbb{R})$ the
values $k=1/2,1,3/2,2,\ldots $. 

 The eigenstates $|k,n\rangle\,,\,n=0,1,\ldots\,,$ of $K_0$
are normalized elements of the associated Hilbert space and  can
 be used as a (infinite) basis. That basis can be generated with the help of
raising and lowering operators
\begin{equation}
  \label{eq:40}
 K_+=K_1+i\,K_2~,~~K_-
=K_1-i\,K_2~,
\end{equation}
with
\begin{equation}
  \label{eq:805}
  K_-|k,n=0\rangle = 0 \,,~~K_0|k,n=0 \rangle = k\,|k,n=0\rangle\,. 
\end{equation}
 For an 
irreducible representation the Casimir operator
\begin{equation}
  \label{eq:41}
L=K_1^2+K_2^2-K_0^2
\end{equation}
has the eigenvalues
\begin{equation}
  \label{eq:42}
 l=k(1-k)\,.
\end{equation}
This
has an immediate surprising quantum physical implication:
Classically we have the trigonometric Pythagorean relation
\begin{equation}
  \label{eq:43}
 (h_1)^2+(h_2)^2 = h_0^2\,,
\end{equation}
but -
because of Eq.\ \ref{eq:41} - quantum theoretically we get for an irreducible
representation
\begin{equation}
  \label{eq:44}
K_1^2+K_2^2=K_0^2+L= K_0^2+k\,(1-k)\,.
\end{equation}

As the eigenvalues \ref{eq:42} of $L$ vanish
only for $k=1$, we see that {\em the quantum effects can
violate Pythagoras' theorem}! 

 In the case of
the harmonic oscillator we have $k=1/2,l=1/4$. So we have
``quantum trigonometrical deviations'' for that system!

{\em Testing the quantum relation \ref{eq:44} is one of the major
experimental challenges of the whole approach discussed here\,}!

Another important point is the following:

 It has been
realized  \cite{mlo} that the generators $K_j$ may be
constructed from the operators \ref{eq:1007} and \ref{eq:1008} in a non-linear
(Holstein-Primakoff type \cite{ho}) manner:
\begin{equation}
  \label{eq:45}
 K_+ =
a^+\sqrt{N+2k}\,,~K_-=
 \sqrt{N+2k}\,a\,,~K_0=N+k\,.
\end{equation}
Here I have - as is usual - dropped the ``hat'' on the
 operators $a,a^+$ and $N$.

 However, it is far  more interesting to turn the
 argument around: 

Given the self-adjoint operators $K_j$ of a
 positive discrete series irreducible unitary representation of
 the group $SU(1,1)\cong SL(2,\mathbb{R}) = Sp(2,\mathbb{R})$, then one
 can define  annihilation and
 creation operators
\begin{equation}
  \label{eq:46}
 a=(K_0+k)^{-1/2}K_-\,,~~a^+
 =K_+(K_0+k)^{-1/2}\,,~~N=K_0-k\,,
\end{equation}
which have the usual properties
\begin{equation}
  \label{eq:47}
 [a,a^+]=1~,~~ N=a^+a~.
\end{equation}
The {\em composite} position and momentum operators
\begin{eqnarray}
  \label{eq:797}
  \tilde{Q} &=& \frac{1}{\sqrt{2}}[(K_0+k)^{-1/2}K_- +K_+(K_0+k)^{-1/2}]\,, \\
\tilde{P} &=& \frac{1}{i\sqrt{2}}[(K_0+k)^{-1/2}K_- -K_+(K_0+k)^{-1/2}]\,,
\end{eqnarray}
are the quantized counterparts of the classical composite coordinates
 \ref{eq:792} and \ref{eq:796}!

{\em Thus, in a sense the operators $K_0, K_1$ and $K_0$ are at least
 as fundamental
as the operators $Q$ and $P$ and it appears possible - at least in principle - 
to base the structures of quantum mechanics on the group $SO^{\uparrow}(1,2)$,
its covering groups and corresponding higher dimensional generalizations\,}!

\subsection{Chapter 3} 
 Ch.\ 3 discusses the problem how the operators $K_1$ and
 $K_2$  may be used in order to ``measure'' the phase $\vp$
 appearing in  physically interesting state vectors.

 The state vectors of that analysis
 are  three  different types of coherent states (Barut-Girardello
 \cite{ba1}, Perelomov \cite{per0,per} and the conventional
 Schr\"odinger-Glauber \cite{schro1,glau1}
coherent
 states) which are associated with the
 Lie algebra of $SO^{\uparrow}(1,2)$ and all of which are
 characterized by a complex number  with a non-vanishing phase. 

The three types can be defined by the relations
 \begin{eqnarray}\label{eq:1516}K_-|k,z\rangle
 &=& z\,|k,z\rangle~,~~ z=|z|\,e^{i\phi} \in \mathbb{C} ~,
 , \\ \label{eq:1037}(K_0+k)^{-1}K_-|k,\lambda \rangle &=& \lambda\,
 |k,\lambda \rangle~,~~ \lambda = |\lambda|\,e^{i\,\theta} \in
 \mathbb{D}~, \\ \label{eq:1038}&& \mathbb{D} = \{\lambda
\in \mathbb{C}~,~ |\lambda| < 1 \}~,  \\\label{eq:1039}(K_0+k)^{-1/2}K_-|k,
\alpha
\rangle &=& \alpha\,|k,\alpha
 \rangle~,~~ \alpha = |\alpha|\,e^{i\,\beta} \in \mathbb{C}~.
  \end{eqnarray}

(As to the definition \ref{eq:1039} see the first of the Eqs.\ \ref{eq:46}.)

Ch.\ 3 contains a large number of matrix elements of the operators
$K_0, K_1$ and $K_2$, as well as functions of them,
 with respect to the coherent states \ref{eq:1516}-\ref{eq:1039}.

The results obtained show very clearly that the operators $K_1$ and $K_2$
are well suited for ``measuring'' the phase content of those states
and that their matrix elements provide a large number of predictions
for experimental tests.
 \subsection{Chapter 4}
That  chapter deals with $SU(1,1)$-related properties of the harmonic
 oscillator. 

For $k=1/2$ we can identify the Hamilton operator $H$
 with the  operator $K_0$ (the frequency $\omega$ being normalized to $1$).

 The group $SU(1,1)$ has
the following explicit irreducible unitary representation for $k=1/2$:
\begin{eqnarray}\label{eq:1507}(f_2,f_1)&=&\frac{1}{2\,\pi}\int_0^{2\,\pi}
d\varphi~\bar{f}_2(\varphi)\,f_1(\varphi)~,\\
 \label{eq:1019}|k=1/2,n \rangle &=& e^{i\,n\,\varphi}~, n=0,1,\ldots ~, \\
 \label{eq:1020}K_0&=&\frac{1}{i}~\partial_{\varphi}+1/2~,\\
 \label{eq:1021}K_-&=&
 e^{-i\,\varphi}~\frac{1}{i}~\partial_{\varphi}~,
 \\\label{eq:1022}K_+&=&e^{i\,\varphi}~(\frac{1}{i}~\partial_{\varphi}+1)~,\\
 \label{eq:1023}H&=& K_0~.  \end{eqnarray} 

  It is possible to describe the whole quantum theory of the
 harmonic oscillator in the Hilbert space with the scalar
 product \ref{eq:1507}\,!

According to the general relations \ref{eq:1516}-\ref{eq:1039} we  get {\em 
three}
different types of coherent states for the harmonic oscillator! In the Hilbert
space above they have the forms
\begin{eqnarray}
  \label{eq:785}
  f_z(\vp) &=& \frac{e^{z\,e^{i\vp}}}{\sqrt{I_0(2|z|)}}\,, \\
&&~~~I_0\,:~\text{modified Bessel function of 1st kind}\,, \nonumber \\
f_{\lambda}(\vp) &=& (1-|\lambda|^2)^{1/2}(1-\lambda\,e^{i\vp})^{-1}\,,
\label{eq:786} \\ f_{\alpha}(\vp) &=& e^{-|\alpha|^2/2} \sum_{n=0}^{\infty}
\frac{(\alpha\,e^{i\vp})^n}{\sqrt{n!}}\,, \label{eq:787}
\end{eqnarray}  and they have  a number of interesting properties!

 The use of this Hilbert space also allows for a critical evaluation
 of the widespread - rather
 controversial - notion of ``phase states'' \cite{phst}.

   Despite the  superficial impression, the phase
 $\vp$ above is {\em not} the {\em quantum mechanical observable
 canonically conjugate
 to the Hamilton operator} $H\,$!  It is merely the mathematical
 variable used in the  Hilbert space with the scalar product \ref{eq:1507} and
the basis \ref{eq:1019}\,!

 The two genuine {\em quantum
 mechanical} observables $K_1$ and $K_2$ here have the form
 \begin{eqnarray} \label{eq:1508}K_1&=& \frac{1}{2}(K_+ +K_-) =
 \cos\vp\,\frac{1}{i}\partial_{\vp} + \frac{1}{2}\,e^{i\,\vp}~, \\
 \label{eq:1024}K_2&=& \frac{1}{2i}(K_+ -K_-) =
 \sin\vp\,\frac{1}{i}\partial_{\vp} + \frac{1}{2i}\,e^{i\,\vp}~.
 \end{eqnarray}
 They are self-adjoint with respect to the scalar
 product \ref{eq:1507} and have a continuous spectrum. 

The rest of that chapter discusses the quantum mechanics of the
 harmonic oscillator in terms of the Hardy space $H^2(\mathbb{R}, d\xi)$
on the real line and the relationship of that Hilbert space to the usual
$L^2(\mathbb{R}, d\xi)\,$.
\subsection{Chapter 5}
That chapter discusses possible attempts to define operators
 $\widehat{\cos\vp}$ and $\widehat{\sin\vp}$  in the sense of London, 
 Susskind and Glogower by combining the operators
 $K_j$ in a non-linear way.

 At first sight  suitable generalizations of the operators
 \ref{eq:1504} and \ref{eq:1010}  appear to be possible. But this is
 only so as long as one uses them in lowest order. Already their squares
are of doubtful use! The reason is the following: 

If one expresses the operators \ref{eq:1504} and \ref{eq:1010} and their
generalizations for $k \neq 1/2$ in terms of the $K_1, K_2$ and $K_0$, then
the {\em cosine}-operator does not only depend on $K_1$ - as one would expect -
but also on $K_2$. Similarly, the {\em sine}-operator not only depends on $K_2$
but also on $K_1$. This is unsatisfactory and even contradictory
in the present framework, where $K_1$ stands for  cosine- and $K_2$ for
 sine-properties!

Actually it is not necessary to define such cosine- and sine-operators at all!
The results of Ch.\ 3 clearly show that
  $K_1$ and $K_2$ themselves serve all the desired purposes!
\subsection{Chapter 6}
 Ch.\ 6 discusses possible applications of the general group theoretical
 framework to quantum optical problems. The group $SU(1,1) \cong 
SL(2,\mathbb{R}) = Sp(2,\mathbb{R})$, its
 Lie algebra and associated coherent states entered quantum optics
 in the middle of the eighties of the last century
 \cite{mil,fish,wod,schu,ger1,bis}. This came in the course of generating
 squeezed states \cite{squ} from the well-known coherent
 states \cite{coh1,coh1a,per,
 coh2,coh3a,coh3,coh4} with their ``minimum uncertainty'' properties:
 the product of the r.m.s.\ fluctuations of the operators $Q$ and
$P$ with repect to the Schr\"odinger-Glauber coherent states satisfies
the {\em equality} in the Heisenberg uncertainty relation. In the case of
squeezed states one of the r.m.s.\ fluctuations is made smaller
 (``squeezed'') at the expense of the other factor in the product.

As a preparation for dealing with squeezing properties Sec.\ 6.1 discusses
 the adjoint representation of the group $SO^{\uparrow}(1,2)$, i.e.\ the
representation in the vector space of its Lie algebra.

 Sec.\ 6.2 deals
with Schwarz's inequality for scalar products as the basis for the different
inequalities discussed in connection with squeezing.

Squeezed states can be generated by
 ``squeezing operators'' which are bilinear in photon
 creation and annihilation operators and it was realized that
 certain combinations obey the Lie algebra of $SO^{\uparrow}(1,2)$: 

 Sec.\ 6.3 discusses the
 case for one mode where 
\begin{equation}
  \label{eq:49}
 K^{(1)}_+ =\frac{1}{2}(a^+)^2\,,~K^{(1)}_-=
\frac{1}{2}a^2\,,~K^{(1)}_0=\frac{1}{2}(a^+a+1/2)\,.
\end{equation}
As
$K^{(1)}_-$ here annihilates $|n=0\rangle$ as well as
$|n=1\rangle$ the representation decomposes into one with states
having an even number of quanta and one with states having an odd number of 
quanta. For even numbers one has $k=1/4$
and for odd numbers $k=3/4$.

Here only the even states are invariant under the gauge transformation
\ref{eq:784} which leads to experimentally interesting selection rules.

 Another  realization of the Lie algebra  is provided by a {\em pair}
$a^+_j\,, a_j\,,\, j=1,2,$ of creation and annihilation operators:
\begin{equation}
  \label{eq:50}
 K^{(2)}_0=\frac{1}{2}(a_1^+a_1+a_2^+a_2+1)~,~~
K^{(2)}_+=a_1^+a_2^+~, ~~K^{(2)}_-=a_1a_2~.
\end{equation}
This is discussed in Sec.\ 6.4 where  the relation of such a system to
the problem of  simultaneous measurements of non-commuting operators \cite{ar}
is indicated, too.

\subsection{Chapter 7}
It has become popular in quantum optics to express quantum mechanical
expectation values
\begin{equation}
  \label{eq:788}
  \langle A \rangle_{\rho} = \tr\rho\,A)\,,~\rho\,:~\text{density operator}\,,
\end{equation}
of  self-adjoint operators $A$ in terms of of a density function $w(q,p)$
for a phase space like \ref{eq:13} and a phase space function $\tilde{A}(q,p)$
corresponding to $A$ such that
\begin{equation}
  \label{eq:48}
  \langle A \rangle_{\rho} =\int_{{\cal S}^2_{q,p}}dqdp\,
 w(q,p)\,\tilde{A}(q,p)\,. 
\end{equation}
 Several such approaches are in use: 

The oldest one is due to Weyl and especially Wigner \cite{wig1}.
 It makes essential
use of Fourier transformation between coordinate and momentum space and is
very closely related to properties of the Weyl-Heisenberg group.

A second approach is due to Husimi \cite{hus} and centers around the
function
\begin{equation}
  \label{eq:51}
  Q(\alpha, \bar{\alpha}) = \langle \alpha |\rho|\alpha \rangle \,,
\end{equation}
where $|\alpha \rangle$ is a conventional Schr\"odinger-Glauber coherent
state.

A third scheme, due to Sudarshan \cite{su} and Gauber \cite{glau2}, postulates
the following ``diagonal P-representation'' for the density operator:
\begin{equation}
  \label{eq:52}
  \rho = \frac{1}{\pi}\int_{\mathbb{C}}d^2\alpha\,P(\alpha,\bar{\alpha})
|\alpha\rangle
\langle \alpha |\,.
\end{equation}
Here the ``density''  $P(\alpha, \bar{\alpha})$ can become negative
 and highly singular!

As we have now two more types of coherent states, namely $|k,z\rangle$ and
$|k,\lambda \rangle$, we can define the corresponding distributions
on the phase space \ref{eq:5}\,: \begin{eqnarray}
  \label{eq:782}
  S_k(z,\bar{z}) &=& \langle k,z|\rho|k,z \rangle\,, \\
T_k(\lambda, \bar{\lambda}) &=& \langle k,\lambda|\rho|k,
\lambda \rangle\,, \label{eq:789}
\end{eqnarray}
and the density operator representations
\begin{eqnarray}
  \label{eq:790}
  \rho&=& \int_{\mathbb{C}}d\mu_k(z)\,F_k(z,\bar{z})\,|k,z\rangle
 \langle k,z|\,,\\
 \rho&=& \int_{\mathbb{D}}d\mu_k(\lambda)\,G_k(\lambda,\bar{\lambda})\,
|k,\lambda\rangle \langle k,\lambda|\,. \label{eq:791}
\end{eqnarray}

The distributions \ref{eq:782} and \ref{eq:789} as well as the representations
\ref{eq:790} and \ref{eq:791} can be discussed and analysed in a very similar
manner as has been done extensively for the distribution \ref{eq:51} and the
representation \ref{eq:52}. 

As we now have  altogether six ``densities'', there exist a large number of
relations between them which might be very useful for future applications.

I have not attempted to introduce distributions corresponding to Wigner's
one, because it is so closely related to the Weyl-Heisenberg group and
its action as a translation group. 
\subsection{Chapter 8}
Ch.\ 8 deals with the symplectic properties of interference patterns,
especially the $SO^{\uparrow}(1,2)$ structure of  the main observable 
quantities:

Consider the sum
\begin{equation}
  \label{eq:53}
A= A_1+A_2 = |A_1|\,e^{-i\,\varphi_1}+|A_2|\,e^{-i\,\varphi_2}
\end{equation}
of two complex
amplitudes $A_j$. 

The quantities $|A_j|$ and $\varphi_j$ may be functions
of other parameters, e.g.\ space or/and time variables etc.,
depending on the concrete experimental situation.

 The absolute
square of $A$ has the form \begin{eqnarray} \label{eq:1509}w_3(I_1,I_2,
\varphi)&=&|A|^2=I_1+I_2+2\,
\sqrt{I_1\,I_2}\, \cos\varphi~,\\ \label{eq:1025}&& I_j =|A_j|^2\,,\,
j=1,2\,,~~ \vp=\vp_1-\vp_2\,.  \nonumber
\end{eqnarray} Phase shifting one of the two amplitudes $A_j$ by
 appropriate devices yields new intensities: \begin{eqnarray}
\label{eq:1510}w_4(I_1,I_2,\varphi) &=& w_3(\varphi +\pi)=
I_1+I_2-2\sqrt{I_1\,I_2}\cos\varphi~,\\ \label{eq:1026}w_5(I_1,I_2,\varphi)
&=&w_3(\varphi +\pi/2)= I_1+I_2-2\sqrt{I_1\,I_2}\sin\varphi~,\\
\label{eq:1027}w_6(I_1,I_2,\varphi)& =& w_3(\varphi- \pi/2)=
I_1+I_2+2\sqrt{I_1\,I_2}\sin\varphi~.\end{eqnarray}

 The typical
quantities for a {\em classical} description of the interference
pattern are then
\begin{eqnarray} \label{eq:1511}4\,h_1(\vp,I_{1,2})&=&w_3-w_4=
4\,I_{1,2}\,\cos\varphi\,,~I_{1,2}=\sqrt{I_1\,I_2}~,\\
 \label{eq:1028}4\,h_2(\vp,I_{1,2}) &=&
w_5-w_6 =-4\,I_{1,2}\,\sin\varphi~,\\ \label{eq:1029}4\,h_0(\vp,I_{1,2})
 &=& 4\,I_{1,2}=
\sqrt{(w_4-w_3)^2+(w_6-w_5)^2}
>0~. \end{eqnarray}
  We see that the essential part of the
interference pattern is characterized by the three classical
observables $h_j(\vp,I_{1,2})$ with their associated
 $\mathfrak{so}(1,2)$ Lie algebra
 structure we know already.

The system has a number of interesting gauge properties, symplectic reductions
and intriguing relations to the symplectic group $Sp(4,\mathbb{R})$ which all
play a role in the quantum theory of the system.

 The quantum version of the classical relation \ref{eq:1029} poses
 critical questions as to the
validity of the ``operational phase'' analysis of the
interesting experiments by Mandel et al.\
\cite{Noh1}, because we know  from Ch.\ 2 that such
a trigonomical Pythagorean relation can be violated on the quantum level.

Sec.\ 8.2.2 discusses how one may test experimentally - at least in principle -
the different predictions for the observable quantities of interference 
patterns mentioned above, e.g.\ by multiport homodyning \cite{homod}.
{\em That section is mainly an appeal to the experts\,}!

 One last general remark in this context: 

It is essential to realize that a group theoretical quantization
does $not$ assume that the generators of the basic Lie algebra
themselves can be expressed by some conventional canonical operators
 $(Q_j,P_j)\,,\, j=1,\ldots,$  or in
terms of  associated annihilation and creation operators $a_j$ 
and $ a_j^+$. 

A well-known similar example is the angular momentum: In the case
of an orbital angular momentum the components
$\hat{l}_j\,,\,j=1,2,3,$ can be expressed in terms of  three pairs
$(Q_j,P_j)$, but this is not essential at all for the quantum
theory. That can be derived from the
single property that the 3 operators $\hat{l}_j$ generate the Lie
algebra of the group $SO(3)$ or that of its covering group $SU(2)$!

The representations of the latter allow for half-integer spins not
expected from semi-classical arguments!
 {\em Classically} the
Poisson brackets of the 3 components $l_1 = q_2p_3-q_3p_2\,,l_2
=\ldots$ fulfill the $SO(3)$ Lie algebra, too. However, this
applies to the {\em orbital} angular momentum only.

 Correspondingly,
though  some $SO^{\uparrow}(1,2)$ Lie
algebra representations may be constructed in terms of creation and
annihilation operators, this is not possible for other interesting ones and
also definitely not necessary!
\section{Apologies}
This article has been written by an outsider and non-specialist
 as to the impressively active and successful quantum
optics community! I have tried rather hard to understand some central topics
and problems of that growing and fascinating field. But I am sure I missed
important contributions and essential papers\,! So I would like to apologize
to all those authors the work of which I failed to appreciate or simply
overlooked. I shall certainly value being informed about any oversight
or worse. 

In addition I apologize to the mathematical physicists for my pedestrian
way of dealing with the mathematics of the problems. But I wanted to focus
on the physical aspects of those. 
\chapter{The $ SO^{\uparrow}(1,2)$ Lie algebra generators $K_0, K_1,$ and
 $K_2$ as quantum ``observables'' and the associated number states}
\section{General structures}
According to the Introduction and Appendix A.2 the basic classical
``canonical observables'' on the phase space 
\begin{equation}
  \label{eq:724}
  {\cal S}^2_{\vp,I} = \{(\vp,I);\, \vp \in \mathbb{R} \bmod{2\pi},\,~I > 0\,
\}
\end{equation}
 are
\begin{equation}
  \label{eq:721}
  h_0(\vp,I) = I\,,~~h_1(\vp,I) = I\,\cos\vp\,,~~h_2(\vp,I) = -I\,\sin\vp\,.
\end{equation}
They obviously are not independent, but obey the quadratic 
Pythagorean
relation
\begin{equation}
  \label{eq:58}
(I\cos\varphi)^2+(I\sin\varphi)^2=I^2\,.
\end{equation}
They also obey the Lie algebra $\mathfrak{so}(1,2)$ of the group
$SO^{\uparrow}(1,2)$, namely
\begin{equation}
  \label{eq:723}
  \{h_0,h_1\}_{\vp,I} = -h_2\,,~~\{h_0,h_2\}_{\vp,I} = h_1\,,~~
\{h_1,h_2\}_{\vp,I} = h_0\,,
\end{equation}
where
\begin{equation}
  \label{eq:725}
   \{h_0,h_1\}_{\vp,I} \equiv \partial_{\vp}h_0\,\partial_I h_1 -
\partial_Ih_0\,\partial_{\vp}h_1\,,\ldots\,.
\end{equation}
The relations \ref{eq:721}-\ref{eq:723} constitute the main structure
properties for any kind of analysis, uses or applications as to the
{\em classical} phase space \ref{eq:724}.

The important fact for a quantization of that phase space now is that
the group $SO^{\uparrow}(1,2)$ acts transitively, effectively and
globally Hamilton-like on it (see the Introduction and Appendix A.2).
This allows for a consistent group theoretical quantization in terms
of the positive discrete series of irreducible unitary representations
of the group $SO^{\uparrow}(1,2)$ or one of its covering groups (Appendix
B.3). 

The quantization is implemented by replacing the classical observables
\ref{eq:721} by the corresponding self-adjoint operators $K_0,K_1$ and
$K_2$ which represent the Lie algebra $\mathfrak{so}(1,2)$ in the
Hilbert space of the unitary representation under consideration:
\begin{equation}
  \label{eq:726}
  h_0 \to K_0\,~~h_1 \to K_1\,,~~ h_2 \to K_2\,.
\end{equation}
The self-adjoint  $K_j$ obey the commutation relations
\begin{align}
\label{eq:55} [K_0,K_1] & = i\,K_2\,,~~[K_0,K_2] = -i\,K_1\,,~~[K_1,K_2] = -i\,
K_0\,, \\
 [K_+,K_-]&=-2K_0\,,~[K_0,K_{\pm}]=\pm
K_{\pm}\,,\end{align} where
\begin{equation}
  \label{eq:54}
 K_+=K_1+iK_2\,,~~K_-=K_1-iK_2\,.
\end{equation}

In order to calculate expectation values and  other matrix elements
 we have
to know the actions of the operators $K_j, j=0,1,2,$ on the
Hilbert spaces associated with the positive discrete series 
 representations of $SO^{\uparrow}(1,2)$ (or
its covering groups). 

In the following I  frequently use properties 
which are discussed in more detail in Appendix B and in the literature
quoted there.

As the eigenfunctions of $K_0$ -- the generator of the compact
subgroup $O(2)$ of $SO^{\uparrow}(1,2)$ -- form a complete basis
of the associated Hilbert spaces, it is convenient to use them as
a starting point. The operators \ref{eq:54}
act as ladder operators. The positive discrete
series is characterized by the property that there exists a state
$|k,0\rangle$ for which
\begin{equation}
  \label{eq:727}
   K_-|k,0\rangle =0\,,~~K_0|k,0\rangle =k\,|k,0\rangle\,.
\end{equation}
  The number $k
>0$ characterizes the representation: 

 For a general normalized
 eigenstate
 $|k,n\rangle$ of $K_0$
we have
 \begin{eqnarray} \label{eq:1512}K_0|k,n\rangle&=&(k+n)|k,n\rangle\,
,~n=0,1,\ldots,~k > 0~,
\label{k3} \\ \label{eq:1030}\label{erz}
 K_+|k,n\rangle&=&\omega_n\,[(2k+n)(n+1)]^{1/2}|k,n+1\rangle~,~~
 |\omega_n|=1~, \\ \label{eq:1031}\label{vern}
 K_-|k,n\rangle&=&\frac{1}{\omega_{n-1}}[(2k+n-1)n]^{1/2}
 |k,n-1\rangle~.\end{eqnarray}

 In irreducible unitary representations the
operator $K_-$ is the adjoint operator of $K_+:\;
(f_1,K_+f_2)=(K_-f_1,f_2)$. The phases $\omega_n$ serve to
guarantee this property. Their choice depends on the concrete
realization of the representations. In the examples discussed in
Ref. \cite{bo} and in Appendix B.3 they have the values
 1 or $i$. In the following I
assume $\omega_n$ to be independent of $n:\omega_n=\omega$. But
then it can be absorbed into the definition of $K_{\pm}$ and we
can ignore it in the following discussions.

The Casimir operator 
\begin{align}
  \label{eq:56}
 L=K_1^2+K_2^2-K_0^2&
=K_+K_-+K_0(1-K_0)= \\&=K_-K_+ - K_0(1+K_0)
\end{align}
is a multiple of the unit operator in an irreducible unitary representation 
and has  the eigenvalues
\begin{equation}
  \label{eq:728}
  l=k(1-k)\,.
\end{equation}

 The allowed values of $k$ depend on the group:

 For
$SO^{\uparrow}(1,2)$ itself one can have $k=1,2,\ldots$ and for the
double coverings $SU(1,1)\cong SL(2,\mathbb{R}) =Sp(2,\mathbb{R})$
the values $k=1/2,1,3/2,\ldots $. For the universal
covering group $k$ may be any real number $>0$ (for details see
Refs.\ \cite{sa1,bo}). The appropriate choice will depend on the
physics to be described.

 In any case, for any unitary representation
the number $k$ has to be non-vanishing and positive!

Writing the Casimir operator for an irreducible representation in the form
\begin{equation}
  \label{eq:57}
 K_1^2 +K_2^2
= K_0^2 + l\,,~l=  k-k^2 =\frac{1}{4} -(k-\frac{1}{2})^2
\end{equation}
and comparing it with the corresponding
classical Pythagorean (trigonomical) relation \ref{eq:58}
one sees immediately
how that important geometrical property {\em may be violated quantum
theoretically\footnote{The additional term $l=k-k^2$ in Eq.\ \ref{eq:57} is a genuine quantum
effect associated with the non-classical properties of the ground state - see Eqs. 
\ref{eq:727} - as shown explicitly by the example of the harmonic oscillator. It is not
to be compared with the  constant $a \neq 0$ in the equation $x_1^2+x_2^2 =
x_0^2 + a$ of a classical hyperboloid!}\,}!

 Only for $k=1$ there is no such violation!

But already for
the important harmonic oscillator where $k=1/2$ (see Ch.\ 4 below)
we have $l =1/4$ and the r.h.\ side of Eq.\ \ref{eq:57} is enlarged
compared to the classical case \ref{eq:58}. This is the largest value
 $l$ may
take.

 We shall later (Ch.\ 6) encounter representations with
 $k=1/4$ and $k=3/4$. In both cases  $l=3/16$.

 On the other hand, if $k >1$
then the r.h.\ side of Eq.\ \ref{eq:57} is reduced. 

The relation \ref{eq:57} implies that we have the following constraint
on the mean-square fluctuations $(\Delta K_j)^2_k$ and the averages
$\langle K_j \rangle_k \equiv \langle k|K_j|k 
\rangle$ of the operators $K_j\,,j=0,1,2,$ with respect
to any state $|k\rangle$ of an irreducible unitary representation with
index $k$:
\begin{equation}
  \label{eq:405}
  (\Delta K_1)^2_k + (\Delta K_2)^2_k- (\Delta K_0)^2_k +\langle K_1
 \rangle_k^2
+\langle K_2 \rangle_k^2 -
\langle K_0 \rangle_k^2 = k-k^2
\end{equation}

{\em The modification \ref{eq:57} of the classical trigonomical relation
\ref{eq:58} is one of the most striking predictions of the present
approach to group quantizing the phase space \ref{eq:724}
It is, of course, extremely important to test such a prediction
experimentally\,}. I shall come back to this point on several occasions.
\section{Matrix elements of the number states}
  Eq.\ \ref{erz} implies
\begin{equation}
  \label{eq:59}
|k,n\rangle
=\frac{1}{\sqrt{(2k)_n\,n!}}
(K_+)^n|k,0\rangle~.
\end{equation}
Here the definition
\begin{equation}
  \label{eq:60}
 2k\,(2k+1)\cdots
(2k+n-1) \equiv (2k)_n = \frac{\Gamma(2k+n)}{\Gamma(2k)}~
\end{equation}
has
been used, where $(a)_n$ denotes Pochhammer's symbol \cite{er1,span}.

The completeness relation for the states $|k,n\rangle$ may formally
be written as
\begin{equation}
  \label{com0}
  \sum_{n=0}^{\infty} |k,n\rangle \langle k,n| = \boldsymbol{1} .
\end{equation}
The expectation values of the self-adjoint operators
\begin{equation}
  \label{eq:729}
    K_1 =(K_+
+K_-)/2\,,~~~ K_2=(K_+-K_-)/2i\,,
\end{equation}
  (which correspond to the
classical observables $I\cos\varphi$ and $-I\sin\varphi$) with
respect to the eigenstates $|k,n\rangle$ and the associated mean-square
fluctuations can be calculated with the help of the relations
\ref{k3}-\ref{vern}:
\begin{equation}
  \label{eq:61}
 \langle
k,n|K_j|k,n\rangle=0~,~~j=1,2\,;~n=0,1,\ldots ~.
\end{equation}
Thus, for the
number states the expectation values of the ``quadrature''
operators $K_1$ and $K_2$ vanish. This is what one expects. 

 The
corresponding mean-square fluctuations are \begin{eqnarray}
 \label{eq:1513}(\Delta
K_j)^2_{k,n}& =& \langle k,n|K_j^2|k,n\rangle = \frac{1}{2}(n^2
+2nk+k)= \\ \label{eq:1032}&=&\frac{1}{2}[(n+k)^2+l]\,,~
 j=1,2\,, \nonumber\\
\label{eq:1033}(\Delta K_j)^2_{k,n=0} &=& \langle k,n=0|K_j^2|k,n=0\rangle
 = \frac{k}{2}~.  \end{eqnarray}

 Because of
$[K_1,K_2]=-iK_0$, the general uncertainty relation
\begin{equation}
  \label{eq:62}
 \Delta
A\;\Delta B \geq \frac{1}{2}|\langle \, |[A,B]|\, \rangle |
\end{equation}
for self-adjoint operators $A$ and $B$ here takes the special form
\begin{equation}
  \label{eq:63}
 (\Delta K_1)_{k,n}\; (\Delta K_2)_{k,n} = \frac{1}{2}(n^2+2k
n+k) \geq \frac{1}{2}|\langle k,n|K_0|k,n\rangle
|=\frac{1}{2}(n+k)~.
\end{equation}
The equality sign holds  for the ground
state $|k,n=0\rangle$ only.

 The Eqs.\ \ref{eq:1513} imply further  that
\begin{eqnarray}
 \label{eq:1514}\langle k,n|K_1^2|k,n\rangle +
\langle k,n|K_2^2|k,n\rangle &=& (\Delta K_1)^2_{k,n}+(\Delta
K_2)^2_{k,n}= \\ \label{eq:1034}= (n+k)^2 + l &=& \label{ncorr}
\langle k,n|K_0^2|k,n\rangle +
l~, \end{eqnarray} which is, of course, just special case of the
Casimir operator relation \ref{eq:57}. But it shows in addition that for
very large $n$ the term $l $ on the r.h.s.\ of Eq.\
\ref{ncorr} is negligible and the {\em correspondence principle} holds. 

We also have 
\begin{equation}
  \label{eq:799}
  \langle k,n|K_1\,K_2 +K_2\,K_1|k,n\rangle =0\,.
\end{equation}
\section{Position and momentum operators as func\-tions of the Lie algebra
 generators $K_0,K_1$ and $K_2$}
 If $f(K_0)$ is
an ``appropriate'' function of the operator $K_0$, then the
commutation relations \ref{eq:55}  imply  relations which are completely
analogous to those in Eq. \ref{eq:16}:
\begin{equation}
  \label{eq:64}
 K_-f(K_0) = f(K_0+\boldsymbol{1})\,
K_-\,,~f(K_0)\,K_+ = K_+f(K_0+\boldsymbol{1})\,.
\end{equation}
``Appropriate'' means that the application of the operators $f(K_0)$
and $f(K_0+\boldsymbol{1})$  to all  number states \ref{eq:59} is
 well-defined,
including the application to the ground state $|k,n=0\rangle $!

The relations \ref{eq:64} imply the following important
observation (see also Sec.\ 1.4): 

According to Eq.\ \ref{k3} the operator $K_0$ is a positive definite
one, in an irreducible unitary representation of the positive
discrete series! Therefore the operator  $(K_0+k)^{-1/2}$ does always
exist in the Hilbert space for such a representation and we can define
\begin{equation}
  \label{eq:65}
 a= (K_0+k)^{-1/2}K_-~,~a^+ =
K_+(K_0+k)^{-1/2}~,~ N=K_0-k~,
\end{equation}
which, according to the
relations \ref{erz} and \ref{vern} have the properties
\begin{equation}
  \label{eq:66}
 a|k,n\rangle =
\sqrt{n}\,|k,n-1\rangle~,~~a^+|k,n\rangle =
\sqrt{n+1}\,|k,n+1\rangle~.
\end{equation}
It follows that
\begin{equation}
  \label{eq:67}
 [a,a^+]= 1~.
\end{equation}

The relation \ref{eq:67} can also be shown in the following way:
 From Eq.\ \ref{eq:56} we
have
\begin{equation}
  \label{eq:68}
 K_-K_+= L+K_0(K_0+1)\,,~~ L=k(1-k)\,,
\end{equation}
so that
\begin{equation}
  \label{eq:69}
 a\cdot a^+ =
\frac{k(1-k)+K_0(K_0+1)}{K_0+k}= K_0-k+1=N+1\,.
\end{equation}

On the other hand,
because of
\begin{equation}
  \label{eq:70}
 K_+K_-= L+K_0(K_0-1)
\end{equation}
and the relations \ref{eq:64}, we
get
\begin{equation}
  \label{eq:71}
 a^+\cdot a = K_+(K_0+k)^{-1}K_- = (K_0+k-1)^{-1}K_+K_-
=K_0-k=N\,.
\end{equation}
Thus, it always possible to associate creation and annilation
operators \ref{eq:65} with a given irreducible unitary representation of the
positive discrete series of $SO^{\uparrow}(1,2)$ or $SU(1,1) \cong SL(2,
\mathbb{R})= Sp(2,\mathbb{R})$  or any other of the covering groups! 

As a consequence we can  introduce the {\em composite self-adjoint
 position and momentum
operators} (with $\omega =1$)
\begin{eqnarray}
  \label{eq:730}
  \tilde{Q}&=& \frac{1}{\sqrt{2}}(a^++a)=\frac{1}{\sqrt{2}}[K_+(K_0+k)^{-1/2}
+(K_0+k)^{-1/2}K_- 
] ~~~ \\ &=& \frac{1}{\sqrt{2}}[(K_0+k)^{-1/2}K_1 +
K_1(K_0+k)^{-1/2}]- \label{eq:731} \\ &&-\frac{i}{\sqrt{2}}[(K_0+k)^{-1/2}K_2 -
K_2(K_0+k)^{-1/2}]\,, \nonumber \\
\tilde{P}&=& \frac{i}{\sqrt{2}}(a^+-a)=\frac{i}{\sqrt{2}}[K_+
(K_0+k)^{-1/2}-(K_0+k)^{-1/2}K_-] 
 ~~~ \label{eq:732} \\ &=& -\frac{i}{\sqrt{2}}[(K_0+k)^{-1/2}K_1 -
K_1(K_0+k)^{-1/2}]-\label{eq:733} \\ && -\frac{1}{\sqrt{2}}[(K_0+k)^{-1/2}K_2 +
K_2(K_0+k)^{-1/2}]\,. \nonumber
\end{eqnarray}
This is possible\footnote{As to to the relationship of the
operators \ref{eq:730}-\ref{eq:733} to the Stone--von Neumann theorem
\cite{sto} see Subsec.\ 4.4.4\,.} for any allowed $k$.

The operators $\tilde{Q}$ and $\tilde{P}$ are the quantum versions of the
classical coordinates \ref{eq:792} and \ref{eq:796}. They are ``observables''
in the sense of Secs.\ 1.4 and 6.3 because they transform according to
the group $SO^{\uparrow}(1,2)$ (adjoint representation from Sec.\ 6.1),
and not according to $Sp(2,\mathbb{R})$ (Sec.\ 6.3)!

As has been stressed in Sec.\ 1.4.1, {\em the possibility to construct position
and momentum operators from them makes the operators $K_1, K_2$ and $K_0$
as least as fundamental as the $Q$ and $P$ themselves\,!}

The following point is important, too: The matrix elements of the operators 
\ref{eq:65}, \ref{eq:730} and \ref{eq:732} in general will depend  on the
index $k$, i.e.\ one the choice of the unitary representation. This will
become evident and explicitly shown in Secs.\ 3.1 and 3.2 in connection
with coherent states. 

\section{A contraction of the Lie algebra $\mathfrak{so}(1,2)$ to the
Weyl-Heisenberg algebra}
There is another relationship between the Lie algebra \ref{eq:55} and the
Lie algebra 
\begin{equation}
  \label{eq:734}
  [a,a]=0\,,~~[a^+,a^+]=0\,,~~[a,a^+]=1\,,
\end{equation} of the Heisenberg-Weyl group \ref{eq:26}
 \cite{ba1}:

 Define
\begin{equation}
  \label{eq:72}
\check{K}_{\pm} = (2k)^{-1/2}K_{\pm}~,~\check{K}_3 = k^{-1}K_0~
\end{equation}
Then
\begin{equation}
  \label{eq:73}
 [\check{K}_+, \check{K}_-] = -\check{K}_3~,~
[\check{K}_3, \check{K}_{\pm}] =\pm k^{-1}\check{K}_{\pm}~.
\end{equation}
If
we now take the limit $k \to \infty$, then - formally - the limit
of the Lie algebra \ref{eq:73} is the Weyl-Heisenberg Lie algebra
\ref{eq:734}! This may be justified by looking at the
matrix elements of the new operators \ref{eq:72}: From Eqs.\ \ref{k3}-
\ref{vern}
we get
\begin{eqnarray} \label{eq:1515}\langle k, n_2| \check{K}_3|k,n_1\rangle&=
&(1+\frac{n_1}{k})\delta_{n_2,n_1},
 \\
 \label{eq:1035}\langle k,n_2|\check{K}_+|k,n_1\rangle&=&[(1+\frac{n_1}{2k})
(n_1+1)]^{1/2}\, \delta_{n_2,n_1+1},~~
  \\
 \label{eq:1036}\langle k,n_2|\check{K}_-|k,n_1\rangle&=&[(1+
\frac{n_1-1}{2k})n_1]^{1/2}\,
 \delta_{n_2,n_1-1}~. \end{eqnarray} Taking the limit $k \to \infty$
  shows that
\begin{equation}
  \label{eq:74}
 \lim_{k \to \infty}\check{K}_3 \to 1~,~\lim_{k \to \infty}\check{K}_+
 \to a^+~,~\lim_{k \to \infty}\check{K}_-
 \to a~.
\end{equation}
However, we have  seen in the previous section that it is not necessary to go
 to the limit $k \to \infty$ in order to arrive at the Lie algebra
\ref{eq:734} if a representation of the Lie algebra \ref{eq:55}
 is given. We may just use the relations \ref{eq:65} or \ref{eq:730} and
\ref{eq:732}!
\chapter{Three types of coherent state matrix elements of the 
 ``observables'' $K_0,K_1$ and
 $K_2$}
  Next I want to discuss  matrix elements of the ``observables'' $K_j,\,
j=0,1,2,$
 with respect to
   coherent states. The purpose is to see how the properties of their classical
 counterparts
 \begin{equation}
   \label{eq:685} 
  h_0 = I\,,~~ h_1 =I\,\cos\vp\,,~~ h_2= -I\,\sin\vp\,, \nonumber
 \end{equation}
  are incorporated
 into those
 matrix elements, especially how the angle $\vp$ is represented.

 The three different types of coherent states discussed below are all
 characterized by a complex number and it is interesting to see
 how the phase of these numbers is ``measured'' by the operators
 $K_1$ and $K_2$. Many of the mathematical results discussed in the following
 are well-known. They are repeated here in order to keep the
 discussion  of the intended aim rather self-contained. Most
 of the related References will be given in the course of the
 chapter.

 The three types of coherent states -- those of Barut-Girardello
 \cite{ba1}, 
 Perelomov \cite{per0,per1,per}
 and Schr\"odinger-Glauber \cite{schro1, glau1} --
 can all be characterized
 as eigenstates of the annihilation operators $K_-, (K_0+k)^{-1}K_-$ and
$(K_0+k)^{-1/2}K_-$, as defined in the Eqs.\ \ref{eq:1516}-\ref{eq:1039}. 
  \section{Barut-Girardello coherent states}
  \subsection{General properties}
 From the relation \ref{eq:59} we get for the Barut-Girardello states
 \ref{eq:1516}
\begin{equation}
  \label{eq:75}
 \langle k,n|k,z\rangle=
\left[\frac{1}{(2k)_n\,n!}\right]^{1/2}z^n\,
\langle k,n=0|k,z\rangle~,
\end{equation}
so that
\begin{eqnarray} \label{eq:1517}\langle k,z|k,z\rangle& =& \sum_{n=0}^{\infty}=
\langle k,z|k,n\rangle \langle k,n|k,z\rangle\\  \label{eq:1040}&=& |\langle
k,n=0|k,z\rangle|^2
\, g_k(|z|^2)~, \nonumber \\ \label{eq:1041}g_k(|z|^2) &=& \sum_{n=0}^{\infty}\frac{|z|^{2n}}{ (2k)_n\,n!} \\
\label{eq:1042}&=&\Gamma(2k) |z|^{1-2k}\,I_{2k-1}(2|z|)~. \nonumber
\end{eqnarray} Here
\begin{equation}
  \label{eq:76}
 I_{\nu}(x) = \left
(\frac{x}{2}\right)^{\nu}\sum_{n=0}^{\infty}
\frac{1}{n!\,\Gamma(\nu+n+1)}\left (\frac{x}{2}\right)^{2n}
\end{equation}
is
the usual modified Bessel function of the first kind \cite{er2}
which has the asymptotic expansion \begin{eqnarray} \label{eq:1518}I_{\nu}(x)
&\asymp& \frac{\displaystyle
e^x}{\sqrt{2\pi\,x}}\left[1-\frac{4\nu^2 -1}{8\,x}+
2\frac{(4\nu^2-1)(4\nu^2-9)}{16^2\,x^2}+ O(x^{-3})\right]\\
\label{eq:1043}&&~~~~~~~~~~~~ \mbox{for}~x\rightarrow +\infty ~. \nonumber
\end{eqnarray} For the function $g_k(|z|^2)$ this implies
the asymptotic behaviour
\begin{eqnarray} \label{eq:1519}g_k(|z|^2)
&\asymp& \frac{\Gamma(2k)}{2\sqrt{\pi}}|z|^{\frac{1}{2}-2k}~e^{
2|z|}\left[1-\frac{4(2k-1)^2 -1}{16\,|z|}+ \right.\\ \label{eq:1044}&+& \left.
\frac{[4(2k-1)^2-1][4(2k-1)^2-9]}{2\,16^2\,|z|^2}+
O(|z|^{-3})\right]\nonumber \\
\label{eq:1045}&&~~~~~~~~~~~~ \mbox{for}~|z| \rightarrow +\infty ~. \nonumber
\end{eqnarray}
  If $\langle
k,z|k,z\rangle=1$ we have
\begin{equation}
  \label{eq:77}
|\langle k,n=0|k,z\rangle|^2  =
\frac{1}{g_k(|z|^2)}~.
\end{equation}
Choosing the phase of $\langle
k,n=0|k,z\rangle$ to be zero we finally get the expansion
\begin{equation}
  \label{eq:78}
|k,z\rangle= \frac{1}{\sqrt{g_k(|z|^2)}}\sum_{n=0}^{\infty}
\frac{z^n}{\sqrt{(2k)_n\,n!}}\,|k,n\rangle ~~.
\end{equation}
Notice that
$|k,z=0 \rangle = |k,n=0\rangle$. \\  In order to avoid explicit
factors like $|z|^{2k-1}$ etc.\ it is convenient to introduce the
function $g_k(w)$ which in addition is an entire holomorphic
function of the complex variable $w$. This last property does not
hold for the function $I_{\nu}$ if $\nu$ is not an integer. \\ In
the general language of hypergeometric series \cite{er1c} we have
the relation
\begin{equation}
  \label{eq:79}
 g_k(w) = {_0}F_1(2k;w)~.
\end{equation}
${_0}F_1(2k;w)$
is an entire function of order $1/2$ and type $2$, i.e.\ it
behaves for large $|w|$ as $\leq O(\exp(2\,|w|^{1/2}))$ (see Eq.\
\ref{eq:1519}). As to the notion of ``order'' and ``type'' of an entire
function see Ref.\ \cite{hil}.

  The function ${_0}F_1(2k;w)$ obeys the differential equation
\begin{equation}
  \label{eq:80}
 w\frac{d^2({_0}F_1)}{dw^2}+2k\,\frac{d({_0}F_1)}{dw} -{_0}F_1
=0~.
\end{equation}
Two different coherent states are not orthogonal:
\begin{eqnarray} \label{eq:1520}\langle k, z_2|k,z_1 \rangle &=&
\sum_{n=0}^{\infty}\langle k, z_2|k,n \rangle  \langle k,n|k,z_1
\rangle =  \\ \label{eq:1046}&=& \frac{g_k(
\bar{z}_2\,z_1)}{\sqrt{g_k(|z_2|^2)\,g_k(|z_1|^2)}}~. \nonumber
\end{eqnarray} They are complete, however, in the sense that, with
$z=|z|\,e^{i\,\phi}$, we have the  relation
\begin{multline} \label{eq:1047}
 \frac{2}{\pi\,\Gamma(2k)}\int_0^{\infty}d|z|\, |z|^{2k}\,
K_{2k-1}(2|z|)\,g_k(|z|^2)\times \\
\times \int_0^{2\pi}d\phi\, \langle k,n_2|k,|z|\,
e^{i\phi}\rangle \langle k,  |z|\, e^{i\phi}|k,n_1 \rangle =
\\  = \langle k,n_2|k,n_1\rangle
=\delta_{n_2n_1}~,\end{multline} because \cite{GR} 
\begin{equation}
  \label{eq:406}
 \int_0^{\infty} d|z|\,
|z|^{2(n+k)}K_{2k-1}(2|z|)=\frac{1}{4}n!\,\Gamma(2k+n)~. 
\end{equation}
  Here
$K_{\nu}(x)$ is the modified Bessel function of the third kind
\cite{er2}. It has the asymptotic expansion \begin{multline}\label{asympk}
 K_{\nu}(x) \asymp  \\ \asymp \sqrt{\frac{\pi}{2\,x}}\,e^{
-x}\,\left[1+\frac{4\nu^2 -1}{8\,x}+
2\frac{(4\nu^2-1)(4\nu^2-9)}{16^2\,x^2}+ O(x^{-3})\right] \\
 \mbox{for}~x\rightarrow +\infty ~ 
\end{multline} and  a singularity for $x \to 0$ like
\begin{eqnarray} \label{eq:1521}K_{\nu}(x) & \to & 
\frac{\Gamma(\nu)}{2}\left(\frac{x}{2}\right)^{-\nu}\,(1 +
O(x^2)\,)~\mbox{
 for } \nu > 0~, \\ \label{eq:1051}
 K_0(x) &\to &-[\gamma +\ln\frac{x}{2}] + O(x^2\,\ln x)~, 
\\ \label{eq:1052}\gamma
 &=& 0.57\ldots~: \mbox{ Euler's constant}\,. \nonumber
  \end{eqnarray} That singularity is, however, removed by the
additional factor $|z|^{2k}$ in the measure of the integral \ref{eq:1047}. 

  Formally we may express the completeness relation \ref{eq:1047}
  (``resolution of the identity'')
as \begin{eqnarray} \label{eq:1522}\label{com1} \int_{\mathbb{C}} d\mu_k(z)
\,|k,z\rangle \langle k, z| &=& \boldsymbol{1}\,, \\
\label{eq:1053}d\mu_k(z) &=& m_k(z)\, d^2z~,\label{meas1} \\
\label{eq:1054}m_k(z)&=&\frac{2}{\pi\,\Gamma(2k)}\,|z|^{2k-1}\,
K_{2k-1}(2|z|)\,g_k(|z|^2)~,
\nonumber\\
\label{eq:1055}&& d^2z = |z|\,d|z|\,d\phi~. \nonumber \end{eqnarray} The
completeness relation implies the following convolution property
for the scalar product \ref{eq:1520}:
\begin{equation}
  \label{eq:81}
 \int_{\mathbb{C}}
d\mu_k(z)\langle k, z_2|k,z\rangle \langle k,z|k, z_1 \rangle = \langle k,
z_2|k,z_1\rangle ~.
\end{equation}
\subsection{Associated Hilbert space of holomorphic functions}
It is worthwhile to have a brief look at the
Hilbert space of holomorphic functions associated with the
coherent states $|k,z\rangle $: The coefficients
\begin{equation}
  \label{eq:82}
\tilde{f}_{k,n}(z) = \frac{z^n}{\sqrt{(2k)_n\,n!}}~,~n=0,1,\ldots
\end{equation}
in the expansion \ref{eq:78} form an orthonormal basis of the Hilbert
space of holomorphic square-inte\-grable functions $\tilde{f}(z)$
with the scalar product
\begin{eqnarray} \label{eq:1523}(\tilde{f}_2,\tilde{f}_1)_{k,\tilde{\mu}} &=&
\int_{\mathbb{C}}d\tilde{\mu}_k(z)\,
\bar{\tilde{f}}_2(z)\,\tilde{f}_1(z)~,\\ \label{eq:1056}d\tilde{\mu}_k(z) &=&
\tilde{m}_k(|z|) \,d^2z~, \\ \label{eq:1057}\tilde{m}_k(|z|) &=& 
\frac{2}{\pi\,\Gamma(2k)}\,
|z|^{2k-1}\, K_{2k-1}(2|z|)~.
\end{eqnarray} The weight function $\tilde{m}_k(|z|)$ of the measure
$d\tilde{\mu}_k(z)$  has the following limiting
properties: \begin{eqnarray}\label{eq:1524}\tilde{m}_k(|z|=0) &=&
 \frac{1}{(2k-1)\,\pi}
\mbox{ for } 2k-1 > 0~, \\ \label{eq:1058}\tilde{m}_{1/2}(|z|) &\to&
-\frac{2}{\pi}[\gamma + \ln|z|]\mbox{ for } |z| \to 0.
\end{eqnarray} Thus, the weight $\tilde{m}_{1/2}(|z|)$ has a logarithmic
singularity for $|z| \to 0$. Like above, this singularity is, however,
suppressed by the additional factor $|z|$ in $d^2z
=|z|\,d|z|\,d\phi$. \\ For $|z| \to \infty $ we get from \ref{asympk}
\begin{equation}
  \label{eq:83}
\tilde{m}_k(|z|) \asymp
\frac{|z|^{2k-3/2}}{\sqrt{\pi}\,\Gamma(2k)}\,e^{-2|z|}\,\left[1+
\frac{4(2k-1)^2
-1}{16\,|z|} + O(|z|^{-2}) \right]~.
\end{equation}
The same arguments which
lead to the relation \ref{eq:1047}
 show that
\begin{equation}
  \label{eq:84}
(\tilde{f}_{k,n_2},\tilde{f}_{k,n_1})_{k,\tilde{\mu}}=\delta_{n_2\,n_1}~.
\end{equation}

The so-called ``reproducing kernel'' \cite{aro,barg2,neeb} for this 
Hilbert space is
given by
\begin{equation}
  \label{eq:85}
 \tilde{\Delta}_k(\bar{z}_2,\,z_1) =
\sum_{n=0}^{\infty}
\bar{\tilde{f}}_{k,n}(z_2)\,\tilde{f}_{k,n}(z_1) =
g_k(\bar{z}_2 z_1)\,.
\end{equation}
Its most essential property
\begin{equation}
  \label{eq:86}
\int_{\mathbb{C}}
\tilde{\mu}_k(z_2)\,\tilde{\Delta}_k(\bar{z}_2,z_1)\,z_2^n =
z_1^n~,~n \geq 0\,,
\end{equation}
follows immediately from the orthogonality
relations \ref{eq:84}. Thus, if
\begin{equation}
  \label{eq:87}
 \tilde{f}(z)= \sum_{n=0}^{\infty}
c_n\,z^n
\end{equation}
is a convergent series, then the relations \ref{eq:86}
imply
\begin{equation}
  \label{eq:88}
 \int_{\mathbb{C}}
\tilde{\mu}(z_2)\,\tilde{\Delta}_k(\bar{z}_2,z_1)\,\tilde{f}(z_2) =
\tilde{f}(z_1)~.
\end{equation}
Therefore, the kernel
$\tilde{\Delta}(\bar{z}_2,z_1)$ here plays the same role as the usual
$\delta$-function in other circumstances. It has the property
\begin{equation}
  \label{eq:89}
\bar{\tilde{\Delta}}_k(\bar{z}_2,z_1) =
\tilde{\Delta}_k(\bar{z}_1,z_2)~,
\end{equation}
so that
\begin{equation}
  \label{eq:90}
\int_{\mathbb{C}}
\tilde{\mu}_k(z_2)\,\tilde{\Delta}_k(\bar{z}_1,z_2)\,(\bar{z}_2)^n
= (\bar{z}_1)^n~.
\end{equation}
Notice, however, that we do not have the
reproducing property \ref{eq:88} if a function $f$ is not holomorphic, i.e.
$\partial_{\bar{z}}f \neq 0\,$. In this case the generalization
of the relation \ref{eq:86} is \begin{eqnarray} \label{eq:1525}
\int_{\mathbb{C}}
\tilde{\mu}_k(z_2)\,\tilde{\Delta}(\bar{z}_2,z_1)\bar{z}_2^{\bar{n}}\,z_2^n
&=& 0 \mbox{ for } \bar{n} > n~,\label{nhol1} \\ &=&
\frac{n!\,(2k)_n}{(n-\bar{n})!\,(2k)_{n-\bar{n}}}\,z_1^{n-\bar{n}}
\mbox{ for } n \geq \bar{n}~. \label{nhol2} \end{eqnarray}
(here $\bar{n} =0,1,\ldots$ denotes an independent natural number, not
the complex conjugate of $n$\,.)

The corresponding
generalization of \ref{eq:90} is obtained by complex conjugation of
the relations \ref{nhol1} and \ref{nhol2}

 The convolution \ref{eq:81} implies
\begin{equation}
  \label{eq:407}
  \int_{\mathbb{C}}d\tilde{\mu}_k(z)\,\tilde{\Delta}_k(\bar{z}_2,z)\,
\tilde{\Delta}_k(\bar{z},z_1) = \tilde{\Delta}_k(\bar{z}_2,z_1)~,
\end{equation}
or
\begin{equation}
  \label{eq:408}
\int_{\mathbb{C}}d\tilde{\mu}_k(z)\, g_k(\bar{z}_2\,z)\,g_k(\bar{z}\,z_1) = 
g_k(\bar{z}_2\,z_1)~.  
\end{equation}

One can also implement a - positive discrete series - irreducible
unitary representations with index $k$ of $SU(1,1)$ etc.\ in the present
Hilbert space. The corresponding Lie algebra generators are
\begin{eqnarray} \label{eq:1526}K_+ &=& z~,\\ \label{eq:1059}K_-
 &=& 2k\frac{d}{dz} +
z\frac{d^2}{dz^2}~, \\ \label{eq:1060}K_0 &=& z\frac{d}{dz} +k~.
\end{eqnarray} They obey the commutation relations \ref{eq:55} and the
relations \ref{k3}-\ref{vern} when applied to the basis functions \ref{eq:82}.

That
$K_-$ is the adjoint operator of $K_+$ may be seen as follows: 

The scalar product of two functions
\begin{equation}
  \label{eq:91}
 \tilde{f}_j(z) =
\sum_{n=0}^{\infty} c_n^{(j)}\tilde{f}_{k,n}(z)~,~j=1,2,
\end{equation}
is
given by the series
\begin{equation}
  \label{eq:92}
 (\tilde{f}_2,\tilde{f}_1)_{\tilde{\mu}} =
\sum_{n=0}^{\infty}\bar{c}_n^{(2)}c_n^{(1)}.
\end{equation}
Applying $K_+$ to
$\tilde{f}_1$, observing the action \ref{erz} of $K_+$ on $\tilde{f}_{k,n}$
 and the
orthonormality \ref{eq:84} gives
\begin{equation}
  \label{eq:93}
(\tilde{f}_2,K_+\tilde{f}_1)_{\tilde{\mu}} =
\sum_{n=0}^{\infty}\sqrt{(2k+n)(n+1)}\,\bar{c}_{n+1}^{(2)}c_{n}^{(1)}~.
\end{equation}
Applying $K_-$ to $\tilde{f}_2$ and calculating
$(K_-\tilde{f}_2,\tilde{f}_1)$ gives the same result:
\begin{equation}
  \label{eq:94}
(K_-\tilde{f}_2,\tilde{f}_1)_{\tilde{\mu}} =
(\tilde{f}_2,K_+\tilde{f}_1)_{\tilde{\mu}}~.
\end{equation}
The operator
$K_-$ has the eigenfunctions $\tilde{f}_{k,\zeta}(z)$:
\begin{equation}
  \label{eq:95}
K_-\tilde{f}_{k,\zeta}(z) = \zeta\,\tilde{f}_{k,\zeta}(z)~,~ \zeta
\in\mathbb{C}.
\end{equation}
The normalizable solution of this differential
equation is (see also Eq.\ \ref{eq:80})
\begin{equation}
  \label{eq:96}
 C\,\sum_{n=0}^{\infty}\frac{\zeta^n}{\sqrt{(2k)_n
\,n!}}\tilde{f}_{k,n}(z) = C\,g_k(\zeta z)~. ~
\end{equation}
Normalization
finally gives
\begin{equation}
  \label{eq:97}
 \tilde{f}_{k,\zeta}(z)=
\frac{1}{g_k(|\zeta|^2)}\,g_k(\zeta z)~.
\end{equation}
Let
\begin{equation}
  \label{eq:98}
\tilde{f}(z) = \sum_{n=0}^{\infty} c_n\,\tilde{f}_{k,n}(z)
\end{equation}
be
a normalized function,
\begin{equation}
  \label{eq:99}
 (\tilde{f},\tilde{f})_{\tilde{\mu}} =
\sum_{n=0}^{\infty}|c_n|^2 =1~.
\end{equation}
Then we have the estimate
\begin{equation}
  \label{eq:100}
|f(z)|^2 \leq \sum_{n=0}^{\infty} |c_n\tilde{f}_{k,n}(z)|^2 \leq
(\sum_{n=0}^{\infty}|c_n|^2)(\sum_{n=0}^{\infty}|\tilde{f}_{k,n}(z)|^2)
= g_k(|z|^2)~.
\end{equation}
Thus
\begin{equation}
  \label{eq:101}
 |\tilde{f}(z)| \leq
\sqrt{g_k(|z|^2)}.
\end{equation}
This describes the allowed upper bound
and the possible growth properties for the functions
$\tilde{f}(z)$.

The regular solutions of the eigenvalue equations
\begin{eqnarray}
  \label{eq:678}
  K_1 \tilde{f}_{k,h_1}(z)&=& h_1\, \tilde{f}_{k,h_1}(z)\,, \\
 K_2 \tilde{f}_{k,h_2}(z)&=& h_2\, \tilde{f}_{k,h_2}(z)\,, \label{eq:679}
\end{eqnarray}
where the operators $K_1 =(K_++K_-)/2$ and $K_2 =(K_+-K_-)/2i$ follow
from the Eqs.\ \ref{eq:1526} and \ref{eq:1059}, may easily be determined
with the help of Ref.\ \cite{er4}:
\begin{eqnarray}
  \label{eq:680}
  \tilde{f}_{k,h_1}(z)&=& \tilde{C}_{k,h_1}\,e^{-iz}\,\Phi(k-ih_1,2k;2iz)\,,
~h_1 \in \mathbb{R}\,, \\
 \tilde{f}_{k,h_2}(z)&=& \tilde{C}_{k,h_2}\,e^{-z}\,\Phi(k-ih_2,2k;2z)\,,
~h_2 \in \mathbb{R}\,.
\label{eq:681}
\end{eqnarray}

 Here $C_{k,h_1}$ and $C_{k,h_2}$ are (normalization) constants
and $\Phi(a,c;z)$ is the at the origin regular basic solution
($\Phi(a,c; z=0)=1$) of the
differential equation for confluent hypergeometric functions \cite{er5}.

The convolution relation \ref{eq:81} shows that the scalar product
\ref{eq:1520} is a reproducing kernel, too:
\begin{equation}
  \label{eq:102}
 \Delta_k(\bar{z}_2,z_1) =
\langle z_2|z_1\rangle =
\sum_{n=0}^{\infty}\bar{f}_{k,n}(z_2)\,f_{k,n}(z_1) =
\frac{g_k(
\bar{z}_2\,z_1)}{\sqrt{g_k(|z_2|^2)\,g_k(|z_1|^2)}}~,
\end{equation}
where the functions
\begin{equation}
  \label{eq:103}
 f_{k,n}(z)
\frac{1}{\sqrt{g_k(|z|^2)}}\,\frac{z^n}{\sqrt{(2k)_n\,n!}}~,~n=0,1,\ldots
\end{equation}
form an orthonormal basis with respect to a scalar product
with the integration measure \ref{meas1}:
\begin{equation}
  \label{eq:104}
(f_{k,n_2},f_{k,n_1})_{k,\mu} =
\int_{\mathbb{C}}d\mu_k(z)\bar{f}_{k,n_2}(z)f_{k,n_1}(z) =
\delta_{n_2n_1}\,.
\end{equation}
Due to the coefficient
$[g_k(|z|^2)]^{-1/2}$ the functions \ref{eq:103} are no longer
holomorphic. Instead of \ref{eq:86} we now have the relations
\begin{equation}
  \label{eq:105}
\int_{\mathbb{C}}d\mu_k(z_2)\Delta_k(\bar{z}_2,z_1) z^n_2
/\sqrt{g_k(|z_2|^2)} =z^n_1/\sqrt{g_k(|z_1|^2)}~.
\end{equation} Notice also that
\begin{equation}
  \label{eq:684}
  |k,n \rangle = \int_{\mathbb{C}} d \mu_k(z) \bar{f}_{k,n}(z)\, |k,z
 \rangle \,.
\end{equation}

For a
normalizable function
\begin{equation}
  \label{eq:106}
 f(z) = \sum_{n=0}^{\infty} c_n
f_{k,n}(z)
\end{equation}
the same arguments as above lead to the bound
\begin{equation}
  \label{eq:107}
|f(z)| \leq 1 ~.
\end{equation}
This appears to impose a severe restriction
on possible allowed functions $f(z)$. But this restriction is
spurious, because the two Hilbert spaces with the different
measures \ref{meas1} and \ref{eq:1056} are unitarily equivalent. Obviously the
Hilbert space with the measure \ref{eq:1056} is mathematically the
``cleaner'' one, because its basic functions are genuinely
holomorphic!

 There is a completely analogous situation for the conventional
(``Schr\"o\-dinger-Glauber'') normalized coherent states \cite{schro1,glau1}
\begin{eqnarray} \label{eq:1527}|\alpha\rangle &=& e^{-|\alpha|^2/2}\sum_{n=0}^{\infty}
\frac{\alpha^n}{\sqrt{n!}}|n\rangle~,~\alpha \in  \mathbb{C}~, \\
\label{eq:1061}\langle \alpha_2|\alpha_1 \rangle &=&
e^{-(|\alpha_2|^2+|\alpha_1|^2)/2}\,e^{\bar{\alpha}_2\,\alpha_1}~.
\end{eqnarray}
 The  integration measure here is
\begin{equation}
  \label{eq:108}
 d^2\alpha/\pi =
d\,\Re(\alpha)\,d\,\Im (\alpha)/\pi ~,
\end{equation}
with respect to which
the functions
\begin{equation}
  \label{eq:109}
 h_n(\alpha) = e^{
-|\alpha|^2/2}\frac{\alpha^n}{\sqrt{n!}}
\end{equation}
form an orthonormal
basis.  

 The completeness relation may be written as
\begin{equation}
  \label{eq:110}
\frac{1}{\pi}\int_{\mathbb{C}}d^2\alpha |\alpha\rangle \langle
\alpha| = \boldsymbol{1}\,.
\end{equation}
The reproducing
kernel is
\begin{equation}
  \label{eq:111}
 \Delta(\alpha_2,\alpha_1) = \langle
\alpha_2|\alpha_1 \rangle =
\sum_{n=0}^{\infty}\bar{h}_n(\alpha_2)h_n(\alpha_1) = e^{
-(|\alpha_2|^2+|\alpha_1|^2)/2}\,e^{
\bar{\alpha}_2\alpha_1}~.
\end{equation}
\\
Square-integrable functions
\begin{equation}
  \label{eq:112}
 f(\alpha)= \sum_{n=0}^{\infty} c_n
h_n(\alpha)~
\end{equation}
are bounded as
\begin{equation}
  \label{eq:113}
 |f(\alpha)| \leq 1~.
\end{equation}
On the
other hand we have the associated Bargmann-Segal Hilbert space
\cite{barg2,seg,coh4} with the integration measure
\begin{equation}
  \label{eq:114}
d\tilde{\mu}(\alpha) = \frac{1}{\pi}d^2\alpha\,e^{ -|\alpha|^2}
\end{equation}
and the orthonormal basis of holomorphic functions
\begin{equation}
  \label{eq:115}
\tilde{h}_n(\alpha) = \frac{\alpha^n}{\sqrt{n!}}~,~n=0,1,\ldots ~.
\end{equation}
Here the reproducing kernel is given by
\begin{equation}
  \label{eq:116}
\tilde{\Delta}(z_2,z_1) = e^{ \bar{\alpha}_2\,\alpha_1} ~.
\end{equation}
It
has the property
\begin{equation}
  \label{eq:117}
 \int_{\mathbb{C}} d\tilde{\mu}(\alpha_2)\,
\tilde{\Delta}(\alpha_2,\alpha_1)\,\alpha_2^n = \alpha_1^n~.
\end{equation}
The growth of square-integrable holomorphic functions
\begin{equation}
  \label{eq:118}
\tilde{h}(\alpha) = \sum_{n=0}^{\infty} c_n \tilde{h}_n(\alpha)
\end{equation}
is restricted by
\begin{equation}
  \label{eq:119}
 |\tilde{h}(\alpha)|
\leq e^{ |\alpha|^2/2}~,
\end{equation}
i.e.\ is of order 2 and type 1/2. 

The above discussions about the different Hilbert space and their
associated reproducing kernels will be of special interest in
Ch.\ 7 when we discuss (quasi)-probability distributions
related to the different types of coherent states associated with
the Lie algebra  $\mathfrak{so}(1,2)$.
\subsection{Expectation values of quantum observables}
 The following expectation values are associated with the
states $|k,z\rangle$:
\begin{eqnarray} \label{eq:1528}\langle K_0\rangle_{k,z} &\equiv& \langle
 k,z|K_0|k,z\rangle=
k+|z|\, \rho_k(|z|)\,, \\ \label{eq:1062}&& \rho_k(|z|)=
\frac{I_{2k}(2|z|)}{I_{2k-1}(2|z|)} <1~,
\\ \label{eq:1063}\langle K_0^2\rangle_{k,z}&=& k^2+|z|^2+ |z|\,
\rho_k(|z|)~,\end{eqnarray} so that
\begin{equation}
  \label{eq:120}
 (\Delta
K_0)^2_{k,z}=|z|^2[1-\rho_k^2(|z|)]+(1-2k) |z|\,\rho_k(|z|)~.
\end{equation}
The inequality $\rho_k < 1$ in Eq.\ \ref{eq:1062} will be justified in
Appendix D.1. 

 For the number operator
\begin{equation}
  \label{eq:121}
 N = K_0 -k
\end{equation}
this
implies \begin{eqnarray}\label{eq:1529}\langle N\rangle_{k,z} &\equiv&
\bar{n}_{k,z} = |z|\rho_k(|z|) ~,\\ \label{eq:1064}\langle N^2 \rangle_{k,z}
 &= &
|z|^2 +(1-2k)|z|\,\rho_k~,\\ \label{eq:1065}(\Delta N)^2_{k,z}&=&
|z|^2(1-\rho^2_k) +(1-2k) |z|\,\rho_k~. \end{eqnarray}
 The quantity
\begin{equation}
  \label{eq:122}
 R = \frac{(\Delta n)^2 - \bar{n}}{\bar{n}^2}~,
\end{equation}
is
used in quantum optical discussions \cite{pau2} as a (rough)  measure for
deviations from Poisson statistics for which $R=0$. It is closely
related to Mandel's $Q$-parameter \cite{mand1}
\begin{equation}
  \label{eq:123}
 Q=\frac{(\Delta
n)^2 - \bar{n}}{\bar{n}} = \bar{n}\, R~.
\end{equation}
$R$ and $Q$ here have the
values
\begin{eqnarray}  \label{eq:1530}R_{k,z} &=& \frac{1}{\rho_k^2} -
\frac{2k}{|z|\,\rho_k}-1~, \\ \label{eq:1066}Q_{k,z}& =& |z|\,
(\frac{1}{\rho_k} -
\rho_k) -2k~. \end{eqnarray}

 It follows from the expansion \ref{eq:78}
that the probability
\begin{equation}
  \label{eq:124}
 w_{k,z\leftrightarrow n} \equiv w(|k,n
\rangle \leftrightarrow |k,z \rangle) = |\langle k,n|k,z \rangle
|^2
\end{equation}
is given by
\begin{equation}
  \label{eq:125}
 w_{k,z\leftrightarrow n} =
\frac{|z|^{2n}}{(2k)_n\,n!\,g_k(|z|^2)} ~.
\end{equation}
The following
limits are of interest: \\
 For $|z| \to 0$ one has
\begin{equation}
  \label{eq:126}
 \rho_k(|z|)
\to
\frac{|z|}{2\,k}\,\left(1-\frac{|z|^2}{2\,k\,(2k+1)}\right)~~\mbox{for}~|z|\to
0\,,
\end{equation}
and for very large $|z|$, the {\em correspondence limit\,}, we get
 from Eq.\ \ref{eq:1518} that
\begin{eqnarray} \label{eq:1531}\rho_k(|z|)&\asymp& 1 -\frac{4k-1}{4|z|}+
\frac{16\,(k^2-k)+3}{32|z|^2}+O(|z|^{-3})~, \\
\label{eq:1067}\rho_k^2(|z|)&\asymp& 1 -\frac{4k-1}{2|z|}+
\frac{8\,k^2-6\,k+1}{4\,|z|^2}+ O(|z|^{-3})~\\ \label{eq:1068}&& \mbox{ for }
 |z|
\to \infty~. \nonumber \end{eqnarray}
 
 This implies
 \begin{eqnarray}  \label{eq:1532}\langle K_0\rangle_{k,z}& \asymp &|z|
+\frac{1}{4}+O(|z|^{-1})~,\\ \label{eq:1069}(\Delta K_0)^2_{k,z} &\asymp&
\frac{1}{2}|z| + O(|z|^{-1})~, \\
\label{eq:1070}\bar{n}_{k,z}&\asymp& |z| +1/4 -k + O(|z|^{-1})~, \\
\label{eq:1071}(\Delta N)^2_{k,z}&\asymp&\frac{1}{2}|z| + O(|z|^{-1})\asymp
\frac{1}{2}\bar{n}_{k,z}~, \\ \label{eq:1072}R_{k,z} &\asymp & -
\frac{1}{2\,\bar{n}_{k,z}}+O(|z|^{-2}) ~, \\
\label{eq:1073}Q_{k,z} &\asymp & -\frac{1}{2}~,
\\\label{eq:1074}w_{k,z\leftrightarrow n} &\asymp &
2\sqrt{\pi}\frac{|z|^{2(k+n)-1/2}}{n!\,\Gamma(2k+n)}\,e^{-2|z|}\,
(1+O(|z|^{-1}))~.
\end{eqnarray}
The last relations show that that distribution does not become
Poisson-like in the classical limit but stays sub-Poisson-like,
even though $R_{k,z}$ tends to $0$ in this limit. 

 For $|z| \to 0$ the limiting behaviour of the expectation values 
\ref{eq:1528} etc.\ can be derived with the help of the
 relation \ref{eq:126}.

For $K_1$ and $K_2$ we have the following expectation values:
\begin{eqnarray} \label{eq:1533}\langle K_1\rangle_{k,z}
&=&\frac{1}{2}(\bar{z}+z)=|z|\,\cos\phi~,\\ \label{eq:1075}\langle
K_2\rangle_{k,z} &=&\frac{1}{2i}(\bar{z}-z)=-|z|\,\sin\phi~,\\
\label{eq:1076}\langle K_1^2\rangle_{k,z}&=& |z|^2\cos^2\phi
 +\frac{1}{2}\langle
K_0\rangle_{k,z}~, \\ \label{eq:1077}\langle K_2^2\rangle_{k,z} &=&
|z|^2\sin^2\phi +\frac{1}{2}\langle K_0\rangle_{k,z}~, \\ \label{eq:1078}
\langle
K_1^2+ K_2^2\rangle_{k,z} & =& \langle K_0^2 \rangle_{k,z}+l =
|z|^2 + \langle K_0\rangle_{k,z}\,,
\end{eqnarray}
\begin{eqnarray}
\label{eq:1079}(\Delta K_1)^2_{k,z}& =&(\Delta K_2)^2_{k,z}
 =\frac{1}{2}\langle
K_0\rangle_{k,z}~,\label{delK} \\
 \label{eq:1080}\langle
 K_1\,K_2+K_2\,K_1 \rangle_{k,z} &=&
 \frac{1}{2i}(\bar{z}^2-z^2) = -|z|^2\sin 2\phi~, \\
\label{eq:1081}\langle S(K_1,K_2)_{k,z}  \rangle_{k,z} &=& 0~,
\\\label{eq:1082}S(K_1,K_2)_{k,z} &=& \frac{1}{2}(K_1\,K_2 + K_2\,K_1)
-\langle K_1 \rangle_{k,z}\langle K_2\rangle_{k,z}\,, ~~~~~~ \\
 \label{eq:1083}\langle(K_0+k)^{-1} \rangle_{k,z} &=& |z|^{-1}\,\rho_k(|z|)~,
 \\\label{eq:1084}\langle a=(K_0+k)^{-1/2} K_- \rangle_{k,z} &=&z \,
 \langle(K_0+k)^{-1/2}
 \rangle_{k,z}~, \\ \label{eq:1085}\langle(K_0+k)^{-1/2} \rangle_{k,z} &=&
 \frac{1}{g_k(|z|^2)}
 \sum_{n=0}^{\infty}\frac{|z|^{2n}}{\sqrt{2k+n}\,(2k)_n\,n!} = \\
 \label{eq:1086}&=& \frac{1}{\sqrt{\pi}\,g_k(|z|^2)} \times \nonumber \\
&& \times \int_0^{\infty}
 dt\,t^{-1/2}\,e^{-2k\,t}
 g_k(|z|^2 e^{-t})\,,~~~ \label{eq:683} \\ \label{eq:1087}\langle(K_0+k)^{-1/2}
 \rangle_{k,z} &\to&
 |z|^{-1/2}\left[1-\frac{2k-1/4}{4|z|} + O(|z|^{-2})\right]~~~~~~ \\
 \label{eq:1088}&&
 \mbox{ for } |z| \to \infty~.  \nonumber
\end{eqnarray} In deriving the relations \ref{eq:1076} and \ref{eq:1077}
 the equality
$K_-K_+= K_+K_-+2K_0$ (Eq.\ \ref{eq:56}) has been used.

 The Eqs\ \ref{eq:1533} and \ref{eq:1075} show clearly that the expectation
values of $K_1$ and $K_2$ have the same form as their
classical counterparts 
\begin{equation}
  \label{eq:682}
  h_1 = I\,\cos\vp\,,~~~h_2 = -I\,\sin\vp\,,
\end{equation}
the difference being that $|z|$ and $\langle K_0\rangle_{k,z}$, see Eq.\ 
\ref{eq:1528}, 
 coincide only for very large $|z|$. 

But  the operators $K_1$ and $K_2$ do
measure the phase of $z$ for {\em all}  values of $|z|$ in
 a straightforward way:
\begin{equation}
  \label{eq:128}
 \tan\phi =
- \frac{\langle K_2\rangle_{k,z}}{\langle K_1\rangle_{k,z}}~.
\end{equation}
Notice also that the expectation values \ref{eq:1533} and \ref{eq:1075} 
of $K_1$ and $K_2$ do not
depend on the index $k$ at all, contrary to that of $K_0$.

We see  that we may identify the complex number $z$ as
\begin{equation}
  \label{eq:129}
 z= h_1-i\,h_2~,
\end{equation}
where $h_1$ and $h_2$ are the classical versions \ref{eq:682} of the
quantum observables $K_1$ and $K_2$.

The situation is analogous to
Schr\"{o}dinger-Glauber coherent states $|\alpha \rangle$ where
the complex number $\alpha$ is interpreted as
$(q+i\,p)/\sqrt{2}\,$. A difference is, however, that in our case
we have a third observable $K_0$ - or $h_0$, respectively - the
role of which is less trivial than that of the identity operator
in the usual quantum mechanical descriptions. 

 Eq.\ \ref{eq:1078} shows how
the deviations from the Pythagorean relation
\ref{eq:58} are determined by the expectation value of $K_0$! 

The relations \ref{delK} imply  that  the general inequality
\begin{equation}
  \label{eq:130}
 (\Delta
K_1)^2 (\Delta K_2)^2 \geq \frac{1}{4}|\langle K_0 \rangle|^2~.
\end{equation}
becomes an equality here, i.e. the states $|k,z\rangle$ are ``
minimal uncertainty'' states, but they are not squeezed (see Ch.\
6 for details, also as to the quantity $S(K_1,K_2)_{k,z}$). 

In turning the series \ref{eq:1085} into an integral the relation
\begin{equation}
  \label{eq:131}
(2k+n)^{-1/2} =
\frac{1}{\sqrt{\pi}}\int_0^{\infty}dt\,t^{-1/2}\,e^{-(2k+n)\,t}
\end{equation}
has been used. The asymptotic behaviour of the integral \ref{eq:683} will
be justified in Appendix D.2.

As the Barut-Girardello coherent states $|k,z\rangle$ are not
squeezed the interest in them grew only slowly - compared to the
Perelomov coherent states discussed below - within the quantum
optical community
\cite{agar1,buz1,agar2,gerr1,ban0,trif1,agar3,ban1,
brif1,agar4,gerr2a,gerr2,brif3,brif4,brif5,trif2,pera1,
wan1,peri1}. It was realized that an appropriate dynamics could
evolve Barut-Girardello states into squeezed states
\cite{buz1,peri1} and that a superposition of 2 of them can be
squeezed, too \cite{ban1,gerr2a}.
\section{Perelomov coherent states}
\subsection{General Properties}  The Perelomov coherent states as
defined by Eq.\ \ref{eq:1037} have always been introduced in a different
way \cite{per0}. I shall discuss the relationship between the two
approaches below. Let me first turn to the implications of the
definition \ref{eq:1037}: 

 Multiplying  Eq.\ \ref{eq:1037} from the left with
$\langle k,n|$ and using the relations \ref{k3} and \ref{erz}
yields the recursion formula
\begin{equation}
  \label{eq:132}
 \langle k, n+1|k,\lambda\rangle =
\left (\frac{2k+n}{n+1}\right)^{1/2}\,\lambda\,\langle
k,n|k,\lambda \rangle~,
\end{equation}
which implies
\begin{equation}
  \label{eq:133}
 \langle
k,n|k,\lambda \rangle = \left (\frac{(2k)_n}{n!}\right
)^{1/2}\,\lambda^n\,\langle k,n=0|k,\lambda \rangle.
\end{equation}
Using the
relation \cite{er3,span1}
\begin{equation}
  \label{eq:134}
  (2k)_n =
\frac{\Gamma(2k+n)}{\Gamma(2k)}=(-1)^n\,n!\,{-2k \choose n}
\end{equation}
we obtain  a summable binomial power series:
\begin{eqnarray} \label{eq:1534}\langle k,\lambda |k, \lambda \rangle &=&
 \sum_{n=0}^{\infty}\langle k, \lambda|k,n\rangle \langle k,n|k,
 \lambda \rangle \nonumber \\ \label{eq:1089}&=&|\langle
k,n=0|k,\lambda \rangle|^2\,\sum_{n=0}^{\infty} {-2k \choose
n}\,(-|\lambda|^2)^n \\ \label{eq:1090}&=& |\langle k,n=0|k,\lambda \rangle|^2
\,(1-|\lambda|^2)^{-2k}~. \nonumber \end{eqnarray} The series \ref{eq:1089}
converges only for $|\lambda|^2 <1\,$.

 From $\langle k, \lambda|k,
\lambda \rangle = 1$ we get
\begin{equation}
  \label{eq:135}
 |\langle k,n=0|k,\lambda \rangle|
= (1-|\lambda|^2)^k ~.
\end{equation}
Choosing the phase of $ \langle
k,n=0|k,\lambda \rangle $ to be zero finally gives
\begin{equation}
  \label{eq:136}
 |k,\lambda
\rangle = (1-|\lambda|^2)^k\, \sum_{n=0}^{\infty} \left (
\frac{(2k)_n}{n!}\right )^{1/2}\,\lambda^n\,|k,n\rangle\,.
\end{equation}

Since their introduction \cite{per0} the existing definitions of
the states \ref{eq:136} are as follows: 

 From the Lie algebra \ref{eq:55} it
follows that \cite{hel0,gil,per1,per}
\begin{equation}
  \label{eq:137}
 U(w) \equiv e^{  (w/2)\,K_+
-(\bar{w}/2)\,K_-} = e^{ \lambda\,K_+}\,e^{
\ln(1-|\lambda|^2)\,K_0}\,e^{ -\bar{\lambda}\,K_-}\,,
\end{equation}
where
\begin{equation}
  \label{eq:138}
w = |w|\,e^{i\,\theta} \in \mathbb{C}~,~ \lambda =
\tanh(|w|/2)\,e^{i\,\theta}\,.
\end{equation}
Notice
that the complex numbers $w$ and $\lambda$ have the same phase! 

For later convenience I collect some elementary relations:
\begin{eqnarray} \label{eq:1535}|w| &=& \ln
\left(\frac{1+|\lambda|}{1-|\lambda|}\right)~,\\ \label{eq:1091}1-|\lambda|^2
 &=&
\frac{1}{\cosh^2(|w|/2)}~, \\ \label{eq:1092}1+|\lambda|^2 &=&
 \frac{\cosh|w|}{\cosh^2(|w|/2)}~, \\
\label{eq:1093}\frac{|\lambda|}{1-|\lambda|^2} &=& \frac{1}{2} \sinh |w| \,.
\end{eqnarray}
Applying the operator relation \ref{eq:137} to the ground state $|k, n=0
\rangle$ and using the Eqs.\ \ref{vern} and \ref{eq:59} then gives
\begin{equation}
  \label{eq:139}
 e^{
(w/2)\,K_+ -(\bar{w}/2)\,K_-}\,|k, n=0 \rangle = |k,\lambda
\rangle \,.
\end{equation}
That this state is an eigenstate of
$(K_0+k)^{-1}K_-$ has been noticed before \cite{wue1,wan}
(as to the general algebraic structure see also the Refs.\
\cite{sud,vog}). When taking the first of the relations in Eq.\ \ref{eq:65}
 into account, the special case $k=1/2$ has been discussed
previously, too, as an eigenstate of the operator $E_-$ from Eq.\ \ref{eq:17},
 without emphazising its property as a Perelomov
coherent state  \cite{lern2,lebl,sha,vour1}.

 In this paper I
use the property \ref{eq:1037}  as the defining one! 

The definition \ref{eq:139} was motivated by the property of the
conventional coherent states \ref{eq:1527} to be definable as
\begin{equation}
  \label{eq:140}
|\alpha\rangle = D(\alpha)|0\rangle~,~D(\alpha) = e^{\alpha\,a^+
-\bar{\alpha}\,a}=
e^{-|\alpha|^2/2}\,e^{\alpha\,a^+}\,e^{-\bar{\alpha}\,a}\,.
\end{equation}

The states $|k, \lambda \rangle $ have
properties similar to those of the states $|k, z \rangle $ from above: \\
They are not orthogonal,
\begin{equation}
  \label{eq:141}
 \langle k, \lambda_2 | k, \lambda_1
\rangle = (1-|\lambda_1|^2)^k\,
(1-|\lambda_2|^2)^k\,(1-\bar{\lambda}_2\, \lambda_1)^{-2\,k}\,,
\end{equation}
they are, however,  complete in the following sense: From
\begin{equation}
  \label{eq:142}
 \int
d\mu_k(\lambda)\, \langle k,n|k, \lambda \rangle \langle k,
\lambda |k,n \rangle = \frac{(2k)_n}{n!} \int d\mu_k(\lambda)
(1-|\lambda|^2)^{2\,k}\, |\lambda|^{2\,n}\,,
\end{equation}
where
\begin{equation}
  \label{eq:143}
d\mu_k(\lambda) = \frac{2k-1}{\pi}(1-|\lambda|^2)^{-2}
|\lambda|\,d|\lambda| \, d\theta \,,
\end{equation}
and \cite{er1a}
\begin{equation}
  \label{eq:144}
\int_0^1 d\,x (1-x)^{2k-2}\,x^n = \mbox{B}(n+1, 2k-1) =
\frac{n!\,\Gamma(2k-1)}{\Gamma(2k+n) }\,,
\end{equation}
we have
\begin{equation}
  \label{eq:145}
\int_{\mathbb{D}} d\,\mu_k(\lambda)\, \langle k,n_2|k, \lambda
\rangle \langle k, \lambda |k,n_1 \rangle = \delta_{n_2n_1} \,,
\end{equation}
where in addition
\begin{equation}
  \label{eq:146}
 \int_0^{2\pi} d\theta e^{i\,(n_1-n_2)} = 0~~
\mbox{ for } n_2 \neq n_1~
\end{equation}
has been used. 

 It follows from
Eq.\ \ref{eq:144} that the integral \ref{eq:142} exists for $k=1/2$,
 too, because
$(2k-1)\,\Gamma(2k-1) = \Gamma(2k)\,$!

In terms of the variable $w$ from Eq.\ \ref{eq:138} the measure \ref{eq:143}
has the form
\begin{equation}
  \label{eq:127}
  d\mu_k = \frac{2k-1}{4\pi}\sinh|w|\,d|w|d\theta\,.
\end{equation}

 Formally we may write the completeness relation \ref{eq:145} as
\begin{equation}
  \label{com2}
\int_{\mathbb{D}} d\,\mu_k(\lambda)\, |\lambda \rangle \langle
\lambda | = \boldsymbol{1}.
\end{equation}
It is obviously
advantageous to introduce the measure
\begin{equation}
  \label{eq:148}
 d\tilde{\mu}_k(\lambda)
= (1-|\lambda|^2)^{2k}\,d \mu_k(\lambda)\,,
\end{equation}
because (see, e.g.\
\cite{bar1,bo})
\begin{equation}
  \label{eq:149}
 (\lambda^{n_2}, \lambda^{n_1})_{\mathbb{D},k}
=\int_{\mathbb{D}}d\tilde{ \mu}_k(\lambda)
\,\bar{\lambda}^{n_2}\,\lambda^{n_1} =
\frac{n_1!}{(2k)_{n_1}}\,\delta_{n_2\,n_1}\,.
\end{equation}
The integral \ref{eq:149} provides the well-known scalar product for a
Hilbert space of holomorphic functions on the unit disc
$\mathbb{D}$, a so-called Hardy space \cite{shiop}. 

 We have the
orthonormal basis
\begin{equation}
  \label{eq:150}
 e_{k,n}(\lambda) =
\sqrt{\frac{(2k)_n}{n!}}\,\lambda^n~
\end{equation}
The role of the usual
$\delta$-function here plays the ``reproducing kernel''
\cite{aro,barg2,coh4,neeb}, too (see subsection 3.1.2):
\begin{equation}
  \label{eq:151}
\Delta_k(\bar{\lambda}_2, \lambda_1) = \sum_{n=0}^{\infty}
\bar{e}_{k,n}(\lambda_2)\,e_{k,n}(\lambda_1) =
(1-\bar{\lambda}_2\,\lambda_1)^{-2\,k}\,,
\end{equation}
with
\begin{equation}
  \label{eq:152}
\int_{\mathbb{D}} d\tilde{\mu}_k(\lambda_2)\,\Delta_k(\bar{\lambda}_2,
\lambda_1)\,\lambda_2^n = \lambda_1^n~.
\end{equation}
In summing the series
\ref{eq:151}  the relation \ref{eq:134} has been used.

 For a given unitary
representation the scalar product between the normalized coherent
states $|k,z\rangle\,$, Eq.\ \ref{eq:78}, and $|k,\lambda\rangle$, Eq.\
\ref{eq:136}, is given by
\begin{equation}
  \label{eq:153}
 \langle k,\lambda|k,z \rangle =
\frac{(1-|\lambda|^2)^k}{
\sqrt{g_k(|z|^2)}}\,e^{\bar{\lambda}\,z}~.
\end{equation}
With
$z=|z|\exp(i\,\phi)$ and $\lambda =|\lambda|\exp(i\,\theta)$ we
have
\begin{equation}
  \label{eq:154}
 \langle k, \lambda|k,z \rangle =
\frac{(1-|\lambda|^2)^k}{
\sqrt{g_k(|z|^2)}}\,e^{|\lambda||z|e^{i\,(\phi-\theta)}}~.
\end{equation}
This shows: The scalar product \ref{eq:154} decreases for (the
classical limit, see below) $|\lambda| \to 1$ and for (the
classical limit) $|z| \to \infty$ (because $|\lambda| < 1$ and
because of the asymptotic behaviour \ref{eq:1519}). The decrease is
maximal for $\phi-\theta = \pm \pi$ and minimal for $\phi =
\theta$.

From  the completeness relations \ref{com1} and \ref{com2} we get
the integral transforms
\begin{equation}
  \label{eq:155}
 |k,z\rangle =
\int_{\mathbb{D}}d\mu_k(\lambda) \langle
k,\lambda|k,z\rangle\,|k,\lambda \rangle\,,~~ |k,\lambda\rangle =
\int_{\mathbb{C}}d\mu_k(z) \langle k,z|k, \lambda \rangle\,|k,z
\rangle\,.~~
\end{equation}
With the kernel \ref{eq:153} these become 
\begin{eqnarray} \label{eq:1536}|k,z \rangle &=&
\frac{1}{\sqrt{g_k(|z|^2)}}\sum_{n=0}^{\infty}
\frac{z^n}{\sqrt{(2k)_n\,n!}}\,|k,n\rangle = \\ \label{eq:1094}&=&
\int_{\mathbb{D}}d\mu_k(\lambda)
\frac{(1-|\lambda|^2)^k\,e^{\bar{\lambda}\,z}}{
\sqrt{g_k(|z|^2)}}\,|k,\lambda\rangle =  \\
\label{eq:1095}&=&\frac{1}{\sqrt{g_k(|z|^2)}}
\int_{\mathbb{D}}d\tilde{\mu}_k(\lambda)
\,e^{\bar{\lambda}\,z}\sum_{n=0}^{\infty}
\left(\frac{(2k)_n}{n!}\right)^{1/2}\lambda^n\,|k,n\rangle\,;  \\
\label{eq:1096}|k,\lambda\rangle &=&  (1-|\lambda|^2)^k \sum_{n=0}^{\infty}
\left(\frac{(2k)_n}{n!}\right)^{1/2}\lambda^n\,|k,n\rangle = \\
\label{eq:1097}&=&\int_{\mathbb{C}}d\mu_k(z)
\frac{(1-|\lambda|^2)^k\,e^{\bar{z}\,\lambda}}{
\sqrt{g_k(|z|^2)}}\,|k,z\rangle =  \\
\label{eq:1098}&=&(1-|\lambda|^2)^k \int_{\mathbb{C}}d\tilde{\mu}_k(z)
\,e^{\bar{z}\,\lambda}\sum_{n=0}^{\infty}\frac{z^n}{\sqrt{(2k)_n\,n!}}
|k,n\rangle\,.  \end{eqnarray} One immediately can read
off the following unitary mappings of the bases \ref{eq:82} and \ref{eq:150}:
\begin{eqnarray} \label{eq:1537}\tilde{f}_{k,n}(z) &=&
 \int_{\mathbb{D}}d\tilde{\mu}_k(\lambda)
\,e^{\bar{\lambda}\,z}\,e_{k,n}(\lambda)~,~ n=0,1,\ldots\,, \\
\label{eq:1099}e_{k,n}(\lambda) &=&\int_{\mathbb{C}}d\tilde{\mu}_k(z)
\,e^{\bar{z}\,\lambda}\,\tilde{f}_{k,n}(z)~. \end{eqnarray} Notice
that the kernels $\exp(\bar{\lambda}\,z)$ and $
\exp(\bar{z}\,\lambda)$ do not depend on the index $k\,$. 

 For a discussion of
related integral transforms see Ref.\ \cite{basu}

Multiplying the relations \ref{eq:1536} and \ref{eq:1096} from the
 left with the states
$\langle k, z_2|$ and $\langle k, \lambda_2|$, respectively, yields
 the following useful integral transforms

\begin{eqnarray} \label{eq:1538}g_k(\bar{z}_2\,z_1) &=&
\int_{\mathbb{D}}d\tilde{\mu}_k(\lambda)\, e^{\bar{z}_2\,\lambda +
\bar{\lambda}\,z_1}~, \\ \label{eq:1101}(1-\bar{\lambda}_2\,\lambda_1)^{-2k}
 &=&
\int_{\mathbb{C}}d\tilde{\mu}_k(z)\,e^{\bar{\lambda}_2\,z +
\bar{z}\,\lambda_1} ~.
\end{eqnarray}
Differentiating these relations with respect to $\bar{z}_2$ etc.\
generates new relations, e.g.
\begin{equation}
  \label{eq:156}
\frac{dg_k(\bar{z}_2\,z_1)}{d\bar{z}_2} =
\frac{z_1}{2k}\,g_{k+1/2}(\bar{z}_2\,z_1)
=\int_{\mathbb{D}}d\tilde{\mu}_k(\lambda)\, \lambda \,
e^{\bar{z}_2\,\lambda + \bar{\lambda}\,z_1}\,,
\end{equation}
where the
relation $(2k)_{n+1} = 2k\,(2k+1)_n $ has been used \cite{span2}.

 The Hilbert spaces with the scalar product \ref{eq:149} ``carry''
irreducible unitary representations of $SU(1,1)$ etc.

 The generators
of the Lie algebra are \cite{bar1,bo}
\begin{equation}
  \label{eq:157}
 K_+ = 2k\,\lambda +
\lambda^2\frac{d}{d\lambda}~,~K_- = \frac{d}{d\lambda}~,~K_0 =
\lambda\frac{d}{d\lambda}+k~.
\end{equation}
The basis functions \ref{eq:150} are the
eigenfunctions of $K_0$. The - unnormalized - eigenfunctions of
$K_-$ are
\begin{equation}
  \label{eq:158}
 f_{k,z}(\lambda)  = C_{k,z} \,e^{z\,\lambda}~,~K_-
f_{k,z}=z\,f_{k,z}~.
\end{equation}
As \begin{eqnarray} \label{eq:1539}(f_{k,z},f_{k,z}) &=&
|C_{k,z}|^2 \int_{\mathbb{D}}d\tilde{\mu}_k(\lambda)\,
|e^{z\,\lambda}|^2 = \\ \label{eq:1102}&=& |C_{k,z}|^2\,g_k(|z|^2)
\int_{\mathbb{D}}d\mu_k(\lambda)\langle z|\lambda\rangle
\langle\lambda|z \rangle = |C_{k,z}|^2\,g_k(|z|^2)~,\nonumber
\end{eqnarray} we have the normalized eigenfunctions
\begin{equation}
  \label{eq:159}
 f_{k,z}(\lambda) =
\frac{1}{\sqrt{g_k(|z|^2)}}\,e^{z\,\lambda}~.
\end{equation}
(Notice that
the integration over $\theta$ in Eq.\ \ref{eq:1539} yields the same result
for the exponents
$z\,\bar{\lambda}+\bar{z}\,\lambda=2\,|z||\lambda|\,\cos(\theta-\phi)$
and $z\,\lambda + \bar{z}\,\bar{\lambda}=2\,|z||\lambda|\,\cos(\theta+\phi)$). 

The unnormalized eigenfunctions of $K_1$ and $K_2$ are
\begin{eqnarray} \label{eq:1540}f_{k,h_1}(\lambda) &=&
C_{k,h_1}\,\left(\frac{1+i\lambda}{1-i\lambda}\right)^{-i\,h_1}(1+
\lambda^2)^{-k}~,
\\ \label{eq:1103}K_1f_{k,h_1}(\lambda)&=& h_1\,f_{k,h_1}(\lambda)~,~ h_1 \in
\mathbb{R}~; \\
\label{eq:1104}f_{k,h_2}(\lambda) &=&
C_{k,h_2}\,\left(\frac{1+\lambda}{1-\lambda}\right)^{-i\,h_2}(1-
\lambda^2)^{-k}~,
\\ \label{eq:1105}K_2f_{k,h_2}(\lambda)&=& h_2\,f_{k,h_2}(\lambda)~,~ h_2 \in
\mathbb{R}~, \end{eqnarray}
where $C_{k,h_1}$ and $C_{k,h_2}$ are (normalization) constants.
\subsection{Expectation values of quantum observables}
Next we come to the expectation values and the fluctuations of the
operators $K_j$ with respect to the states \ref{eq:136}. We have
\cite{wod,buz1,trif1} (using the relations \ref{eq:1535}-\ref{eq:1093}):
\begin{eqnarray} \label{eq:1541}\langle K_0 \rangle_{k, \lambda} &\equiv&
 \langle
k, \lambda|K_0| k, \lambda \rangle =
k\,\frac{1+|\lambda|^2}{1-|\lambda|^2}= k\, \cosh|w|~, \\
\label{eq:1106}\langle K_0^2 \rangle_{k, \lambda}&=&
k^2\,\frac{(1+|\lambda|^2)^2}{(1-|\lambda|^2)^2} + 2\,k\,
\frac{|\lambda|^2}{(1-|\lambda|^2)^2} = \\ \label{eq:1107}& =&
k^2\,\cosh^2|w| + \frac{k}{2}\, \sinh^2|w|~, \nonumber \\
\label{eq:1108}(\Delta K_0)^2_{k, \lambda} &=& 2\,k
\,\frac{|\lambda|^2}{(1-|\lambda|^2)^2}= \frac{k}{2}\,
\sinh^2|w|~, \nonumber \\ &=& \frac{1}{2}(\frac{\langle K_0 \rangle_{k,
 \lambda}^2}{k}
-k)\,. \nonumber
\end{eqnarray} Here $|\lambda| = \tanh|w/2|$ (see Eq.\ \ref{eq:138}). 

For the number operator \ref{eq:121}, the  parameter $R$ defined in Eq.\ 
\ref{eq:122}
and Mandel's Q-parameter \ref{eq:123} we get
\begin{eqnarray} \label{eq:1542}\langle N \rangle_{k,\lambda}&\equiv
 &\bar{n}_{k,\lambda} =
k\,(\cosh|w|-1)~, \\
\label{eq:1109}(\Delta N)^2_{k,\lambda} &=& \frac{k}{2}\,\sinh^2|w| =
\bar{n}_{k,\lambda}\,(1+\bar{n}_{k,\lambda}/2k)~, \\
\label{eq:1110}R_{k,\lambda} &=& \frac{1}{2k}~, \\ 
\label{eq:1111}Q_{k,\lambda} &=&
\frac{\bar{n}_{k,\lambda}}{2k}~. \end{eqnarray} The probability
\begin{equation}
  \label{eq:160}
 w_{k,\lambda \leftrightarrow n} \equiv w(|k,n\rangle
\leftrightarrow |k,\lambda \rangle ) = |\langle k,n|k, \lambda
\rangle |^2
\end{equation}
is given by
\begin{equation}
  \label{eq:161}
 w_{k,\lambda \leftrightarrow n} =
(1-|\lambda|^2)^{2k}\, \frac{(2k)_n\,|\lambda|^{2n}}{n!}~.
\end{equation}
For
$k= 1/2$ this distribution represents the Bose - statistics
\cite{agar4a,vour1} of free quanta with energies $E_{\nu}\,, \nu =
0,1,\ldots ,$ in a heat bath of (inverse) temperature $\beta =
1/k_BT$ and chemical potential $\mu$:
\begin{equation}
  \label{eq:162}
 w_{1/2,\lambda
\leftrightarrow n} = (1-|\lambda|^2)\,
|\lambda|^{2n}~,\,|\lambda|^2 = e^{-\beta\,(E_{\nu}-\mu)}
\end{equation}
is
the probability  to find $n$ quanta in a state with energy
$E_{\nu}$.

 For the expectation
values and fluctuations of $K_1$ and $K_2$ we get:
\begin{eqnarray} \label{eq:1543}\langle K_1 \rangle_{k, \lambda}& = &
2\,k\, \frac{|\lambda|}{1-|\lambda|^2}\,\cos\theta  =
k\,\sinh|w| \,\cos\theta ~,
\\ \label{eq:1112}&=& \sqrt{2\,k}\, (\Delta\,K_0)_{k,\lambda}
\,\cos\theta = \frac{2\,|\lambda|}{1+|\lambda|^2} \langle K_0 \rangle_{k,
 \lambda} \cos \theta  ~,
 \nonumber \\
\label{eq:1113}\langle K_2 \rangle_{k, \lambda} & =& - 2\,k\, 
\frac{|\lambda|}{1-|\lambda|^2}\,\sin\theta  = - k \,\sinh|w|\,\sin\theta~,
~~~~~\\
\label{eq:1114}&=& - \sqrt{2\,k}\, (\Delta\,K_0)_{k,\lambda}\,
\sin\theta~, \nonumber \\
\label{eq:1115}\langle K_1^2 \rangle_{k, \lambda} &=& 4\,k^2\,\cos^2\theta\,
\frac{|\lambda|^2}{(1-|\lambda|^2)^2}+\frac{k}{2}
\frac{|1+\lambda^2|^2}{(1-|\lambda|^2)^2}~,
\\ \label{eq:1116}&& |1+\lambda^2|^2 = 1 + 2\,|\lambda|^2\,\cos2\theta +
|\lambda|^4~, \\ \label{eq:1117}(\Delta K_1)^2_{k, \lambda} &=&
\frac{k}{2}\,\frac{|1+\lambda^2|^2}{(1-|\lambda|^2)^2}~, \\
 \label{eq:1118}\langle K_2^2 \rangle_{k, \lambda} & =&
4\,k^2\,\sin^2\theta\,
\frac{|\lambda|^2}{(1-|\lambda|^2)^2}+\frac{k}{2}
\frac{|1-\lambda^2|^2}{(1-|\lambda|^2)^2}~,
\\ \label{eq:1119}&& |1-\lambda^2|^2 = 1 - 2\,|\lambda|^2\,\cos2\theta +
|\lambda|^4~, \\ \label{eq:1120}(\Delta K_2)^2_{k, \lambda} &=&
\frac{k}{2}\,\frac{|1-\lambda^2|^2}{(1-|\lambda|^2)^2}~, \\
\label{eq:1121}(\Delta K_1)^2_{k, \lambda}\,(\Delta K_2)^2_{k, \lambda} & = &
      \frac{k^2}{4}\, \frac{|1+\lambda^2|^2\,|1-\lambda^2|^2}{
      (1-|\lambda|^2)^4}\, \ge
      \frac{1}{4}\,|\langle K_0\rangle_{k, \lambda}|^2 = \\ \label{eq:1122}&=&
      \frac{k^2}{4}\,\frac{(1+|\lambda|^2)^2}{(1-|\lambda|^2)^2}~,
      \nonumber \\ \langle K_1 \rangle^2_{k, \lambda} +
 \langle K_2 \rangle^2_{k,
       \lambda}&=& \langle K_0\rangle^2_{k, \lambda} -k^2\,, \label{eq:686} \\
 \label{eq:1123}(\Delta K_1)^2_{k, \lambda} +
 (\Delta K_2)^2_{k,
       \lambda}&=& (\Delta K_0)^2_{k, \lambda}+k~. \end{eqnarray}

Here again - like in the cases \ref{eq:1533} and
 \ref{eq:1075} - the expectation
values \ref{eq:1543} and \ref{eq:1113} have the simple structure
\begin{equation}
  \label{eq:798}
  r\,\cos\theta\,,~~-r\,\sin\theta\,,~~r=k\,\sinh|w|\,.
\end{equation}
The modulus $r$ approaches the expectation
 value $\langle K_0\rangle_{k,\lambda}\,$,\, \ref{eq:1541}\,, for large $|w|$.
 The Eqs.\ \ref{eq:1543} and \ref{eq:1113}
show again how the operators
       $K_1$ and $K_2$ ``measure'' the phase of the complex
       numbers $\lambda$ or $w$. We have
\begin{equation}
  \label{eq:171}
       \tan\theta= - \frac{\langle K_2 \rangle_{k, \lambda}}{
       \langle K_1 \rangle_{k, \lambda}}~.
\end{equation}

      Sometimes the following relations are useful:
      \begin{eqnarray}\label{eq:1544}\langle K_+^2 \rangle_{k, \lambda} & =&
      2k\,(2k+1)\,\frac{\bar{\lambda}^2}{(1-|\lambda|^2)^2}~,\\
      \label{eq:1124}\langle K_+K_- \rangle_{k, \lambda} &
      =&2k\,|\lambda|^2\,\frac{2\,k+|\lambda|^2}{(1-|\lambda|^2)^2}~,\\
      \label{eq:1125}\langle S_{k,\lambda}(K_1,K_2) \rangle_{k, \lambda} & =&
      \frac{k}{2i}\,\frac{\bar{\lambda}^2-\lambda^2}{(1-|\lambda|^2)^2}
=-k\frac{|\lambda|^2\sin 2\theta}{(1-|\lambda|^2)^2}~,\\
      \label{eq:1126}S_{k,\lambda}(K_1,K_2)&=& \frac{1}{2}(K_1K_2+K_2K_1)-
\langle K_1 \rangle_{k,
      \lambda}\,\langle K_2 \rangle_{k, \lambda}~. \end{eqnarray}
      We note the equality
\begin{equation}
  \label{eq:163}
 (\Delta K_1)^2_{k, \lambda}\,(\Delta K_2)^2_{k,
      \lambda} = \frac{1}{4}\,|\langle K_0\rangle_{k, \lambda}|^2
      +|\langle S_{k,\lambda}(K_1,K_2) \rangle_{k, \lambda}|^2~,
\end{equation}
and shall discuss the background of this relation in Ch.\ 6 in more
      detail.

      As to the fluctuations  the following  $\theta$-values
      are of special interest:
       \begin{eqnarray}\label{eq:1545}(\Delta K_1)^2_{k, \lambda}(|\lambda| >0,
 \theta = 0,\pi)
       &=&
       \frac{k}{2}\,\frac{(1+|\lambda|^2)^2}{(1-|\lambda|^2)^2} >
       \\ \label{eq:1127} > (\Delta K_2)^2_{k, \lambda}(\theta = 0,\pi)& =&
       \frac{k}{2} 
        < \frac{1}{2}\,\langle K_0\rangle_{k, \lambda}\,, \\
       \label{eq:1129}(\Delta K_2)^2_{k, \lambda}(|\lambda| > 0, \theta =
       -\pi/2\,,\pi/2)
&=& \frac{k}{2}\,\frac{(1+|\lambda|^2)^2}{(1-|\lambda|^2)^2} >
       \\ \label{eq:1130} > (\Delta K_1)^2_{k, \lambda}(\theta =
 -\pi/2\,,\pi/2) &=&
       \frac{k}{2}  
       < \frac{1}{2}\,\langle K_0\rangle_{k, \lambda}
        \,.  \end{eqnarray} For all 4 cases we
       get the equality sign in the ``uncertainty'' inequality \ref{eq:130}.
The same follows from Eq.\ \ref{eq:163} because the correlations \ref{eq:1125}
vanish for those $\theta$-values.

       In addition we have squeezing - for $|\lambda| > 0$ - in
       these cases because of the last inequalities in the relations 
\ref{eq:1127} and \ref{eq:1130}!
Squeezing will be discussed in more detail in Ch.\ 6. 

Because of the squeezing properties the Perelomov coherent states attracted
the attention of the quantum optics community earlier than the Barut-Girardello
ones. Here is a selection of the early papers:
 \cite{mil,fish,wod,schu,ger1,buz1}
 
 For the
expectation value of the annihilation operator $a= (K_0
+k)^{-1/2}K_-$ we get \begin{eqnarray}  \label{eq:1546}\langle a= (K_0
+k)^{-1/2}K_- \rangle_{k,\lambda} &=& \lambda\, \langle
(K_0+k)^{1/2}\rangle_{k,\lambda}~, \\
\label{eq:1132}\langle (K_0+k)^{1/2}\rangle_{k,\lambda}&=&
(1-|\lambda|^2)^{2k}\,\sum_{n=0}^{\infty}\frac{
(2k)_n\,(2k+n)^{1/2}}{n!}|\lambda|^{2\,n}~. ~~~~ \nonumber
\end{eqnarray} Observing that $(2k+n)^{1/2} =(2k+n)(2k+n)^{-1/2}$
and again using the relation \ref{eq:131} gives for the sum in Eq.\
 \ref{eq:1546}
\begin{eqnarray} \label{eq:1547}\sum_{n=0}^{\infty}\frac{
(2k)_n\,(2k+n)^{1/2}}{n!}|\lambda|^{2\,n}&=&
|\lambda|^2\,\frac{du_{2k}(|\lambda|^2)}{d\,|\lambda|^2} +
2k\,u_{2k}(|\lambda|^2)~,~~~~ \\ \label{eq:1133}u_{2k}(|\lambda|^2) &=&
\frac{1}{\sqrt{\pi}}\int_0^{\infty}dt\,
t^{-1/2}(e^t-|\lambda|^2)^{-2k}~. \end{eqnarray} Notice that
\begin{equation}
  \label{eq:164}
\frac{d^m\,u_{2k}}{d(|\lambda|^2)^m} =
(2k)_m\,u_{2k+m}~,~m=1,2,\ldots~.
\end{equation}
As $2k$ in general is a
positive integer we may generate the necessary $u_{2k}$ from
\begin{equation}
  \label{eq:165}
u_1(|\lambda|^2) = \Phi(|\lambda|^2,1/2,1) =
(|\lambda|^2)^{-1}\,F(|\lambda|^2,1/2) =
(|\lambda|^2)^{-1}\,\mbox{Li}_{\frac{1}{2}}(|\lambda|^2)\,,
\end{equation}
where
\begin{equation}
  \label{eq:166}
 \Phi(z,s,a) = \sum_{n=0}^{\infty}\frac{z^n}{(a+n)^s}
\end{equation}
is Lerch's function \cite{er1b} and
\begin{equation}
  \label{eq:167}
 \mbox{Li}_s(z) = F(z,s) = z\,\Phi(z,s,a=1)=
\sum_{n=1}^{\infty} \frac{z^n}{n^s}
\end{equation}
 the so-called
``polylogarithm'' of index $s$ \cite{lew}. 

 Expectation values of the type 
\begin{eqnarray} \label{eq:1548}\langle (K_0+b)^{-1}
\rangle_{k,\lambda} &=& s_{k}^{(b)}(|\lambda|^2) =(1-|\lambda|^2)^{2k} \,
v_{2k,b}(|\lambda|^2)~,
\\ \label{eq:1134}v_{2k,b}(|\lambda|^2)&=& \sum_{n=0}^{\infty}
\frac{(2k)_n}{(k+b+n)\,n!}\,|\lambda|^{2\,n}~,~b \geq 0\,,
\end{eqnarray} may be calculated in a similar fashion:

 With 
\begin{equation}
  \label{eq:168}
 \frac{1}{k+b+n} = \int_0^1 du\,u^{k+b+n-1}
\end{equation}
we get
\begin{equation}
  \label{eq:169}
v_{2k,b}(|\lambda|^2)  =
\frac{1}{|\lambda|^{2(k+b)}}\,\int_0^{|\lambda|^2}dx\,x^{k+b-1}(1-x)^{-2k}
~.
\end{equation}
The sum $v_{2k,b}(|\lambda|^2) $ may be expressed by a
hypergeometric function \cite{er3a}:  $\,
_{2}F_1(2k,k+b;k+b+1;|\lambda|^2)/(k+b)\,$, but in practice it
will be more convenient to (partially) integrate:
\begin{eqnarray} \label{eq:1549}\hat{v}_{2k,b}(|\lambda|^2) &\equiv&
\int_0^{|\lambda|^2} dx\,x^{k+b-1}(1-x)^{-2k} = \\ \label{eq:1135}&=&
\frac{|\lambda|^{2(k+b-1)}}{2k-1}(1-|\lambda|^2)^{-2k}-
\frac{k+b-1}{2k-1}\,\hat{v}_{2k-1,b-1}(|\lambda|^2)\,, ~~~~~~~\\
 \label{eq:1136}&& k
> 1/2\,,\,k+b-1 > 0~. \end{eqnarray} (The index $2k$ of $v_{2k,b}$
refers to the integrand $(1-x)^{-2k}$ only!) \\ For, e.g.\ $k=1/2$
and $b=0,1/2$ we can integrate directly:
  \begin{eqnarray}
    \label{eq:688}
 \hat{v}_{1,0}(|\lambda|^2)& =& \ln
\left(\frac{1+|\lambda|}{1-|\lambda|}\right) =|w|\,, \\
\hat{v}_{1,\frac{1}{2}}(
|\lambda|^2)
&=& -\ln (1-|\lambda|^2)= 2\,\ln\cosh(|w|/2)\,,   
  \end{eqnarray} (see the relations \ref{eq:1535} and \ref{eq:1091})
so that \begin{eqnarray} \label{eq:1550}\langle
(K_0)^{-1} \rangle_{k=1/2,\lambda}&=&
\frac{1-|\lambda|^2}{|\lambda|}\,\ln
\left(\frac{1+|\lambda|}{1-|\lambda|}\right)= \frac{2\,|w|}{\sinh|w|}~, \\
\label{eq:1137}\langle (K_0+1/2)^{-1} \rangle_{k=1/2,\lambda}&=&
-\frac{1-|\lambda|^2}{|\lambda|^2}\,\ln(1-|\lambda|^2) \\ & =&
\frac{2\, \ln\cosh(|w|/2)}{\sinh^2(|w|/2)}\,. \nonumber
\end{eqnarray}
 The {\em classical
(correspondence) limits} of the above formulae are obtained for
$|w| \to \infty$ or $|\lambda| \to 1$. As to the behaviour of the
function \ref{eq:166} in this limit see the Refs.\ \cite{barn3,er1b}

\section{Schr\"{o}dinger-Glauber coherent states}
As the properties of these states are very well-known I confine
myself to those properties which are of special interest in the
context of the groups $SU(1,1)$ etc.\  and their irreducible unitary
representations.
  \subsection{Some general properties}
  We know already from the discussion of the Eqs. \ref{eq:65} in Ch.\ 2
 that we may
 define annihilation and creation operators
 \begin{equation}
   \label{eq:689}
   a=(K_0+k)^{-1/2}K_-\,,~~~a^+ = K_+(K_0+k)^{-1/2}\,,
 \end{equation} for any given self-adjoint operators $K_0,K_1$ and $K_2$
from an irreducible unitary representation of the positive discrete series
of $SU(1,1)$. It follows from the definition \ref{eq:1039} and the 
 relations \ref{vern} and \ref{k3}
that
\begin{equation}
  \label{eq:172}
 |k,\alpha \rangle = e^{-|\alpha|^2/2}\, \sum_{n=0}^{\infty}
 \frac{\alpha^n}{\sqrt{n!}}\,|k,n\rangle~,~\alpha =
 |\alpha|\,e^{i\,\beta}~,
\end{equation} is an eigenstate of $a$ from Eq.\ \ref{eq:689}. It has
all the general properties of the usual Schr\"odinger-Glauber coherent 
states! Because of the additional dependence  on the index $k$, we now get
a whole family of such states!
 One has to realize,
 however, that for $k\neq 1/2$ the states $|k, n \rangle $ are not
 the usual Hermite functions, eigenfunctions of the harmonic
 oscillator. I shall discuss the  case $k=1/2$ in the
 next chapter. 

 We have the usual completeness relation
 \begin{equation}
   \label{com3}
 \frac{1}{\pi}\, \int_{ \mathbb{C}}d^2\alpha \,|k,\alpha\rangle \langle k,\alpha| = 
\boldsymbol{1}\,.
 \end{equation}
Of special interest are the following scalar products between  the coherent
states $|k,z\rangle\,$, $|k,\lambda \rangle $ and $ |k,\alpha\rangle $:

  From their number state representations \ref{eq:78},
 \ref{eq:136}, and \ref{eq:172} 
 we obtain 
\begin{eqnarray} \label{eq:1551}\langle k,\alpha|k,z\rangle &=& \label{caz}
\frac{e^{-|\alpha|^2/2}}{\sqrt{g_k(|z|^2)}}\,C_k(\bar{\alpha};z)\,, \\
\label{eq:1138}C_k(\bar{\alpha};z) &=& \sum_{n=0}^{\infty}
\frac{1}{\sqrt{(2k)_n}}\,\frac{(\bar{\alpha}\,z)^n}{n!}=\bar{C}_k(\bar{z};
\alpha)~, \end{eqnarray} \begin{eqnarray}
\label{eq:1139}\langle k,\alpha|k,\lambda \rangle &=&
e^{-|\alpha|^2/2}(1-|\lambda|^2)^k\,D_k(\bar{\alpha}; \lambda)\,, \\
\label{eq:1140}D_k(\bar{\alpha}; \lambda)&=&\sum_{n=0}^{\infty}
\sqrt{(2k)_n}\,\frac{(\bar{\alpha}\,\lambda)^n}{n!}=\bar{D}_k(\bar{\lambda};
\alpha)\,.
\end{eqnarray}

 From the completeness relations \ref{com1}, \ref{com2} and 
\ref{com3}
 one gets a number of mappings in terms of integral
transforms
\begin{eqnarray} \label{eq:1552}|k,z\rangle &=&
\frac{1}{\pi}\int_{\mathbb{C}}d^2\alpha \, \langle k,
\alpha|k,z\rangle\,|k,\alpha\rangle = \\\label{eq:1141}&=&
\frac{1}{\sqrt{g_k(|z|^2)}}\int_{\mathbb{C}}d\tilde{\mu}
(\alpha)\,C_k(\bar{\alpha};z)\,e^{|\alpha|^2/2}|k,\alpha \rangle~; \nonumber \\
\label{eq:1142}|k,\alpha\rangle &=& \int_{\mathbb{C}}d\mu_k(z)\, \langle
k,z|k,\alpha \rangle\,|k,z\rangle = \\ \label{eq:1143}&=&
e^{-|\alpha|^2/2}\int_{\mathbb{C}}d\tilde{\mu}_k(z)\,
C_k(\bar{z};\alpha)\sqrt{g_k(|z|^2)}\,|k,z\rangle~; \nonumber \\
\label{eq:1144}|k,\lambda\rangle &=&\frac{1}{\pi}\int_{\mathbb{C}}d^2\alpha
\,\langle k,
\alpha|k,\lambda \rangle\,|k,\alpha\rangle = \\
\label{eq:1145}&=&(1-|\lambda|^2)^k\,\int_{\mathbb{C}}d\tilde{\mu}
(\alpha)\,D_k(\bar{\alpha};\lambda)\,e^{|\alpha|^2/2}|k,\alpha
\rangle~; \nonumber \\\label{eq:1146}|k,\alpha\rangle &=&
\int_{\mathbb{D}}d\mu_k(\lambda)\, \langle k,\lambda|k,\alpha
\rangle\,|k,\lambda \rangle = \\ \label{eq:1147}&=&
e^{-|\alpha|^2/2}\int_{\mathbb{D}}d\tilde{\mu}_k(\lambda)\,
D_k(\bar{\lambda};\alpha)(1-|\lambda|^2)^{-k}\,|k,\lambda\rangle~;
\nonumber
\end{eqnarray} These mappings imply the following unitary
transformations of the basis functions \begin{eqnarray}
\label{eq:1553}\tilde{f}_{k,n}(z) &=&\int_{\mathbb{C}}d\tilde{\mu}
(\alpha)\,C_k(\bar{\alpha};z)\,\tilde{h}_n(\alpha)~, \\
\label{eq:1148}\tilde{h}_n(\alpha) &=&\int_{\mathbb{C}}d\tilde{\mu}_k(z)\,
C_k(\bar{z};\alpha)\tilde{f}_{k,n}(z)~; \\
\label{eq:1149}e_{k,n}(\lambda)&=&\int_{\mathbb{C}}d\tilde{\mu}
(\alpha)\,D_k(\bar{\alpha};\lambda)\,\tilde{h}_n(\alpha)~, \\
\label{eq:1150}\tilde{h}_n(\alpha)&=&\int_{\mathbb{D}}d\tilde{\mu}_k(\lambda)\,
D_k(\bar{\lambda};\alpha)\,e_{k,n}(\lambda)~. \end{eqnarray} Like
in subsection 3.2.1 we get additional interesting integral
transforms by multiplying the relations \ref{eq:1552}-\ref{eq:1146} from
 the left with
an appropriate $\langle k,\cdot|\,$. I list only a few examples:
Generalizations are straightforward.
\begin{eqnarray} \label{eq:1554}e^{\bar{\alpha}_2\,\alpha_1}&=&
\int_{\mathbb{C}}d\tilde{\mu}_k(z)\,C_k(\bar{\alpha}_2;z)\,
C_k(\bar{z};\alpha_1)=
\\ \label{eq:1151}&=& \int_{\mathbb{D}}d\tilde{\mu}_k(\lambda)
\,D_k(\bar{\alpha}_2;\lambda)\,
D_k(\bar{\lambda};\alpha_1)~, \\ \label{eq:1152}e^{\bar{\lambda}\,z} &=&
\int_{\mathbb{C}}d\tilde{\mu}_k(\alpha)\,D_k(\bar{\lambda};
\alpha)\,C_k(\bar{\alpha};z)~, \\ \label{eq:1153}C_k(\bar{\alpha};z)&=&
\int_{\mathbb{D}}d\tilde{\mu}_k(\lambda)\,D_k(\bar{\alpha};\lambda)\,
e^{\bar{\lambda}\,z}~, \\ \label{eq:1154}D_k(\bar{\alpha};\lambda) &=&
\int_{\mathbb{C}}d\tilde{\mu}_k(z)\,C_k(\bar{\alpha};z)\,e^{\bar{z}
\,\lambda}~.
\end{eqnarray}
\subsection{Expectation values of the  observables
 $K_1$, $K_2$ and $K_0$ }Again using the relations \ref{k3}-\ref{vern} we
  get \begin{eqnarray} \label{eq:1555}\langle K_1\rangle_{k,\alpha} &=&
 |\alpha|\,\cos\beta\, \langle  k, \alpha|\sqrt{N+2k}|k, \alpha\rangle
 \,, \\
  \label{eq:1155}\langle K_2\rangle_{k,\alpha}
 &=&
- |\alpha|\, \sin\beta\,\langle k,  \alpha|\sqrt{N+2k}|k, \alpha\rangle\,, \\
\label{eq:1156}\langle K_0\rangle_{k,\alpha} &=& \langle N\rangle_{k,\alpha}+k
=|\alpha|^2 +k = \bar{n}_{k,\alpha}+k\,,
\end{eqnarray} where
\begin{equation}
  \label{eq:173}
 \langle k, \alpha|\sqrt{N+2k}|k, \alpha\rangle
= h_1^{(k)}(|\alpha|) =
e^{-|\alpha|^2}\sum_{n=0}^{\infty}\sqrt{2k+n}\,\frac{|\alpha|^{2n}}{n!}\,.
\end{equation}
 Notice that the variable  $|\alpha|^2$ here, too,  equals the
observable quantity $\bar{n}_{k,\alpha}\,$, the average number of
quanta associated with the state $|k, \alpha\rangle\,$! 

Contrary to the mean numbers $\bar{n}_{k,z}$ and
$\bar{n}_{k,\lambda}$ which depend on the index $k$ (Eqs.\ \ref{eq:1529} and
\ref{eq:1542}) {\em the  average $\bar{n}_{k,\alpha}$ does not depend on
$k$} and we can omit the label $k$:
\begin{equation}
  \label{eq:174}
 |\alpha|^2 =
\bar{n}_{\alpha}\,.
\end{equation}
In addition, we see again that the
operators $K_1$ and $K_2$ measure the phase of the complex number
$\alpha$ associated with the state $|k, \alpha \rangle $ in a very
similar way as in the previous two cases $|k,z\rangle$ and $|k, \lambda
 \rangle\,$. Like there we have
\begin{equation}
  \label{eq:175}
\tan\beta = - \frac{\langle K_2 \rangle_{k, \alpha}}{ \langle K_1
\rangle_{k, \alpha}}~.
\end{equation}
We further get
\begin{eqnarray} \label{eq:1556}\langle K_1^2 \rangle_{k, \alpha}&=&
|\alpha|^2\,\cos^2\beta \, h^{(k)}_2(|\alpha|)+ h^{(k)}(|\alpha|)~,\\
\label{eq:1157}h^{(k)}_2(|\alpha|) &=& e^{-|\alpha|^2}\, \sum_{n=0}^{\infty}
\sqrt{(2k+n)(2k+n+1)}\,\frac{|\alpha|^{2n}}{n!}~,~~~~~~~~ \\
\label{eq:1158}h^{(k)}(|\alpha|)&=&
\frac{|\alpha|^4}{2}-\frac{1}{2}(h_2^{(k)}-2k-1)\,|\alpha|^{2}
+\frac{k}{2}\, ~,\\
\label{eq:1159}(\Delta K_1)^2_{k,\alpha}&=&
|\alpha|^2\,\cos^2\beta\,[h_2^{(k)}-(h_1^{(k)})^2] + h^{(k)}~;
 \\ \label{eq:1160}\langle K_2^2 \rangle_{k, \alpha}&=&
|\alpha|^2\,\sin^2\beta \, h^{(k)}_2(|\alpha|)+ h^{(k)}(|\alpha|)~,\\
\label{eq:1161}(\Delta K_2)^2_{k,\alpha}&=&
|\alpha|^2\,\sin^2\beta\,[h_2^{(k)}-(h_1^{(k)})^2] + h^{(k)}~; \\
\label{eq:1162}(\Delta K_1)^2_{k,\alpha}\,(\Delta K_1)^2_{k,\alpha}&=&
\frac{|\alpha|^4\,\sin^2 2\beta}{4}\,[h_2^{(k)}-(h_1^{(k)})^2]^2 +
\\ \label{eq:1163}&& + |\alpha|^2\,[h_2^{(k)}-(h_1^{(k)})^2]\,h^{(k)} +
 (h^{(k)})^2~,
\nonumber \\\label{eq:1164}(\Delta K_1)^2_{k,\alpha}+(\Delta K_2)^2_{k,
\alpha}&=&
|\alpha|^2\,[h_2^{(k)}-(h_1^{(k)})^2]+2h^{(k)}~; \\
\label{eq:1165}S(K_1,K_2)_{k,\alpha} &=& -\frac{|\alpha|^2}{2}\,\sin 2\beta
\,[h_2^{(k)}-(h_1^{(k)})^2]~; \\
 \label{eq:1166}\langle
K_0^2 \rangle_{k, \alpha}&=& (|\alpha|^2 + k)^2 + |\alpha|^2~, \\
\label{eq:1167}(\Delta K_0)^2_{k, \alpha} &=& |\alpha|^2 = \langle N\rangle_{k,
\alpha}= \bar{n}_{\alpha}~. \end{eqnarray}

Using the methods described in Appendix D.3 we obtain the following
asymptotic expansions for the functions $h_1^{(k)}(|\alpha|)\,$ and
$h_2^{(k)}(|\alpha|)$ and certain combinations of them in the case
of large $|\alpha|$:
\begin{eqnarray} \label{eq:1557}h_1^{(k)}(|\alpha|) &\asymp&
|\alpha|\,{[}1+(k-\frac{1}{8})\,|\alpha|^{-2} +
c_{1;-4}\,|\alpha|^{-4}+ \\ \label{eq:1168}&&+ O(|\alpha|^{-6}){]}\,,~
 c_{1;-4} =
-\frac{1}{2}\,k^2 -\frac{3}{8}\,k +\frac{7}{128}~,
\nonumber \\
\label{eq:1169}(h_1^{(k)})^2 & \asymp &
|\alpha|^2\,[1+(2k-\frac{1}{4})|\alpha|^{-2}-(k-\frac{1}{8})|\alpha|^{-4}
+ ~~~~~~~~\\ \label{eq:1170}&& + O(|\alpha|^{-6}\,]~, \nonumber \\
\label{eq:1171}h_2^{(k)}(|\alpha|) &\asymp&
|\alpha|^2\,{[}1+(2k+1/2)\,|\alpha|^{-2}-\frac{1}{8}\,|\alpha|^{-4}+
\\ \label{eq:1172}&& +O(|\alpha|^{-6}){]}\,, \nonumber
\\\label{eq:1173}h_2^{(k)}(|\alpha|)-(h_1^{(k)})^2(|\alpha|) &\asymp&
 \frac{3}{4}
+(k-\frac{1}{4})|\alpha|^{-2} +O(|\alpha|^{-4})~, \\
\label{eq:1174}h^{(k)}(|\alpha|) &\asymp& \frac{|\alpha|^2}{4}\,[1
+(2k+\frac{1}{4})|\alpha|^{-2} +O(|\alpha|^{-4})\,]\,.
\end{eqnarray} This gives the following asymptotic approximations
for large $|\alpha|$
\begin{eqnarray} \label{eq:1558}\langle K_1 \rangle_{k,\alpha} &\asymp&
|\alpha|^2\,\cos\beta\, {[}1+(k-\frac{1}{8})\,|\alpha|^{-2} +
c_{1;-4}\,|\alpha|^{-4}+ O(|\alpha|^{-6}){]}\,, ~~~~~~~~~~~~\\
 \label{eq:1175}(\Delta K_1)^2_{k,\alpha} &\asymp&
\frac{|\alpha|^2\,\cos^2\beta}{4}\,{[}3 + (4k-1)|\alpha|^{-2} +
O(|\alpha|^{-4}) {]}+\\ \label{eq:1176}&&+
\frac{|\alpha|^2}{4}\,[1+(2k+\frac{1}{4})
|\alpha|^{-2} + O(|\alpha|^{-4})]; \nonumber \\
\label{eq:1177}\langle K_2 \rangle_{k,\alpha} &\asymp& -|\alpha|^2\,\sin\beta\,
{[}1+(k-\frac{1}{8})\,|\alpha|^{-2} +
c_{1;-4}\,|\alpha|^{-4}+ O(|\alpha|^{-6}){]}\,,~~~~~~~~~ \\
\label{eq:1178}(\Delta K_2)^2_{k,\alpha} &\asymp&
\frac{|\alpha|^2\,\sin^2\beta}{4}\,{[}3 + (4k-1)|\alpha|^{-2} +
O(|\alpha|^{-4}) {]}+\\ \label{eq:1179}&&+
\frac{|\alpha|^2}{4}\,[1+(2k+\frac{1}{4}) |\alpha|^{-2} +
O(|\alpha|^{-4})]~. \nonumber
 \end{eqnarray}
 Whereas the operators \[ Q=\frac{1}{\sqrt{2}}(a^+ + a)~,~P
 =\frac{i}{\sqrt{2}}(a^+ -
 a)~,~[Q,P]=i~, \] have very simple properties with respect to the
 states $|k,\alpha \rangle\,$, namely
\begin{equation}
  \label{eq:176}
 (\Delta Q)^2_{k,\alpha} = (\Delta P)^2_{k,\alpha}
 = \frac{1}{2}~,~(\Delta Q)^2_{k,\alpha}(\Delta P)^2_{k,\alpha}= \frac{1}{4}
 |\langle k,\alpha| [Q,P]|k,\alpha \rangle|^2~,
\end{equation}
the operators $K_1$ and $K_2$ obviously have not.

 Because
 $[K_1,K_2]= -i\,K_0$ we have for arbitrary states $|\psi\rangle$ 
\begin{equation}
  \label{eq:177}
 (\Delta
 K_1)^2_{\psi}\,(\Delta K_2)^2_{\psi} \geq \frac{1}{4}\,|\langle
 \psi |K_0|\psi \rangle |^2 ~.
\end{equation}
One speaks of ``squeezing'' (see Ch.\ 6) if
\begin{equation}
  \label{eq:178}
 (\Delta K_1)^2_{\psi} < \frac{1}{2}|\langle
 \psi |K_0|\psi \rangle | \mbox{ or }(\Delta K_2)^2_{\psi} < \frac{1}{2}|\langle
 \psi |K_0|\psi \rangle |~.
\end{equation}
Because
\begin{equation}
  \label{eq:179}
 \langle k,\alpha |K_0|k, \alpha \rangle = |\alpha|^2
+k = |\alpha|^2(1+k|\alpha|^{-2})~,
\end{equation}
we see from Eqs.\ \ref{eq:1175} and \ref{eq:1178}
that for large $|\alpha|$ the squared uncertainties \ref{eq:1159} or 
\ref{eq:1161} are
squeezed in leading order  if either
\begin{equation}
  \label{eq:180}
 \cos^2\beta < \frac{1}{3}
\mbox{ or }\sin^2\beta < \frac{1}{3}~.
\end{equation}
\chapter{Harmonic oscillator}
There are many interrelations between  the operators $a$ and $a^+$
(or $Q$ and $P$) of the harmonic oscillator the Lie algebra
 generators $K_j,\,j=0,1,2,$ of
the group $SU(1,1)$ etc. Several of those relations we encountered
already in previous chapters. 

 Of special interest are the
non-linear ones \cite{mlo}
\begin{equation}
  \label{eq:181}
 K_+ = a^+\,\sqrt{N+2k}~,~K_- =
\sqrt{N+2k}\,a~,~ K_0 = N+k~,
\end{equation}
which allows the construction of
self-adjoint $K_j$, once the $a\,,a^+$ are given. 

 But  even
more important is the inversion (see Secs.\ 1.5.1 and 2.3): 
Given self-adjoint $K_j$, an
irreducible unitary representation of the positive discrete series 
with index $k$ and the
associated Hilbert space, we can define
\begin{equation}
  \label{eq:182}
 a =
(K_0+k)^{-1/2}K_-~,~a^+ =K_+(K_0+k)^{-1/2}~,~ N=K_0-k=a^+a~.
\end{equation}

Another interesting  relationship between the $a^+\,,a$ and
the $K_j$ is the following: The bilinear expressions
\begin{equation}
  \label{eq:183}
 K_+ =
\frac{1}{2}(a^+)^2~,~K_- =\frac{1}{2}a^2~,~K_0=\frac{1}{2}(a^+a+
1/2)~
\end{equation}
obey the Lie algebra \ref{eq:55}. As the operator $K_-$ here
annihilates the states $|n=0\rangle$ {\em and} $|n=1\rangle$, we get two
different representations of the Lie algebra, one with $k=1/4$ and
one with $k=3/4$. Details of these will be discussed Sec.\ 6.3, also in
Sec.\ 6.4  realizations
of that Lie algebra by a pair $a_1,a_2$ of annihilation operators
and the related creation operators.

As the harmonic oscillator has the Hamiltonian $H =
\omega\,(N+1/2)$ the third of the relations \ref{eq:181} strongly
 suggests to identify 
 \begin{equation}
   \label{eq:170}
   H_{osc} =  K_0~\text{for}~k=\frac{1}{2}
 \end{equation} (frequency $\omega$ scaled to $=1$) and to
implement the 
quantum mechanics of the harmonic oscillator in a Hilbert space
 which contains a unitary
irreducible representation of $SU(1,1)$ or $SL(2,\mathbb{R})=Sp(2,\mathbb{R})$
 with index $k=1/2$. 

{\em All the results and predictions of the last chapter, derived for
 general indi\-ces $k$,
apply, of course, for $k=1/2$, too,} and will therefore  not be listed here
again. They contain, to be sure, a wealth of new informations about the
quantum mechanical harmonic oscillator! One just has to take $k=1/2$
in all the formulae of Ch.\ 3! In some cases one has to take the limit
$k \to 1/2$, starting from $k > 1/2$

The present chapter focuses on properties which are special
for unitary representations with $k=1/2$.
\section{Quantum mechanics of the harmonic \\ oscillator in the Hardy
space of the unit circle} In subsection 3.2.1  we  encountered the
scalar product
\begin{equation}
  \label{eq:184}
 (f,g)_{\mathbb{D},\, k}
=\frac{2k-1}{\pi}\int_{\mathbb{D}}
\bar{f}(\lambda)g(\lambda)\,(1-|\lambda|^2)^{2k-2}
  |\lambda|\,d|\lambda|\,d\theta ~,
\end{equation}
for the important (Bargmann) Hilbert space
   $ \mbox{$\cal H$}_{\mathbb{D},\, k}$
   of holomorphic functions on the unit
  disc $\mathbb{D}$ (Eq.\ \ref{eq:1038}).
   It can be used for any real $k > 1/2$ and also - as we have seen - in the
  limiting case $k\rightarrow 1/2$. \\
  Any holomorphic function in $\mathbb{D}$ can be
  expanded in powers of $\lambda$ and we saw that the functions
 
\begin{equation}
  \label{eq:185}
 e_{k,n}(\lambda)= \sqrt{\frac{(2k)_n}{n!}}\,
  \lambda^n~~,~~ n = 0,1,2, \ldots ~~,
\end{equation}
form an orthonormal basis of  $\mbox{$ \cal H$}_{\mathbb{D},\, k}$.

 If we have  the functions
\begin{equation}
  \label{eq:186}
f(\lambda)=\sum_{n=0}^{\infty}a_n\,\lambda^n~,~~g(\lambda)=
\sum_{n=0}^{\infty}b_n\,\lambda^n~,
\end{equation}
then, according to Eq.\ \ref{eq:149},
 their scalar product $(f,g)_{\mathbb{D},\, k}$ is given by
the series
\begin{equation}
  \label{eq:187}
 (f,g)_{\mathbb{D},\, k}=\sum_{n=0}^{\infty}
\frac{n!}{(2k)_n}\,
 \bar{a}_n \,b_n~~.
\end{equation}
This series can be used  to extend the
definition of the scalar product for Hilbert spaces
 $\mbox{$\cal H$}_{\mathbb{D},\,k}$ to all real $k>0\,$!

For the special case $k=1/2$ the coefficient in front of
$\bar{a}_n\, b_n$ in Eq.\ \ref{eq:187} has the value 1. This allows for an
interesting reinterpretation of the Hilbert space $ \mbox{$\cal
H$}_{\mathbb{D},\frac{1}{2}} $:

 Consider the Hilbert space $L^2(S^1,d
\vp)$ on the unit circle with the scalar product
\begin{equation}
  \label{eq:188}
(f_2,f_1)=\frac{1}{2\pi}\int_0^{2\pi}d\varphi \;
\bar{f}_2(\varphi)\,f_1(\varphi)~,
\end{equation}
an orthonormal
 basis of which is given by the
functions $\exp(i\,n \,\varphi)\;,~n \in \mathbb{Z}$.

That subspace of functions $f(\varphi) \in  L^2(S^1,d\vp)$ which have only
non-negative Fourier coefficients, i.e.\  $ a_n =0 
\mbox{ for } n<0\,$,
 is being called the ``Hardy space $H^2(S^1,d\vp)$ of the unit disc''
$\mathbb{D}$ or ``unit cirle'' ($S^1 = \partial \mathbb{D}$) \cite{hard2},
  and the corresponding scalar pro\-duct
 will be denoted by $(f_1,f_2)_+$. \\ The Hilbert space
  $H^2(S^1,d \vp)$  has the orthonormal
 basis
\begin{equation}
  \label{eq:189}
 e_n(\varphi) = e^{i\,n\,\varphi}\,,~n = 0,1,2,\ldots ~.
\end{equation}

 Hardy spaces  have a number of interesting
 properties and are closely related to Hilbert spaces of holomorphic
 functions \cite{bar1,hard2} because the unit circle is the
 boundary $\partial \mathbb{D}$ of $\mathbb{D}\,$! Notice that the
 eigenfunctions \ref{eq:189} may be considered as limits of those in Eq.\ 
\ref{eq:185}:
\begin{equation}
  \label{eq:190}
 e_n(\varphi) = \lim_{|\lambda| \to 1}e_{k=1/2,
 n}(|\lambda|\,e^{i\,\varphi})~,~ \lambda \in \mathbb{D}~.
\end{equation}
(Mathematically the limit should be taken in terms of the
appropriate norm \cite{hard2}!)

  If we have  two Fourier series $\in H^2(S^1\,,d\vp)$,
\begin{equation}
  \label{eq:191}
 f_1(\varphi)=\sum_{n=0}^{\infty}a_n\,e^{i\,n\,\varphi}~,~
~f_2(\varphi)= \sum_{n=0}^{\infty}b_n\,e^{ i\,n\,\varphi}~~,
\end{equation}
they have the scalar product
\begin{equation}
  \label{eq:192}
 (f_1,f_2)_+=
\frac{1}{2\pi}\int_0^{2\pi} d\varphi\,
\bar{f}_1(\varphi)f_2(\varphi)=\sum_{n=0}^{\infty}\bar{a}_n
\,b_n~~.
\end{equation} Comparing with \ref{eq:187} for $k=1/2$ we see that
 we may realize the Hilbert space $ \mbox{$\cal
H$}_{\mathbb{D},\, 1/2}$ by using the Hardy space $H^2(S^1,\,d\vp)$!

The $SU(1,1)$ Lie algebra generators \ref{eq:157} for $k=1/2$ are now \cite{bo}
\begin{eqnarray} \label{eq:1559}K_+ &=&
e^{i\,\varphi}\,(\frac{1}{i}\partial_{\varphi}+1)~, \\
\label{eq:1180}K_- & =& e^{-i\,\varphi}\,\frac{1}{i}\partial_{\varphi}~, \\
\label{eq:1181}K_0 &=& \frac{1}{i}\partial_{\varphi} + \frac{1}{2}~,
\end{eqnarray} for which  the relations \ref{k3}-\ref{vern} take the form
\begin{eqnarray}
\label{eq:1560}K_0\,e_n(\varphi) &=& (n+\frac{1}{2})\,e_n(\varphi)~, \\
\label{eq:1182}K_+\,e_n(\varphi)&=& (n+1)\,e_{n+1}(\varphi)~, \\
\label{eq:1183}K_-\,e_n(\varphi)&=& n\,e_{n-1}(\varphi)~. \end{eqnarray}
The operators
\begin{equation}
  \label{eq:834}
  a = (K_0+1/2)^{-1/2}K_-\,,~~ a^+=K_+(K_0+1/2)^{-1/2}
\end{equation}
have the usual properties
\begin{equation}
  \label{eq:835}
  a\,e_{n}(\vp)= \sqrt{n}\,e_{n-1}(\vp)\,,~~a^+\,e_n(\vp) = \sqrt{n+1}\,
e_{n+1}(\vp)\,,
\end{equation}
and have the usual matrix elements \cite{mesa}. The same applies, of
course, to those of the operators $Q$ and $P$:
\begin{equation}
  \label{eq:917}
  Q=\frac{1}{\sqrt{2}}\,(a^+ +a)\,,~~~ P=\frac{i}{\sqrt{2}}\,(a^+ -a)\,.
\end{equation}
Before I turn to the harmonic oscillator itself let me add a few
mathematical remarks: 

 The reproducing kernel (see Eqs.\ \ref{eq:151} and
\ref{eq:152}) here has the form
\begin{equation}
  \label{eq:193}
 \Delta(\varphi_2, \varphi_1) =
\sum_{n=0}^{\infty} \overline{e_n(\varphi_2)}\,e_n(\varphi_1) =
(1-e^{i\,(\vp_1-\vp_2)})^{-1}~,
\end{equation}
\[ \frac{1}{2\,\pi}
\int_0^{2\,\pi} d\vp_2 \, \Delta(\varphi_2, \varphi_1)\ e_n(\vp_2)
= e_n(\vp_1)~.
\]The kernel has a singularity (pole) for $\varphi_2 = \varphi_1$.
In calculations one has to replace
$\exp(i\,(\varphi_1-\varphi_2))$ by
$(1-\epsilon)\,\exp(i\,(\varphi_1-\varphi_2))$ and then take the
limit $\epsilon \to 0$ at the end. 

 The operators $ (K_0 +
1/2)^{-s}\,, s = 1/2,1$ etc., as needed, e.g.\ in Eqs.\ \ref{eq:797} and
\ref{eq:1037}, are represented by integral kernels
\begin{eqnarray} \label{eq:1561}(K_0 + \frac{1}{2})^{-s}(\varphi_2, \varphi_1)
&=& \sum_{n=0}^{\infty} (n+1)^{-s}
(e^{i\,(\varphi_1-\varphi_2)})^n = \\ \label{eq:1184}&=& \Phi(e^{i\,
(\varphi_1-\varphi_2)}, s, 1)= \\
\label{eq:1185}&=& e^{-i\,(\varphi_1-\varphi_2)}\sum_{n=1}^{\infty} n^{-s}
(e^{i\,(\varphi_1-\varphi_2)})^n \\ \label{eq:1186}&=&
e^{-i\,(\varphi_1-\varphi_2)}
\mbox{Li}_s(e^{i\,(\varphi_1-\varphi_2)})~,
\end{eqnarray}
 where $
\Phi(z,s,a)$  is Lerch's function and $\mbox{Li}_s(z)$  the
polylogarithm (see Eqs.\  \ref{eq:166} and \ref{eq:167} of subsection 3.2.2).

 For $ s>0$, $a=1$ and
$\varphi_2 \neq \varphi_1 $ one has the integral representation
\cite{er1b}
\begin{equation}
  \label{eq:194}
 (K_0 + \frac{1}{2})^{-s}(\varphi_2, \varphi_1) =
\frac{1}{\Gamma(s)} \int_0^{\infty} dt\, \frac{t^{s-1}}{e^t -
e^{i\,(\varphi_1-\varphi_2)}} ~.
\end{equation}

According to the Eqs.\ \ref{eq:170} and \ref{eq:1181} we now get for
the Hamiltonian of the harmonic oscillator
\begin{equation}
  \label{eq:198}
 H_{osc} =
(\frac{1}{i}\,\partial_{\varphi} + \frac{1}{2})~.
\end{equation}

As here $|k=1/2, n \rangle = \exp(i\,n\,\varphi)$, we can sum the
coherent state series \ref{eq:78}, \ref{eq:136} and \ref{eq:172} 
  as functions of $\varphi$:

 For the coherent state \ref{eq:78} we get 
\begin{equation}
  \label{eq:199}
 f_z(\varphi) =
\frac{1}{\sqrt{I_0(2\,|z|)}}\,e^{\displaystyle z\,e^{i\,\varphi}}
= \frac{1}{\sqrt{I_0(2\,|z|)}}\,e^{\displaystyle
|z|\,e^{i\,(\varphi +\phi)}}~.
\end{equation}

Thus, $f_z(\vp)$ is essentially the generating function for the basis
\ref{eq:189}.

Applying the time evolution
operator
\begin{equation}
  \label{eq:200}
 U(t) = e^{-iHt} = e^{-iK_0t}
\end{equation}
to
the function $f_z(\varphi)$ (in their number basis \ref{eq:78}) yields
\begin{equation}
  \label{eq:201}
 U(t)\cdot f_z(\varphi) =
e^{-it/2}\,f_{z(t)}(\varphi)~,~z(t) =
z\,e^{-it}~.
\end{equation}
Thus, the time evolution
essentially translates the phase $\phi$ of $z$ by $-t$ (recall 
that the frequency $\omega$ has been scaled to the value $1$). 

Similarly we get for the coherent states \ref{eq:136}:
\begin{equation}
  \label{eq:202}
f_{\lambda}(\varphi) = (1-|\lambda|^2)^{1/2}\, (1-\lambda\, e^{i\,
\varphi})^{-1} = (1-|\lambda|^2)^{1/2}\, (1-|\lambda|\, e^{i\,
(\varphi + \theta)})^{-1}
\end{equation}
and
\begin{equation}
  \label{eq:203}
 U(t)\cdot
f_{\lambda}(\varphi) =
e^{-it/2}\,f_{\lambda(t)}(\varphi)~,~\lambda(t) =
\lambda\,e^{-it}~.
\end{equation}

For the coherent states \ref{eq:172} the
resulting series cannot be summed in an elementary way:
\begin{equation}
  \label{eq:204}
f_{\alpha}(\varphi) =
e^{-|\alpha|^2/2}\,\sum_{n=0}^{\infty}\frac{(\alpha\,e^{i\,\varphi})^n
}{\sqrt{n!}} =
e^{-|\alpha|^2/2}\,\sum_{n=0}^{\infty}\frac{(|\alpha|\,e^{i\,(\varphi
+\beta)})^n }{\sqrt{n!}}~,
\end{equation}
\begin{equation}
  \label{eq:205}
 U(t)\cdot f_{\alpha}(\varphi) =
e^{-it/2}\,f_{\alpha(t)}(\varphi)~,~\alpha(t) =
\alpha\,e^{-it}\,.
\end{equation}
(Recall that the series \ref{eq:172} can be
summed to a Gaussian function if one takes for $|n\rangle$ the
Hermite functions \ref{eq:806}.)

 The above results for the time evolution of the 3 different coherent states
 show that the phases of $z\,,\lambda$ and $\alpha$ are
 the dynamical variables, {\em not} the mathematical auxiliary phase
 $\varphi$!

 An asymptotic expansion of the  function \ref{eq:204} has been given in Ref.\
 \cite{garr} as
\begin{equation}
  \label{eq:206}
 f_{\alpha}(\varphi) \asymp
 (2\,\pi)^{1/4}\,(2\,|\alpha|)^{1/2}\,e^{\displaystyle -[|\alpha|^2\,
 (\varphi+\beta)^2 - i \,(|\alpha|^2 -1/2)\,(\varphi + \beta)]}~.
\end{equation}

 The above discussions show explicitly that we can associate {\em three}
 different
coherent states with the harmonic oscillator, all of which stay coherent
with time! They have different ``squeezing'' and many different other
 properties, already discussed in Ch.\ 3 in a more general context. 

I would like to stress again that - contrary to all possible appearances - the
 phase $\varphi$ of the Hilbert space $H^2(S^1, d\vp)$ with
 the basis \ref{eq:189}
 and the scalar product \ref{eq:192} is {\em not} the quantum mechanical
 canonically conjugate observable with respect to the operator
 $K_0$ of Eq.\ \ref{eq:1181}. The reason is that $\varphi$ as a
 multiplication operator is not self-adjoint with respect to the
 scalar product \ref{eq:192} \cite{reed1}. 

The quantity $\vp$ is merely a mathematical auxiliary variable which
parame\-trizes the Hilbert space. The information about the physical phases
of the states \ref{eq:199}, \ref{eq:202} and \ref{eq:204} has to be extracted
by means of the operators $K_1$ and $K_2$ as discussed in detail in the
previous Ch.\ 3.

In addition, the
 multiplication operator $ \exp(i\,\varphi)$ is not unitary on
 $H^2(S^1,d\vp)$, because it acts on the eigenstate basis as an isometric 
 shift operator \cite{shiop}:
\begin{equation}
  \label{eq:207}
 e^{i\,\varphi}\, e_n(\varphi) = e_{n+1}(\varphi)~,
\end{equation}
where the inverse transformation
\begin{equation}
  \label{eq:208}
 e^{-i\,\varphi}\, e_n(\varphi) =
 e_{n-1}(\varphi)
\end{equation}
is, however, not defined for $e_{n=0}\,$! This implies again that
$\vp$ cannot be a self-adjoint operator, because otherwise $\exp(i\,\vp)$
would be unitary!

In the present approach to the quantum theory of the harmonic oscillator
the basic observables are the self-adjoint operators $K_0,\,K_1$ and $K_2$.
Even the position and momentum operators $Q$ and $P$ are functions of them
 (see Eq.\ \ref{eq:917})!
\section{Some critical remarks on ``phase states''}
 At this point a few remarks about the so-called ``phase states''
 \cite{phst} may be appropriate: 

In the search for a possible
 phase operator it was surmised \cite{suss} that the following
 (``phase'') state
 might be a candidate for an eigenstate of the yet to be found
 phase operator:
\begin{equation}
  \label{eq:209}
 |\varphi\rangle
 =\sum_{n=0}^{\infty}e^{i\,n\,\varphi}\,|n\rangle ~,
\end{equation}
where $|n\rangle$
 are the usual eigenstates (Hermite functions) of the harmonic
 oscillator and the $e_n(\varphi) = \exp(i\,n\,\varphi)$
 constitute a basis for the Hilbert space
 with the scalar product \ref{eq:192}. The situation is the same as in the
case of the three types \ref{eq:78}, \ref{eq:136} and \ref{eq:172} of
coherent states which are introduced as series in the number states the 
coefficients of which form a basis in an associated Hilbert space!

In the literature, however, the functions
 $\exp(i\,n\,\vp)$ are treated as mere coefficients multiplying the
 basis vectors
 $|n\rangle$. The norm of the state \ref{eq:209} is then
 obviously infinite.

 For the formal scalar product with
 respect to the basis $|n\rangle $ of two such
 states one gets
\begin{equation}
  \label{eq:210}
 \langle \varphi_2|\varphi_1 \rangle =
 \sum_{n=0}^{\infty}e^{i\,n\,(\varphi_1-\varphi_2)}\,,
\end{equation}
which is
 just the reproducing kernel \ref{eq:193}, and not a delta-function!
 The latter property usually is then interpreted as an indication that there
 is no such self-adjoint phase operator which has the state $|\varphi
 \rangle$ as an eigenstate, because states $|\varphi_1\rangle$ and
$|\varphi_2\rangle$ are not orthogonal for $\varphi_2 \neq
\varphi_1\,$! The argument, however, is due to  misunderstandings:

 First, let me rewrite the expression \ref{eq:209} in a more symmetrical
way: 
\begin{equation}
  \label{eq:211}
 |\vp \rangle \to F(\varphi,x)
 =\sum_{n=0}^{\infty}e_n(\varphi)\,u_n(x)\,,
\end{equation}
 where $|n\rangle = u_n(x)$ are Hermite's functions
\cite{mesa} with the scalar product
\begin{equation}
  \label{eq:836}
  (g_2,g_1) = \int_{-\infty}^{\infty}dx\,\bar{g}_2(x)\,g_1(x)\,.
\end{equation}
Eq.\ \ref{eq:211} is the - formal
 - sum over the products $e_n(\varphi)\,u_n(x)$ of the
 eigenfunctions of the oscillator Hamilton operator $H_{osc}$,
 represented in two different Hilbert spaces, namely $H^2(S^1,\vp)$ and
 $L^2(\mathbb{R},dx)$! Thus, the products in the sum are just the
 ``diagonal'' elements of the the basis $\{e_{n_1}(\varphi)\,u_{n_2}(x)\,,\,
 n_1\,,n_2 = 0,1,2,\ldots \}$ for the
 tensor product of the two Hilbert spaces! 

 Instead of taking
 the formal ``scalar product'' of two ``states'' $F_j(\varphi_j,x_j)\,$,
$j=1,2 $ with respect
 to the basis $u_n(x)$ -- i.e. $ x_2 = x_1 = x $ and integration over $x$,
 (Eq.\ \ref{eq:195}),
 -- with the $e_n(\varphi_j)$ as coefficients, like what is being done in Eq.\
\ref{eq:210}, we may as well
 do it  the other way round and obtain
\begin{equation}
  \label{eq:212}
 \frac{1}{2\,\pi}
 \int_0^{2\,\pi}d\,\varphi\,
 \bar{F}_2(\varphi,x_2)\,F_1(\varphi, x_1) =
 \sum_{n=0}^{\infty} \bar{u}_n(x_2)\,u_n(x_1) = \delta(x_2-x_1)~.
\end{equation}
Here the $\delta$-function is the reproducing kernel for the
 basis $\{u_n(x)\}\,$! 

 One of the misunderstandings is the
 following: 

 A reproducing kernel represents the properties of the
 {\em completeness relation} for the functions of the basis in a
 concrete Hilbert space. The completeness does not have to be
 expressed by a $\delta$-function as the examples of the coherent
 states $|k,z\rangle\,,|k,\lambda \rangle $ and $|k,\alpha \rangle$ in the
 last chapter
 clearly show. They also show that the completeness relation
  is independent of the orthogonality of
 the associated eigenfunctions. 

  Another well-known example for such a situation
 is the Hilbert space of positive frequency solutions of the free
 Klein-Gordon equation \cite{jost}. They have the Fourier
 representation \begin{eqnarray} \label{eq:1563}\psi^{(+)}(x)& =&
 \frac{1}{i\,(2\,\pi)^{3/2}} \int_{p^0>0}
 d^4p\, e^{-i\,p\cdot x}\, \delta(p^2-m^2)\, a(p)~,~ \\ \label{eq:1190}&& x =
 (x^0,\vec{x})~,~p\cdot x = p^0x^0-\vec{p}\cdot\vec{x}\,, \nonumber
 \end{eqnarray}
 and their scalar product is \begin{eqnarray}  \label{eq:1564}(\psi_2^{(+)},
\psi_1^{(+)}) &=& i\,\int_{x^0=t}d^3x\,
 \bar{\psi}_2^{(+)}\partial_0\psi_1^{(+)} -(\partial_0\bar{\psi}_2^{(+)})\,
\psi_1^{(+)}\\ \label{eq:1191}&
 =&
 \int_{p^0 > 0} d^4p\,\delta(p^2-m^2)\bar{a}_2(p)\,a_1(p)~.
 \end{eqnarray}
 For an orthonormal basis $\{f_{n}\}\,,\,
 (f_{n_2},f_{n_1}) = \delta_{n_2\,n_1 }\,$, one has the
 completeness relation
\begin{equation}
  \label{eq:213}
 \sum_{n =0}^{\infty}
 \bar{f}_{n}(x_2)\,f_{n}(x_1) = i\,\Delta_+
 (x_2-x_1)~,
\end{equation}
where $\Delta_+$ is the distribution (generalized function)
\begin{equation}
  \label{eq:214}
 \Delta_+
 (x_2-x_1) = \frac{1}{i\,(2\,\pi)^{3}} \int_{p^0
 >0}d^4p\,\delta(p^2-m^2)\,e^{-i\,p\cdot x} ~,
\end{equation}
which is not a
 $\delta$-function, either.

 In order to avoid the (pseudo-) problems  mentioned above it has been
 proposed \cite{lern2,barn,rev6} to truncate the sum \ref{eq:209} at
 some finite
 $n=s$ and start with a
 finite dimensional phase state space with a discretized phase variable,
 where everything is under
 mathematical control. Barnett and Pegg suggested to
  do all  required calculations in the finite dimensional space first and let
 the dimension $s+1$ go to $\infty$ at the very end. This proposal
 has led to a large number of follow-up papers \cite{phst}. But it
 is unsatisfactory for several reasons: In finite dimensional vector spaces all
 Hermitian operators are also self-adjoint, i.e.\ have a complete
 set of eigenfunctions and therefore a spectral representation.
 This is no longer true in infinite dimensional Hilbert spaces
 (see, e.g.\ Ref.\ \cite{reed1}) and therefore one has to expect
 problems for the limiting theory, which may be reached - if at
 all - by weak convergence only. As to a discussion of the
 mathematical problems involved see  Ref.\ \cite{dub}.

 Actually it is not necessary at all to employ the additional
 oscillator basis $|n\rangle$ in Eq.\ \ref{eq:209}. We have seen that it
 suffices to work with the functions \ref{eq:189} and the associated
 Hilbert space $H^2(S^1,d\vp)$ alone! {\em All} the usual quantum
 theory of the harmonic oscillator can be described by means of
 that space.

Furthermore, we have seen in the previous section, that the phase
$\vp$ is a mere auxiliary mathematical variable and {\em not} a
canonical quantity!

As to the relationship between the Hilbert space $H^2(S^1,d\vp)$ with
its basis $e_n(\vp)$ and the usual Hilbert space $L^2(\mathbb{R},dx)$
with its basis of Hermite functions $u_n(x)$ see Sec.\ 4.5.
\section{Eigenfunctions of $K_1$ and $K_2$}
 In the present framework the quantum theoretical properties of
 the phase are incorporated into the operators $K_1$ and $K_2$
 which, according to the Eqs.\ \ref{eq:1559} and \ref{eq:1180},
 here have the form
 \begin{eqnarray} \label{eq:1565}K_1&=& \frac{1}{2}(K_+ +K_-) =
 \cos\vp\,\frac{1}{i}\partial_{\vp} + \frac{1}{2}\,e^{i\,\vp}~, \\
 \label{eq:1192}K_2&=& \frac{1}{2i}(K_+ -K_-) =
 \sin\vp\,\frac{1}{i}\partial_{\vp} + \frac{1}{2i}\,e^{i\,\vp}~.
 \end{eqnarray} The determination of their eigenfunctions
 $f_{h_1}(\vp)$ and $f_{h_2}(\vp)$ is straightforward. Let me
 start with $K_2$. The case $K_1$ can be dealt with by a shift
 $\vp \to \vp+\pi/2$. 

Integrating the differential equation
\begin{equation}
  \label{eq:215}
 K_2f_{h_2}(\vp) = h_2 f_{h_2}(\vp)
\end{equation}
yields \cite{grad}
\begin{equation}
  \label{eq:216}
 f_{h_2}(\vp) = C\,(\sin\vp)^{-1/2}(\tan(\vp/2))^{i\,h_2}\,e^{-i\,\vp/2}~,~
 h_2 \in
 \mathbb{R}~,~ \vp \in (0,\pi)~,
\end{equation}
where $C$ is a normalization
 constant. In the interval $\vp \in (\pi, 2\pi)$ the functions
 $\sin\vp$ and $\tan(\vp/2)$ are negative. Here we get
\begin{equation}
  \label{eq:217}
 f_{h_2}(\vp) = C\,|\sin\vp|^{-1/2}|\tan(\vp/2)|^{i\,h_2}\,e^{-i\,\vp/2}~,~
 h_2 \in
 \mathbb{R}~,~ \vp \in (\pi,2\pi)~,
\end{equation}
Thus we have
 
\begin{equation}
  \label{eq:218}
 f_{h_2}(\vp) = C\,|\sin\vp|^{-1/2}|\tan(\vp/2)|^{i\,h_2}\,e^{-i\,\vp/2}~,~
 h_2 \in
 \mathbb{R}~,~ \vp \in (0,2\pi)~.
\end{equation}
The normalization constant
 $C$ is determined from
\begin{equation}
  \label{eq:219}
 \frac{1}{2\,\pi} \int_0^{\pi}d\vp\,
 \bar{f}_{h_2'}(\vp)\,f_{h_2}(\vp) =
 \frac{|C|^2}{2\,\pi}\int_0^{\pi}\frac{d\,\vp}{\sin\vp}
(\tan(\vp/2))^{i\,(h_2-h_2')}~.
\end{equation}
The substitution
\begin{equation}
  \label{eq:220}
 u=\ln\tan(\vp/2)~,~du = \frac{d\vp}{\sin\vp}~,~
  u(\vp \to 0^+, \pi^-) \to -\infty\,, + \infty,
\end{equation}
then gives
\begin{equation}
  \label{eq:221}
  \frac{1}{2\,\pi} \int_0^{\pi}d\vp\,
 \bar{f}_{h_2'}(\vp)\,f_{h_2}(\vp) =
 |C|^2\,\delta(h_2'-h_2)~.
\end{equation}
Taking the integral \ref{eq:221} from $\pi$ to
 $2\,\pi$  gives the same contribution, so that
 $|C|^2 =1/2$. Thus, we finally have
\begin{equation}
  \label{eq:222}
 f_{h_2}(\vp) = |2\sin\vp|^{-1/2}|\tan(\vp/2)|^{i\,h_2}\,e^{-i\,\vp/2}~,~ h_2 \in
 \mathbb{R}~,~ \vp \in (0,2\pi)~.
\end{equation}

The substitution $\vp \to \vp + \pi/2$ transforms the operator
 $K_2$ of Eq.\ \ref{eq:1192} into the operator $K_1$ of Eq.\ \ref{eq:1565}. Its
 normalized
 eigenfunctions therefore are
\begin{equation}
  \label{eq:223}
 f_{h_1}(\vp) = |2\cos\vp|^{-1/2}|\tan(\vp/2 + \pi/4)|^{i\,h_1}\,
e^{-i\,\vp/2}~,~ h_1 \in
 \mathbb{R}~,~ \vp \in (0,2\pi)~.
\end{equation}

 The hypothetical ansatz
\begin{equation}
  \label{eq:224}
 f_{h_1}(\vp)
 = \sum_{n=0}^{\infty} c_n^{(1)}\,e_n(\vp)
\end{equation} for the eigenfunctions of $K_1$ from \ref{eq:1565}
leads to the
 recursion formula
\begin{equation}
  \label{eq:225}
 c_{n+1}^{(1)} =
 \frac{2\,h_1}{n+1}\,c_n^{(1)} - \frac{n}{n+1}\,c_{n-1}^{(1)}~,
 n=0,1,2, \ldots ~.
\end{equation}
The first few terms are the following
 \begin{eqnarray} \label{eq:1566}c_1^{(1)}/c_0^{(1)} &=& 2\,h_1  ~,\\
\label{eq:1193}c_2^{(1)}/c_0^{(1)} &=& 2\,h_1^2 - \frac{1}{2}~, \nonumber \\
\label{eq:1194}c_3^{(1)}/c_0^{(1)} &=& \frac{2^3}{3!}\,h_1^3 - \frac{5}{3}\,h_1~,
\nonumber \\ \label{eq:1195}c_4^{(1)}/c_0^{(1)} &=& \frac{2^4}{4!}\,h_1^4
-\frac{7}{3}\,h_1^2 + \frac{3}{8}~. \nonumber \end{eqnarray} For
 $f_{h_2}(\vp) $ we get accordingly:
\begin{equation}
  \label{eq:226}
 c_{n+1}^{(2)} =
 \frac{2i\,h_1}{n+1}\,c_n^{(2)} + \frac{n}{n+1}\,c_{n-1}^{(2)}~,
 n=0,1,2, \ldots ~.
\end{equation}
\begin{eqnarray} \label{eq:1567}c_1^{(2)}/c_0^{(2)} &=& 2i\,h_2  ~,\\
\label{eq:1196}c_2^{(2)}/c_0^{(2)} &=& -2\,h_2^2 + \frac{1}{2}~, \nonumber \\
\label{eq:1197}c_3^{(2)}/c_0^{(2)} &=& -\frac{2^3i}{3!}\,h_2^3 -
\frac{5i}{3}\,h_2~, \nonumber \\ \label{eq:1198}c_4^{(2)}/c_0^{(2)} &=&
\frac{2^4}{4!}\,h_2^4 -\frac{7}{3}\,h_2^2 + \frac{3}{8}~.
\nonumber
\end{eqnarray}
These formulae allow nothing to say about the possible convergence of the
series.
\section{The harmonic oscillator in the Hardy space of the complex
upper half-plane}
There is  the obvious question to be asked:

 The
usual quantum mechanical description of the harmonic oscillator is
in terms of the Hilbert space $L^2(\mathbb{R}, d\xi)$ with the
scalar product
\begin{equation}
  \label{eq:195}
 (g_2,g_1) =
\int_{-\infty}^{\infty}d\xi\,\bar{g}_2(\xi)\,g_1(\xi)\,,\, \xi = \beta x\,,\,
\beta = \sqrt{m\,\omega/\hbar}\,,
\end{equation}
and the
Hermite functions - the oscillator eigenfunctions of the stationary
Schr\"odinger equation - as an orthonormal basis (see, e.g.\ Ref.\
 \cite{mesa}):
\begin{equation}
  \label{eq:806}
 u_n(\xi)= \frac{e^{-\xi^2/2}}{2^{n/2}\,\sqrt{n!}\,\pi^{1/4}}\,H_n(\xi)\,, 
\end{equation}
where $H_n(\xi)$ is the $n$-th Hermite polynomial.

 If the same
quantum mechanics is to be described by the Hilbert space $H^2(S^1,d\vp)$
with the scalar product \ref{eq:192} and the eigenfunctions \ref{eq:189},
 what is the
relationship between the two spaces? 

 The answer is somewhat subtle and not quite straightforward\footnote{This
was, unfortunately, obscured in the first (even printed) version of the
 present paper, though it is 
indicated in the frequently quoted Ref.\ \cite{bo}.}: I shall first
discuss the relationship between the Hardy space $H^2(S^1,d\vp)$ of the
circle and the unitarily equivalent Hardy space $H^2(\mathbb{R},d\xi)$
on the real line.

 The  space $H^2(\mathbb{R},d\xi)$
 consists of that closed subspace of functions
$g(\xi) \in L^2(\mathbb{R},d\xi)$ wich are limits (in an appropriate 
topology) 
\begin{equation}
  \label{eq:837}
 g(\xi) = \lim_{\eta \to 0} g(\xi +i\,\eta) 
\end{equation}
of functions $g(z=\xi+i\,\eta)$ which are holomorphic in the upper
complex half-plane. Afterwards I shall explain how the elements
of $L^2(\mathbb{R},d\xi)$ can be projected on the subspace
 $H^2(\mathbb{R},d\xi)$ and and how the two spaces are unitarily related.
 The space $H^2(\mathbb{R},d\xi)$ is of physical
interest by itself.
\subsection{The relationship between the Hardy space of the circle and
that of the real line}
 The transformation  \cite{sa1,hard2}:
\begin{equation}
  \label{eq:196}
 \xi = \frac{1}{i}\, \frac{e^{i\,\varphi}
-1}{e^{i\,\varphi} +1}= \tan\frac{\varphi}{2}~,~~ e^{i\,\varphi} =
\frac{1+i\,\xi}{1-i\,\xi}~,~ \varphi = 2\arctan\xi\,,
\end{equation}
maps the
unit circle $S^1 =\partial\mathbb{D}$ onto the real line $\mathbb{R}$
and vice versa. 

Given  a function $f(\vp) \in H^2(S^1, d\vp)$, we can define
a function $g^{(+)}(\xi) \in H^2(\mathbb{R},d\xi)
 \subset L^2(\mathbb{R},d\xi)$
 and vice versa:
\begin{eqnarray} \label{eq:1562}g^{(+)}(\xi) &=&
\frac{1}{\sqrt{\pi}\,(1-i\,\xi)}\,f\left(e^{i\,\varphi} =
\frac{1+i\,\xi}{1-i\,\xi}\,,\, \varphi = 2 \arctan \xi\right )\,,
\\ \label{eq:1187}f(\varphi) &=& \frac{2\,\sqrt{\pi}}{1+e^{i\,\varphi}}\,
g^{(+)}(\,\xi=
\tan (\varphi/2)\,)~, \\ \label{eq:1188}&& (1+e^{i\,\varphi})(1-i\,\xi) = 2~.
\end{eqnarray} The mapping is unitary because
\begin{equation}
  \label{eq:197}
(g_2^{(+)},g_1^{(+)})=\int_{-\infty}^{\infty}d\xi\,\bar{g}_2^{(+)}(\xi)\,
g_1^{(+)}(\xi) =
\frac{1}{2\pi}\int_0^{2\pi} d\varphi\,
\bar{f}_2(\varphi)f_1(\varphi)~,~\frac{d\varphi}{2} =
\frac{d\xi}{1+\xi^2}~.
\end{equation}
The unitary transformation \ref{eq:1562} and \ref{eq:1187} maps
 the basis \ref{eq:189}
onto an orthonormal basis
\begin{equation}
  \label{eq:687}
  v_n(\xi) = \frac{1}{\sqrt{\pi}(1-i\,\xi)}\left(\frac{1+i\,\xi}{1-i\,\xi}
\right)^n\,,\,n=0,1,\ldots,
\end{equation}
 of $H^2(\mathbb{R},d\xi)$.

Except for an irrelevant phase, the basis \ref{eq:687} coincides with
the basis \ref{eq:528} for $k=1/2,\, \Im(w) = v =0$, and the scalar
product \ref{eq:529}. 

If we replace the real variable $\xi$ in Eq.\ \ref{eq:687} by the complex
$z=\xi + i\,\eta$ one, then the functions $v_n(z)$ are holomorphic in the
upper half-plane and have a pole of order $n+1$ in the lower on.

A formal transformation of the type \ref{eq:1562}
 was already discussed by London
 \cite{lon2}, using methods of Jordan and Pauli!

 The functions \ref{eq:687}
 are  eigenfunctions, with eigenvalues $n+1/2$,  of the (Hamilton) 
operator
\begin{eqnarray}
  \label{eq:809}
  \tilde{K}_0=\tilde{H}_{osc}& =&\frac{1}{2i}[\xi +(\xi^2+1)
\frac{d}{d\xi}\,]\,, \\
\tilde{H}_{osc}\,v_n(\xi) &=& (n+\frac{1}{2})\,v_n(\xi)\,. \label{eq:820}
\end{eqnarray}
This follows from Eq. \ref{eq:1684} of Appendix B. The associated Eqs.\
\ref{eq:1486} give - after an appropriate redefinition of the phases -
\begin{eqnarray}
  \label{eq:810}
  \tilde{K}_+ &=& \frac{1+i\xi}{2}\,[1-i(1+i\xi)\frac{d}{d\xi}]\,, \\
  \tilde{K}_+v_n(\xi) &=& (n+1)\,v_{n+1}(\xi)\,, \label{eq:819} \\
\tilde{K}_- &=& -\frac{1-i\xi}{2}\,[1+i(1-i\xi)\frac{d}{d\xi}]\,, 
 \label{eq:818} \\
\tilde{K}_- v_n(\xi) &=& n\,v_{n-1}\,, \label{eq:822}
\end{eqnarray}
which imply
\begin{eqnarray}
  \label{eq:811}
 \tilde{K}_1 &=& \frac{i}{2}\,[\xi+(\xi^2-1)\,\frac{d}{d\xi}]\,. \\
\tilde{K}_2 &=& \frac{1}{i}\,(\frac{1}{2}+\xi\,\frac{d}{d\xi})\,,\label{eq:821}
\end{eqnarray}
The eigenfunctions of $\tilde{K}_1$ and $\tilde{K}_2$ are discussed below.

The eigenfunctions \ref{eq:806} have a remarkable property:
Their density
\begin{equation}
  \label{eq:812}
  \rho_{v;n}(\xi) = |v_n(\xi)|^2 = \frac{1}{\pi(1+\xi^2)}
\end{equation}
 is independent of $n\,$: $\rho_{v;n} = \rho_v$! 

This is a remarkable difference compared to the densities $|u_n(\xi)|^2$
of the eigenfunctions \ref{eq:687} which depend strongly on $n\,$! The
property \ref{eq:812} is, of course, a consequence of the fact that the
eigenfunctions \ref{eq:189} have the constant density $1$\,!

In probability theory and statistics the density \ref{eq:812} is
called the ``density of the Cauchy distribution'' \cite{cau1} (in physics:
``Lorentz'' \cite{vank} or ``Breit-Wigner'' distribution).

One property of the Cauchy distribution is that its moments
\begin{equation}
  \label{eq:807}
  \langle \xi^m \rangle = \int_{-\infty}^{\infty}d\xi\,\xi^m
\rho_v(\xi)
\end{equation}
do not exist for $m \geq 1$. (For odd $m$ the integral is formally $=0$ because
the integrand is an odd function. However, the characteristic function
\begin{equation}
  \label{eq:838}
  \phi(t) = \int_{-\infty}^{\infty}d\xi\,\rho_v(\xi)\,e^{i\,t\,\xi} =e^{-|t|}
\end{equation}
 of the Cauchy distribution is not differentiable at $t=0\,$!)

 The divergence of the moment integrals means especially that
 the functions \ref{eq:806}
do not belong to the domain of definition of the multiplication operator
$\xi$ because $\xi\,v_n(\xi)$ is not square integrable\,!
 On the other hand, they do belong to the domain of the operator
 $(1/i)d/d\xi$\,, because
 \begin{equation}
   \label{eq:808}
   \frac{1}{i}\frac{dv_n}{d\xi}=\frac{(1+i\xi+2n)(1+i\xi)^{n-1}}{\sqrt{\pi}\,
(1-i\xi)^{n+2}}\,.
 \end{equation}

If we introduce the variable x of Eq.\ \ref{eq:195} - which has the
dimension of a length, whereas $\xi$ is dimensionless -  and the measure $dx$,
 then the density \ref{eq:812}
takes the form
\begin{equation}
  \label{eq:813}
  \rho_{v}(x;\lambda) = \frac{\beta}{\pi(1+\beta^2x^2)}=
\frac{\lambda}{\pi(\lambda^2+x^2)}\,,\,\lambda =1/\beta\,.
\end{equation}
It has its maximum for $x=0$, with the value
\begin{equation}
  \label{eq:814}
  \rho_{v}(x=0;\lambda) = \frac{\beta}{\pi}=\frac{1}{\pi\,\lambda}\,.
\end{equation}

If we denote by $x_{(1/2)}$ the arguments where $\rho_{v}(x;\lambda)$ takes
half its maximum value \ref{eq:814}, then
\begin{equation}
  \label{eq:815}
  x_{\pm(1/2)} = \pm \frac{1}{\beta}= \pm \lambda\,.
\end{equation}
The points $x_{(\pm i)}$ of inflexion of the function \ref{eq:813}
are
\begin{equation}
  \label{eq:910}
  x_{(\pm i)} =\pm \frac{\lambda}{\sqrt{3}}\,.
\end{equation}
In the (classical!) limits $\omega \to \infty$ or $\hbar \to 0$ we
have $\beta \to \infty\,,\,\lambda \to 0$, where the density
 \ref{eq:813} approaches the
delta-function (see, e.g.\ Ref.\ \cite{gel}, vol.\ I):
\begin{equation}
  \label{eq:816}
  \lim_{\lambda \to 0} \rho_{v}(x;\lambda) = \delta(x)\,.
\end{equation}
Shifting the center of the distribution density from $x=0$ to $x=a$ 
yields the density (I drop the index ``v'' in the following)
\begin{equation}
  \label{eq:839}
  \rho(x;\lambda,a) = \frac{1}{\pi}\frac{\lambda}{\lambda^2 + (x-a)^2}\,,
\end{equation} which has the characteristic function
\begin{equation}
  \label{eq:840}
  \phi(t) = e^{i\,a\,t-\lambda\,|t|}\,.
\end{equation}
The characteristic function \ref{eq:840} has an important property which
relates the Cauchy density \ref{eq:839} to an associated Markovian Cauchy
 process \cite{caup} with
respect to the parameter $\lambda$: 

If $\rho_i= \rho(x_i;\lambda_i,a_i)\,,\,i=1,2,$ are the Cauchy densities of two
independent random variables $x_i$ with corresponding characteristic functions
$\phi_i(t)$, then the product of those functions is given by
\begin{equation}
  \label{eq:841}
  \phi_{3}(t) = \phi_2(t)\,\phi_1(t) = e^{i(a_2+a_1)-(\lambda_2+\lambda_1)
|t|}\,,
\end{equation}
i.e.\ $\phi_{3}$ is again the characteristic function of a Cauchy distribution
which is a convolution of the two original ones:
\begin{equation}
  \label{eq:842}
  \rho(x_3;\lambda_3=\lambda_1+\lambda_2\,,\, a_3=a_1+a_2) =
\int_{-\infty}^{\infty}dx\,\rho(x_3-x;\lambda_2,a_2)\,\rho(x;\lambda_1,
a_1)\,.
\end{equation}
This important ``Chapman-Kolmogorov'' property here is a consequence
of the fact that the inverse Fourier
transform of the product \ref{eq:841} yields the convolution \ref{eq:842}.
Important is the resulting semi-group property in $\lambda$. The $a_i$
may be put to zero.

As the moments of the Cauchy distribution do not exist one need some
other means in order to characterize salient features of the distribution.
One possibility is to use so-called {\em ``quantils of order $p$''}
 \cite{cau2}:

The Cauchy distribution function $F(x)$ of the density $\rho$ is given by
\begin{equation}
  \label{eq:843}
  F(x) = \int_{-\infty}^x du\,\rho(u;\lambda,a) = \frac{1}{2}+
\frac{1}{\pi}\arctan\left(
\frac{x-a}{\lambda}\right)\,.
\end{equation}
The $p$th quantil is defined as the value $x=x_p$ for which
\begin{equation}
  \label{eq:844}
  F(x_p) = p\,,\, 0<p<1\,.
\end{equation}
$x_{1/2}$ is called the ``{\em median}''. Here we have
\begin{equation}
  \label{eq:845}
  x_{1/2} = a\,.
\end{equation}
For the lower and upper ``{\em quartils}'' $x_{1/4}$ and $x_{3/4}$ we get
\begin{equation}
  \label{eq:846}
  x_{1/4}= a-\lambda\,,~x_{3/4} =a+\lambda\,,~\lambda =\frac{1}{2}(x_{3/4}-
x_{1/2})\,,~a=\frac{1}{2}(x_{1/4}+x_{3/4})=x_{1/2}\,.
\end{equation}
Thus, $(x_{3/4}-x_{1/4})/2$ is a measure for the width of the distribution.

Whereas the conventional eigenfunctions \ref{eq:806} are eigenfunctions
of the parity operation $\xi \to -\xi$, the eigenfunctions \ref{eq:687}
are not, but they obey the relation
\begin{equation}
  \label{eq:823}
  \bar{v}_n(-\xi) = v_n(\xi)\,.
\end{equation}
This corresponds to the property
\begin{equation}
  \label{eq:824}
  \overline{\tilde{H}}_{osc}(-\xi) = \tilde{H}_{osc}(\xi)\,.
\end{equation}

The self-adjoint position and momentum operators \ref{eq:730} and
\ref{eq:732} here take the form
\begin{eqnarray}
  \label{eq:817}
  \tilde{Q} &=& \frac{1}{\sqrt{2}}[\tilde{K}_+(\tilde{K}_0+1/2)^{-1/2}
+ (\tilde{K}_0 +1/2)^{-1/2}\tilde{K}_-]\,, \\
\tilde{P} &=& \frac{i}{\sqrt{2}}[\tilde{K}_+(\tilde{K}_0+1/2)^{-1/2}
- (\tilde{K}_0 +1/2)^{-1/2}\tilde{K}_-]\,. \label{eq:911}
\end{eqnarray}
These operators have the same matrix elements with respect to the
basis \ref{eq:687} as the operators $Q=x$ and $P=(1/i)d/dx$ have with
respect to the basis \ref{eq:806}!
\subsection{The eigenfuctions of $\tilde{K}_1$ and $\tilde{K}_2$}
I briefly discuss the eigenfunctions $\tilde{f}_{h_1}(\xi)$ and 
$\tilde{f}_{h_2}(\xi)$ of the operators \ref{eq:811}
and \ref{eq:821}. The procedure is similar as in Sec.\ 4.3: 
For $\xi^2 >1$ we have
\begin{equation}
\label{eq:852}
  \tilde{f}_{h_1}(\xi) = C_+\, \left(\frac{\xi+1}{\xi-1}\right)^{i\,h_1}\,
(\xi^2-1)^{-1/2}
 \text{ for } \xi^2 >1\,, h_1 \in \mathbb{R}\,,\,C_+= \text{const}.\,,
\end{equation}
and for $\xi^2 <1$ we obtain analogously
\begin{equation}
  \label{eq:848}
  \tilde{f}_{h_1}(\xi) =C_-\, \left(\frac{1+\xi}{1-\xi}\right)^{i\,h_1}\,
(1-\xi^2)^{-1/2}\,\text{ for } \xi^2<1\,, h_1 \in \mathbb{R}\,.
\end{equation}
Setting $\xi=1+\epsilon\,,\,\epsilon >0 $  for the relation \ref{eq:852}
 and $\xi =1-\epsilon$
for \ref{eq:848} near $\xi =1$ and requiring continuity of the eigenfunctions
yields $C_-=C_+ =C$.

In order to determine the constants $C$ it is useful to
introduce a new variable:
\begin{equation}
  \label{eq:849}
  e^u =\frac{\xi+1}{\xi-1}\,,\,\xi^2>1\,,\,du=-2(\xi^2-1)^{-1}d\xi\,, 
\end{equation} where $
 u\to -\infty \text{ for } \xi \to -1^-\,,\, u\to 0^- \text{ for }
\xi \to -\infty\,;\, u \to \infty \text{ for } \xi \to 1^+\,,\,
u \to 0^+ \text{ for } \xi \to \infty\,, $ so that 
\begin{equation}
  \label{eq:850}
 \int_{-\infty}^{-1}d\xi \bar{\tilde{f}}_{h_1'}(\xi)\,\tilde{f}_{h_1}(\xi)
 = \frac{|C|^2}{2}
\int_{-\infty}^0du\,e^{i(h_1-h_1')u}
\end{equation} and
\begin{equation}
  \label{eq:851}
 \int^{\infty}_{1}d\xi\, \bar{\tilde{f}}_{h_1'}(\xi)\tilde{f}_{h_1}(\xi)
 = \frac{|C|^2}{2}
\int^{\infty}_0du\,e^{i(h_1-h_1')u}\,.
\end{equation} Putting
\begin{equation}
  \label{eq:853}
  e^u =\frac{1+\xi}{1-\xi}\,,\,du =2(1-\xi^2)d\xi
\end{equation}
for $\xi^2 <1$ gives correspondingly
\begin{equation}
  \label{eq:854}
 \int^{1}_{-1}d\xi\, \bar{\tilde{f}}_{h_1'}(\xi)\tilde{f}_{h_1}(\xi)
 = \frac{|C|^2}{2}
\int_{-\infty}^{\infty}du\,e^{i(h_1-h_1')u}\,. 
\end{equation} Adding the three contributions yields
\begin{equation}
  \label{eq:855}
  \int_{-\infty}^{\infty}d\xi\,\bar{\tilde{f}}_{h_1'}(\xi)\tilde{f}_{h_1}(\xi)
=2\pi\,|C|^2\,\delta (h_1'-h_1)\,, 
\end{equation}
so that 
\begin{equation}
  \label{eq:856}
  C=\frac{1}{\sqrt{2\pi}}
\end{equation}
gives the appropriate normalization:
\begin{equation}
  \label{eq:857}
\tilde{f}_{h_1}(\xi) =\left\{\begin{array}{l} {\displaystyle
  \frac{1}{\sqrt{2\pi}}\left(
\frac{\xi+1}{\xi-1}\right)^{ih_1}}(\xi^2-1)^{-1/2} \text{ for } \xi^2 >1\,, \\
{\displaystyle \frac{1}{\sqrt{2\pi}}\left(
\frac{1+\xi}{1-\xi}\right)^{ih_1}}(1-\xi^2)^{-1/2} \text{ for } \xi^2 < 1\,.
 \end{array} \right. 
\end{equation}
In a similar manner we get
\begin{equation}
  \label{eq:847}
 \tilde{f}_{h_2}(\xi) =\left\{\begin{array}{l} {\displaystyle
  \frac{1}{2\sqrt{\pi}}\,\xi^{ih_2-1/2}} \text{ for } \xi >0\,, \\
{\displaystyle \frac{1}{2\sqrt{\pi}}\,(-\xi)^{ih_2-1/2}} \text{ for } \xi <0\,,
 \end{array} \right.  
\end{equation} with the normalization
\begin{equation}
  \label{eq:858}
  (\tilde{f}_{h_2'},\tilde{f}_{h_2}) = \delta(h_2'-h_2)\,.
\end{equation}
\subsection{The Fourier transform of $g(\xi) \in H^2(\mathbb{R},\,d\xi)$}
The Fourier transforms\footnote{It is convenient here to have a different
sign convention in the exponent compared to  the one used in quantum
 mechanics.}
\begin{equation}
  \label{eq:859}
  \hat{g}(p) =\frac{1}{\sqrt{2\pi}}\,\int_{-\infty}^{\infty}d\xi
\,g(\xi)\,e^{-i\,p\,\xi}\,,~g(\xi) \in H^2(\mathbb{R},\,d\xi)
\end{equation}
have the important property (Paley-Wiener \cite{pal}) that
\begin{equation}
  \label{eq:860}
  \hat{g}(p) = 0 \text{ for } p < 0\,,
\end{equation}
and we have the inversion
\begin{equation}
  \label{eq:861}
  g(\xi) = \frac{1}{\sqrt{2\pi}}\,\int_{0}^{\infty}dp
\,\hat{g}(p)\,e^{i\,\xi\,p}\,.
\end{equation} Furthermore
\begin{equation}
  \label{eq:867}
 \int_0^{\infty}dp\,|\hat{g}(p)|^2 = \int_{-\infty}^{\infty}d\xi\,
|g(\xi)|^2\,. 
\end{equation}
The Fourier transforms of the basis functions \ref{eq:687} are \cite{ertrI}
\begin{equation}
  \label{eq:862}
  \hat{v}_n(p) = \sqrt{2}(-1)^n\,e^{-p}L_n(2p)\,,\,n=0,1,\ldots\,,
\end{equation}
where the $L_n$ are Laguerre's polynomials \cite{lag}.
The $\hat{v}_n(p)$ are eigenfunctions of the operator
\begin{eqnarray}
  \label{eq:863}
  \hat{K}_0& =& \frac{1}{2}\left(p-\frac{d}{dp}-p\frac{d^2}{dp^2}\right)\,, \\
\hat{K}_0\,\hat{v}_n(p) &=&(n+1/2)\,\hat{v}_n(p)\,. \label{eq:864}
\end{eqnarray}
The other generators have the form
\begin{eqnarray}
  \label{eq:865}
  \hat{K}_1 & =& \frac{1}{2}\left(p+\frac{d}{dp}+p\frac{d^2}{dp^2}\right)\,, \\
\hat{K}_2 &=& i\,\left(\frac{1}{2}+p\,\frac{d}{dp}\right)\,. \label{eq:866}
\end{eqnarray} Using the properties of $L_n$ \cite{lag} one verifies that
\begin{eqnarray}
  \label{eq:868}
  \hat{K}_+\hat{v}_n(p)&=&(n+1)\,\hat{v}_{n+1}(p)\,, \\
\hat{K}_-\hat{v}_n(p)&=&n\,\hat{v}_{n-1}(p)\,.
\end{eqnarray}
The eigenvalue equation
\begin{equation}
  \label{eq:869}
  \hat{K}_1\hat{f}_{h_1}(p) = h_1\,\hat{f}_{h_1}(p)
\end{equation}
has the regular solution \cite{er4}
\begin{equation}
  \label{eq:870}
 \hat{f}_{h_1}(p) = C\,e^{-ip}\,_{1}F_1(1/2-ih_1,1; 2ip)\,,\,h_1
 \in \mathbb{R}\,, 
\end{equation}
where $_{1}F_1(a,c;z)$ is the usual regular confluent hypergeometric
(``Kummer's'') function \cite{confl}.
The normalization constant $C$ can be obtained by comparing the relation
 \cite{mo}
\begin{equation}
  \label{eq:871}
  \lim_{\epsilon \to 0} \frac{1}{\sqrt{2\pi}}\int_0^{\infty}dp\,e^{-ip}
\,_{1}F_1(1/2-ih_1,1;2ip)\,
e^{i\,p\,(\xi +i\epsilon)}=\frac{i}{\sqrt{2\pi}}\left(\frac{\xi+1}{\xi-1}
\right)^{ih_1}(\xi^2-1)^{-1/2}\, \text{ for } \xi > 1\, 
\end{equation}
whith Eq.\ \ref{eq:857}. The result is
\begin{equation}
  \label{eq:872}
  \hat{f}_{h_1}(p) = -i\,e^{-ip}\,_{1}F_1(1/2-ih_1,1;2ip)\,,\,h_1 \in
\mathbb{R}\,.
\end{equation}
For the eigenfunctions of $\hat{K}_2$ we get (compare Eq.\ \ref{eq:847})
\begin{equation}
  \label{eq:873}
  \hat{f}_{h_2}(p)= \frac{1}{\sqrt{2\pi}}\,p^{-ih_2-1/2}\,,\,h_2 \in
 \mathbb{R}\,.
\end{equation}
\subsection{Relationships between $L^2(\mathbb{R},d\xi)$ and 
$H^2(\mathbb{R},d\xi)$} I briefly mention two essential relationships
between the spaces   $L^2(\mathbb{R},d\xi)$ and 
$H^2(\mathbb{R},d\xi)$ which are of interest here: $H^2(\mathbb{R},d\xi)$
as a closed subspace of  $L^2(\mathbb{R},d\xi)$ \cite{hard2}
 and the unitary correspondence
between the two. 
\subsubsection{$H^2(\mathbb{R},d\xi)$ as a subspace of  $L^2(\mathbb{R},d\xi)$}
Let me start with $H^2(\mathbb{R},d\xi)$ as a subspace: it consists of those
elements $g^{(+)}(\xi) \in L^2(\mathbb{R},d\xi)$ which are limits
 $\lim_{\eta \to 0}
g(z=\xi+i\,\eta)$ of functions $g(z)$ which are holomorphic in the
 upper complex
half-plane ($\eta >0$). 

If $g(\xi) \in  L^2(\mathbb{R},d\xi)$ its projection $g^{(+)}(\xi)$ 
onto the subspace $H^2(\mathbb{R},d\xi)$ is obtained by (double) 
Fourier transformation \cite{h2l2}, namely
\begin{eqnarray}
  \label{eq:874}
  \hat{g}(p) &=& \frac{1}{\sqrt{2\pi}}\int_{-\infty}^{\infty}d\xi\,g(\xi)\,
e^{-i\,p\,\xi}\,,\\ g^{(+)}(\xi) &=&   \frac{1}{\sqrt{2\pi}}
\int_{0}^{\infty}dp\,\hat{g}(p)\,
e^{i\,\xi\,p}\,.\label{eq:875}
\end{eqnarray}
We know already that $\hat{g}(p) = 0$ for $p \leq 0$ if $g(\xi) \in 
 H^2(\mathbb{R},d\xi)$
(Paley--Wiener theorem). So Eq.\ \ref{eq:875} is the appropriate projection.
The complementary subspace  
\begin{equation}
  \label{eq:876}
\bar{H}^2(\mathbb{R},d\xi) = L^2(\mathbb{R},d\xi)\setminus 
 H^2(\mathbb{R},d\xi)   
\end{equation}
consists of the functions
\begin{equation}
  \label{eq:877}
   g^{(-)}(\xi) =   \frac{1}{\sqrt{2\pi}}
\int_{-\infty}^{0}dp\,\hat{g}(p)\,
e^{i\,\xi\,p}
\end{equation}
They are limits of functions $g(z)$ which are holomorphic in the {\em lower}
half-plane.

 A basis of $\bar{H}^2(\mathbb{R},d\xi)$ is given by
 $\{\bar{v}_n(\xi)\}$\,.

Obviously we have
\begin{equation}
  \label{eq:890}
  g(\xi) =g^{(+)}(\xi)+g^{(-)}(\xi)\,.
\end{equation}

For the basis functions \ref{eq:806} we have the Fourier transforms \cite{fher}
\begin{equation}
  \label{eq:878}
  \hat{u}_n(p) = \frac{(-i)^n}{(2^n\,n!\,\sqrt{\pi})^{1/2}}\,
 e^{-p^2/2}\,H_n(p)\,, 
\end{equation}
so that
\begin{equation}
  \label{eq:879}
  u_n^{(+)}(\xi) = \frac{1}{\sqrt{2\pi}}\int_0^{\infty}dp\,\hat{u}_n(p)\,
e^{i\,\xi\,p}\,.
\end{equation}
The integrals \ref{eq:879} may be evaluated with the help of the generating
function \cite{lag}
\begin{equation}
  \label{eq:880}
  \sum_{n=0}^{\infty}\frac{t^n}{n!}\,H_n(p) = e^{-t^2+2\,t\,p}\,,
\end{equation}
which yields
\begin{equation}
  \label{eq:881}
 F(t;\xi)= \sum_{n=0}^{\infty}\frac{t^n}{\sqrt{n!}}u_n^{(+)}(\xi) = 
\frac{e^{t^2/2}}{2^{1/2}\,\pi^{3/4}}\int_0^{\infty}dp\,e^{-p^2/2
-i(\sqrt{2}t-\xi)
\,p}\,.
\end{equation}
Evaluation of integral in the last Eq.\ follows from \cite{laplp2}
\begin{equation}
  \label{eq:882}
  \int_0^{\infty} dp\,e^{-p^2/2-z\,p} = \sqrt{\frac{\pi}{2}}\,e^{z^2/2}\,
[1-\text{erf}(z/\sqrt{2})] \equiv \sqrt{\frac{\pi}{2}}\,e^{z^2/2}\,
\text{erfc}(z/\sqrt{2}) \,,
\end{equation}
where $\text{erf}(w)$ is the ``error function'' \cite{errf}
\begin{equation}
  \label{eq:883}
  \text{erf}(w)=\frac{2}{\sqrt{\pi}}\int_{0}^{w}du\,e^{-u^2}
\,
\end{equation}
and 
\begin{equation}
  \label{eq:894}
  \text{erfc}(w) =1-\text{erf}(w) = \frac{2}{\sqrt{\pi}}\int_{w}^{\infty}du
\,e^{-u^2}\,
\end{equation}
the ``complementary error function''.

For imaginary arguments $w=i\,v\,,\,v \in \mathbb{R}\,,$ $
\text{erf}(i\,v)\,$ is
 purely imaginary and generally $\text{erf}(-w) = -\text{erf}(w)$.

 For \ref{eq:881} we obtain
\begin{equation}
  \label{eq:884}
F(t;\xi) = \frac{1}{2\pi^{1/4}}\,e^{-(t^2/2-\sqrt{2}\,t\,\xi+\xi^2/2)}\,
\text{erfc}[i(t-\xi/\sqrt{2})]\,.  
\end{equation}

The functions
\begin{equation}
  \label{eq:885}
  u^{(+)}_n(\xi) = \frac{1}{\sqrt{n!}}\,\left.
\frac{d^nF(t;\xi)}{dt^n}\right|_{\displaystyle
 t=0}
\end{equation}
may be calculated with the help of the relation \cite{errf}
\begin{equation}
  \label{eq:886}
  \frac{d^{n+1}\text{erf}(b\,t+a)}{dt^{n+1}} =-\frac{2}{\sqrt{\pi}}\,
(-b)^{n+1}e^{-(b\,t+a)^2} H_n(b\,t+a)\,,\,n=0,1,\ldots\,,
\end{equation}
where here $b=i$ and $a =-i\,\xi/\sqrt{2}$. Thus
\begin{equation}
  \label{eq:888}
  \left.\frac{d^{n+1}\text{erf}[i(t-\xi/\sqrt{2})]}{d\,t^{n+1}}\right|_{
\displaystyle t=0}
 =\frac{2}{\sqrt{\pi}}\,
i^{n+1}e^{\xi^2/2}H_n(i\xi/\sqrt{2})\,. 
\end{equation}
As the $H_n(i\xi/\sqrt{2})$ are real for even $n$ and imaginary for
odd $n$ the r.h.\ side of Eq.\ \ref{eq:888} is always purely imaginary.

Examples:
\begin{eqnarray}
  \label{eq:887}
  u_0^{(+)}(\xi) &=& \frac{1}{2\,\pi^{1/4}}\,e^{-\xi^2/2}\,[1+
\text{erf}(i\,\xi/\sqrt{2})]\,, \\
 u_1^{(+)}(\xi) &=& \frac{1}{2\,(2\sqrt{\pi})^{1/2}}\,e^{-\xi^2/2}\,H_1(\xi)
\,[1+\text{erf}(i\,\xi/\sqrt{2})]-\pi^{-3/4}\,i\,. \label{eq:889}
\end{eqnarray}
The {\em real} parts of these functions are just one half of the original
$u_0(\xi)$ and $u_1(\xi)$. The other half comes from $u^{(-)}_0(\xi)$ and
$u^{(-)}_1(\xi)$ (see Eq.\ \ref{eq:890}).

Other interesting quantities are the ``transition'' amplitudes
\begin{eqnarray}
  \label{eq:891}
  c_{m,n} &=& (v_m,u^{(+)}_n) = (\hat{v}_m,\hat{u}_n) \\
&=& (-1)^m\,(-i)^n\,\frac{\sqrt{2}}{\pi^{1/4}\,(2^n\,n!)^{1/2}}
\int_0^{\infty}dp\,e^{-p^2/2-p}\,L_m(2p)\,H_n(p) \,.
\end{eqnarray}
They can be calculated with the help of the generating function \ref{eq:880}
and \cite{lag}
\begin{equation}
  \label{eq:893}
  \sum_{m=0}^{\infty}s^m\,L_m(u) = \frac{e^{\displaystyle 
u\,s/(s-1)}}{1-s}\,,|s|<1\,,
\end{equation}
which yield
\begin{eqnarray}
  \label{eq:895}
  c_{m,n} &=& \frac{1}{m!\,\sqrt{n!}}\,\left.\frac{d^{m+n}G(s,t)}{ds^m\,dt^n}
\right|_{\displaystyle s=0,\,t=0}\,, \\
G(s,t)&=& \sum_{m=0,n=0}^{\infty}c_{m,n}\,s^m \,\frac{t^n}{\sqrt{n!}} \\& =&
\frac{\pi^{1/4}}{s+1}\exp{\left[\frac{1}{2}\left(\frac{1-s}{1+s}\right)^2 +
\sqrt{2}\,\frac{1-s}{1+s}\,t\,i -\frac{1}{2}\,t^2\right]}\,\text{erfc}
\left(\frac{1}{
\sqrt{2}}\,\frac{1-s}{1+s} +t\,i\right),~~~\label{eq:896}
\end{eqnarray}
where again the relation \ref{eq:882} has been used.

Another possibility in order to evaluate the integrals \ref{eq:891}
is the use of the relation 
\begin{equation}
  \label{eq:897}
  \int_0^{\infty}dp \,p^n\,e^{p^2/2-p} = \frac{(-1)^n}{n!}\sqrt{\frac{\pi}{2}}
\left.\frac{d^n}{dz^n}\,\left[e^{z^2/2}\,\text{erfc}(\frac{z}{\sqrt{2}})\right]
\right|_{z=1}\,,
\end{equation}
which follows from \ref{eq:882}.

Examples:
\begin{eqnarray}
  \label{eq:898}
  c_{0,0} &=& \pi^{1/4}\,\sqrt{e}\,\text{erfc}(\frac{1}{\sqrt{2}}) =
 0.6965\,, \\
|c_{0,0}|^2 &=& 0.4851\,; \label{eq:900}\\
c_{1,0} &=& \pi^{1/4}\,\left[2\,\sqrt{\frac{2}{\pi}}-3\,\sqrt{e}\,
\text{erfc}(\frac{1}{\sqrt{2}})\right] = 0.0351\,, \label{eq:901}\\
|c_{1,0}|^2 &=& 0.0012\,; \label{eq:902}\\
c_{0,1} &=& \pi^{1/4}\,\left[\sqrt{2\,e}\,\text{erfc}(\frac{1}{\sqrt{2}})
-\frac{2}{\sqrt{\pi}}\right]\,i = -0.5173\,i\,, \label{eq:903} \\
|c_{0,1}|^2 &=& 0.2676\,; \\
c_{0,2}&=& \pi^{-1/4}\,\left[1-\sqrt{\frac{e\,\pi}{2}}\,\text{erfc}(\frac{1}{
\sqrt{2}})\right] = 0.2586\,, \\
|c_{0,2}|^2 &=& 0.0669\,.
\end{eqnarray}
Here the numerical value \cite{errf1}
\begin{equation}
  \label{eq:899}
  \text{erfc}(\frac{1}{\sqrt{2}}) = 0.31731\ldots
\end{equation}
has been used.

The relationship between the Laguerre polynomials $L_m(p)$ and the
Hermite polynomials $H_n(p)$ for $p \geq 0$ may briefly sketched as
follows:

The Laguerre polynomials form an orthonormal basis of the Hilbert
 space $L_{(+)}^2(p \geq 0, dp)$ with weight function $\exp(-p)$ \cite{lag}:
 \begin{equation}
   \label{eq:892}
   \int_0^{\infty}dp\,e^{-p}\,L_m(p)\,L_n(p) = \delta_{m\,n}\,.
 \end{equation}
The situation with respect to the Hermite functions \ref{eq:806} is somewhat
more complicated: 

From $H_n(-p)=(-1)^n\,H_n(p)$ it follows that
\begin{equation}
  \label{eq:904}
  \int_{-\infty}^{\infty}dp\,e^{-p^2}H_m(p)\,H_n(p) = [1+(-1)^{m+n}]\,
\int_{0}^{\infty}dp\,e^{-p^2}H_m(p)\,H_n(p)= 2^m\,m!\,\sqrt{\pi}
\,\delta_{m\,n}\,,
\end{equation}
which shows that the functions $H_m(p)\, \exp{(-p^2/2)}$ form an orthogonal
set on $L_{(+)}^2(p \geq 0, dp)$ for either $m$ and $n$ both even or $m$ and
$n$ both odd. Each of the sets $\{H_{2n}(p)\,,\,n=0,1,\ldots\}$ and 
$\{H_{2n+1}(p)\,,\,n=0,1,\ldots\}$ provide a basis in the following sense:
They are related to the special generalized Laguerre polynomials
\cite{lag} 
\begin{eqnarray}
  \label{eq:905}
  H_{2n}(\sqrt{u}) &=& (-1)^n\,2^{2n}\,n!\,L_n^{-1/2}(u)\,, \\
H_{2n+1}(\sqrt{u}) &=& (-1)^n\,2^{2n+1}\,n!\,\sqrt{u}\,L_n^{1/2}(u)\,,
 \label{eq:906}
\end{eqnarray}
where
\begin{equation}
  \label{eq:907}
  L_n^{\alpha}(u) = \begin{pmatrix} n+\alpha \\ n\end{pmatrix}\,
_{1}F_1(-n,\,\alpha+1;u)\,,\,\alpha > -1\,,
\end{equation}
and
\begin{equation}
  \label{eq:908}
  \int_0^{\infty}du\,e^{-u}\,u^{\alpha}\,L_m^{\alpha}(u)L_n^{\alpha}(u)
=\Gamma(\alpha+1)\, \begin{pmatrix} n+\alpha \\ n\end{pmatrix}\,\delta_{m\,n}
\,.
\end{equation}
For fixed $\alpha >-1$ the sets $\{ L_n^{\alpha}(u)\,,\,n=0,1,\ldots \}$
together with the weight function $u^{\alpha}\,\exp{(-u)}$ form an orthonormal
basis for the Hilbert space $L_{(+)}^2(u \geq 0, du)$\,.
With the help of the relations \ref{eq:905} and \ref{eq:906} one finds
\cite{erdht2}
\begin{equation}
  \label{eq:909}
  \int_0^{\infty}dp\,e^{-p^2}\,H_{2m}(p)\,H_{2n+1}(p) = 
2^{2(m+n)}\,(m-n+1/2)_n\,
(n-m+3/2)_m\,,
\end{equation}
which means that the functions $H_{2m}(p)\,\exp{(-p^2/2)}$ and
 $H_{2n+1}(p)\,\exp{(-p^2/2)}$ are no longer orthogonal on 
$L_{(+)}^2(p \geq 0, dp)$\,!
\subsubsection{The unitary equivalence of  $L^2(\mathbb{R},d\xi)$ and 
$H^2(\mathbb{R},d\xi)$}
We have seen that the operators \ref{eq:817} and \ref{eq:911}
have the same matrix elements with respect to the basis \ref{eq:687}
as the usual operators $\xi$ and $-i\,d/d\xi$ do have with respect to
the basis \ref{eq:806}. Thus, in view of the Stone--von Neumann theorem
(see Subsec.\ 1.3.1) the question of their unitary equivalence
arises\footnote{I thank K.\ Fredenhagen for clarifying remarks concerning this
problem.}. The answer is surprisingly simple: 

The genuine subspace  $H^2(\mathbb{R},d\xi) \subset L^2(\mathbb{R},d\xi)$
can be mapped unitarily onto $L^2(\mathbb{R},d\xi)$ by the prescription
\begin{equation}
  \label{eq:912}
  v_n(\xi) \Longleftrightarrow u_n(\xi)\,.
\end{equation}
Then any two functions
\begin{equation}
  \label{eq:913}
  g^{(+)}_j(\xi)= \sum_{n=0}^{\infty}c^{(j)}_n\,v_n(\xi) \in
 H^2(\mathbb{R},d\xi)\,,\,j=1,\,2,
\end{equation}
are mapped unitarily onto
\begin{equation}
  \label{eq:914}
  g_j(\xi)= \sum_{n=0}^{\infty}c^{(j)}_n\,u_n(\xi) \in
 L^2(\mathbb{R},d\xi)\,,\,j=1,\,2, 
\end{equation}
and vice versa. The mapping is unitary because
\begin{equation}
  \label{eq:915}
  (g_2^{(+)},g_1^{(+)}) = \sum_{n=0}^{\infty}\bar{c}_n^{(2)}\,c_n^{(1)} =
(g_2,g_1)\,.
\end{equation}
This implies the unitary correspondences
\begin{equation}
  \label{eq:916}
  \tilde{Q}  \Longleftrightarrow \xi\,,~~\tilde{P}  \Longleftrightarrow 
\frac{1}{i}\frac{d}{d\xi}\,.
\end{equation}
This is an explicit example of the Stone--von Neumann theorem \cite{sto}.

The above arguments may actually be generalized to representations with
$k \neq 1/2\,$: In that case the correspondence \ref{eq:912} is replaced
by
\begin{equation}
  \label{eq:825}
  |k,n\rangle \Longleftrightarrow u_n(\xi)
\end{equation}
and the operators $\tilde{Q}$ and $\tilde{P}$ of Eq.\ \ref{eq:916} by the
operators \ref{eq:730} - \ref{eq:733}. 

The unitary equivalence of the position and momentum operators does, however,
not imply that the associated generators $K_j$ for different $k$
 are unitarily equivalent!
They are not, of course, because they belong to different inequivalent 
irreducible unitary representations of the group $SO^{\uparrow}(1,2)$
or one of its covering groups!

\section{A few generalizations for $k \neq 1/2$}
Several of the above results related to the Hardy space $H^2$ of
the unit circle may be generalized to unitary representations with
$k \neq 1/2\,$ \cite{bo,boy,sa2}: 

 The idea again is to implement the
scalar product \ref{eq:184} first in terms of a series expansion and then
realize that series expansion by means of $H^2$. 

 One  starts by defining the self-adjoint operator $A_k$ on $H^2$
 which is diagonal
in the basis \ref{eq:189} of $H^2$ and which acts on it as
\begin{equation}
  \label{eq:227}
A_k\,e_n(\vp)= \frac{n!}{(2k)_n}\, e_n(\vp)\,.
\end{equation}
Then one can
define a $H^2$ related Hilbert space $H^2_{A_{k}}$ with the
scalar product
\begin{equation}
  \label{eq:228}
 (f_1,f_2)_k \equiv (f_1,A_k\,f_2)=
\sum_{n=0}^{\infty} \frac{n!}{(2k)_n}\,
 \bar{a}_n\,b_n~~
\end{equation}
for the functions \ref{eq:191}. The series \ref{eq:228}
 representing the scalar product of $H^2_{A_k}$ is obviously
 the same as \ref{eq:187} which represents the scalar product for
$\mbox{$\cal H$}_{\mathbb{D},\, k}$. This exhibits the very close
relationship between the two Hilbert spaces. More explicitly this means:

 An orthonormal basis for $H^2_{A_k}$ is given by
\begin{eqnarray}
 \label{eq:1568}\chi_{k,n}(\vp)&=& \sqrt{\frac{(2k)_n}{n!}}\,
  e_n(\vp),\, n =0,1,\ldots, \\
  \label{eq:1199}(\chi_{k,n_1},\chi_{k,n_2})_k &=&
  \delta_{n_1n_2}~.
   \end{eqnarray}
  The  operators $K_0,\,K_+,\,K_-$ now have the form \cite{boa}
  \begin{eqnarray} \label{eq:1569}K_0&=&\frac{1}{i}\,\partial_{\vp}+k~, \\
\label{eq:1200}K_+&=& e^{
  i\,\vp}(\frac{1}{i}\partial_{\vp}+2\,k) ~, \\ \label{eq:1201}K_-&=&
 e^{-i\,\vp}\frac{1}{i}\,\partial_{\vp}~~.
  \end{eqnarray} Their action on the basis functions \ref{eq:1568} is given by
  \begin{eqnarray} \label{eq:1570}K_0\chi_{k,n}&=&(k+n)\chi_{k,n}~~, \\
 \label{eq:1202}K_+\chi_{k,n}&=& [(2k+n)(n+1)]^{1/2}\chi_{k,n+1}~~, \\
 \label{eq:1203}K_-\chi_{k,n}&=& [(2k+n-1)n)]^{1/2}\chi_{k,n-1}~~.
\end{eqnarray}

It is important to realize that the  operators $K_0,\, K_+,\, K_-$
belong to a  representation which is unitary only with respect to
the scalar product \ref{eq:228}, not with respect to the scalar product
\ref{eq:192}!
This may be seen explicitly as follows: Applying the operators
$K_+$ and $K_-$ to the series
\begin{equation}
  \label{eq:229}
f_1(\vp)=\sum_{m=0}^{\infty}a_m\,\chi_{k,m}(\vp)~,~ ~f_2(\vp)=
\sum_{n=0}^{\infty}b_n\,\chi_{k,n}(\vp)~,
\end{equation}
using the relations \ref{eq:1202} and \ref{eq:1203} and the orthonormality
 \ref{eq:1199} yields
\begin{equation}
  \label{eq:230}
 (f_2,K_+f_1)_k = \sum_{n=0}^{\infty}
  [(2k+n)(n+1)]^{1/2}\,\bar{b}_{n+1}\,a_n = (K_-f_2,f_1)_k~~,
\end{equation}
which says that $K_-$ is the adjoint operator of $K_+$ with respect
 to the scalar product \ref{eq:228}. But one sees immediately that
 this is not so
 with respect to the scalar product \ref{eq:192}! 

 The reproducing kernel here is
\begin{equation}
  \label{eq:231}
 \Delta_k(\vp_2,\vp_1)
 =\sum_{n=0}^{\infty} \bar{\chi}_{k,n}(\vp_2)\,\chi_{k,n}(\vp_1)=
 (1-e^{i\,(\vp_1-\vp_2)})^{-2k}~.
\end{equation}
Because
\begin{equation}
  \label{eq:232}
 \lim_{n \to \infty}
 \left( \frac{(2k)_n}{n!}\right)^{1/n} = 1\,,
\end{equation}
the radius of convergence of the series \ref{eq:231} is 1 and the
 kernel becomes singular for $\vp_2=\vp_1$. In the
 calculations one has to replace the basic functions $e_n(\vp)$ here too by
 $(1-\epsilon)\,e_n(\vp)$ and take the limit $\epsilon \to 0$ at
 the end. 

For the coherent states \ref{eq:78}, \ref{eq:136}  and \ref{eq:172}
 one gets here
 from the basis \ref{eq:1568} the functions \begin{eqnarray} 
\label{eq:1571}f_{k,z}(\vp)
 &=&
 \frac{1}{\sqrt{g_k(|z|^2)}}\,e^{\displaystyle z\,e^{i\,\vp}}~,\\
\label{eq:1204}f_{k,\lambda}(\vp) &=&
(1-|\lambda|^2)^k\,(1-\lambda\,e^{i\,\vp})^{-2k}~, \\ \label{eq:1205}f_{k,
\alpha}(\vp)
&=&\sum_{n=0}^{\infty}\sqrt{(2k)_n}\frac{(\alpha\,e^{i\,\vp})^n}{n!}~.
\end{eqnarray}
From the expressions \ref{eq:1200} and \ref{eq:1201} we obtain the operators
 \begin{eqnarray} \label{eq:1572}K_1 &=& \frac{1}{2}(K_+-K_-) =
\cos\vp\frac{1}{i}\partial_{\vp} + k\,e^{i\,\vp}\,, \\ \label{eq:1206}K_2
&=&\frac{1}{2i}(K_+ -K_-) = \sin\vp \frac{1}{i}\partial_{\vp}
+\frac{k}{i}\,e^{i\,\vp}~. \end{eqnarray} They have the
eigenfunctions \begin{eqnarray} \label{eq:1573}f_{k,h_1}(\vp) &=& C_1 \,
|\cos\vp|^{-k}|\tan(\vp/2 + \pi/4)|^{i\,h_1}\,e^{-i\,k\,\vp}~,~ \\
\label{eq:1207}&& h_1 \in
 \mathbb{R}~,~ \vp \in (0,2\pi)~, \nonumber \\
 \label{eq:1208}f_{k,h_2}(\vp) &=& C_2\, |\sin\vp|^{-k}|
\tan(\vp/2)|^{i\,h_2}\,e^{-i\,k\,\vp}~, \\
 \label{eq:1209}&& h_2 \in
 \mathbb{R}~,~ \vp \in (0,2\pi)~. \end{eqnarray}
\chapter{Operators for $\mathbf{\cos\vp}$ and $\mathbf{\sin\vp}\,$?}
We have seen in Ch.\ 3 that the operators $K_1$ and $K_2$ by themselves are
well suited in order to determine the phase content of a state.
One may nevertheless ask whether there are ``reasonable''
operators for $\cos\vp$ and $\sin\vp$  in the present
framework
and how the London-Susskind-Glogower operators \ref{eq:1504} and 
\ref{eq:1010} fit in.  

In view of the relations \ref{eq:65} the following  operators in
 an irreducible unitary representation with Bargmann index $k$ are a
generalization of the ones in Eq.\ \ref{eq:17}:
\begin{eqnarray} \label{eq:1574} E_{k,-} &=&
(K_0+k)^{-1}K_- ~,
\\ \label{eq:690} E_{k,+} &=& K_+(K_0+k)^{-1} = (E_{k,-})^+~.
\end{eqnarray} For $k=1/2$ we get back the the operators \ref{eq:17}.

 Let us look at some  properties of the operators \ref{eq:1574} and
\ref{eq:690}: 

Using the relations \ref{eq:68} and \ref{eq:70} with $L=k(1-k)$ we get
 \begin{eqnarray}
\label{eq:1575}E_{k,-}\cdot E_{k,+} &=& 1 - \frac{2k-1}{K_0+k}~, \\
\label{eq:1210}E_{k,+}\cdot E_{k,-} &=& 1 - \frac{2k-1}{K_0+k-1}~,\\
\label{eq:1211}\langle k,0|E_{k,+}\cdot E_{k,-}|k,0\rangle &=& 0~\forall~k~.
\end{eqnarray} We see that the operators $E_{k,-}$ and $E_{k,+}$ are {\em not}
isometric for $k \neq 1/2\,$! The case $k=1/2$ is a very special one
 and not generic! 

The relation \ref{eq:1211} follows from \ref{eq:1210}
 even for $k=1/2$ if
one takes the limit $k \to 1/2$ after forming the expectation value
of \ref{eq:1210} with respect to $|k,0 \rangle$. \\
The operators \ref{eq:1574} and \ref{eq:690} obey the commutation relation
\begin{eqnarray} \label{comE} [E_{k,-}, E_{k,+}] &=&
 \frac{2k-1}{(K_0+k)(K_0+k-1)}~, \\ \label{eq:1212}\langle k,0|
 [E_{k,-}, E_{k,+}]|k,0 \rangle &=& \frac{1}{2k}~. \end{eqnarray}
In the case $k=1/2$ the expectation values of the commutator \ref{comE}
 vanish for all states, except for the ground state: \ref{eq:1212}. This
is well-known and follows immediately from the relations \ref{eq:18},
but the situation is obviously different for $k\neq 1/2$!

 The obvious generalizations of the $\cos\vp$ - and
$\sin\vp$ - operators \ref{eq:1504} and \ref{eq:1010} are
\begin{eqnarray}\tilde{C}_k &=& \frac{1}{2}(E_{k,+}+E_{k,-})
 \label{cosk}
 ~, \\ \tilde{S}_k &=& \frac{1}{2i}(E_{k,-}-E_{k,+})~. \label{sink}
 \end{eqnarray}
 These have the properties  \begin{eqnarray} \label{eq:1578}[K_0,\tilde{C}_k ]
 &=&
 -i\,\tilde{S}_k~,~[K_0,\tilde{S}_k ] =
 i\,\tilde{C}_k~,\\
\label{eq:1214}{[} \tilde{C}_k \, , \, \tilde{S}_k {]}
&=&\frac{(2k-1)\, i}{2}\,\frac{1}{(K_0+k)(K_0+k-1)}~, \\
\label{eq:1215}\langle k,0|[\hat{C}_k\,,\,\hat{S}_k]|k,0\rangle &=&
\frac{i}{4k}~,
\\ \label{eq:1216}\tilde{C}_k^2+\tilde{S}_k^2&=& 1- \frac{2k-1}{2}\left(
\frac{1}{K_0+k-1}+ \frac{1}{K_0+k}\right)~, \\
\label{eq:1217}\langle k,0|\tilde{C}_k^2+\tilde{S}_k^2|k,0\rangle &=&
\frac{1}{4k}~.
\end{eqnarray}
In view of the following properties with respect to the number states and
the coherent states \ref{eq:78} and \ref{eq:136}
these operators may appear to be appealing:
 \begin{eqnarray}  \label{eq:1579}\langle
k,n|\tilde{C}_k|k,n\rangle &=& 0~, ~\langle
k,n|\tilde{S}_k|k,n\rangle =0~, \\
 \label{eq:1218}\langle \tilde{C}_k \rangle_{k,z} &=&\cos\phi\, \rho_k(|z|)~, ~
\langle \tilde{S}_k \rangle_{k,z} = \sin\phi\, \rho_k(|z|)~, \\
   \label{eq:1220}\langle \tilde{C}_k \rangle_{k,\lambda} &=&
\cos\theta\, |\lambda|~, ~
\langle \tilde{S}_k \rangle_{k,\lambda} = \sin\theta\, |\lambda|~, 
\end{eqnarray} where the relations \begin{eqnarray} \label{eq:1580}\langle
k,z|E_{k,-}|k,z\rangle &=& \frac{z}{|z|}\,\rho_k(|z|)~, \\
\label{eq:1222}\langle k, \lambda|E_{k,-}|k,\lambda\rangle &=& \lambda
\end{eqnarray} have been used (see Eqs.\ \ref{eq:1083} and \ref{eq:1037}).
 Notice that the coherent states $|k,\lambda
 \rangle$ are eigenstates of $E_{k,-}$!

However, problems appear for higher powers
 of the operators
$\tilde{C}_k$ and $\tilde{S}_k\,$, like $\tilde{C}_k^2$ etc.:

 If we express $\tilde{C}_k$ and $\tilde{S}_k\,$ in terms of the
observables $K_1$ and $K_2$, instead of $K_+$ and $K_-$, we get
 \begin{eqnarray} \label{eq:1581}\tilde{C}_k
&=& \frac{1}{2} [(K_0+k)^{-1}K_1 + K_1(K_0+k)^{-1}] \\ &&
-\frac{i}{2} [(K_0+k)^{-1}K_2 - K_2(K_0+k)^{-1}]~,\nonumber \\ \label{eq:1224}
\tilde{S}_k
&=& -\frac{1}{2} [(K_0+k)^{-1}K_2 + K_2(K_0+k)^{-1}] \\ &&
+\frac{1}{2\,i} [(K_0+k)^{-1}K_1 - K_1(K_0+k)^{-1}]~. \nonumber
\end{eqnarray}
Thus, {\em the $\cos$-operator $\tilde{C}_k$ contains contributions
from the $\sin$-observable $K_2$ and the $\sin$-operator
$\tilde{S}_k$ contains contributions from the $\cos$-observable
$K_1\,$}. 

These contributions do not matter for the expectation
values \ref{eq:1579}-\ref{eq:1220} which are linear in $\tilde{C}_k$ and
$\tilde{S}_k$, but they will in general matter for higher powers
of these operators. That  can be seen by comparing the expectation values
 $ \langle
k,n|\tilde{C}_k^2|k,n\rangle $ or $
 \langle \tilde{C}_k ^2\rangle_{k,z}$ etc. with corresponding expectation
values of suitable variants of the operators \ref{eq:1581} and \ref{eq:1224}:

 The actions of the operators
$\tilde{C}_k$ and $\tilde{S}_k$ on the number states $|k,n\rangle$
are \begin{eqnarray} \label{eq:1582}\tilde{C}_k|k,n\rangle &=&
\frac{1}{2}(\tilde{f}^{(k)}_n|k,n-1\rangle
+\tilde{f}^{(k)}_{n+1}|k,n+1\rangle)~, \\
\label{eq:1226}\tilde{S}_k|k,n\rangle &=&
\frac{1}{2i}(\tilde{f}^{(k)}_n|k,n-1\rangle
-\tilde{f}^{(k)}_{n+1}|k,n+1\rangle)~, \\
\label{eq:1227}\tilde{f}^{(k)}_n &=&
\frac{[n\,(2k+n-1)]^{1/2}}{2k+n-1}~,~\tilde{f}^{(k)}_{n=0} =0~.
 \end{eqnarray} This gives the following expectation values for the squared
operators
\begin{eqnarray} \label{eq:1583}\langle k,n|\tilde{C}_k^2|k,n\rangle
 &=&\langle k,n|\tilde{S}_k^2|k,n\rangle
=  \frac{1}{4}((\tilde{f}^{(k)}_n)^2
+(\tilde{f}^{(k)}_{n+1})^2~, 
\\ \label{eq:1229}\langle \tilde{C}_k^2\rangle_{k,n=0} &=&\langle
\tilde{S}_k^2\rangle_{k,n=0} =\frac{1}{8\,k}~.
\end{eqnarray}

  In order to compare these fluctuations with those of
 $\cos$- and $\sin$-operators which
 are ``pure'' ones,  the expressions \ref{eq:1581} and \ref{eq:1224}
 suggest to define
  \begin{eqnarray} \label{eq:1584}\hat{C}_k &=&
\frac{1}{2} [(K_0+k)^{-1}K_1 + K_1(K_0+k)^{-1}] \\ \label{eq:1231}&=&
\frac{1}{4}[E_{k,-} +E_{k+1,-}+E_{k,+}+E_{k+1,+}]~, \\
\label{eq:1232}\hat{S}_k &=& -\frac{1}{2} [(K_0+k)^{-1}K_2 +
 K_2(K_0+k)^{-1}] \\
\label{eq:1233}&=& \frac{1}{4i}[E_{k,-} +E_{k+1,-}-E_{k,+}-E_{k+1,+}]~.
\end{eqnarray}
Because  \begin{eqnarray}\label{eq:1585}\langle k,z|E_{k+1,-}|k,z \rangle &=&
z\,\langle k,z|(K_0+k+1)^{-1}|k,z \rangle \\ \label{eq:1234}&=&
\frac{z}{|z|}\,\rho_k(|z|) - \frac{z}{|z|^2}
+\frac{2k\,z}{|z|^3}\rho_k(|z|)~,\nonumber \end{eqnarray} and (see
the relation \ref{eq:1548})
\begin{equation}
  \label{eq:380}
  \langle k,\lambda|E_{k+1,-}|k,\lambda \rangle =
  \lambda\, s_{k}^{(b=k+1)}(|\lambda|^2)\,,
\end{equation}
 we have now 
\begin{eqnarray}  \label{eq:1587}\langle
k,n|\hat{C}_k|k,n\rangle &=& 0~, ~\langle
k,n|\hat{S}_k|k,n\rangle =0~, \\
 \label{eq:1236}\langle \hat{C}_k \rangle_{k,z} &=&\cos\phi\,
 \left( \rho_k(|z|)
 -\frac{1}{2|z|}+ \frac{k}{|z|^2}\rho_k(|z|)\right)~,
 \\
\label{eq:1237}\langle \hat{S}_k \rangle_{k,z} &=& \sin\phi\,\left(\rho_k(|z|)
 -\frac{1}{2|z|}+ \frac{k}{|z|^2}\rho_k(|z|)\right)~, \\
  \label{eq:1238}\langle \hat{C}_k \rangle_{k,\lambda} &=&
\cos\theta\, |\lambda| \,[1+s_{k}^{(b=k+1)}(|\lambda|^2)]/2\,, \\
\label{eq:1239}\langle \hat{S}_k \rangle_{k,\lambda} &=& \sin\theta\,
|\lambda| \,[1+s_{k}^{(b=k+1)}(|\lambda|^2)]/2\,.
\end{eqnarray}
 The expectation values of $\hat{C}_k$ and
$\hat{S}_k$ with respect to the conventional coherent states
$|k,\alpha \rangle $ may be dealt with in the same way as with the other ones
above. 

 From the relations
\begin{eqnarray} \label{eq:1588}\hat{C}_k|k,n\rangle &=&
\frac{1}{4}(\hat{f}^{(k)}_n|k,n-1\rangle
+\hat{f}^{(k)}_{n+1}|k,n+1\rangle)~, \\
\label{eq:1240}\hat{S}_k|k,n\rangle &=& \frac{1}{4i}(\hat{f}^{(k)}_n|k,n-1
\rangle
-\hat{f}^{(k)}_{n+1}|k,n+1\rangle)~, \\
\label{eq:1241}\hat{f}^{(k)}_n &=&
[n\,(2k+n-1)]^{1/2}\left(\frac{1}{2k+n-1}+\frac{1}{2k+n}\right)~,\\
\label{eq:1242}&&~\hat{f}^{(k)}_{n=0} =0~. \nonumber
 \end{eqnarray}  we get  the fluctuations
\begin{eqnarray} \label{eq:1589}\langle k,n|\hat{C}_k^2|k,n\rangle
 &=&\langle k,n|\hat{S}_k^2|k,n\rangle
=\\ &=& \frac{1}{16}((\hat{f}^{(k)}_n)^2
+(\hat{f}^{(k)}_{n+1})^2)~, \nonumber
\\ \label{eq:1244}\langle\hat{C}_k^2\rangle_{k,n=0} &=&\langle
\hat{S}_k^2\rangle_{k,n=0} =
\frac{(4\,k+1)^2}{32\,k\,(2k+1)^2}~. 
\end{eqnarray} In order to see now the difference in the consequences of the
different definitions of the operators
 $\tilde{C}_k\,,\tilde{S}_k$
and $\hat{C}_k\,,\hat{S}_k\,$, respectively, let us look at the special 
but important case of
 the ground state expectation values of their squares for $k=1/2$:
\begin{eqnarray}
\label{eq:1590}\langle \tilde{C}_{k=1/2}^2\rangle_{k=1/2,n=0} &=&\langle
\tilde{S}_{k=1/2}^2\rangle_{k=1/2,n=0} = \frac{1}{4}= 0.25~, \\
\label{eq:1247}\langle \hat{C}_{k=1/2}^2\rangle_{k=1/2} &=&\langle
\hat{S}_{k=1/2} ^2\rangle_{k=1/2,n=0} = \frac{9}{64} \approx 0.14~.
\end{eqnarray}

In view of the operators $\hat{C}_k$ and $\hat{S}_k$ one may ask,
why divide by the operator $K_0+k$ and not by $K_0$ itself? This
leads to the operators \cite{ka3}
\begin{equation}
  \label{eq:691}
  \check{C}_k=\hat{C}_{k-1}~~ \text{and}~~
\check{S}_k=\hat{S}_{k-1} 
\end{equation}
 and their actions
\begin{eqnarray} \label{eq:1591}\check{C}_{k}|k,n\rangle &=&
\frac{1}{4}(\check{f}^{(k)}_n|k,n-1\rangle
+\check{f}^{(k)}_{n+1}|k,n+1\rangle)~, \\
\label{eq:1249}\check{S}_{k}|k,n\rangle &=&
\frac{1}{4i}(\check{f}^{(k)}_n|k,n-1\rangle
-\check{f}^{(k)}_{n+1}|k,n+1\rangle)~, \\
\label{eq:1250}\check{f}^{(k)}_n &=&
[n\,(2k+n-1)]^{1/2}\left(\frac{1}{k+n-1}+\frac{1}{k+n}\right)~,
\\ \label{eq:1251}&& \check{f}^{(k)}_{n=0} =0~.
 \end{eqnarray}
It follows that
\begin{eqnarray} \label{eq:1592}\langle k,n|\check{C}_{k}^2|k,n\rangle
 &=&\langle k,n|\check{S}_{k}^2|k,n\rangle
=\\ &=& \frac{1}{16}((\check{f}^{(k)}_n)^2
+(\check{f}^{(k)}_{n+1})^2)~, \nonumber
\\ \label{eq:1253}\langle \check{C}_k^2\rangle_{k,n=0} &=&\langle
\check{S}_{k}^2\rangle_{k,n=0} =
\frac{(2\,k+1)^2}{8\,k\,(k+1)^2}~, \end{eqnarray}
For  $k=1/2$ we get from \ref{eq:1253} 
\begin{equation}
  \label{eq:692}
 \langle \check{C}_{k=1/2}^2\rangle_{k=1/2,n=0}=
 \langle \check{S}_{k=1/2}^2\rangle_{k=1/2,n=0} =
\frac{4}{9}\approx 0.44~. 
\end{equation}
In the special case $k=1/2$ the operators \ref{eq:691} were already
discussed some time ago \cite{lern3,lyn1}.

 Comparing the different results \ref{eq:1590}, \ref{eq:1247}
and \ref{eq:692}
 one realizes the problems as to  finding an appropriate definition of
 suitable operators $\widehat{\cos}$ and $\widehat{\sin}\,$!

  However, such a search is not necessary in the present
 $SO^{\uparrow}(1,2)$ framework, because here the primary quantum observables
incorporating the properties of $\cos\vp$ and $\sin\vp$ are the
operators $K_1$ and $K_2$, {\em not}  $\tilde{C}_k$ and $\tilde{S}_k$ or
$\hat{C}_k$ and $\hat{S}_k$ or
$\check{C}_k$ and $\check{S}_k$ as discussed above. 

 In any case:
In view of the analysis  made  above in connection with the
 expressions \ref{eq:1581}
and \ref{eq:1224} we  see that the London Susskind-Glowgower operators
$\tilde{C}_k$ and $\tilde{S}_k$ are {\em not} appropriate for measuring
angle properties of a state!

\chapter{The group $SO^{\uparrow}(1,2)$ as a framework for  applications  in
quantum optics}
 Structures of the group $SU(1,1) \cong SL(2,\mathbb{R})=Sp(2,\mathbb{R})$
 and $SO^{\uparrow}(1,2) = SU(1,1)/Z_2$ - especially its
Lie algebra $\mathfrak{so}(1,2)$ - appear to be around ``all over
 the place'' in
quantum optics and seem to ``loom'' behind many corners! It is the
purpose of the present and the next  chapters to put those
structures into the perspective of the present approach. Before I
give a selection of references from the quantum optics literature
let me start with some general remarks:
\section{Adjoint representation}
Much can be learnt from the adjoint representation associated with
any unitary representation of the group $SU(1,1)$, i.e.\ the
representation of the group $SO^{\uparrow}(1,2)$ in the
3-dimensional vector space of the Lie algebra spanned by the basis
$K_j\,,\,j=0,1,2,$ or $K_0,K_+,K_-$:

 If \begin{eqnarray}
\label{eq:1594}U(w)&=&e^{\displaystyle(w/2)\,K_+ -(\bar{w}/2)\,K_-}
=e^{\displaystyle i\,w_2K_1+i\,w_1K_2}~,\\ \label{eq:1256}&& w=w_1+i\,w_2
=|w|\,e^{i\,\theta}~, \nonumber \end{eqnarray} then it follows
from the general formula (see, e.g.\ Ref.\ \cite{mer})
\begin{equation}
  \label{eq:234}
e^A\,B\,e^{-A}= B +
[A,B]+\frac{1}{2!}[A,[A,B]]+\frac{1}{3!}[A,[A,[A,B]]]~\cdots
\end{equation}
and the commutation relation \ref{eq:55} that \begin{eqnarray}
\label{eq:1595}U(-w)\,K_+\,U(w) &=& \frac{1}{2}(\cosh|w|+1)\,K_+ + \\
\label{eq:1257}&& +\frac{1}{2}e^{-2\,i\,\theta}(\cosh|w| -1)\, K_- +
e^{-i\,\theta}\,\sinh|w|\,K_0 \nonumber \\
\label{eq:1258}U(-w)\,K_-\,U(w) &=& \frac{1}{2}(\cosh|w|+1)\,K_- + \\
\label{eq:1259}&& +\frac{1}{2}e^{2\,i\,\theta}(\cosh|w| -1)\, K_+
 + e^{i\,\theta}\,\sinh|w|\,K_0 \nonumber \\
\label{eq:1260}U(-w)\,K_0\,U(w) &=& \cosh|w|\,K_0 + \\ \label{eq:1261}&&
+\frac{1}{2}\,\sinh|w|(e^{i\,\theta}\, K_+ +
e^{-i\,\theta}\,K_-)~,\nonumber \\ \label{eq:1262}U(-\tau)\,K_+\,U(\tau) &=&
e^{-i\,\tau}\,K_+~,
\\
\label{eq:1263}U(-\tau)\,K_-\,U(\tau) &=& e^{i\,\tau}\,K_-~, \\
 \label{eq:1264}&&
U(\tau)=e^{i\,\tau\,K_0}~. \end{eqnarray} 

For the
operators $K_1$ and $K_2$ this means \begin{eqnarray}
\label{eq:1596}U(-w)\,K_1\,U(w)
&=& [1 +\cos^2\theta(\cosh|w|-1)]\,K_1 - \\
\label{eq:1265}&& -\sin\theta\,\cos\theta\,(\cosh|w| -1)\,K_2 +
\cos\theta\,\sinh|w|\,K_0~, \nonumber \\
\label{eq:1266}U(-w)\,K_2\,U(w) &=& [1+ \sin^2\theta(\cosh|w|-1)]\,K_2 -\\
\label{eq:1267}&& -\sin\theta\,\cos\theta\,(\cosh|w| -1)\,K_1  -
\sin\theta\,\sinh|w|\,K_0~,\nonumber\\
\label{eq:1268}U(-w)\,K_0\,U(w) &=& \sinh|w|\,(\cos\theta\,K_1-
\sin\theta\,K_2) +
\cosh|w|\,K_0~, \\ \label{eq:1269}U(-\tau)\,K_1\,U(\tau) &=& \cos\tau\,K_1 +
\sin\tau\, K_2~, \\
\label{eq:1270}U(-\tau)\,K_2\,U(\tau) &=& -\sin\tau\,K_1 + \cos\tau\, K_2~.
\end{eqnarray}

 The transformations leave the
quadratic Killing form $L = K_1^2+K_2^2-K_0^2 $ invariant. This
property reflects the fact that the above transformations of the
3-dimensional Lie algebra vector space are (1+2)-dimensional
Lorentz transformations with $K_0$ playing the role of a ``time
variable''.  That can be made more explicit in the following
way:

 Let us define the two ``spatial'' vectors
\begin{equation}
  \label{eq:235}
 \vec{K} =
(K_1,K_2)~,~\vec{n} = (-\cos\theta,\sin\theta)~.
\end{equation}
Then the
transformation formulae \ref{eq:1596}-\ref{eq:1268} can be written as
\begin{eqnarray} \label{eq:1597}U(-w)\vec{K}\,U(w) &=& \vec{K}+(\cosh|w|-1)
(\vec{n}\cdot\vec{K})\,\vec{n}- \sinh|w|\,\vec{n}\,K_0\,, \\
\label{eq:1271}U(-w)K_0\,U(w)& = &\cosh|w|\,K_0 -\sinh|w|\,(\vec{n}\cdot
\vec{K})~.
\end{eqnarray} These equations describe Lorentz transformations (``boosts'')
in the direction $\vec{n}\,$! 

Notice that all the above relations are independent of the index $k$.
The $k$-dependence will appear in the matrix elements with respect
to a given Hilbert space.
The reason for using the
transformations $U(-w)\,K_1\,U(w)$ etc.\ instead of the usual
$U(w)\,K_1\,U(-w)$ etc.\ is the following: In applications it is
frequently useful to generate a (coherent) state $|\psi\rangle$ by
applying one of the above unitary operators $U$ to a given
reference state $|\psi_0 \rangle$, namely $|\psi\rangle =
U|\psi_0\rangle$. The best-known examples are the  coherent
states $|\alpha \rangle$ (Eq.\ \ref{eq:140}) and $|k,\lambda \rangle$
 (Eq.\ \ref{eq:139}) where the reference
state $|\psi_0\rangle$ is the ground state $|k, n=0\rangle$. 

 For the Perelomov state \ref{eq:139}
the generating unitary operator is the operator $U(w)$ from above. The
expectation value of, e.g.\ $K_1$, is then given by
\begin{equation}
  \label{eq:693}
  \langle
k,\lambda|K_1|k,\lambda \rangle = \langle k,0|U(-w)\,K_1\,U(w)|k,0
\rangle\,. 
\end{equation}
 According to Eq.\ \ref{eq:1596} this expectation value may by
calculated by taking the ground state expectation value of the
r.h.\ side of \ref{eq:1596} which immediately gives $k\,\cos\theta\,
\sinh|w|$, in agreement with Eq.\ \ref{eq:1543}. The other cases can be
dealt with accordingly.

The unitary operator $U(w)$ is not a very natural one from a group
theoretical point of view, because the vector subspace spanned by
 the operators $K_1$ and $K_2$ does not form a Lie subalgebra,
i.e.\ the unitary transformations $U(w)$ do not form a subgroup.

From a group theoretical aspect one would decompose a general
unitary transformation in an irreducible representation either
according to Cartan's polar
decomposition or according to Iwasawa's decomposition (see Appendix B): 

The polar decomposition  of a general unitary $SU(1,1)$
transformation is given by
\begin{equation}
  \label{eq:236}
 U(\tau_2,w_2,\tau_1) =
e^{i\,\tau_2\,K_0}\cdot e^{i\,w_2\,K_1}\cdot e^{i\,\tau_1\,K_0}~,
\end{equation}
and the Iwasawa decomposition by
\begin{equation}
  \label{eq:237}
 U(\tau,w_2,\nu)=
e^{i\,\tau\,K_0}\cdot e^{i\,w_2\,K_1}\cdot e^{i\,\nu\,N}~,~ \nu
\in \mathbb{R}~,
\end{equation}
where
\begin{eqnarray} \label{eq:1598}N&=& K_2+K_0~, \\ \label{eq:1272}{[}
 N,K_1{]}=i \, N~,~{[}N,K_0{]}&=&
i\,K_1\,,~{[}K_1,K_0{]}= i\,(K_0-N)\,. \end{eqnarray} $N$ is the
generator of a nilpotent subgroup. The Iwasawa decomposition
appears rarely in quantum optical papers \cite{mu}.
\\ \label{eq:1273}From the commutation relations \ref{eq:1272}
 we get for the adjoint
representation with respect to the basis $K_1,\,K_0$ and $N$:
\begin{eqnarray} \label{eq:1599}e^{-i\,\nu\,N}K_1e^{i\,\nu\,N} &=& K_1
 +\nu\,N~,
\\\label{eq:1274}e^{-i\,\nu\,N}K_0e^{i\,\nu\,N} &=& K_0 + \nu\,K_1
+\frac{\nu^2}{2}\,N~, \\
\label{eq:1275}e^{-i\,w_2\,K_1}Ne^{i\,w_2\,K_1} &=& e^{-w_2}N ~, \\
\label{eq:1276}e^{-i\,w_2\,K_1}K_0e^{i\,w_2\,K_1} &=& (\cosh w_2 + \sinh
w_2)\,K_0 - \sinh w_2\, N ~. \end{eqnarray} The rest follows
immediately from Eqs.\ \ref{eq:1269} and \ref{eq:1270}.
\section{Schwarz's inequality and $SO^{\uparrow}(1,2)$ squee\-zing}
\subsection{Uncertainty relations} 
One of the main purposes of the operators $U(w)$ in
quantum optics is to generate ``squeezed'' states
\cite{stol,lu,yue,hol,wod,rev8}. I recall the main features of their
definition: Let $A$ and $B$ two self-adjoint operators with the
commutator $[A,B]=i\,C$, where $C$ is again self-adjoint. Then the
lower limit for the product of the ``uncertainties''
$(\Delta\,A)_{\psi}$ and $(\Delta\,B)_{\psi}$ in the state
$|\psi\rangle\,$, where
\begin{equation}
  \label{eq:695}
  (\Delta\,A)_{\psi}^2 = \langle
\psi|(A-\langle \psi|A|\psi\rangle)^2|\psi \rangle\,, 
\end{equation}
 is
derived from Schwarz's inequality
\begin{equation}
  \label{eq:238}
\langle\psi_2|\psi_2\rangle\,\langle \psi_1|\psi_1\rangle \geq
|\langle\psi_2|\psi_1\rangle|^2
\end{equation}
for scalar products as follows
\cite{rob,schro2,jack,jor,mer1,dod}\cite{coh3}: Taking for $|\psi_2\rangle$ and
$|\psi_1\rangle$ the states 
\begin{equation}
  \label{eq:694}
 \tilde{A}|\psi\rangle \equiv
(A-\langle A\rangle_{\psi})|\psi\rangle~~\text{and}~~
\tilde{B}|\psi\rangle \equiv (B-\langle
B\rangle_{\psi})|\psi\rangle\,, 
\end{equation}
 we get
\begin{equation}
  \label{eq:239}
 (\Delta A)_{\psi}^2\cdot
(\Delta B)_{\psi}^2 \geq |\langle \psi |\tilde{A}\cdot \tilde{B} =
\frac{1}{2}[\tilde{A},\tilde{B}]+
\frac{1}{2}(\tilde{A}\tilde{B}+\tilde{B}\tilde{A})|\psi \rangle
|^2.
\end{equation}
As the expectation value of the commutator $[A,B]$ is
purely imaginary and that of
\begin{equation}
  \label{eq:240}
 S_{\psi}(A,B) =
\frac{1}{2}(AB+BA)-\langle A \rangle_{\psi}\langle
B\rangle_{\psi}
\end{equation}
purely real we can write Eq.\ \ref{eq:239} as
\begin{equation}
  \label{eq:241}
 (\Delta
A)_{\psi}^2\cdot (\Delta B)_{\psi}^2 \geq \frac{1}{4}|\langle \psi
| [A,B]|\psi\rangle|^2+ |\langle \psi|S_{\psi}(A,B)|\psi \rangle
|^2 \geq  \frac{1}{4}|\langle \psi | [A,B]|\psi\rangle|^2 ~.
\end{equation}
If $\langle \psi|S_{\psi}(A,B)|\psi \rangle$ vanishes  then the
inequality \ref{eq:241} reduces to the usual Heisenberg uncertainty
relation. If $\langle \psi|S_{\psi}(A,B)|\psi \rangle \neq 0$ then
the first of the inequalities \ref{eq:241} is stronger than the second
one.

One now says that `` the state $|\psi \rangle $ is squeezed with
respect to the operator $A\,$'' if
\begin{equation}
  \label{eq:242}
(\Delta A)_{\psi}^2 < [
|\langle \psi | [A,B]|\psi\rangle|^2/4+ |\langle
\psi|S_{\psi}(A,B)|\psi \rangle |^2]^{1/2}~,
\end{equation}
(``Robertson-Schr\"odinger squeezing''), or
\begin{equation}
  \label{eq:243}
 (\Delta
A)_{\psi}^2 < \frac{1}{2} |\langle \psi | [A,B]|\psi\rangle|~
\end{equation}
(``Heisenberg squeezing''). 

 The latter condition is more
restrictive. 

 In order to preserve the inequalities \ref{eq:241} the
second uncertainty $(\Delta B)_{\psi}$ has to be enlarged or ``stretched''
accordingly. 

 Depending on the state $|\psi\rangle $ the r.h.\
side of the inequalities might be quite large. Then it appears
appropriate \cite{trif1,mann1} to sharpen the criteria \ref{eq:242}
 and \ref{eq:243} to
\begin{equation}
  \label{eq:244}
(\Delta A)_{\psi}^2 < \min_{\{|\psi\rangle\}}[ |\langle \psi
| [A,B]|\psi\rangle|^2/4+ |\langle \psi|S_{\psi}(A,B)|\psi \rangle
|^2]^{1/2} \equiv \Delta_{RS}^2~,
\end{equation}
(``absolute
Robertson-Schr\"odinger squeezing''), or
\begin{equation}
  \label{eq:245}
 (\Delta A)_{\psi}^2 <
\min_{\{|\psi\rangle\}}\frac{1}{2} |\langle \psi |
[A,B]|\psi\rangle| \equiv \Delta_H^2~,
\end{equation}
(``absolute Heisenberg
squeezing''). Here $\{|\psi \rangle\}$ means a given set of
states, e.g.\ a basis of the Hilbert space. 

Example:

 If $A=K_1\,,\,B=K_2$
and $|\psi\rangle = |k,n\rangle$ then $\langle
k,n|S_{k,n}(K_1,K_2)|k,n \rangle = 0$ and the minimum $k/2$ of the r.h.\
side in  \ref{eq:245} is obtained for the ground state
$|k,n=0\rangle$.
\subsection{Group theoretical generation of squeezed states}
In order to see the squeezing properties of the operators $U(w)$
let us specialize to pure special Lorentz transformations in the
$(K_1,K_0)-$ and $(K_2,K_0)-$ subspaces, respectively:
\begin{eqnarray} \label{eq:1600}U(-w_1)\,K_1\,U(w_1)& = &
\cosh w_1\, K_1 + \sinh w_1\,K_0~, \\
\label{eq:1277}U(-w_1)\,K_0\,U(w_1)& = & \sinh w_1\, K_1 + \cosh w_1\,K_0~,
\\ \label{eq:1278}U(w_1)&=& e^{i\,w_1\,K_2}~; \\
\label{eq:1279}U(-i\,w_2)\,K_2\,U(i\,w_2)& = &
\cosh w_2\, K_2 - \sinh w_2\,K_0~, \\
\label{eq:1280}U(-i\,w_2)\,K_0\,U(i\,w_2)& = & -\sinh w_2 \, K_2 + \cosh
w_2\,K_0~,
\\ \label{eq:1281}U(i\,w_2)&=& e^{i\,w_2\,K_1}~. \end{eqnarray} If we
 now define  the
operators
\begin{equation}
  \label{eq:246}
 K_{1\pm} = K_1 \pm K_0~,~ K_{2\pm} = K_2 \pm K_0~,
\end{equation}
where
\begin{equation}
  \label{eq:247}
 [K_{1+},K_{1-}]= 2\,i\,K_2~,~[K_{2+},K_{2-}]=
-2\,i\,K_1~,
\end{equation}
then we get \begin{eqnarray}
\label{eq:1601}U(-w_1)\,K_{1\pm}\,U(w_1) &\equiv & \hat{K}_{1\pm}= e^{\pm
w_1}K_{1\pm}
 ~,\\ \label{eq:1282}U(-w_1)\,K_2 \,U(w_1) &\equiv& \hat{K}_2 = K_2 ~,
 \\ \label{eq:1283}{[} \hat{K}_{1+},\hat{K}_{1-}{]}&=& 2\,i\,K_2~;\\
\label{eq:1284}U(-i\,w_2)\,K_{2\pm}\,U(i\,w_2) &\equiv & \hat{K}_{2\pm}= e^{\mp
w_2}K_{2\pm}
 ~,\\ \label{eq:1285}U(-i\,w_2)\,K_1 \,U(i\,w_2) &\equiv& \hat{K}_1 = K_1 ~,
 \\ \label{eq:1286}{[}\hat{K}_{2+},\hat{K}_{2-}{]}&=& -2\,i\,K_1~.
 \end{eqnarray}
We have the following correspondences of the operators $K_{1\pm}$ and 
$K_{2\pm}$ and the basic classical quantities $I,I\cos\vp$ and $I\sin\vp$:
\begin{equation}
  \label{eq:696}
  K_{1\pm} \leftrightarrow I\,(\cos\vp\pm1)\,,~~~~
 K_{2\pm} \leftrightarrow -I\,(\sin\vp\pm1)\,.
\end{equation}
The operator $K_{2+}$ is identical with the generator $N$ of Eq.\
 \ref{eq:1598}, $K_{2-}$ is the generator $\bar{N}$ of another nilpotent
group (see Eqs.\ \ref{eq:503} and \ref{eq:504}).

The operators $U(w_1)$ and $U(i\,w_2)$ generate special Perelomov
coherent states \ref{eq:139} from a ground state $|k,0\rangle$. 

(If $w_2=0\,,$ we have $w_1
> 0,$ for  $\theta = 0$, and  $w_1 <0$ for  $\theta = \pi$, and
if  $w_1= 0\,,$ then $w_2>0$, for $\theta =\pi/2$, and $w_2 < 0$
for $\theta = -\pi/2\,$, i.e.\ the variable
\begin{equation}
  \label{eq:697}
 \lambda = \lambda_1
+i\,\lambda_2 
\end{equation}in $|k, \lambda \rangle$ 
 is purely real in the
first case and purely imaginary in the second.) 

 When calculating expectation values with respect to these states
we can do so by just calculating the ground state expectation
values of the r.h. sides of the above Eqs.\ \ref{eq:1601}-\ref{eq:1286}:
\begin{eqnarray} \label{eq:1602}\langle K_{1\pm}\rangle_{k,\lambda =\lambda_1}
&=& \langle \hat{K}_{1\pm} \rangle_{k,\lambda =0} = e^{\pm w_1}
\langle K_{\pm 1} \rangle_{k,\lambda=0} = \pm k\,e^{\pm w_1}~, \\
\label{eq:1287}\langle K_{1\pm}^2\rangle_{k,\lambda =\lambda_1} &=& \langle
\hat{K}_{1\pm}^2 \rangle_{k,\lambda =0} = e^{\pm 2\,w_1}
\langle K_{\pm 1}^2 \rangle_{k,\lambda=0}\\\label{eq:1288}& =& e^{\pm
 2\,w_1}(k/2+k^2)~,\nonumber \\
\label{eq:1289}(\Delta K_{1\pm})^2_{k,\lambda= \lambda_1} &=&
 \frac{k}{2}\,e^{\pm
2\,w_1}~; \\ \label{eq:1290}\langle K_{2\pm}\rangle_{k,\lambda
 =i\,\lambda_2} &=&
\langle \hat{K}_{2\pm} \rangle_{k,\lambda =0} = e^{\mp w_2}
\langle K_{\pm 2} \rangle_{k,\lambda=0} = \pm k\,e^{\mp w_2}~, \\
\label{eq:1291}\langle K_{2\pm}^2\rangle_{k,\lambda =i\,\lambda_2} &=& \langle
\hat{K}_{2\pm}^2 \rangle_{k,\lambda =0} = e^{\mp 2\,w_2}
\langle K_{\pm 2}^2 \rangle_{k,\lambda=0}\\\label{eq:1292}& =& e^{\mp
 2\,w_2}(k/2+k^2)~,\nonumber \\
\label{eq:1293}(\Delta K_{2\pm})^2_{k,\lambda=i\,\lambda_2} &=&
\frac{k}{2}\,e^{\mp 2\,w_2}~.
 \end{eqnarray} For the r.h.\ side of
the inequality \ref{eq:241} we get in case of the pair $K_{1\pm}\,$:
\begin{eqnarray} \label{eq:1603}|\langle [K_{1+},K_{1-}] \rangle_{k,\lambda =
\lambda_1}| &=& |\langle [K_{1+},K_{1-}] \rangle_{k,\lambda =
0}| \\ \label{eq:1294}& =& 2|\langle K_2 \rangle_{k,\lambda = 0}| = 0~;
 \nonumber \\
\label{eq:1295}|\langle S_{k,\lambda=\lambda_1}(K_{1+},K_{1-})
 \rangle_{k,\lambda
= \lambda_1}| &=& |\langle
S_{k,\lambda=0}(K_{1+},K_{1-})\rangle_{k,\lambda = 0}|
 = \frac{k}{2}~. \end{eqnarray}
Combined with Eqs.\ \ref{eq:1289} we obtain the equality
\begin{equation}
  \label{eq:248}
 (\Delta K_{1+})_{k,\lambda=\lambda_1}
 (\Delta K_{1-})_{k,\lambda=\lambda_1}= \frac{k}{2} =
 |\langle
S_{k,\lambda=0}(K_{1+},K_{1-})\rangle_{k,\lambda = 0}|~.
\end{equation}
That such an equality has to hold we know already from the general
result \ref{eq:163}. Here we learn in addition from \ref{eq:1289}
  that one can make
 one of the uncertainties as small as possible by a corresponding
 choice of $w_1$, at the expense of the other uncertainty.

The corresponding discussion for the operators $K_{2\pm}$ gives
completely analogous results. 

 We have here - Eqs.\ \ref{eq:1603} and \ref{eq:1295} - the somewhat unusual
case that the expectation value of the commutator in the
inequality \ref{eq:241} vanishes whereas the expectation value of
$S_{\psi}(A,B)$ is non-vanishing!  A deeper reasons for
the equality \ref{eq:248} will be discussed below.

The examples just discussed may be generalized immediately: Take
any normalizable state vector $|k,\sigma \rangle$ of the Hilbert
space of a representation with Bargmann index $k$ and apply one of
the operators $U(w_1),\,U(i\,w_2)$ to it. Then the states
\begin{equation}
  \label{eq:249}
|k,\sigma;w_1\rangle = U(w_1)|k,\sigma \rangle~,~
|k,\sigma;w_2\rangle = U(i\,w_2)|k,\sigma \rangle~,
\end{equation}
are
squeezed as to the operators $K_{1\pm}$ and $K_{2\pm}$,
respectively: We have, e.g. from Eqs.\ \ref{eq:1601}-\ref{eq:1283}
\begin{eqnarray} \label{eq:1604}\langle K_{1\pm}\rangle_{k,\sigma;w_1}
&=& \langle \hat{K}_{1\pm} \rangle_{k,\sigma} = e^{\pm w_1}
\langle K_{\pm 1} \rangle_{k,\sigma}~, \\
\label{eq:1296}\langle K_{1\pm}^2\rangle_{k,\sigma;w_1} &=& \langle
\hat{K}_{1\pm}^2 \rangle_{k,\sigma} = e^{\pm 2\,w_1}
\langle K_{\pm 1}^2 \rangle_{k,\sigma}~, \\
\label{eq:1297}(\Delta K_{1\pm})^2_{k,\sigma;w_1} &=& e^{\pm 2\,w_1}(\Delta
K_{1\pm})^2_{k,\sigma}~.
\end{eqnarray}
Because the transformations \ref{eq:249} are unitary and because of the
relations \ref{eq:1601}-\ref{eq:1283} we get 
\begin{eqnarray} \label{eq:1605}|\langle [K_{1+},K_{1-}] \rangle_{k,\sigma;w_1
}| &=& |\langle [K_{1+},K_{1-}] \rangle_{k,\sigma}
|= 2\,|\langle K_2 \rangle_{k,\sigma}|~,\\
\label{eq:1298}|\langle S_{k,\sigma;w_1}(K_{1+},K_{1-}) \rangle_{k,\sigma;w_1}|
&=& |\langle S_{k,\sigma}(K_{1+},K_{1-})\rangle_{k,\sigma}|
 ~. \end{eqnarray}
The inequality \ref{eq:241} holds, of course, for the original state
$|k,\sigma \rangle $ and the operators $A=K_{1+}$ and $B=K_{1-}$. The Eqs.\
\ref{eq:1297}, \ref{eq:1605} and \ref{eq:1298} then show that the same
inequality holds for the transformed state $|k,\sigma;w_1 \rangle $:
\begin{equation}
  \label{eq:698}
 (\Delta K_{1+})^2_{k,\sigma;w_1}\cdot(\Delta K_{1-})^2_{k,\sigma;w_1} 
\geq |\langle K_2 \rangle_{k,\sigma}|^2 +|\langle S_{k,\sigma}(K_{1+},K_{1-})
\rangle_{k,\sigma}|^2\,,
\end{equation} where the uncertainties on the l.h.\ side are now squeezed
and stretched according to Eqs.\ \ref{eq:1297}.
 
 The present
squeezing procedure can, of course, applied especially to the
states $|k,n\rangle, |k,z\rangle, |k,\lambda\rangle$ and
$|k,\alpha\rangle$ discussed  in Ch.\ 3. How this may
be done even experimentally will be indicated below in Sec.\ 6.5.
\subsection{Schwarz's equality!}
We have just seen that we get interesting additional information if
the inequality \ref{eq:241} becomes an equality. This can be exploited further:

 A {\em necessary and sufficient} condition
for Schwarz's inequality \ref{eq:238} to become an equality  is the linear
 dependence of the states \ref{eq:694} \cite{hew}.

Let $|\psi_0 \rangle$ be a state for which the equality holds.
Then
\begin{equation}
  \label{eq:250}
 \tilde{B}|\psi_0\rangle =
\gamma\,\tilde{A}|\psi_0\rangle\,,~\gamma \in \mathbb{C}\,,
\end{equation}
i.e.\ we have the ``eigenvalue'' equation
\begin{equation}
  \label{eq:251}
(B-\gamma\,A)|\psi_0\rangle = (\langle
B\rangle_{\psi_0}-\gamma\,\langle
A\rangle_{\psi_0})|\psi_0\rangle~,
\end{equation}
where the complex parameter
$\gamma$ may be calculated as \cite{mer1} \begin{eqnarray}
 \label{eq:1606}\gamma
&=& \frac{2\langle S_{\psi_0}(A,B)\rangle_{\psi_0} + \langle
[A,B]\rangle_{\psi_0}}{2\,(\Delta A)^2_{\psi_0}}~, \\
\label{eq:1299}\frac{1}{\gamma} &=&\frac{2\langle S_{\psi_0}(A,B)
\rangle_{\psi_0}
- \langle
[A,B]\rangle_{\psi_0}}{2\,(\Delta B)^2_{\psi_0}}~, \\
\label{eq:1300}|\gamma| &=& \frac{(\Delta B)_{\psi_0}}{(\Delta
A)_{\psi_0}}~,~\arg{\gamma} = \arctan\left( -i \frac{\langle
[A,B]\rangle_{\psi_0}}{2\langle
S_{\psi_0}(A,B)\rangle_{\psi_0}}\right)~.
\end{eqnarray} Thus, $\gamma$ will in general be complex. 

As a first example take the coherent states $|k,z\rangle$ for which the
``Schwarz equality'' holds (Eqs.\ \ref{delK} and \ref{eq:1081})
 Here the relation \ref{eq:250} takes the form
 
\begin{equation}
  \label{eq:252}
 (K_1 - \langle K_1
\rangle_{k,z})|k,z\rangle = i\,(K_2-\langle
K_2\rangle_{k,z})|k,z\rangle~,~\langle K_j \rangle_{k,z} = z_j\,,j=1,2\,;\,z =
z_1+i\,z_2\,,
\end{equation}
which is just another version of the defining equation $K_-|k,z \rangle =
z |k,z \rangle$. 

As a second example consider the operators $K_1$
and $K_2$ acting on the coherent state $|k,\lambda\rangle$.
According to Eq.\ \ref{eq:163} the uncertainty inequality is an equality.
This implies that
\begin{equation}
  \label{eq:253}
 (K_1- \gamma\,K_2)|k,\lambda\rangle =
(\langle K_1\rangle_{k,\lambda}- \gamma\,\langle
K_2\rangle_{k,\lambda})|k,\lambda\rangle~,
\end{equation}
where $\gamma$ may
be calculated, according to Eqs.\ \ref{eq:1300}, from the relations
 \ref{eq:1120}, \ref{eq:1117}, \ref{eq:1541} (because $[K_1,K_2]=-i\ K_0\,$), 
 and \ref{eq:1125}:
 \begin{equation}
   \label{eq:699}
 |\gamma| = \frac{|1-\lambda^2|}{|1+\lambda^2|}\,,~\arg \gamma = \arctan \left
( \frac{1-|\lambda|^4}{2\,|\lambda|^2\,\sin (2\theta)} \right)\,.  
 \end{equation}
 
\section{The Lie algebra $\mathfrak{so}(1,2)$ in terms of 
 pro\-ducts of 1-mode operators $a$ and $a^+$}
 The Lie algebra $\mathfrak{so}(1,2)$ of the group $SU(1,1)$
 entered quantum
optics - unrecognized as such - in connection with squeezing
\cite{stol,lu,yue,hol}. The group theoretical background was realized
later \cite{mil,fish,wod,schu}. 

 A realization of the
 Lie algebra $\mathfrak{so}(1,2)$ in terms of annihilation and
creation operators $a$ and $a^+$ is given by \cite{lip,barg3a,nied,mosh3}
\begin{equation}
  \label{eq:261}
 K_+ =
\frac{1}{2}(a^+)^2~,~K_- = \frac{1}{2}\,a^2~,~K_0 =
\frac{1}{2}(a^+ a + 1/2)~.
\end{equation} An alternative form is (we put the frequency $\omega=1$ in the
following)
\begin{equation}
  \label{eq:257}
  K_0 = \frac{1}{4}(Q^2+P^2)\,,~~K_1 =\frac{1}{4}(Q^2-P^2)\,,~~K_2=-\frac{1}{4}
(QP+PQ)\,.
\end{equation}
As $K_-$ annihilates the states
$|n_{osc} =0\rangle$ {\em and} $|n_{osc} =1\rangle$,

\begin{equation}
  \label{eq:258}
  K_-|n_{osc}=0 \rangle =0\,,~~  K_-|n_{osc}=1 \rangle =0\,,
\end{equation}
we get 2 different irreducible
representations of the Lie algebra $\mathfrak{so}(1,2)$, one which is given by
states with even numbers of oscillator quanta and one with odd
numbers. Because
\begin{equation}
  \label{eq:262}
 K_0 |n_{osc}\rangle
=\frac{1}{2}\,(n_{osc}+1/2)|n_{osc}\rangle\,,~n_{osc} =0,1,2,\ldots,
\end{equation}
we see that $K_0$
has the eigenvalues
\begin{equation}
  \label{eq:259}
   (2\,n_{osc} + 1/2)/2= n+\frac{1}{4}~~\text{and}~~ (2\,n_{osc}+1 +
1/2)/2= n+\frac{3}{4}\,,~n=0,1,\ldots,\,
\end{equation}
 in the cases of even and odd numbers of quanta, respectively. 

That is to say, we get one irreducible unitary representation
with $k=1/4$ and one with $k=3/4$. 

As to the related group these are true
representations of a 2-fold covering of $SU(1,1)\cong SL(2,\mathbb{R})=
Sp(2,\mathbb{R})$ and a 4-fold covering
of $SO^{\uparrow}(1,2)$. Those 2-fold covering groups of the symplectic
groups $Sp(2n,\mathbb{R})$ in $2n$ dimensions are called ``metaplectic''
 \cite{wei,foll} (see Appendices B and C for more details). 
 
The two representations with $k=1/4$ and $k=3/4$  may be realized
 not only in the 2 subspaces ${\cal H}_+$ and ${\cal H}_-$ of
the Fock space of the harmonic oscillator just mentioned, but also
in the Hilbert space with the series inner product \ref{eq:187} of Ch.\
4 or in the associated Hilbert space with the scalar product \ref{eq:228}.
In the case of $k=3/4$ - but not for $k=1/4$ - one can also use
the Bargmann Hilbert space  with the scalar product \ref{eq:184}.

If one uses  for the harmonic oscillator the conventional Hilbert space
with the scalar product \ref{eq:195} (with the variable $\xi$ replaced
by $q$ and Hermite's functions (see \ref{eq:806})
\begin{equation}
  \label{eq:260}
  f_{n_{osc}}(q) = C_{n_{osc}}\,e^{-q^2/2}\,H_{n_{osc}}(q)\,,~C_{n_{osc}} = 
\text{ const.}\,,
\end{equation}
as basis, then
the subspace ${\cal H}_+$ for the unitary representation with $k=1/4$
 is spanned by 
the Hermite functions with even Hermite polynomials $H_{n_{osc}}$ 
(they are invariant
under the reflection $q \to -q$) and the subspace ${\cal H}_-$ for the
 representation
with $k=3/4$ is spanned by the Hermite functions with odd Hermite polynomials.

In the ``even'' subspace ${\cal H}_+$ the Hamiltonian
\begin{equation}
  \label{eq:254}
  H_{osc}=2\,K_0
\end{equation}
has the eigenvalues 
\begin{equation}
  \label{eq:255}
  (2n_{osc}+ 1/2)\,,
\end{equation} and in the ``odd'' subspace ${\cal H}_-$ its eigenvalues are
\begin{equation}
  \label{eq:256}
  (2n_{osc}+1 +1/2)\,.
\end{equation}

The two subsystems may be given the following model interpretation:

As the odd Hermite functions do all vanish at $q=0$ one may interprete them
as the energy eigenfunctions of a system with the potential
\begin{equation}
  \label{eq:700}
  V(q)=\frac{1}{2}q^2~\text{for}~q\geq 0\,,~V(q)=\infty~\text{for}~q <0\,.
\end{equation} A more mathematical description is that we obtain a system
confined to the half plane $ q \geq 0$ by identifying $-q$ with $q$. The
resulting (classical) system has a phase space 
\begin{equation}
  \label{eq:701}
 \mathbb{R}^2/Z_2[\cdot q]=\{(q,p) \in
 \mathbb{R}^2, (-q,p) \equiv (q,p) \}\,. 
\end{equation}
(The notion $Z_2[\cdot q]$ is to indicate that the group $Z_2 =\{e,-e\}$
acts on the variable $q$.) 
Such a space is called an ``orbifold'' \cite{cush1} (see also Appendix A.3)
 Note that here we do {\em not} identify $-p$ with $p$\,! 

The model interpretation for the even Hermite functions and their Hilbert
space ${\cal H}_+$ I take from Appendix A.3: If we identify the points
$(-q,-p)$ and $(q,p)$, then the resulting orbifold
\begin{equation}
  \label{eq:702}
  \mathbb{R}^2/Z_2[\cdot (q,p)] =\{(q,p) \in
 \mathbb{R}^2,\, (-q,-p) \equiv (q,p) \}\,. 
\end{equation} is a cone with its tip at $(q,p)=(0,0)$. 

We now  may interpret the group $Z_2$ as a kind of ``gauge'' group the action
of which leaves ``observables'' like the expressions \ref{eq:257} and the
even Hermite functions invariant. 

The coordinates $q$ and $p$ are {\em not} observables in this sense, only
the equi\-valence classes
\begin{equation}
  \label{eq:738}
  \{(q,p) \equiv (-q,-p)\}\,.
\end{equation}

The crucial point may also be seen as follows:
With help of the relation \ref{eq:234} one can show that 
 the operators \ref{eq:261}
 generate the following rotations:
\begin{eqnarray}
  \label{eq:735}
   e^{-i\,\tau\, K_0}a\,e^{i\,\tau\, K_0} &=& e^{i\,\tau/2}\,a\,, \\
 e^{-i\,\tau\, K_0}a^+\,e^{i\,\tau\, K_0} &=& e^{-i\,\tau/2}\,a^+\,,
\label{eq:736}
\\ \label{eq:1313}e^{-i\,\tau\,K_0}Q\,e^{i\,\tau\,K_0}& =&
\cos\frac{\tau}{2}\,Q-\sin\frac{\tau}{2}\,P\,,
\\ \label{eq:1314}e^{-i\,\tau\,K_0}P\,e^{i\,\tau\,K_0} &=&
 \sin\frac{\tau}{2}\,Q+ \cos\frac{\tau}{2}\,P\,,
\\ e^{-i\,\tau\,K_0}K_1\,e^{i\,\tau\,K_0}& =&
\cos\tau\,K_1+\sin\tau\,K_2\,,\label{eq:737}
\\e^{-i\,\tau\,K_0}K_2\,e^{i\,\tau\,K_0} &=&
 -\sin\tau\,K_1+ \cos\tau\,K_2.
\end{eqnarray}

These formulae show that $Q$ and $P$ transform according to the maximal compact
subgroup \ref{ka1} of the symplectic group $Sp(2,\mathbb{R})$, whereas
$K_1$ and $K_2$ transform according to the maximal compact subgroup $O(2)$
of $SO^{\uparrow}(1,2)\cong Sp(2,\mathbb{R})/Z_2$, where $Z_2$ denotes the 
center of $Sp(2,\mathbb{R})$!

 The difference is expressed by the property that for $\tau = 2\pi$ the
operators $Q$ and $P$ change sign whereas $K_1$ and $K_2$ remain invariant!
See also the closely related discussion in Sec.\ 1.3\,!
 Gauge transformations like
 \begin{equation}
   \label{eq:739}
   Z_2[\cdot(q,p)]:~~(q,p) \to (-q,-p)
 \end{equation}
 have been
discussed in a number of papers by Prokhorov and Shabanov \cite{prok1} and
the subject was reviewed recently by Shabanov \cite{sha1}.
 There is one crucial difference, however,
between one of their quantum mechanical conclusions and the above result
\ref{eq:255}   concerning the energy spectrum:

 They also get a doubling of
the energy levels of the harmonic oscillator, but {\em including} the
ground state energy, whereas the ground state energy in Eq.\ \ref{eq:255}
is the {\em same as that of the harmonic oscillator}. However, they derive
the energy spectrum of the orbifold \ref{eq:702} by means of a semi-classical
Bohr-Sommerfeld procedure  which is known to be quantitatively unreliable
 as far as ground state energies are concerned.

What does all this mean for quantum optics? The crucial point is that
the orbifold \ref{eq:702}, with the tip deleted, is diffeomorphic to the
phase space \ref{eq:5} we started from (see also Appendix A.3). 
So the above unitary representation
with $k=1/4$ may be a possible candidate for a quantization of that phase
space!

Squeezing properties of the even and odd oscillator states discussed here
are analyzed in Ref.\ \cite{pari}.

\subsection{Squeezing of $Q$ or $P$}
The special forms \ref{eq:261} or \ref{eq:257} of $K_0, K_1$ and $K_2$
have all the properties discussed in general in the previous section,
especially those as to squeezing from Sec.\ 6.2. For completeness I 
add here their action on the canonical operators 
\begin{equation}
  \label{eq:263}
 Q
=\frac{1}{\sqrt{2}}(a^+ +a)\,,~P=
\frac{i}{\sqrt{2}}(a^+ - a)\,,
\end{equation}
namely
\begin{eqnarray}
\label{eq:1609}e^{-i\,w_2\,K_1}Q\,e^{i\,w_2\,K_1} &=& \cosh(w_2/2)\,Q
+ \sinh(w_2/2)\,P ~, \\
\label{eq:1310}e^{-i\,w_2\,K_1}P\,e^{i\,w_2\,K_1} &=& \cosh(w_2/2)\,P
+  \sinh(w_2/2)\,Q~, \\
\label{eq:1311}e^{-i\,w_1\,K_2}Q\,e^{i\,w_1\,K_2} &=& e^{w_1/2}\,Q\,, \\
\label{eq:1312}e^{-i\,w_1\,K_2}P\,e^{i\,w_1\,K_2} &=& e^{-w_1/2}\,P\,.
 \end{eqnarray}
 All these
transformations and the rotations \ref{eq:1313}-\ref{eq:1314} leave 
the commutation relations
\begin{equation}
  \label{eq:264}
 Q\,P-P\,Q=i
\end{equation}
invariant. This is a consequence of the property that the
group $SU(1,1)$ is isomorphic to the symplectic group $Sp(2,\mathbb{R})$ 
in $1+1$ dimensions (Appendix B). 

 In quantum optics the operators $Q$ and
$P$ are frequently denoted by $\hat{X}_1$ and $\hat{X}_2$ or
$\hat{X}$ and $\hat{Y}$ and called ``quadratures'', reflecting the
two orthogonal directions in the associated phase space
$\mathbb{R}^2$: \begin{eqnarray} \label{eq:1610}Q&=&\hat{X}_1 =
\hat{X}~, \\ \label{eq:1315}P&=& \hat{X}_2 = \hat{Y}~. \end{eqnarray}

 The
relations \ref{eq:1311} and \ref{eq:1312} show explicitly the squeezing
(``Lorentz boost'') generated by $K_2$.

The operators $K_{1\pm}$ and $K_{2\pm}$ from the  section 6.2.2
here have the form
\begin{eqnarray} \label{eq:1611}K_{1+}=K_1+K_0&=& \frac{1}{4}\,(a^++a)^2 =
\frac{1}{2}\,Q^2~, \\
\label{eq:1316}K_{1-}=K_1-K_0&=& \frac{1}{4}\,(a^+-a)^2 =
-\frac{1}{2}\,P^2~, \\
\label{eq:1317}K_{2+}=K_2+K_0&=& \frac{1}{4i}\,(a^++ia)^2 =\\
\label{eq:1318}&=&\frac{1}{4}(\,Q^2+P^2-Q\,P-P\,Q)\,
,  \\
\label{eq:1319}K_{2-}=K_2-K_0&=& \frac{1}{4i}\,(a^+-ia)^2 =\\
\label{eq:1320}&=&-\frac{1}{4}(\,Q^2+P^2+Q\,P+P\,Q) ~,
\nonumber \end{eqnarray} They have, of course, all the
algebraic properties listed in that section. 
\subsection{The $\mathfrak{so}(1,2)$ structure of squared hermitian 
amplitudes}
 The $\mathfrak{so}(1,2)$ Lie
algebra realization \ref{eq:261} also occurs in  so-called
``amplitude-squared squeezing'' \cite{hill1,ger2,hill2}. The starting
observation is of general interest:

  Let
\begin{equation}
  \label{eq:265}
E=\lambda\,(\, a\,e^{-i\,\omega\,t} + a^+\,e^{i\,\omega\,t})\,,~~
\lambda \in \mathbb{R}\,,
\end{equation}
be the hermitian quantized amplitude
of a 1-mode field. Then its square has the form \begin{eqnarray}
\label{eq:1612}E^2&=& \lambda^2\,\{[(a^+)^2+(a)^2]\,\cos(2\,\omega\,t)+ \\
&&+i\,[(a^+)^2-(a)^2]\,\sin(2\,\omega\,t) + \label{eq:1321} +
 2\,a^+a+1\}  \nonumber \\
\label{eq:1322}&=& 4\,\lambda^2\,[K_1\,\cos(2\,\omega\,t)-K_2\,
\sin(2\,\omega\,t)
+K_0]~.  \end{eqnarray} Thus, the Lie algebra \ref{eq:261}
generates the squared quantized amplitude $E^2$! 
We shall come back to the more general case involving interfering amplitudes
in Ch.\ 8. 
\section{The Lie algebra  $\mathfrak{so}(1,2)$ in terms of  2-mode operators}
The realization \ref{eq:261} of the  Lie algebra $\mathfrak{so}(1,2)$
 by a single pair
of annihilation and creation operators leads only to two irreducible unitary
representations with the Bargmann indices $k = 1/4\,,3/4$.

 The
outcome is much richer if one takes two pairs
\cite{bied,bar2,sol,wod,schu}: The operators
\begin{equation}
  \label{eq:266}
 K_+=a_1^+a_2^+~,
~K_-=a_1a_2~,~K_0=\frac{1}{2}(a_1^+a_1+a_2^+a_2+1)~
\end{equation}
obey the
commutation relations \ref{eq:55}. 

 The tensor product $ \mbox{$\cal
H$}^{osc}_1\otimes \mbox{$\cal H$}^{osc}_2$ of the two harmonic
oscillator Hilbert spaces contains all the irreducible unitary
representations of the group \\
 $SU(1,1)\cong SL(2,\mathbb{R})=Sp(2,\mathbb{R})$
 (for which
$k=1/2,1,3/2,\ldots$) in the following way: 

 Let $|n_j\rangle_j,~n_j =0,1,\ldots\,, \, j=1,2,$ be the eigenstates
of the number operators $N_j$, generated by $a^+_j$ from the
oscillator ground states. 

 Then each of those two subspaces of
 $\mbox{$\cal H$}^{osc}_1\otimes \mbox{$\cal H$}^{osc}_2= \{|n_1
 \rangle_1\otimes
 |n_2\rangle_2\}$ with fixed $|n_1-n_2|\neq 0$  contains an irreducible
  representation
 with
\begin{equation}
  \label{eq:267}
 k=1/2+|n_1-n_2|/2=1,3/2,2,\ldots
\end{equation}
(the operator $N_1-N_2$ commutes
  with all 3 operators in Eqs.\ \ref{eq:266}) and for which the number
  $n$ in the eigenvalue $n+k$ of $K_0$ is given by
\begin{equation}
  \label{eq:268}
  n=\min\{n_1,n_2\}~.
\end{equation}
For the ``diagonal'' case $n_2=n_1$ one gets the unitary
   representation with $k=1/2$. 

Actually the operators \ref{eq:266} are only 3 of 10 independent 
(hermitian) bilinear operators one can build from the two pairs $a_1,a_1^+$
and $a_2,a_2^+$. Those 10 operators generate the Lie algebra $\mathfrak{sp}(4)$
of the real symplectic group $Sp(4,\mathbb{R})$ which plays a major role
in our analysis, too (see Ch.\ 8 and Appendix C).

 Using the realizations one finds \cite{vour4} that the
 two pairs $a_1\,,a_1^+$ and
   $a_2\,,a_2^+$ are $SU(1,1)$- or $Sp(2,\mathbb{R})$- transformed as follows:
\begin{eqnarray}
\label{eq:1613}e^{-i\,w_2\,K_1}a_1\,e^{i\,w_2\,K_1} &=& \cosh(w_2/2)\,a_1 +
i\,\sinh(w_2/2)\,a^+_2\,,~~~ \\
\label{eq:1323}e^{-i\,w_2\,K_1}a^+_1\,e^{i\,w_2\,K_1} &=& \cosh(w_2/2)\,a^+_1 -
i\,\sinh(w_2/2)\,a_2\,,~~~ \\
\label{eq:1324}e^{-i\,w_1\,K_2}a_1\,e^{i\,w_1\,K_2} &=& \cosh(w_1/2)\,a_1 +
\sinh(w_2/2)\,a^+_2\,, \\
\label{eq:1325}e^{-i\,w_1\,K_2}a^+_1\,e^{i\,w_1\,K_2} &=& \cosh(w_1/2)\,a^+_1 +
\sinh(w_2/2)\,a_2\,, \\
\label{eq:1326}e^{-i\,\tau\,K_0}a_1\,e^{i\,\tau\,K_0} = e^{i\,
\tau/2}\,a_1\,,&~&
e^{-i\,\tau\,K_0}a^+_1\,e^{i\,\tau\,K_0} =
e^{-i\,\tau/2}\,a^+_1\,,
\end{eqnarray}
The other half of the resulting relations is obtained by exchanging the
indices $1$ and $2$. 

 As to squeezing the following implications
are of special interest: \begin{eqnarray}
\label{eq:1614}e^{-i\,w_1\,K_2}(a_1+a_2^+)\,e^{i\,w_1\,K_2} &=&
e^{w_1/2}\,(a_1+a_2^+)~,\\
\label{eq:1327}e^{-i\,w_1\,K_2}(a_1-a_2^+)\,e^{i\,w_1\,K_2} &=&
e^{-w_1/2}\,(a_1-a_2^+)~. \end{eqnarray} The rest is obtained by
hermitian conjugation. 

Introducing 
\begin{equation}
  \label{eq:704}
  Q_j =\frac{1}{\sqrt{2}}(a^+_j +a_j)\,,~~P_j =\frac{i}{\sqrt{2}}(a^+_j -
a_j)\,,\,j=1,2, 
\end{equation}
  and \begin{eqnarray} \label{eq:1615}Q_{\pm} &=& Q_1
\pm Q_2~, \\
\label{eq:1328}P_{\pm} &=& P_1 \pm P_2~, \\
\label{eq:1329}{[}Q_+,P_+{]} &=& 2\,i~, \\
\label{eq:1330}{[}Q_-,P_-{]} &=& 2\,i~, \\
\label{eq:1331}{[}Q_+,P_-{]} &=& 0~, \\ \label{eq:1332}{[}Q_-,P_+{]} &=& 0~,
\end{eqnarray} implies the squeezing
relations
\begin{eqnarray}
\label{eq:1616}e^{-i\,w_1\,K_2}Q_+\,e^{i\,w_1\,K_2} &=&
e^{w_1/2}\,Q_+~,\\
\label{eq:1333}e^{-i\,w_1\,K_2}P_+\,e^{i\,w_1\,K_2} &=&
e^{-w_1/2}\,P_+~,\\
\label{eq:1334}e^{-i\,w_1\,K_2}Q_-\,e^{i\,w_1\,K_2} &=&
e^{-w_1/2}\,Q_-~,\\
\label{eq:1335}e^{-i\,w_1\,K_2}P_-\,e^{i\,w_1\,K_2} &=& e^{w_1/2}\,P_-~.
\end{eqnarray}

The combinations $Q_{\pm}\,,P_{\pm}$ of the two original pairs
$(a_1\,,a_1^+)\,,(a_2,a_2^+)$ occur in the discussions \cite{ar} of
the problem how to measure the values of a non-commuting pair of
canonically conjugate observables like, e.g.\ $Q_1$ and $P_1$. The
second pair $(Q_2\,,P_2)$ is related to properties of the
measuring device. The discussions exploit the fact that the
operators $Q_-$ and $P_+$ (or $Q_+$ and $P_-$) commute and may be
measured simultaneously with arbitrary precision. The squeezing
relations \ref{eq:1616}-\ref{eq:1335} underline this interpretation. 

 For the product
coherent state
\begin{equation}
  \label{eq:269}
 |\alpha_1,\alpha_2\rangle = |\alpha_1\rangle
\otimes |\alpha_2 \rangle
\end{equation}
we get the squared uncertainties
\begin{equation}
  \label{eq:270}
( \Delta Q_-)^2_{\alpha_1,\alpha_2} = 1~,~(\Delta
P_+)^2_{\alpha_1,\alpha_2} = 1~,
\end{equation}
According to the relations \ref{eq:1334} and \ref{eq:1333}
 we may ``squeeze'' these uncertainties and make them {\em
simultaneously} arbitrary small by an appropriate choice of
the factor $ \exp(-w_1)$! 

The value $1$ of the squared fluctuations \ref{eq:270} instead of $1/2\,$,
which one has for the basic 1-mode case, is a consequence of the
definitions \ref{eq:1615} and \ref{eq:1328}  of $Q_{\pm}$
 and $P_{\pm}$. A factor
$1/\sqrt{2}$  would have yielded $1/2$ instead of $1$ for the
squared uncertainties \ref{eq:270}. Such a ``renormalization'' may appear
possible for the coordinates $Q_{\pm}$ but as $P_+=P_1+P_2$ is the 
the total momentum of the two modes one should not change its
normalization arbitrarily.
\section{Related applications}
There are a number of other applications of
 the groups $SU(1,1) \cong SL(2,\mathbb{R})=Sp(2,\mathbb{R})$ which
I merely mention here without going into details:
\begin{enumerate}
\item The Lie algebra realization \ref{eq:45} in order to construct
multi-boson squeezed states \cite{kat}.
\item The group $SU(1,1)$ plays also a prominant role in discussions
of Mach-Zehnder type interferometers \cite{yur}.
\end{enumerate}
\section{A few remarks on $SU(1,1)/SO^{\uparrow}(1,2)$ dynamics}
 Up to now I
have discussed only ``kinematical'' aspects of the structure group
$S(1,1)$ etc.\ in quantum optics. The  relations discussed in the previous
sections become
physically much more interesting if the group parameters
$w_j\,,\,j=1,2\,$, $\tau$ etc.\ become dynamical, i.e.\ functions
of time determined by an appropriate Hamiltonian which then
generates squeezed or other states. This can be implemented as follows:

The  Lie algebra $\mathfrak{so}(1,2)$ realizations  \ref{eq:261} and
 \ref{eq:266} appear in physically important 
effective interaction Hamiltonians for nonlinear quantum
optical processes \cite{nonl}.

 Those interaction Hamiltonians
in general take the form 
\begin{equation}
  \label{eq:409}
  H_I = \chi\,K_+ + \bar{\chi}K_-~,
\end{equation}
where the complex quantity $\chi$ contains the time-dependence of $H_I$
and the amplitude(s) and non-linear susceptibilities 
 of the classical source(s) (``pump'') to which those
 quanta (photons) couple which are described by the  creation and annihilation
operators occuring in the Lie algebra operators \ref{eq:261}
and \ref{eq:266}. The free Hamiltonian $H_0$ in $H=H_0+H_I$ is proportional
to $K_0$.

The case \ref{eq:261} is called ``degenerate'' because
both quanta appearing in $K_+$ etc.\ have the same frequency $\omega$ .
 They may either be annihilated,
resulting in {\em one} quantum (better: classical wave) with 
 frequency $2\omega$ (``harmonic generation''),
or created by a classical wave of frequency $2\omega$
 (``parametric down-conversion''). 

The effective interaction Hamiltonian has the same ``degenerate'' structure
 \ref{eq:409} if one has two incoming or outgoing classical waves (``four-wave
mixing'') both with frequency $\omega$ which interact with the two quanta. The
relevant interaction  properties of the two classical waves and the 
nonlinear medium are again incorporated into the quantity $\chi$. 

Qualitatively one encounters the same physical situations in the 
non-degenerate case \ref{eq:266} with the essential difference that
one has now 2 possible modes which can have different frequencies $\omega_1$
and $\omega_2$, so that e.g.\ in non-degenerate parametric down-conversions
a wave of frequency $2\omega$ generates 2 quanta with frequencies $\omega_1$
and $\omega_2$ such that $\omega_1+\omega_2 =2\omega$. 

What is especially important in all these processes is the property of
the self-adjoint operator \ref{eq:409} to generated squeezed states as we have
seen in detail in Sec.\ 6.2. The crucial point is that the
interaction Hamiltonian \ref{eq:409} is an element of the Lie algebra
 $\mathfrak{so}(1,2)$. In addition, the free Hamiltonian $H_0$ in general will
be built from $K_0$. Thus, the Lie algebra $\mathfrak{so}(1,2)$  plays
an essential role in the {\em dynamics} of quantum optical processes, too!

\chapter{(Pseudo-) Probability distributions on the phase space 
${\cal S}^2_{\vp,I}$}
\section{Preliminaries}
A lot of efforts has
been put into attempts to find ways for expressing quantum
mechanical expectation values
\begin{equation}
  \label{eq:271}
 \langle A \rangle_{\rho} =
\tr\rho\,A)
\end{equation}
of a quantum observable (self-adjoint operator)
$A$ with respect to a state - characterized by the density
operator $\rho$ - in terms of a classical density function
$w(q,p)$ on, e.g.\  the phase space
\begin{equation}
  \label{eq:272}
 {\cal S}^2_{q,p} = \{(q,p) \in
\mathbb{R}^2 \}\,,
\end{equation}
and a phase space function $\tilde{A}(q,p)$
(corresponding to the operator $A$) such that
\begin{equation}
  \label{eq:273}
 \langle A
\rangle_{\rho} = \int_{{\cal S}_{(q,p)}} dq\,dp\,
w(q,p)\,\tilde{A}(q,p)~.
\end{equation}
The oldest proposals are those of Weyl
\cite{wey} and  especially Wigner \cite{wig1}. The corresponding Wigner
function
\begin{equation}
  \label{eq:274}
 w(q,p) =
\frac{1}{2\pi}\,\int_{-\infty}^{\infty}\,dx\,e^{-ipx}\,\langle
q+x/2|\rho|q-x/2 \rangle
\end{equation}
is widely used in the modern quantum
optical literature (see the Refs. 
\cite{texb1,texb2,texb3,texb4,texb5,texb6,texb7,texb8,texb9}, especially
Ref.\ \cite{texb8}; additional reviews are
 \cite{tat,wig2,folla,kim3,lee,gad,drag}).

 Because of its intensive use of Fourier
transformations that approach is closely related to the phase
space \ref{eq:272}, its global structure and the associated harmonic analysis
in terms of the translation groups in coordinate and momentum space.
  As the density \ref{eq:274} may become
negative, it is actually not a genuine  probability density. Its negative
values in certain regions of the phase space \ref{eq:272} are usually 
attributed to typical quantum effects. 

Inspired by - among others - a paper by Uhlenbeck \cite{uhl} Husimi in 1940
published a long and very interesting article on various  properties
of the density matrix \cite{hus} in which he suggested to use the ``diagonal''
matrix elements 
\begin{equation}
  \label{eq:275}
 Q(\alpha,\bar{\alpha}) = \langle \alpha |\rho| \alpha \rangle~.
\end{equation}
of the 
 ``coherent'' states $|\alpha = (q+ip)/\sqrt{2}\rangle $ for a
reconstruction of the density operator. Husimi used Gaussian wave packets
and recognized the non-orthogonality \ref{eq:1061} of different such states.

The distribution function \ref{eq:275} is always non-negative. It was later
rediscovered \cite{kan}.

 Of special interest
has been another  highly singular and non-positive ``distribution function''
associated with the density operator,  the so-called
 ``Sudarshan-Glauber diagonal P-representation'' 
 \cite{su,glau2} in terms of
 the coherent states $|\alpha
\rangle \,$:
\begin{equation}
  \label{eq:276}
 \rho = \frac{1}{\pi}\int_{\mathbb{C}}d^2\alpha
\,P(\alpha,\bar{\alpha})\,|\alpha\rangle \langle \alpha |~.
\end{equation}
If it exists
then the expectation values of appropriate operators $B$ are given
by the convenient expression
\begin{equation}
  \label{eq:277}
 \tr\rho\,B)=
\int_{\mathbb{C}}d^2\alpha\,P(\alpha,\bar{\alpha})\,\langle \alpha|B|\alpha
\rangle ~.
\end{equation}
In general $P(\alpha, \bar{\alpha})$ can be negative in certain
regions of the phase space and highly singular
 \cite{su,glau2,kla2,kla3,mill}, namely a
generalized function (linear functional) belonging to the space
$Z'$, which contains, e.g.\ infinite series where the n-th term is
proportional to the n-th derivative of a delta-function \cite{gel,zem}!

 Expressing the Wigner function $w(q,p)$ in terms of the complex variables
$\alpha, \bar{\alpha}$,
\begin{equation}
  \label{eq:278}
 w(\alpha, \bar{\alpha}) = \frac{1}{\pi}
\int_{\mathbb{C}}d^2\alpha_1\ \langle \alpha-\alpha_1|\rho|\alpha
+ \alpha_1 \rangle
\,e^{\alpha\,\bar{\alpha}_1-\bar{\alpha}\,\alpha_1}~,
\end{equation}
one gets
the following relations between the densities $w(\alpha, \bar{\alpha}),
Q(\alpha, \bar{\alpha})$ and $P(\alpha, \bar{\alpha})$:
\begin{eqnarray} \label{eq:1617}w(\alpha, \bar{\alpha}) &=&
\frac{2}{\pi}\,\int_{\mathbb{C}}d^2\alpha_1
P(\alpha_1, \bar{\alpha}_1)\,e^{-2|\alpha_1-\alpha|^2}~, \\
\label{eq:1336}Q(\alpha, \bar{\alpha}) &=& \frac{1}{\pi}\,
\int_{\mathbb{C}}d^2\alpha_1
P(\alpha_1, \bar{\alpha}_1)\,e^{-|\alpha_1-\alpha|^2}~. \end{eqnarray}

 Before I
generalize essential properties of the $|\alpha \rangle$-related
 densities $Q(\alpha,\bar{\alpha})$ and
$P(\alpha, \bar{\alpha})$ to those associated with the
 states $|k,z\rangle$ and $|k,\lambda
\rangle $ - a corresponding generalization of the Wigner function
is not obvious, at least not to me - let me make some general
remarks:

We have seen - already in the Introduction - that the three
appropriate basic ``canonical'' functions on the phase space \ref{eq:5}
are \[h_0 =I\,,~ h_1 = I\cos\vp\,,~h_2=-I\sin\vp\,. \] Thus, one
would start by considering classical probability densities and
``observables'' like
\begin{equation}
  \label{eq:279}
 w(h_0,h_1,h_2)~, ~G(h_0,h_1,h_2)~.
\end{equation}
However, as the three variables $h_j$ are not independent,
 \[ h_0^2 -h_1^2-h_2^2
= 0~, \] the corresponding integration measure is
\begin{equation}
  \label{eq:280}
 \delta(h_0^2
-h_1^2-h_2^2)dh_0dh_1dh_2~.
\end{equation}
As
\begin{equation}
  \label{eq:281}
 \delta(h_0^2
-h_1^2-h_2^2) = \frac{1}{2
\sqrt{h_1^2+h_2^2}}\,[\delta(h_0=\sqrt{h_1^2+h_2^2})+ \delta(h_0=
-\sqrt{h_1^2+h_2^2})]~
\end{equation}
and because $h_0 > 0$ only the first
delta-function contributes. In this way the measure \ref{eq:280} 
reduces to (see Eq.\ \ref{eq:3} )
\begin{equation}
\label{eq:381}
\frac{1}{2\sqrt{h_1^2+h_2^2}}dh_1dh_2 =\frac{1}{2}\,d\vp dI =
\frac{1}{2}dqdp~.
\end{equation}
One just has to replace $h_0$  in $w$ and $G$ of Eq.\ \ref{eq:279} by $
\sqrt{h_1^2+h_2^2}$.

 Obviously,  the measure \ref{eq:381} is
equivalent to the canonical measure $dq dp =d\vp dI $ on phase space. 

The situation is more involved for the quantized theories: For an
irreducible unitary representation with index $k$ we have - as
mentioned several times before - the operator relation
\begin{equation}
  \label{eq:282}
 K_1^2 +
K_2^2 = K_0^2 + l\,,\, l = k(1-k)~.
\end{equation}
Thus, except for the case
$k=1$, the quantum fluctuations ($ k$ characterizes the
non-classical ground state)  modify the classical Pythagorean
relation $h_1^2+h_2^2 = h_0^2$ . 

 If one treats the relation \ref{eq:282}
as a constraint - together with $\tr\rho)=1$ - in order to
determine the density operator $\rho$ by maximizing the associated
entropy \cite{jay}
\begin{equation}
  \label{eq:283}
 S  = - \tr\rho\,\ln\rho)~,
\end{equation}
subject to the
two constraints, then one gets
\begin{equation}
  \label{eq:284}
 \rho =
\frac{e^{-\gamma\,(K_1^2+K_2^2 -K^2_3)}}{Z(\gamma)}~,~Z(\gamma)=
-\ln(\mbox{tr}[e^{-\gamma\,(K_1^2+K_2^2 -K^2_3)}])~.
\end{equation}
In possible
applications the relation 
\begin{equation}
  \label{eq:285}
 K_1^2+K_2^2 -K_0^2 = K_+K_- -K_0^2
+K_0
\end{equation}
will be useful. 

 I will not pursue this ansatz further
here - if $l$ is fixed, a ``microcanonical'' approach
will be more appropriate - and turn to the indicated generalizations
of the expressions \ref{eq:275} and \ref{eq:276}.
 By doing so I
shall just scratch the surface of the underlying structures and
the substance of the many problems. The main purpose is to point
out possible directions for future research. What is needed is a thorough
investigation along the lines of Ref.\ \cite{wolf2} for
 the conventional coherent states. Several points of the present chapter
appear to be related to those of Ref.\ \cite{fan}.
\section{Pseudo-probability distributions asso\-cia\-ted with \\
Barut-Girar\-del\-lo coherent states} 
 Many concepts developed in
connection with pseudo-probability distributions (densities) related to 
the conventional coherent states $|\alpha \rangle
$ may be carried over to the states $|k,z\rangle$. 
There are,
however, at least two important differences: 

 First, contrary to
the states $|\alpha\rangle = D(\alpha)|0\rangle$ which can be
generated form the ground state by a unitary (translation)
operator - Eq.\ \ref{eq:140} -, this appears not to be the case for the
states $|k,z\rangle $. 

 Second, whereas the measure $d^2\alpha$ is
invariant under rotations and translations, the measure
$d\mu_k(z)$ - Eq.\ \ref{meas1} - is invariant only under rotations. The
same applies to the measure $\tilde{\mu}_k(z)$ from Eq.\ \ref{eq:1057}.
 
In the following I use the notations and definitions from Sec.\
3.1. 

 Let
\begin{equation}
  \label{eq:286}
 A = \sum_{m,n =0}^{\infty} |k,m\rangle
A_{k;m\,n}\langle k,n|
\end{equation}
be a self-adjoint operator in the
number state representation \ref{k3}. Then \begin{eqnarray}
 \label{eq:1618}\langle
k,z_2|A|k,z_1 \rangle &=& \label{mael}
[g_k(|z_2|^2)\,g_k(|z_1|^2)]^{-1/2}\,A(k;\bar{z}_2,z_1)~, \\
\label{eq:1337}A(k;\bar{z}_2,z_1)&=&\sum_{m,n =0}^{\infty}
\bar{\tilde{f}}_{k,m}(z_2)\,A_{k;m\,n}\,\tilde{f}_{k,n}(z_1) =  \\
\label{eq:1338}&=& \sum_{m,n
=0}^{\infty}\frac{A_{k;m\,n}}{\sqrt{m!(2k)_m\,n!(2k)_n}}\bar{z}_2^m\,z_1^n
~. \nonumber
\end{eqnarray} Self-adjointness of $A\,,A^+ = A $, implies
\begin{equation}
  \label{eq:287}
\bar{A}(k;\bar{z}_2,z_1) = A(k;\bar{z}_1,z_2)~.
\end{equation}
The expression
\ref{eq:1337} shows that $A(k;\bar{z}_2,z_1)$ is a double series, holomorphic
in $z_1$ and anti-holomorphic in $\bar{z}_2$. Because of the
additional factors $\sqrt{(2k)_m\,(2k)_n}$ in the denominator
which for large $m\,,n$ behave like $\sqrt{m!\,n!}$,  the
convergence of the series \ref{eq:1337} is better than  the corresponding one
for the states $|\alpha \rangle$ \cite{su,glau2}. 

Using the orthonormality \ref{eq:84} of the functions $ \tilde{f}_{k,m}$ we
get the inversion
\begin{equation}
  \label{eq:288}
 A_{k;m\,n} =
\int_{\mathbb{C}}\int_{\mathbb{C}}d\tilde{\mu}_k(z_2)\,d\tilde{\mu}_k(z_1)
\tilde{f}_{k,m}(z_2)\,A(k;\bar{z}_2,z_1)\,\bar{\tilde{f}}_{k,n}(z_1)~.
\end{equation}
From the completeness relation \ref{com1} we have
 \begin{eqnarray} \label{eq:1619}A
&=& \int_{\mathbb{C}}\int_{\mathbb{C}}d\mu_k(z_2)\,d\mu_k(z_1)\,
|k,z_2\rangle \langle k, z_2|A|k,z_1\rangle \langle k,z_1|= \\
\label{eq:1339}&=&
\int_{\mathbb{C}}\int_{\mathbb{C}}d\tilde{\mu}_k(z_2)\,d\tilde{\mu}_k(z_1)
A(k;\bar{z}_2,z_1)\,|k,z_2\rangle \langle k,z_1|~. \end{eqnarray}
For the product $A_2\cdot A_1$ of two operators we obtain
\begin{equation}
  \label{eq:289}
(A_2\cdot A_1)(k;\bar{z}_2,z_1) =
\int_{\mathbb{C}}d\tilde{\mu}_k(z)\,A_2(k;\bar{z}_2,z)\,A_1(k;\bar{z},z_1)~.
\end{equation}
Let us apply these relations to the density operator
\begin{equation}
  \label{eq:290}
 \rho
=  \sum_{m,n =0}^{\infty} |k,m\rangle \rho_{k;m\,n}\langle k,n|
\end{equation}
with its special properties. \begin{eqnarray} \label{eq:1620}\tr\rho) =
\sum_{n=0}^{\infty}\rho_{k;n\,n} &=&
\int_{\mathbb{C}}\int_{\mathbb{C}}d\tilde{\mu}_k(z_2)\,d\tilde{\mu}_k(z_1)\,
\tilde{\Delta}_k(z_2,\bar{z}_1)\,\rho(k;\bar{z}_2,z_1)= ~~~~~\\
\label{eq:1340}&=& \int_{\mathbb{C}}d\tilde{\mu}_k(z_1) \rho(k;\bar{z}_1,z_1) =
1~.
\end{eqnarray} Eqs.\ \ref{eq:288} with $A=\rho$ and \ref{eq:85} were used here
and in the last step use was made of the fact that
$\tilde{\Delta}_k(z_2,\bar{z}_1)$ is a reproducing kernel (Eq.\ \ref{eq:90})
 and
$\rho(\bar{z}_2,z_1)$ an anti-holomorphic function in $\bar{z}_2$.

  We may get the same result more directly from
Eq.\ \ref{mael}:
\begin{equation}
  \label{eq:291}
\tr\rho) = \int_{\mathbb{C}}d\mu_k(z) \langle k,z|\rho|k,z\rangle
=\int_{\mathbb{C}}d\tilde{\mu}_k(z)\,\rho(k;\bar{z},z)~.
\end{equation}
From the positivity of the (Husimi-type) function
\begin{equation}
  \label{eq:292}
 S_k(z,\bar{z}) \equiv \langle k,z|\rho|k,z \rangle \geq 0
\end{equation}
and 
\ref{mael} it follows further that
\begin{equation}
  \label{eq:293}
 \rho(k;\bar{z},z) \geq 0~.
\end{equation}
The
expectation value of a self-adjoint operator $A$ is given by
\begin{equation}
  \label{eq:294}
\langle A \rangle_{\rho;k,z} = \tr\rho\,A) =
\int_{\mathbb{C}}\int_{\mathbb{C}}d\tilde{\mu}_k(z_2)\,d\tilde{\mu}_k(z_1)
\rho(k;\bar{z}_2,z_1)\,A(k;\bar{z}_1,z_2)~.
\end{equation}
Examples:
\begin{equation}
  \label{eq:295}
 \rho
=|k,n\rangle \langle k,n|~.
\end{equation}
For this density operator we get
\begin{eqnarray} \label{eq:1621}\langle k,z_2|\rho|k,z_1\rangle &=& \frac{ \rho(k;\bar{z}_2,z_1)}{
 \sqrt{g_k(|z_2|^2)\,g_k(|z_1|^2)}} \\ \label{eq:1341}\rho(k;\bar{z}_2,z_1)
 &=& \label{rokn}
\bar{\tilde{f}}_{k,n}(z_2)\,\tilde{f}_{k,n}(z_1) =
\frac{(\bar{z}_2\,z_1)^n}{(2k)_n\,n!}~. \end{eqnarray} For
\begin{equation}
  \label{eq:296}
\rho = (1-a)\,\sum_{n=0}^{\infty} a^n\,|k,n\rangle \langle k,n|
~,~ 0 < a <1~,
\end{equation}
we have
\begin{equation}
  \label{eq:297}
 \langle k,z_2|\rho|k,z_1 \rangle =
\frac{1-a}{\sqrt{g_k(|z_2|^2)\,g_k(|z_1|^2)}}\,g_k(a\,\bar{z}_2\,z_1)~.
\end{equation}
This implies that
\begin{equation}
  \label{eq:298}
 S_k(z,\bar{z}) \equiv \langle k,z|\rho|k,z \rangle = 
(1-a)\,\frac{g_k(a\,|z|^2)}{g_k(|z|^2)}~.
\end{equation}
We further
note the relation
\begin{equation}
  \label{eq:299}
 \int_0^{2\pi}d\phi\ |k,|z|e^{i\,\phi}\rangle
\langle k,|z|e^{i\,\phi}| =
\frac{2\pi}{g_k(|z|^2)}\,\sum_{n=0}^{\infty}
\frac{|z|^{2\,n}}{(2k)_n\,n!}\,|k,n\rangle \langle k,n|~.
\end{equation}

Next let us suppose that a certain class of density operators
allows for a so-called ``diagonal representation'' \cite{su,glau2,kla3}
\begin{equation}
  \label{eq:300}
\rho = \int_{\mathbb{C}}d\mu_k(z)\, F_k(z,\bar{z})|k,z\rangle
\langle k,z|~.
\end{equation}
Such a representation will in general require $F_k$ to be
 a generalized function of
type $Z'$, see below.

 Because $\rho$ is self-adjoint, $F_k$ has to
be real. Furthermore, in general $F_k$ will depend on both
variables $z$ and $\bar{z}$. 

(Note the frequent change of the measures $d\mu_k(z)$ and $d\tilde{\mu}_k(z)$
in the following!)

 If the representation \ref{eq:300} is
available, it has a number of interesting properties, e.g.
\begin{eqnarray} \label{eq:1622}\tr\rho\,A)&=&
\int_{\mathbb{C}}d\mu_k(z_1)\int_{\mathbb{C}}d\mu_k(z)\,F_k(z,\bar{z})\,\langle
k,z_1|k,z\rangle \langle k,z|A|k,z_1 \rangle = ~~~~\\ \label{eq:1342}&=&
\int_{\mathbb{C}}d\tilde{\mu}_k(z)\,F_k(z,\bar{z})\,A(k,\bar{z},z)~,
\end{eqnarray} where the completeness relation \ref{com1}
 has been used. 

 If we
put $A=\boldsymbol{1}\,$, then we get the normalization 
\begin{equation}
  \label{eq:301}
 \tr\rho) =
\int_{\mathbb{C}}d\mu_k(z)\, F_k(z,\bar{z}) =1~.
\end{equation}
The diagonal
representation \ref{eq:300} of the density operator  is especially convenient
as to operators of the type
\begin{equation}
  \label{eq:302}
 B_N = \sum_{n_+,n_- =0}^{N_+,N_-}
b_{n_+,n_-}\,K_+^{n_+}\,K_-^{n_-}~,
\end{equation}
because
\begin{equation}
  \label{eq:303}
 \tr \rho B_N)
= \int_{\mathbb{C}}d\mu_k(z)\,F_k(z,\bar{z})\,\sum_{n_+,n_- =0}^{N_+,N_-}
b_{n_+,n_-}\,\bar{z}^{n_+}\,z^{n_-}~.
\end{equation}
This is in complete
analogy to the usual normal-ordered polynomials of the operators
$a^+$ and $a$ applied to the states $|\alpha \rangle\,$. 

However,
normal-ordering is more complicated in the framework of our Lie
algebra, because we have a third operator $K_0\,$! Here
``normal-ordering'' may be defined \cite{per,ban0}  to mean that all the
$K_-$ are to be put to the right in a product, all the $K_+$ to
the left and the $K_0$ in the middle. In rearranging a given
product the following relations have to be observed:
\begin{eqnarray}
\label{eq:1623}K_-\,K_+ &=& K_+\,K_- + 2\,K_0 = \\ &=& l
 + K_0\,(K_0+1)~,\label{noro1} \\
\label{eq:1344}K_0\,K_+ &=& K_+\,(K_0+1)~, \\ \label{eq:1345}K_-\,K_0 &=&
 (K_0+1)\,K_-~.
\end{eqnarray} They follow from the Eqs.\ \ref{eq:55} and \ref{eq:56}
 The relation \ref{noro1} holds only within an irreducible
representation with a value $l=k(1-k)$ of the Casimir operator. \\An
example for normal ordering  is the relation \ref{eq:137}. Its
anti-normally ordered version is \cite{per,ban0}
\begin{equation}
  \label{eq:304}
 U(w) =
e^{-\bar{\lambda}\,K_-}\,e^{-\ln(1-|\lambda|^2)K_0}\,e^{\lambda\,K_+}~.
\end{equation}
Anti-normal ordering here obviously means to put all the $K_+$ to
the right, the $K_-$ to the left and the $K_0$ again in the
middle. 

 Normal-ordering of more general operators like 
\begin{equation}
  \label{eq:147}
  e^{a_+K_+ +a_-K_- +a_3K_0} 
\end{equation} have also been discussed in the literature 
\cite{san,trua,ban0,wue1}, 
the most extensive discussion of normal and anti-normal ordering 
in the present context being contained in Ref.\ \cite{ban0}.

 The expectation value of anti-normally ordered operators like,
e.g.
\begin{equation}
  \label{eq:305}
 B_A = \sum_{n_+,n_- =0}^{N_+,N_-}
b_{n_+,n_-}\,K_-^{n_-}\,K_+^{n_+}~,
\end{equation}
can be expressed in terms
of the Husimi type function $S_k(z,\bar{z})$ from Eq.\ \ref{eq:292}:
 \begin{eqnarray}
 \label{eq:1624}\tr\rho B_A) &=&
\int_{\mathbb{C}}d\mu_k(z)\,\langle k,z|\rho\,B_A|k,z \rangle = \\
\label{eq:1346}&=& \int_{\mathbb{C}}d\mu_k(z)\,\langle k,z|\rho|k,z \rangle
\sum_{n_+,n_- =0}^{N_+,N_-}
b_{n_+,n_-}\,\bar{z}^{n_+}\,z^{n_-}~. \nonumber \end{eqnarray}
Here the cyclic property of the trace has been used. 

 If a
density operator has the anti-normally ordered form
\begin{equation}
  \label{eq:306}
 \rho^{(A)}
= \sum_{n_+,n_- =0}^{N_+,N_-}
\rho_{n_+,n_-}^{(A)}\,K_-^{n_-}\,K_+^{n_+}~,
\end{equation}
then inserting the
 completeness relation \ref{com1} between the two types of operators gives
\begin{equation}
  \label{eq:307}
 \rho^{(A)} =
\int_{\mathbb{C}}d\mu_k(z)\,\sum_{n_+,n_- =0}^{N_+,N_-}
\rho_{n_+,n_-}^{(A)}\,\bar{z}^{n_+}\,z^{n_-}|k,z\rangle \langle
k,z|~,
\end{equation}
i.e.\ for such density operators the diagonal
representation \ref{eq:300} exists, at least formally. 

Again, formally the density operator
\begin{equation}
  \label{eq:308}
 |k,z_0\rangle \langle k,
z_0 |
\end{equation}
may be expressed in terms of the relation
\begin{equation}
  \label{eq:309}
m_k(|z|)\,F_k(z,\bar{z}) = \delta(z-z_0)~,
\end{equation}
where
$\delta(z-z_0)$ is the 2-dimensional delta-function with respect
to the measure $d^2z\,$. 

 For $k=1/2$ one has to take care of
the logarithmic singularity \ref{eq:1051}. One has \cite{magn}
\begin{equation}
  \label{eq:310}
 K_0(2|z|) = =
-(\gamma + \ln|z|)\,I_0(2|z|)+2
\sum_{n=1}^{\infty}\frac{1}{n}\,I_{2n}(2|z|)~,
\end{equation}
so that
\begin{equation}
  \label{eq:311}
\hat{K}_0(|z|) \equiv -(\,K_0(2|z|)/(\gamma + \ln|z|)
\to 1 \mbox{ for } |z| \to 0
\end{equation}

Similarly one can make use of more singular generalized
functions in order to find a $F_k(z,\bar{z})$ which yields a
diagonal representation for a given density operator. Following
Sudarshan \cite{su} one can be tempted to postulate the following ``duality''
between powers of $r =|z|$ and derivatives of the radial
$\delta$-functions $\delta(r)$:
\begin{equation}
  \label{eq:312}
(-1)^{n_1}\int_{0}^{\infty}dr\,r^{n_2}\,\delta^{(n_1)}(r) =
n_2!\,\delta_{n_1n_2}~.
\end{equation}
Here $\delta^{(n)}(r)$ is the n-th
derivative with respect to $r$. The relation is quite formal
because $r^{n_2}$ is not a test function of any of the spaces
 $D\,, S $ or $Z $ \cite{gel,zem} . Actually,
$r^{\lambda}$ is a generalized  function $ \in S'$ or
$ D'$
itself \cite{gel,zem}! 

 As such a generalized function the expression $r^{n_2}\,\delta^{(n_1)}(r)$ has
the property \cite{zem1}
\begin{equation}
  \label{eq:393}
r^{n_2}\,\delta^{(n_1)}(r)   = \left\{
\begin{array}{r@{\quad:\quad}l} 0 & n_1 < n_2  ~,
\\ (-1)^{n_2}\,n_2!\,\delta(r)& n_2 =n_1 ~, \\ (-1)^{n_2}\,
\frac{n_1!}{(n_1-n_2)!}\,\delta^{(n_1-n_2)}(r)& n_1 > n_2
 ~. \end{array}\right. 
\end{equation}
This is compatible with the postulate \ref{eq:312} if we apply 
the generalized function \ref{eq:393} to test functions which have the
constant value $1$  in a compact interval $ 0 \leq r \leq a >0$ and vanish
for $r \geq b > a$. Such functions exist \cite{gel1}. They should
 have vanishing derivatives of arbitrary order at $r=0$! 

As to  different approaches to a mathematical ``solidification'' of the
heuristic ansatz \ref{eq:312} see the remarks below. At the moment I just
use it: 
 
From the number state representation \ref{eq:78} we get the formal expansion
\begin{equation}
  \label{eq:315}
 |k,z\rangle \langle k,z| =
 \frac{1}{g_k(|z|^2)}\sum_{\bar{n},n=0}^{\infty}\frac{|z|^{\bar{n}+n}
 \,e^{i\,(n-\bar{n})\phi}}{\sqrt{(2k)_{\bar{n}}\,(2k)_n\,\bar{n}!\,n!}}\,
 |k,\bar{n}\rangle \langle k,n|\,.
\end{equation}
It may be reproduced by the ansatz
\begin{multline} \label{diaz}
 |z|\,m_k(|z|)\,F_k(z,\bar{z}) = \frac{1}{2\pi}\,\sum_{\bar{m},m
 =0}^{\infty}\frac{\langle k,\bar{m}|\rho|k,m\rangle
 }{(\bar{m}+m)!}\sqrt{(2k)_{\bar{m}}\,(2k)_m\,\bar{m}!\,m!}\\(-1)^{\bar{m}+m}\,
 \delta^{(\bar{m}+m)}(|z|)\,e^{i\,(\bar{m}-m)\phi}\,,
\end{multline}
 which formally leads to the desired result:
\begin{equation}
  \label{eq:316}
 \rho = \int_{\mathbb{C}}d\mu_k(z)F_k(z,\bar{z})\,|k,z\rangle
 \langle k,z| = \sum_{\bar{m},m =0}\langle
 k,\bar{m}|\rho|k,n\rangle\,|k,\bar{m}\rangle \langle k,m|~.
\end{equation}
Here
 the integration over $\phi$ impose the condition $m+\bar{n}=
 \bar{m}+n$ and the integration over $|z|$, according to Eq.\ \ref{eq:312},
  $\bar{m}+m=\bar{n}+n$ . Combined this gives $m = n, \bar{m} =
 \bar{n}$.

 (Here $\bar{m}$ is an independent natural number like $m$, 
not the complex conjugate of $m$\,!) 

 If the series \ref{diaz} does not terminate, the resulting
 generalized function will belong to the type $Z'$ \cite{gel,zem}
the
 associated test functions  of which are Fourier transforms of the smooth
 functions with compact support, i.e.\ they are elements of the space
 $  D$. 

 If
 $\rho$ is diagonal in the number state basis, 
 \begin{equation}
   \label{eq:703}
  \langle
 k,\bar{m}|\rho|k,m \rangle = \rho_{k,m}\,\delta_{\bar{m}\,m}\,, 
 \end{equation} then
\begin{equation}
  \label{eq:317}
 |z|\,m_k(|z|)\,F_k(z,\bar{z}) = \frac{1}{2\pi}\,\sum_{m
 =0}^{\infty}\frac{\rho_{k,m}
 }{(2m)!}\,(2k)_m\,m!\,
 \delta^{(2m)}(|z|)~.
\end{equation}
Especially for
\begin{equation}
  \label{eq:318}
 \rho = |k,n\rangle
 \langle k,n|
\end{equation}
we have
\begin{equation}
  \label{eq:319}
 |z|\,m_k(|z|)\,F_k(z,\bar{z}) = \frac{1}{2\pi}
 \frac{(2k)_m\,m!}
 {(2m)!}\,
 \delta^{(2m)}(|z|)~.
\end{equation}

{\em Remarks:}

The first paper of Sudarshan \cite{su} with its formal use of the relation
\ref{eq:312} led to a number of articles which discussed appropriate
 mathematical approaches:

As generalized functions may be considered as limits of continuous functions
 (I leave out the technical details!), Klauder et al.\ \cite{kla2,kla3}
 and Rocca \cite{roc} clarified this possibility for
 a solid mathematical framework.
Miller and Mishkin \cite{mill} discussed the meaning of such infinite
series like \ref{diaz} as generalized functions of the space $ Z'$.

An interesting regularization in terms of Laguerre polynomials with appropriate
 limits was introduced by Pe\v{r}ina and Mi\v{s}ta \cite{per4}.
The possibility of using polynomials as testfunctions was discussed by
Luk\v{s} \cite{luks}, following another paper by Pe\v{r}ina \cite{per5}.
That approach later was also briefly pointed out by Sudarshan \cite{su1}.
The subject has been analyzed  recently again by W\"unsche \cite{wue3}
and Richter \cite{rich}.

Another tool for finding the quantities
 $F_k$ and $S_k$ is Fourier transformation \cite{su2,cahi}. The 2-dimensional
 Fourier transform for complex variables $w$ and $z$ can be formulated as
 follows   \begin{eqnarray}\label{eq:1625}\label{cfour1} f(w)& =&
 \frac{1}{\pi}\int_{\mathbb{C}}d^2z\,e^{w\,\bar{z}-\bar{w}\,z}\,\tilde{f}(z)~,
 \\ \label{eq:1350}\tilde{f}(z)& =&\label{cfour2}
 \frac{1}{\pi}\int_{\mathbb{C}}d^2w\,e^{-w\,\bar{z}+\bar{w}\,z}\,f(w)~; \\
\label{eq:1351}\delta(w)&=& \delta(\Re(w))\,\delta(\Im(w))=
 \frac{1}{\pi^2}\,\int_{\mathbb{C}}
d^2z\,e^{w\,\bar{z}-\bar{w}\,z}~, \label{cdel} \\ \label{eq:1352}w\,
\bar{z}-\bar{w}\,z&=& 
2i[\Re(z)\,\Im(w)-\Re(w)\,\Im(z)]~.
 \end{eqnarray}
One may define a ``normally ordered''  characteristic function $
\chi_N(w,\bar{w})$ by
\begin{equation}
  \label{eq:320}
 \chi_N(w,\bar{w}) =
\tr\rho\,e^{w\,K_+}\,e^{-\bar{w}\,K_-})~.
\end{equation}
Implementing the trace in terms of number states $|k,n\rangle$, 
inserting for $\rho$
the diagonal representation \ref{eq:300}  and using the completeness
relation \ref{com0}  gives
\begin{equation}
  \label{eq:321}
 \chi_N(w,\bar{w}) =
 \int_{\mathbb{C}}d\mu_k(z)F_k(z,\bar{z})\, e^{w\,\bar{z}-\bar{w}\,z}~.
\end{equation}
Fourier transforming yields 
\begin{equation}
  \label{eq:382}
m_k(|z|)\,F_k(z,\bar{z})=\frac{1}{\pi^2}\int_{\mathbb{C}}d^2w\,
 \chi_N(w,\bar{w})\, e^{-w\,\bar{z}+\bar{w}\,z}~. 
\end{equation}
Without the use of the diagonal representation the characteristic
function \ref{eq:320} can also be evaluated by employing the completeness
relation \ref{com1}:
\begin{equation}
  \label{eq:397}
  \chi_N(w,\bar{w})=\int_{\mathbb{C}}\int_{\mathbb{C}}d\tilde{\mu}_k(z_2)
\,d\tilde{\mu}_k(z_1)\,\rho(k;\bar{z}_2,z_1)\,g_k(\bar{z}_1z_2)\,
e^{w\,\bar{z}_1-\bar{w}\,z_2}~.
\end{equation}

 The anti-normally ordered characteristic function is defined
 correspondingly:
\begin{equation} \label{eq:383}
 \chi_A(w,\bar{w}) =
\tr\rho\,e^{-\bar{w}\,K_-}\,e^{w\,K_+})~.
\end{equation}

Again inserting the completeness relation \ref{com1} between the last two
operators inside the trace gives
\begin{equation}
  \label{eq:384}
  \chi_A(w,\bar{w}) =  \int_{\mathbb{C}}d\mu_k(z)\,S_k(z,\bar{z})\,
 e^{w\,\bar{z}-\bar{w}\,z}~.
\end{equation} By Fourier transforming the last equation 
one can express $S_k$ in terms
of $\chi_A$.

 From the diagonal representation \ref{eq:300} we get the following
 relation between the quantities $F_k$ and $S_k$:
\begin{eqnarray}
  \label{eq:1626}\label{eq:386}
  S_k(z,\bar{z})& \equiv& \langle k,z|\rho|k,z\rangle = 
  \int_{\mathbb{C}}d\mu_k(z_1)\,F_k(z_1,\bar{z}_1)\,|\langle k,z|k,z_1
 \rangle|^{2} \\ \label{eq:1353}&=& \frac{1}{g_k(|z|^{2})}\, 
 \int_{\mathbb{C}}d\tilde{\mu}_k(z_1)\,F_k(z_1,\bar{z}_1)\,
|g_{k}(\bar{z}_1z)|^{2} ~.
\end{eqnarray}
The kernel $|g_k(\bar{z}_2\,z_1)|^2= g_k(\bar{z}_2z_1)\,g_k(z_2\bar{z}_1)$
may be rewritten as \cite{er3c} 
\begin{equation}
  \label{eq:389}
  |g_k(\bar{z}_2z_1)|^2 = \sum_{n=0}^{\infty}\frac{(\bar{z}_2z_1)^n}{(2k)_n
\,n!}\,{_{2}F_1}(-n,-n-(2k-1);2k;\frac{\bar{z}_2z_1}{z_2\bar{z}_1})~.
\end{equation}
The hypergeometric functions in the sum
are essentially Gegenbauer polynomials $C_n^{2k-1/2}$ \cite{er4a}:
\begin{equation}
  \label{eq:390}
 {_{2}F_1}(-n,-n-(2k-1);2k;b)= \frac{n!}{(4k-1)_n}(1-b)^{n}\,
C^{2k-1/2}_n(\frac{1+b}{1-b})~  
\end{equation}

 Whether this is of any use remains to be seen! 

From the representation \ref{eq:300} we further get (see Eq.\ \ref{caz})
\begin{eqnarray}
  \label{eq:1627}\label{eq:385}
 Q_k(\alpha, \bar{\alpha})&=& \langle k,\alpha|\rho|k, \alpha \rangle  =  
\int_{\mathbb{C}}d\mu_k(z)\,F_k(z,\bar{z})\,|\langle k, \alpha|k, z 
\rangle|^2  \\ \label{eq:1354}&=&e^{-|\alpha|^2}\,
 \int_{\mathbb{C}}d\tilde{\mu}_k(z)\,
F_k(z,\bar{z})\,
|C_k(\bar{\alpha};z)|^2 ~. \nonumber
\end{eqnarray}

 Analogously one derives
\begin{eqnarray}
  \label{eq:1628}\label{eq:388}
 S_k(z.\bar{z})&=& \langle k,z|\rho|k, z \rangle  =  
\int_{\mathbb{C}}d^2\,P_k(\alpha,\bar{\alpha})\,|\langle k, z|k, \alpha 
\rangle|^2  \\ \label{eq:1355}&=& \frac{1}{g_k(|z|^2)}
\int_{\mathbb{C}}d^2\alpha \,P_k(\alpha,\bar{\alpha})\,
e^{-|\alpha|^2}\, |C_k(\bar{\alpha};z)|^2~. \nonumber  
\end{eqnarray}

 Additional relations may be obtained from the equality
\begin{equation}
  \label{eq:387}
  \rho = \int_{\mathbb{C}}d\mu_k(z)\,F_k(z,\bar{z})\,|k,z\rangle \langle k,z|
= \int_{\mathbb{C}}d\mu_k(\alpha)\,P_k(\alpha,\bar{\alpha})\,|k,\alpha
\rangle \langle k, \alpha |~.
\end{equation}

In the case of the conventional coherent states $|\alpha\rangle$ the
 ``symmetric'' characteristic function 
 \begin{equation}
   \label{eq:391}
   \chi_S(\alpha,\bar{\alpha}) =\tr\rho\,e^{\alpha\,a^+
-\bar{\alpha}\,a})~
 \end{equation}
is the Fourier transform of the Wigner function $w(\alpha, \bar{\alpha})$. 

 Here the 
situation is more complicated: If we define 
\begin{equation}
  \label{eq:392}
  \chi_S(w,\bar{w}) =\tr\rho\,e^{w\,K_+ -\bar{w}\,K_-})~,
\end{equation}
the trace may be evaluated with the help of the relations \ref{eq:137}
and \ref{com1}  as 
\begin{eqnarray}\label{eq:1629}\label{chiN}
 \chi_S(w,\bar{w})&=&\chi_S(\lambda, \bar{\lambda})= \nonumber \\
 \label{eq:1356}&=&
(1-|\lambda|^2)^k\,\int_{\mathbb{C}}\int_{\mathbb{C}}d\tilde{\mu}_k(z_2)
\tilde{\mu}_k(z_1)\,e^{\lambda\,\bar{z}_1-\bar{\lambda}\,z_2}\times \\
\label{eq:1357}&& ~~~~~~~ \times g_k[(1-|\lambda|^2)\,\bar{z}_1z_2]\,\langle
 k,z_2|\rho|k,z_1\rangle~,
\nonumber \\  \label{eq:1358}\lambda& =&(w/|w|)\,\tanh|w|\,,~~1-
|\lambda|^2=1/\cosh^{2}|w|\,,~~~
 \end{eqnarray}

 It does not seem to be obvious which quantity
 would correspond
to the Wigner function in the present framework. Progress may come from
recent proposals for generalizations of that concept to more general
Lie groups \cite{mann4}.

In any case one has to deal with the following mathematical problem:
 As $\lambda \in \mathbb{D}$,
the integral transform \ref{chiN} represents a mapping from $\mathbb{C}^2$
onto the unit disc $\mathbb{D}$. The associated Fourier transformation
is more sophisticated than the usual one \cite{helg} and, unfortunately,
beyond the scope of the present paper! 

Using the representation \ref{eq:300} the chacteristic function \ref{chiN}
may also be expressed as
\begin{multline}
  \label{eq:314}
 \chi_S(\lambda, \bar{\lambda})=
(1-|\lambda|^2)^k\,\int_{\mathbb{C}}\int_{\mathbb{C}}d\tilde{\mu}_k(z_2)
\tilde{\mu}_k(z_1)\,e^{\lambda\,\bar{z}_1-\bar{\lambda}\,z_2}\cdot\\
\cdot F_k(z_1,\bar{z}_1)\,g_k(\bar{z}_2z_1)\, g_k[(1-|\lambda|^2)\,
\bar{z}_1z_2]~.  
\end{multline}
\section{Pseudo-probability distributions asso\-cia\-ted with \\
Perelomov coherent states}
It is clear from Sec.\ 3.2 that the Perelomov coherent states have
some qualitatively different properties compared to the ones just discussed.
One of the main differences is that the states $|k,\lambda\rangle$ may be
generated by the unitary operator \ref{eq:137}. Another is that the
complex numbers $\lambda$ take only values in the unit disc $\mathbb{D}$.
Third,  the states $|k,\lambda \rangle $ are eigenstates of the
operators $E_{k,-}=(K_0+k)^{-1} K_-$. Together with $E_{k,+}= (E_{k,-})^+$
these operators now play the role the operators $K_-$ and $K_+$ 
had in the last section and the operators $a$ and $a^+$ play
in connection with the states $|\alpha \rangle$. 

Partially one can go through the same routine as in the last section 
(in the following the notions of Sec.\ 3.2 are being used): 

 From the representation \ref{eq:286} and the relations \ref{eq:136} and
\ref{eq:150} we get now
\begin{eqnarray} \label{eq:1630}\langle
k,\lambda_2|A|k,\lambda_1 \rangle &=& \label{maell}
(1-|\lambda_2|)^k\,(1-|\lambda_1|^2)^k\,A(k;\bar{\lambda}_2,\lambda_1)~, \\
\label{eq:1360}A(k;\bar{\lambda}_2,\lambda_1)&=&\sum_{m,n =0}^{\infty}
\bar{e}_{k,m}(\lambda_2)\,A_{k;m\,n}\,e_{k,n}(\lambda_1) = \nonumber \\
\label{eq:1361}&=& \sum_{m,n
=0}^{\infty}A_{k;m\,n} \sqrt{\frac{(2k)_m\,(2k)_n}{m!\,n!}}
\,\bar{\lambda}_2^m\,\lambda_1^n
~. \label{ser2}
\end{eqnarray}
Compared to the series \ref{eq:1337} one has to realize that
 $|\lambda_j| <1,\,j=1,2$, and
that the factor $\sqrt{(2k)_m\,(2k)_n}$ is now in the numerator instead
in the nominator! Otherwise one may proceed formally as in the last section:
replace the variables $z_2,z_1$ by $\lambda_2,\lambda_1$, the complex plane
$\mathbb{C}$ by the unit disc $\mathbb{D}$, the eigenfunctions
 $\tilde{f}_{k,n}(z)$ by $e_{k,n}(\lambda)$ and the integration measures
accordingly. 

 One can also define a Husimi type density
\begin{equation}
  \label{eq:394}
  T_k(\lambda, \bar{\lambda}) \equiv \langle k, \lambda|\rho|k,\lambda
 \rangle \geq 0~,
\end{equation}
and  a diagonal representation
\begin{equation}
  \label{eq:313}
  \rho = \int_{\mathbb{D}}d\mu_k(\lambda)\,G_k(\lambda,\bar{\lambda})\,
|k,\lambda \rangle \langle k, \lambda|~.
\end{equation}
Their relationship is
\begin{eqnarray} \label{eq:1631}T_k(\lambda, \bar{\lambda}) &=& 
\int_{\mathbb{D}}
d\mu_k(\lambda_1)\,
G_k(\lambda_1, \bar{\lambda}_1)\, |\langle k, \lambda|k, \lambda_1
 \rangle|^2 = \\ \label{eq:1362}&=& (1-|\lambda|^2)^{2k}
\int_{\mathbb{D}} d\mu_k(\lambda_1)
(1-|\lambda_1|^2)^{2k}\, G_k(\lambda_1, \bar{\lambda}_1) \times \nonumber \\
&&~~~~~~~~~~~~~~~~~~~~~~ \times [(1-\bar{\lambda}\,
\lambda_1)(1-\lambda\,\bar{\lambda}_1)]^{-2k}\,. \nonumber \end{eqnarray}
With 
\begin{equation}
  \label{eq:395}
  \lambda = |\lambda|e^{i\theta}\,,\,\lambda_1 = |\lambda_1|e^{i\theta_1}\,,
\,\cos(\theta-\theta_1)=t\,,\,|\lambda|\,|\lambda_1|=x~,
\end{equation}
we have 
\begin{equation}
  \label{eq:396}
 [(1-\bar{\lambda}\,
\lambda_1)(1-\lambda\,\bar{\lambda}_1)]^{-2k} = (1-2\,t\,x +x^2)^{-2k} =
\sum_{n=0}^{\infty}C_n^{2k}(t)\,x^n~,  
\end{equation}
where the $C_n^{2k}(t)$ are again  Gegenbauer polynomials \cite{er5a}. 

As we have three different coherent states now in a given unitary
 representation
with index $k$, a large variety of relations may be established (see the
integral transforms in subsections 3.2.1 and 3.3.1). I briefly mention
 only a few
 examples: 
 \begin{equation}
   \label{eq:398}
   S_k(z,\bar{z}) = \frac{1}{g_k(|z|^2)}\int_{\mathbb{D}}
d\tilde{\mu}_k(\lambda)\,
G_k(\lambda,\bar{\lambda})\,e^{\bar{\lambda}\,z+\lambda\,\bar{z}}~,
 \end{equation}
and the ``inverse'' relation
\begin{equation}
  \label{eq:399}
  T_k(\lambda, \bar{\lambda}) = (1-|\lambda|^2)^{2k}\,
\int_{\mathbb{C}}d\tilde{\mu}_k(z)\,
F_k(z,\bar{z})\,e^{\bar{\lambda}\,z+\lambda\,\bar{z}}~.
\end{equation}

As to possible Fourier transforms one may use the trick \cite{meh}
\begin{equation}
 \label{eq:400}
  \langle k,-z|\rho|k,z\rangle = 
 \frac{1}{g_k(|z|^2)}\int_{\mathbb{D}}d\tilde{\mu}_k(\lambda)\,
G_k(\lambda,\bar{\lambda})\,e^{\bar{\lambda}\,z-\lambda\,\bar{z}}~,
\end{equation}
where the exponential in the integrand is now that of a complex Fourier
transform (see  Eq.\ \ref{cfour1}).
Formally we get from  \ref{cdel} 
\begin{eqnarray}
  \label{eq:1632}\label{eq:401}
\frac{2k-1}{\pi}\,(1-|\lambda|^2)^{2k-2}\,G_k(\lambda,\bar{\lambda})& =&
\frac{1}{\pi^2}\,\int_{\mathbb{C}}d^2z\,g_k(|z|^2)\,\langle k,-z|\rho|k,z
\rangle \,e^{\lambda\,\bar{z}-\bar{\lambda}\,z}~ = \nonumber \\
\label{eq:1363}&=&  \frac{1}{\pi^2}\,\int_{\mathbb{C}}d^2z\,\rho(k;-\bar{z},z)
 \,e^{\lambda\,\bar{z}-\bar{\lambda}\,z}~, \label{fourg}
\end{eqnarray}
where in the last step the relation \ref{mael} has been used. \\
Thus, we obtain  the quantity $G_k(\lambda,\bar{\lambda})$
by   Fourier transforming $\rho(k;-\bar{z},z)$,
 a quantity which in general
may be calculated readily. For instance, for the density operator
 \ref{eq:295} we have from \ref{rokn} that
 \begin{equation}
   \label{eq:402}
   \rho(k;-\bar{z},z)= (-1)^n\,\frac{|z|^{2n}}{(2k)_n\,n!}~. 
 \end{equation}
Inserting this into the integral \ref{fourg} yields 
\begin{eqnarray}
\frac{2k-1}{\pi}\,(1-|\lambda|^2)^{2k-2}\,G_k(\lambda,
\bar{\lambda})
&=& \frac{1}{2^{2n}\,(2k)_n\,n!} (\Delta_{\lambda})^n \delta(\lambda)~,
\label{deng} \\
\label{eq:1364}\Delta_{\lambda} &=& \partial^2/\partial\sigma_1^2+ 
  \partial^2/\partial\sigma_2^2\,, \\ && \lambda = \sigma_1+i\,\sigma_2~. 
\nonumber \end{eqnarray}

 As
\begin{equation}
  \label{eq:404}
(\Delta_{\lambda})^n|\lambda|^{2n+2m} = 2^{2n}\,[(m+1)_n]^2\,
|\lambda|^{2m}~,m=0,
1,\ldots~,
\end{equation}
we see: For the number state projection operator \ref{eq:295} the
 pseudo-density $G_k$ has a very similar singularity structure as the
 pseudo-density
$F_k$ (see Eq.\ \ref{eq:319}) though the ``derivations'' have been 
different. The latter one shows: in order to obtain the projection operator
\ref{eq:295} from the relations \ref{deng} and \ref{eq:404} by means
of ``integrating by parts'' under the integral \ref{eq:313}  one has  
 to treat the powers of $|\lambda|^2$ like test functions (with compact
support) when shifting the differential operator $\Delta_{\lambda}$
from the $\delta$-functions to those powers. This is in line with the
 duality postulate \ref{eq:312}. 

 An important point is, of course , that $ |\lambda| <1$ which makes
the Fourier analysis more involved, as was already pointed out at the
end of the last section. Take, e.g. instead of Eq.\ \ref{eq:399} the relation
\begin{equation}
  \label{eq:403}
   \langle k,-\lambda|\rho|k, \lambda \rangle 
 = (1-|\lambda|^2)^{2k}\,\int_{\mathbb{C}}d\tilde{\mu}_k(z)\,
F_k(z,\bar{z})\,e^{-\bar{\lambda}\,z+\lambda\,\bar{z}}~.
\end{equation}
This Fourier integral cannot be inverted in the conventional
manner but requires the more sophisticated means already
mentioned before \cite{helg}. 

As the states $|k,\lambda \rangle$ are eigenstates of the operators
$E_{k,-}$ one may deal with them (and $E_{k,+}$) similarly as with
the operators $K_-$ and $K_+$ in the last section. As to possible
normal orderings one has to take into account the commutators
\ref{comE} etc. I shall skip the details here. 

If the operators
$E_{k,-}$ and $E_{k,+}$ would yield convincing $\cos$- and $\sin$-
operators \ref{cosk} and \ref{sink} 
as discussed critically in Ch.\ 5, the applications
of them would be quite interesting. The discussion in Ch.\ 5 shows, however,
that such an  interpretation has its severe problems. 

\chapter{The $SO^{\uparrow}(1,2)$-structure of interferences}
\section{Classical theory}
 The analysis and 
mathematical descriptions of interference pattern play a very essential
role in classical optics \cite{born}. Consider the following generic
 example which will show all the essential problems of an appropriate
 quantization to be discussed in the next section:
 
 The intensity 
\begin{equation}
  \label{eq:410}
  I = |A_1 + A_2|^2 
\end{equation}
of two ``interfering'' complex amplitudes $A_j\,,j=1,2,$ may be 
expressed in different ways, depending on the coordinates one
uses, cartesian or polar ones:
\begin{equation}
  \label{eq:411}
  A_j = \frac{1}{\sqrt{2}}(q_j +i\,p_j) =|A_j|\,e^{-i\,\vp_j}\,,\,j=1,2~.
\end{equation}
 The two reprensations are classically equivalent, except for the points
$(q_j,p_j) = (0,0)$ where the functional determinants
\begin{equation}
  \label{eq:720}
  \frac{\partial(q_j,\,p_j)}{\partial(\vp_j,|A_j|)} = 2\,|A_j|\,,\,j=1,2,
\end{equation}
 for the mutual
transformations vanishes or its inverse becomes singular!

 For the intensity \ref{eq:410}
we get 
\begin{eqnarray}
  \label{eq:1634}\label{eq:413}
 I = w_3 &=& (\bar{A}_1 + \bar{A}_2)(A_1 + A_2) = \\
 \label{eq:1365}&=&I_1 + I_2 +
 \bar{A}_1\,
A_2 + A_1\,\bar{A}_2  ~, \nonumber \\ 
&& I_j = |A_j|^2\,,\,j=1,2, \\
 \label{eq:1366}\bar{A}_1\,
A_2 + A_1\,\bar{A}_2 &=& 2\, g_1(q_1,p_1;q_2,p_2) \equiv
 q_1\,q_2 + p_1\,p_2  \label{ge1}\\
\label{eq:1367}&=& 2\,h_1(|A_1|,|A_2|,\vp = \vp_1-\vp_2) \equiv \label{ha1} \\ 
\label{eq:1368}&\equiv&  2\,|A_1|\,|A_2|\, \cos\vp
\nonumber \end{eqnarray}
Employing beam splitters and $\lambda/4$ phase shifters (compensators)
allows to shift the phase of one amplitude by $\pm \pi$
 or $\pm \pi/2$,
respectively, relative to that of the other. The result of these modifications
are new intensities:
\begin{eqnarray}
  \label{eq:1635}\label{eq:414}
 w_4 &=& (\bar{A}_1 - \bar{A}_2)(A_1 - A_2) = I_1 + I_2 -
 \\ \label{eq:1369}&& -
 (\bar{A}_1\,A_2 + A_1\,\bar{A}_2)  ~,\nonumber  \\  
 \label{eq:1370}w_5 &=& (\bar{A}_1 -i\, \bar{A}_2)(A_1 +i\, A_2)
 = I_1 + I_2 + \\ \label{eq:1371}&&+i
 (\bar{A}_1\,A_2 - A_1\,\bar{A}_2)  ~,\nonumber \\
i(\bar{A}_1\,A_2 - A_1\,\bar{A}_2)&=& -
 2\,g_2(q_1,p_1;q_2,p_2) \equiv
 -q_1\,p_2
+q_2\,p_1 \label{ge2} \\ &=& 2\,h_2(|A_1|,|A_2|,\vp =
 \vp_1-\vp_2) \equiv
\label{ha2}\\ \label{eq:1374}&\equiv& -2\,|A_1|\,|A_2|
\,\sin\vp\,, \nonumber \\
 \label{eq:1375}w_6 &=& (\bar{A}_1 +i\, \bar{A}_2)(A_1 -i\,
 A_2) = I_1 + I_2 -\\ \label{eq:1376}&&-i
 (\bar{A}_1\,A_2 - A_1\,\bar{A}_2)  ~, \nonumber
\end{eqnarray}
  From the differences of the intensities
\begin{eqnarray}
  \label{eq:1636}\label{eq:415}
  w_3-w_4 &=& 2\,(\bar{A}_1\,A_2 + A_1\,\bar{A}_2) ~, \\
\label{eq:1377}w_5-w_6 &=&  2\,i\,(\bar{A}_1\,A_2 - A_1\,\bar{A}_2)
 \label{we56}
\end{eqnarray}
 one gets the important
 functions $g_j,\, j=1,2,$ or (and)
$ h_j,\,j=1,2,$ which determine the interference pattern. 

I shall discuss the experimental methods (multi-port homodyning etc.)
for determining the densities $w_j$ and their differences later (see
Sec.\ 8.2.2 and the literature quoted there). 
 
Before passing to any quantum theory let me first discuss several
subtleties of the classical canonical structures because they are
important for the quantum theory: 

The functions $g_j$ and $h_j$ are functions on different phase (symplectic)
spaces. The $g_j$ are functions on the 4-dimensional space
\begin{equation}
  \label{eq:416}
  {\cal S}^{4}_{q,p;0,0} = \{ (q_1,p_1;q_2,p_2) \in \mathbb{R}^2 \times 
\mathbb{R}^2\,,
(q_1,p_1) \neq (0,0) \neq (q_2,p_2)\, \}
\end{equation} with the
 {\em local} symplectic form
\begin{equation}
  \label{eq:417}
  \omega_{q,p} = dq_1\wedge dp_1 + dq_2\wedge dp_2~.
\end{equation}
(Though it does not appear to be necessary at this stage, the origin
 of ${\cal S}^{4}_{q,p}$ has been deleted in order to have a one-to-one
relationship between cartesian and polar coordinates. The importance
of deleting the origin has been iscussed before in previous chapters and 
will become evident in this one, too!)

The $h_j\,$, however, are functions on the 2-dimensional space
\begin{equation}
  \label{eq:418}
  {\cal S}^{2}_{\vp,I_{1,2}} = \{(\vp,I_{1,2}=|A_1|\,|A_2|)
 \in S^1 \times \mathbb{R}^+\}\,,
\end{equation}
with the symplectic form
\begin{equation}
  \label{eq:419}
  \omega_{\vp,I_{1,2}} = d\vp\wedge dI_{1,2}\,,
\end{equation} 
that is, here we have the same symplectic manifold we started from in the
very beginning! 

 The deeper relationship between $ {\cal S}^{4}_{q,p;0,0}$ and
${\cal S}^{2}_{\vp,I_{1,2}}$ has its origin in the following ``gauge''
 symmetries:
  
All the densities $w_j,\,j=3,4,5,6,$ are invariant against a simultaneous
phase transformation of the complex amplitudes $A_j$:
\begin{equation}
  \label{eq:420}
  A_j \to e^{i\,\alpha}\,A_j\,,~j=1,2~.
\end{equation}
For $\alpha = \pm \pi$ this gives the reflection
\begin{equation}
  \label{eq:705}
  R:~~(q_1,p_1,q_2,p_2) \to (-q_1,-p_1,-q_2,-p_2)\,,
 ~R^2 = \boldsymbol{1}\,.
\end{equation}
The reflection symmetry \ref{eq:705} is a generalization of the
 $Z_2[\cdot(q,p)]$-symmetry encountered in Ch.\ 6 (Eq.\ \ref{eq:702}).
Though it is contained in the group of phase transformations \ref{eq:420},
it will be important in the following to consider it for itself!

I therefore shall deal with the continuous symmetry \ref{eq:420} and certain
generalizations of it and the discrete
symmetry \ref{eq:705} separately.
\subsection{Continuous gauge transformations and associated symplectic
reductions} 
For the coordinates $q_j$ and $p_j$ the transformations \ref{eq:420}
 imply the following rotations
\begin{align}
  \label{eq:422}
  q_j & \to \tilde{q}_j(\alpha) = \cos\alpha\,q_j -\sin\alpha\,p_j~, \\
\label{eq:1378}p_j & \to \tilde{p}_j(\alpha) = \sin\alpha\,q_j +
 \cos\alpha\,p_j~. \nonumber
\end{align}
All observable (physical) quantities occuring in the relations 
\ref{eq:413}-\ref{eq:1375} should be invariant under the gauge
transformations \ref{eq:420} or \ref{eq:422} respectively !
 
 The moduli $|A_j|$ and the phase difference
$\vp= \vp_1-\vp_2$ are, of course, invariant!. Thus, the phase space
${\cal S}^{2}_{\vp,I_{1,2}}$ represents the $U(1)$ gauge invariant part of
 ${\cal S}^{4}_{q,p;0,0}$, 
with the $U(1)$-gauge dependent part factored out: 

 The transformations
\ref{eq:422} generate circles around the (deleted!) origin in each factor
$\mathbb{R}^2-(0,0)$ of ${\cal S}^{4}_{q,p;0,0}$.
 Any two points on a given circle are physically 
equivalent. Two such circles - one in each factor $\mathbb{R}^2-(0,0)$ -
represent just one point in the (gauge invariant) phase space
 ${\cal S}^{2}_{\vp,I_{1,2}}\,
$! 

 The transformations \ref{eq:422} induce the following vector field 
on ${\cal S}^{4}_{q,p;0,0}$ : 
\begin{equation}
  \label{eq:412}
  X_{2g_0}= p_1\,\partial_{q_1} -q_1\,\partial_{p_1}+p_2\,\partial_{q_2}
-q_2\,\partial_{p_2} ~.
\end{equation}
This follows immediately from the relation
\begin{equation}
  \label{eq:421} 
 \partial f(\tilde{q}_1(\alpha), \tilde{p}_1(\alpha), 
\tilde{q}_2(\alpha), \tilde{p}_2(\alpha)\,)/\partial \alpha(\alpha =0)
 = -X_{2g_0}f~,
\end{equation}
where $f$ is any smooth function on ${\cal S}^{4}_{q,p;0,0}$. (For this and the
following see the notions briefly introduced in Appendix A.1 and the literature
mentioned there!) 

 The vector field 
field $X_{2g_0}$ is associated with the hamiltonian function
\begin{equation}
  \label{eq:423} 
2\, g_0(q_1,p_1,q_2,p_2) = \frac{1}{2}\,(q_1^2+p_1^2+q_2^2+p_2^2)=
I_1 + I_2~.
\end{equation}
This may be seen as follows: If we denote by $i_X\rho$ the interior
product  of a vector field $X$ with a differential form $\rho$  
(see, e.g.\cite{mar1,cho}) then we get for the 2-form \ref{eq:417}
\begin{equation}
  \label{eq:424}
  i_{X_{2g_0}}\omega_{q,p} =p_1\,dp_1+q_1\,dq_1+p_2\,dp_2+q_2\,dq_2 =2\,dg_0~.
\end{equation}
(Note 
that
 $i_{\partial_{q_1}}(dq_1 \wedge dp_1)= dq_1(\partial_{q_1}) \wedge
 dp_1 =dp_1\,,  i_{\partial_{p_1}}(dq_1 \wedge dp_1)= - dp_1(\partial_{p_1})
 \wedge dq_1 =-dq_1\,$ etc.) 

A more familiar - but also more ``hand-waving'' -  argument might
 be the following:

 Consider variations $\delta L$ of the Lagrangean 
\begin{equation}
  \label{eq:425}
  L = \frac{1}{2}(\dot{q}_1^2 + \dot{q}_2^2 -q_1^2 - q_2^2)\,,
\end{equation}
generated by variations $\delta q_j$. Then (by partial differentiation)
\begin{equation}
  \label{eq:426}
  \delta L = \sum_{j=1}^2(\partial L/\partial q_j - \frac{d}{dt}\partial L/
\partial \dot{q}_j)\delta q_j + \frac{d}{dt}\sum_{j=1}^{2} (\partial L/
\partial \dot{q}_j)\delta q_j)~.
\end{equation}

If the curves $q_j(t)$ are solutions of the Euler-Lagrange eqs.\ of
motion, the first term on the r.h.\ side of the relation \ref{eq:426}
vanishes and we have
\begin{equation}
  \label{eq:427}
  \delta L =  \frac{d}{dt}\sum_{j=1}^{2} (\partial L/
\partial \dot{q}_j)\,\delta q_j~.
\end{equation}
If the variations $\delta q_j$ are infinitesimal transformations of
a 1-parameter group which leaves the Lagrangean invariant -- $\delta L =0$ --,
then we have a conservation law (Noether's first theorem). More generally,
it suffices that $L$ is invariant up to a total time derivative,
\begin{equation}
  \label{eq:428}
  \delta L = \frac{d}{dt}C(q,\dot{q})\,,
\end{equation}
because we now have the conservation law
\begin{equation}
  \label{eq:429}
  \frac{d}{dt}[\sum_{j=1}^{2} p_j
\,\delta q_j -C(q,p)]=0\,,~ p_j = \partial L/\partial\dot{q}_j=\dot{q}_j\,. 
\end{equation}
The last situation occcurs in the context of the transformations 
\ref{eq:422}, where 
\begin{equation}
  \label{eq:430}
  \delta q_j = -p_j\,\alpha\,,~\delta p_j = q_j\,\alpha\,,\,|\alpha| \ll 1~.
\end{equation}
This gives
\begin{equation}
  \label{eq:431}
  C= (q_1^2+q_2^2)\,\alpha ~.
\end{equation}
Therefore we have the conservation law
\begin{equation}
  \label{eq:432}
  \frac{d}{dt} (q_1^2+q_2^2+p_1^2+p_2^2) = 0~~ \mbox{ or } \frac{d}{dt}\,
g_0 =0~.
\end{equation}
Thus we get the energy conservation law from the invariance - up to a total
time derivative - under the rotations \ref{eq:422} ! 

The relation \ref{eq:432} implies the {\em constraint}
\begin{equation}
  \label{eq:433}
  \phi_0(q_1,p_1,q_2,p_2) \equiv g_0 -E/2 = 0\,,~ E>0~.
\end{equation}
Before discussing this constraint let us first analyse the group
structures associated with the system: 

For smooth functions on ${\cal S}^{4}_{q,p}$ we have the Poisson brackets
\begin{equation}
  \label{eq:434}
  \{f_1,f_2\}_{q,p} = \partial_{q_1}f_1\,\partial_{p_1}f_2 -
\partial_{p_1}f_1\,\partial_{q_1}f_2 +\partial_{q_2}f_1\,\partial_{p_2}f_2
-\partial_{p_2}f_1\,\partial_{q_2}f_2~.
\end{equation}
For the functions $g_1$ and $g_2$ from \ref{ge1} and \ref{ge2} we get
\begin{equation}
  \label{eq:435}
  \{g_1,g_2\}_{q,p} = \frac{1}{4}(q_1^2+p_1^2-q_2^2-p_2^2)\equiv g_3
= \frac{1}{2}(I_1-I_2)\,.
\end{equation}
The three functions $g_j, j=1,2,3,$ generate the Lie algebra of the
group $SO(3)$ or of its covering group $SU(2)$:
\begin{equation}
  \label{eq:436}
  \{g_j,g_k\}_{q,p} = \epsilon_{jkl}\,g_l~.
\end{equation}
Note that the function $g_2$, Eq.\ \ref{ge2}, has the form of the usual
angular momentum in the plane. 

The function $g_0$ from Eq.\ \ref{eq:423} Poisson commutes with all $g_j, 
j=1,2,3,$
\begin{equation}
  \label{eq:437}
   \{g_j,g_0\}_{q,p} =0\,,\,j=1,2,3~.
\end{equation}
Whereas the functions $g_j,j=1,2,3,$ generate the group $SU(2)$ or $SO(3)$, the
function $g_0$ generates a $U(1)$ or $O(2)$ which is an invariance (gauge)
 group of the
interference observables $g_j,\,j=1,2,3,$ and $g_0$ itself.

 These 4 functions are not independent:
\begin{equation}
  \label{eq:438}
  g_1^2+g_2^2 = I_1 I_2 = g_0^2-g_3^2~.
\end{equation}
On the other hand, let $k_a(\vp,I_{1,2}),\,a=1,2$,
be functions on the phase space \ref{eq:418} with the Poisson
bracket
\begin{equation}
  \label{eq:439}
  \{k_1,k_2\}_{\vp,I_{1,2}} = \partial_{\vp}k_1\,\partial_{I_{1,2}}k_2-
\partial_{I_{1,2}}k_1\,
\partial_{\vp}k_2~.
\end{equation}
With 
\begin{equation}
  \label{eq:440}
   I_{1,2}=\sqrt{I_1\,I_2}=|A_1|\,|A_2|
\end{equation}
we have for the functions $h_a,a=1,2,$ from \ref{ha1} and \ref{ha2}
and $I_{1,2}$:
\begin{equation}
  \label{eq:441}
  \{h_1,h_2\}_{\vp,I_{1,2}} = I_{1,2}\,,~\{h_1,I_{1,2}\}_{\vp,I_{1,2}}
 = h_2\,,~\{h_2,I_{1,2}\}_{\vp,I_{1,2}}
 =-h_1\,,
\end{equation}
which, as we know, constitutes the Lie algebra of the group
 $SO^{\uparrow}(1,2)\,$!

 An immediate question is, of course, how these qualitatively different
group structures (compact $SO(3)$ and non-compact $SO^{\uparrow}(1,2)$)
 are related!

One key lies in the constraint \ref{eq:433}. Before we exploit it, let
me observe that
\begin{eqnarray} \label{eq:1637}\label{Ige1}
 \{I_1I_2,g_1\}_{q,p}& =&\{(g_0^2-g_3^2),g_1\} =-2\,g_3
\{g_3,g_1\}\\\label{eq:1379}& =& 2\,g_3\,g_2\,, \nonumber
 \\  \label{eq:1380}\{I_1I_2,g_2\}_{q,p}& =&-2\,g_3\,g_1\,, \label{Ige2} \\
 \label{eq:1381}\{I_1I_2,g_3\}_{q,p}& =& 0~. \nonumber
\end{eqnarray}
Expressing the real variables $q_j$ and $p_j$ in terms of the 
amplitudes $A_j$ and their complex conjugates we get for the
symplectic form \ref{eq:417}
\begin{equation}
  \label{eq:442}
  \omega_{q,p} = i\,(dA_1 \wedge d\bar{A_1} + dA_2 \wedge d\bar{A}_2)
= d\vp_1 \wedge dI_1 + d\vp_2 \wedge dI_2~. 
\end{equation}
The constraint \ref{eq:433} implies
\begin{equation}
  \label{eq:443}
  dI_2 =-dI_1\,,~d(I_1I_2) = (I_2-I_1)dI_1 = -2\,g_3\,dI_1\,,
\end{equation}
so that
\begin{equation}
  \label{eq:444}
   d\vp_1 \wedge dI_1 + d\vp_2 \wedge dI_2 = d\vp
 \wedge dI_1= -\frac{1}{2\,g_3}\,d\vp \wedge d(I_1I_2)\,,\vp=\vp_1-\vp_2\,.
\end{equation}
As $I_1I_2 = I_{1,2}^2$ we get
\begin{equation}
  \label{eq:445}
  \omega_{q,p} = -\frac{I_{1,2}}{g_3}\,d\vp \wedge dI_{1,2}~.
\end{equation}
The symplectic form  \ref{eq:445}  implies for the
corresponding (``dual'')  Poisson brackets (see Appendix A.1)
\begin{equation}
  \label{eq:446}
  \{\cdot,\cdot\}_{\vp,I_{1,2}} = - \frac{I_{1,2}}{g_3}\,\{\cdot,\cdot\}_{q,p}\,.
\end{equation}
We now see that the relations \ref{eq:435}, \ref{Ige1} and \ref{Ige2}
are equivalent to the relations \ref{eq:441}! (Recall that $h_2 \leftrightarrow
-g_2$.)
We also observe that the symplectic reduction of the
 symplectic space \ref{eq:416}
to the symplectic space \ref{eq:418} is accompanied by a transition
from the structure group $SO(3)$ to the structure group
 $SO^{\uparrow}(1,2)\,$! 

That symplectic reduction needs some more comments: 
According to Dirac's classification \cite{dirac} the constraint
 $\phi_0$ of Eq.\ \ref{eq:433} is
``first class'',  because it commutes with the $g_j,j=1,2,3,$ and (the
Hamiltonian) $g_0$. It generates a $U(1)$ - gauge transformation which
induces a reduction of the original 4-dimensional phase space to a
2-dimensional one with the symplectic form \ref{eq:444}. This comes about
as follows: 
 The constraint \ref{eq:433} reduces the original 4-dimensional
symplectic space ${\cal S}^{4}_{q,p}$ - without deletion of the origin -
 to a 3-dimensional (non-symplectic) one, corresponding to a 3-dimensional
sphere $S^3$. 
  Factoring out the $U(1)$ - gauge transformations
yields a 2-dimensional subspace \cite{ab1}. Mathematically we are
 dealing with the
so-called ``Hopf fibration'' \cite{ab2,ste,cush2,thur1}\footnote{I came across
the textbook \cite{cush1} when this article was practically completed. Ch.\
$I$ of that monograph contains a lot of mathematical material which is very
closely related to that of the present section! See also the
 papers \cite{ler2}.}. 

 For convenience let us put $E=1$. Then
\begin{equation}
  \label{eq:451}
  S^3 = \{(A_1,A_2) \in \mathbb{C}^2; |A_1|^2+|A_2|^2 =1\}~.
\end{equation}
Furthermore 
\begin{equation}
  \label{eq:452}
  S^2 = \{(w,t) \in \mathbb{C}\times \mathbb{R}; |w|^2+t^2 =1, w=u+iv\,\}~.
\end{equation}
The manifold (``chart'') $S^2 - \{(0,0,1)\}$ may be mapped stereographically
onto the complex plane:
\begin{equation}
  \label{eq:453}
  p: S^2 - \{(0,0,1)\} \to \mathbb{C}\,;~ p(w,t) = z=x+iy= \frac{w}{1-t}~,
\end{equation}
with the inversion 
\begin{equation}
  \label{eq:454}
  w=z(1-t)\,,~(1-t)= \frac{2}{1+x^2+y^2}~.
\end{equation}
The mapping (projection) $h: S^3 \to S^2$ is implemented by
\begin{equation}
  \label{eq:455}
  h(A_1,A_2) = (w = 2\,\bar{A}_1 A_2, t= |A_2|^2-|A_1|^2)~.
\end{equation}
Notice that $h(e^{i\alpha}A_1, e^{i\alpha}A_2)= h(A_1,A_2)$. The orbits
of these $U(1) \backsimeq S^1$ transformations which are projected to
points on $S^2$, the so-called ``fibers over $(w,t) \in S^2$'', are 
given by
\begin{align}
  \label{fib} 
  h^{-1}(w,t)&= \{(A_1,A_2) \in \mathbb{C}^2; \\ \label{eq:1382}& |A_1|^2
 =\frac{1-t}{2},\,
|A_2|^2=\frac{|w|^2}{2(1-t)}=\frac{1+t}{2},\,
 w = 2 A_2\bar{A}_1\,\}~. \nonumber
\end{align}
Combining the projection $h$ with a consecutive stereographic one, $p$, 
yields the mapping $p\circ h: S^3 \to \mathbb{C}\,$:
\begin{equation}
  \label{eq:456}
  (p\circ h)(A_1,A_2) = \frac{2\bar{A}_1 A_2}{1-(|A_2|^2-|A_1|^2)} =
\frac{\bar{A}_1 A_2}{|A_1|^2} =z = |z|e^{i\vartheta} \in \mathbb{C}~.
\end{equation}
The relations \ref{eq:456} and \ref{eq:451} imply
\begin{eqnarray}
  \label{eq:1638}\label{eq:457}
  |A_1|& =& (1+|z|^2)^{-1/2}~, \\
\label{eq:1383}|A_2| &=& |z|\,(1+|z|^2)^{-1/2}~, \\
\label{eq:1384}z + \bar{z} = 2\,|z|\cos \vartheta &=&
\frac{A_1 \bar{A}_2 + \bar{A}_1 A_2}{|A_1|^2}=2 \frac{|A_2|}{|A_1|}\,\cos\vp\,
,\,
\vp=\vp_1-\vp_2\,,~~\label{56} \\
\label{eq:1385}z -\bar{z} =2i\,|z|\sin\vartheta &=& 2 i\,
\frac{|A_2|}{|A_1|}\,\sin\vp~.
 \label{57}
\end{eqnarray}
The last Eqs.\ show that $\vartheta = \vp$ . 

Up to now we have ignored the topological fine structures of the manifolds
involved: By implementing the stereographic projection \ref{eq:435} we have
taken out the north pole of the sphere $S^2$, reducing the sphere topologically
to a  (complex) plane. By introducing polar coordinates we in addition  have
 to take out the points $|A_1|=0$ and $|A_2|=0$. This means that $|z| \neq 0$,
too. 
 
The relations \ref{56} and \ref{57} imply
\begin{equation}
  \label{eq:458}
  i\,dz \wedge d\bar{z} = \frac{1}{I_1^2}\,d\vp \wedge dI_1~,
\end{equation}
showing the connection between the Hopf fibration and the symplectic structure
\ref{eq:444}. 

The above reduction \ref{eq:444}  of a 4-dimensional symplectic space
 to a 2-dimenional one by using a single constraint is very special, 
because in general a first class constraint reduces a symplectic space by just
one dimension leaving the additional reduction by an additional dimension
to gauge fixing \cite{dirac},
 i.e.\ selecting  unique representatives from the
equivalence classes formed by the orbits of the gauge transformations. 

A mathematically possible gauge fixing in our case is $\vp_2 =0$, i.e.\
 we take $A_2$ to be
real ($p_2 =0$) and $q_2 >0$. 

 If we express this constraint
 on ${\cal S}^{4}_{q,p}$
by
\begin{equation}
  \label{eq:447}
  \chi_2(q,p) \equiv \arctan\left(\frac{p_2}{q_2}\right)=0~,
\end{equation}
then we have to restrict the $\arctan$ to the interval $(-\pi/2,+\pi/2)$
in order to have a unique solution of the Eq.\ \ref{eq:447}. Otherwise
one would also get the solutions $\vp_2 =\pm \pi$. 

Because
\begin{equation}
  \label{eq:448}
  \{g_0,\chi_2\}_{q,p} = \{I_2,\chi_2\}/2 = 1/2\,,
\end{equation}
we now have two second class constraints \cite{dirac}. 

As to the physics, however, the gauge constraint \ref{eq:447} is too
restrictive, because it requires $A_2$ not only to be real, but to be
positive, too. This could at most be an initial condition, but cannot hold
for a finite time interval which is longer than the time period of the
oscillator. The ``weaker'' constraint $p_2 =0$ which would allow for negative
$A_j$ is unsuitable, too: it would entail the secondary constraint $\{g_0,
p_2\}_{q,p} = q_2/2 =0$, so that $A_2 =0$. In addition it would violate
the postulate $(q_2,p_2) \neq (0,0)$ and the line $p_2 =0$ would cut all
gauge orbits - the circles around  $(q_2,p_2) = (0,0)$ - twice.

A less restrictive gauge fixing is
\begin{equation}
  \label{eq:459}
  \chi_0 = \phi_1+\phi_2 -\beta =0\,,~\beta = \mbox{ const. } ~.
\end{equation}
If one tries to express this again by the original coordinates $q_j$ and 
$p_j$, namely
\begin{equation}
  \label{eq:460}
  \chi_0 = \arctan \frac{p_1}{q_1} + \arctan\frac{p_2}{q_2} -\beta =
\arctan\left(\frac{p_1q_2+q_1p_2}{q_1q_2+p_1p_2}\right) -\beta =0~, 
\end{equation}
with
\begin{equation}
  \label{eq:461}
  \{\phi_0,\chi_0\}_{q,p} =1~,
\end{equation}
then one encounters (Gribov) ambiguities, e.g. $(q_1,p_1) \to (-q_1,
-p_1)$ and $(q_2,p_2) \to (-q_2,-p_2)$ give the same $\chi_0$ in 
\ref{eq:460}. Such ambiguities are allowed, however, if we take the discrete
gauge symmetry \ref{eq:705} into account, which will be discussed below.

In the  case of second class constraints the symplectic
structure on the reduced space is given by the Dirac brackets \cite{dirac}: 

Suppose that we start from a  4-dimensional manifold ${\cal S}^{4}_{q,p}$
 with the
the (local) symplectic form \ref{eq:417}, the associated 
Poisson brackets \ref{eq:434} and two second class constraints
$\phi(q,p)$ and $\chi(q,p)$ which obey 
\begin{equation}
  \label{eq:449}
   \{\phi,\chi\}_{q,p} = \delta = \mbox{ const. }\neq 0~,
\end{equation}
then the Dirac brackets are given by
\begin{equation}
  \label{eq:450}
  \{f_1,f_2\}^* = \{f_1,f_2\}_{q,p} + \frac{1}{\delta}\,\{f_1,\phi\}_{q,p}
\{\chi,f_2\}_{q,p} -\frac{1}{\delta} \,\{f_1,\chi\}_{q,p}
\{\phi,f_2\}_{q,p}~.
\end{equation}
For $\phi = g_0$ and the special functions $f_1,f_2 = g_1,g_2$ or $g_3$ 
the Dirac brackets
are equal to the ordinary Poisson brackets because $g_0$ Poisson
commutes with all the $g_j, j=1,2,3$.

 The Dirac brackets play a non-trivial role, however, in the following
 modi\-fication
of the above canonical framework: 
 The ``gauge'' $\vp_2=0$ 
implies the weaker assumption $A_2 =\bar{A}_2 $.

 A weak form of
this restriction is to replace the amplitude $A_2$ in \ref{eq:410} and
\ref{eq:411} by its complex conjugate $\bar{A}_2$:
\begin{equation}
  \label{eq:462}
  \tilde{I} = |A_1 +\bar{A}_2 |^2~,
\end{equation}
with the corresponding properties
\begin{eqnarray}
  \label{eq:1639}\label{tili}
 \tilde{I}= \tilde{w}_3 &=& (A_1 + \bar{A}_2)(\bar{A}_1 + A_2) =
 \\ \label{eq:1386}&=& I_1 + I_2 +
 \bar{A}_1\,
\bar{A}_2 + A_1\,A_2  ~, \nonumber \\ 
 \label{eq:1387}\bar{A}_1\,
\bar{A}_2 + A_1\,A_2 &=& 2\, g_4(q_1,p_1;q_2,p_2) \equiv
 q_1\,q_2 - p_1\,p_2  \label{ge4}\\
\label{eq:1388}&=& 2\,\tilde{h}_1(|A_1|,|A_2|,\tilde{\vp} =
 \vp_2+\vp_1) \equiv \label{ha4} \\ 
\label{eq:1389}&\equiv&  2\,|A_1|\,|A_2|\, \cos\tilde{\vp}
\nonumber \\
 \label{eq:1390}\tilde{w}_4 &=& (A_1 - \bar{A}_2)(\bar{A}_1 - A_2)
 = I_1 + I_2 -
 \\ \label{eq:1391}&& -
 (\bar{A}_1\,\bar{A}_2 + A_1\,A_2)  ~,\nonumber \\  
 \label{eq:1392}\tilde{w}_5 &=& (A_1 -i\, \bar{A}_2)(\bar{A}_1 +i\, A_2) =
 I_1 + I_2 + \\ \label{eq:1393}&&+i
 (A_1\,A_2 - \bar{A}_1\,\bar{A}_2)  ~,\nonumber \\
\label{eq:1394}i (A_1\,A_2 - \bar{A}_1\,\bar{A}_2)&=& -2\,g_5(q_1,p_1;q_2,p_2)
 \equiv -q_1\,p_2
-q_2\,p_1 \label{ge5} \\ \label{eq:1395}&=& 2\,\tilde{h}_2(|A_1|,|A_2|,
\tilde{\vp}
 = \vp_2+\vp_1)
 \equiv
\label{ha5}\\ \label{eq:1396}&\equiv& -2\,|A_1|\,|A_2|
\,\sin\tilde{\vp}\,, \nonumber \\
 \label{eq:1397}\tilde{w}_6 &=& (A_1 +i\, \bar{A}_2)(\bar{A}_1 -i\, A_2) = 
I_1 + I_2 -\\ \label{eq:1398}&&-i
 (A_1\,A_2 - \bar{A}_1\,\bar{A}_2)  ~, \nonumber
\end{eqnarray}

The functions $g_4,g_5,g_0$ and $g_3$ have the Poisson brackets
\begin{equation}
  \label{eq:463}
  \{g_4,g_5\}_{q,p}= -g_0 \,,~\{g_0,g_4
\}_{q,p}= -g_5\,,~\{g_0,g_5\}_{q,p}=g_4~,
\end{equation}
i.e.\ the $g_4,g_5$ and $g_0$ generate the Lie algebra
of the group $SO^{\uparrow}(1,2)\,$! (Replace $g_4$ or $g_5$ by
 its negative and compare
with \ref{eq:441}.) These 3 functions Poisson commute with
the function $g_3$:
\begin{equation}
  \label{eq:464}
  \{g_j,g_3\}_{q,p}= 0\,,~j=0,4,5\,.
\end{equation}
The function $2g_3$ generates a group $U(1)$ which is the 
invariance (gauge) group of the observables $\tilde{w}_j, j=3,4,5,6$:
\begin{equation}
  \label{eq:465}
  A_1 \to e^{i\alpha}\,A_1\,,~\bar{A}_2 \to e^{i\alpha}\,\bar{A}_2~.
\end{equation}
The consequences can be discussed in a completely analogous way as
 in the case of the
transformations \ref{eq:420}: 

Instead of the transformations \ref{eq:422}
we now have
\begin{eqnarray} \label{eq:1640}\label{brevtra}
q_1 \to \breve{q}_1(\alpha)&=& \cos\alpha\, q_1 - \sin \alpha\,p_1~, \\
\label{eq:1399}p_1 \to \breve{p}_1(\alpha)&=& \cos\alpha\, p_1 + \sin 
\alpha\,q_1~, 
\nonumber \\
\label{eq:1400}q_2 \to \breve{q}_2(\alpha)&=& \cos\alpha\, q_2 +
 \sin \alpha\,p_2~, 
\nonumber \\
\label{eq:1401}p_2 \to \breve{p}_2(\alpha)&=& \cos\alpha\, p_2 -
 \sin \alpha\,q_2~,
\nonumber
\end{eqnarray}
The difference to the rotations \ref{eq:422} lies in the property
that the simultaneous rotations in the $(q_1,p_1)$- and $(q_2,p_2)$ - 
planes have opposite directions now. 

 Instead of the vector field
\ref{eq:412} we here have
\begin{equation}
  \label{eq:466}
  X_{2g_3} = p_1\partial_{q_1}-q_1\partial_{p_1} -
p_2\partial_{q_2}+q_2\partial_{p_2}~,
\end{equation}
with the property
\begin{equation}
  \label{eq:467}
  i_{X_{2g_3}}\omega_{q,p} = 2dg_3\,,~2g_3 =I_1-I_2~.
\end{equation}
The conservation law \ref{eq:432} now is replaced by
\begin{equation}
  \label{eq:468}
  \frac{d}{dt}\,g_3 =0
\end{equation}
and we have the constraint
\begin{equation}
  \label{eq:469}
  \phi_3 \equiv g_3 - \tilde{\epsilon} =0\,,~\tilde{\epsilon} = 
\mbox{ const. }~.
\end{equation}
This constraint  implies
\begin{equation}
  \label{eq:470}
  dI_1 = dI_2\,,~dI_{1,2}^2 = d(I_1I_2) =2\, g_0\,dI_1~,
\end{equation}
leading to a new reduction of the symplectic form \ref{eq:417}:
\begin{equation}
  \label{eq:471}
 \omega_{q,p} = d\tilde{\vp} \wedge dI_1 = \frac{I_{1,2}}{g_0}d\tilde{\vp}
\wedge dI_{1,2}\,, ~ \tilde{\vp} = \vp_1+\vp_2~,
\end{equation}
which is to be compared with the reduced form \ref{eq:445}. \\
For the functions $\tilde{h}_j(\tilde{\vp} =\vp_2+\vp_1,I_{1,2})$ we have
the Poisson brackets
\begin{equation}
  \label{eq:472}
  \{\tilde{h}_1,\tilde{h}_2\}_{\tilde{\vp},I_{1,2}}= I_{1,2}\,,~
\{\tilde{h}_1,I_{1,2}\}_{\tilde{\vp},I_{1,2}} =\tilde{h}_2\,,~
\{\tilde{h}_2,I_{1,2}\}_{\tilde{\vp},I_{1,2}} =-\tilde{h}_1\,,~
\end{equation}
which  form the Lie algebra  $\mathfrak{so}(1,2)$, too. 

 They are
functions on the phase space
\begin{equation}
  \label{eq:480}
   {\cal S}^{2}_{\tilde{\vp},I_{1,2}} =
 \{(\tilde{\vp},I_{1,2}=|A_1|\,|A_2|) \in S^1
 \times \mathbb{R}^+\,\}
\end{equation}
Again the first class constraint \ref{eq:469} becomes a second class one
if one recurs to gauge fixing, e.g.\ $\vp_2 =0$, or
the gauge \ref{eq:447} which has the property
\begin{equation}
  \label{eq:473}
  \{\phi_3,\chi_2\} = -1/2~.
\end{equation}

As  we have a pair of second class constraints now, we have to replace ordinary
Poisson brackets by Dirac brackets \ref{eq:450}. 
 
Because of the relations \ref{eq:464} all the additional terms in \ref{eq:450}
vanishes and thus the Poisson brackets \ref{eq:463}
can be replaced by Dirac brackets $\{\cdot,\cdot\}^*$ without changes! 

This is different for the relations \ref{eq:436} and \ref{eq:437}. Here
we have
\begin{equation}
  \label{eq:474}
  \{g_1,g_2\}^* =- g_0\,,~\{g_0,g_1\}^* =g_2\,,~
\{g_0,g_2\}^* =-g_1~
\end{equation}
and
\begin{equation}
  \label{eq:475}
  \{g_3,g_j\}^* = 0\,,~j=0,1,2~.
\end{equation}
What is surprising is that the ``reduced'' relations \ref{eq:474}
form the Lie algebra  $\mathfrak{so}(1,2)$, whereas the unreduced
ones \ref{eq:436} constitute the Lie algebra  $\mathfrak{so}(3)$! 

Similar to the relation \ref{eq:438} we have for $g_4$ and $g_5$:
\begin{equation}
  \label{eq:476}
  g_4^2+g_5^2 = I_1I_2 = g_0^2-g_3^2~.
\end{equation}
This relation may be used in order to obtain from \ref{eq:474}
\begin{equation}
  \label{eq:477}
  \{I_{1,2},g_1\}^* = \frac{g_0}{I_{1,2}}\,g_2\,,~\{I_{1,2},g_2\}^* =
 -\frac{g_0}{I_{1,2}}\,g_1\,.~
\end{equation}
Rewriting the first of the relations \ref{eq:474} as
\begin{equation}
  \label{eq:478}
  \{g_1,g_2\}^* = -\frac{g_0}{I_{1,2}}\,I_{1,2}
\end{equation}
and comparing with the expression \ref{eq:471} we see that we have
a completely analogous situation as discussed in connection with the
relations \ref{Ige1}, \ref{Ige2}, \ref{eq:445} and \ref{eq:446} if we
replace $\vp=\vp_2-\vp_1 $ by $\tilde{\vp}=\vp_2+\vp_1$, the factor
$-I_{1,2}/g_3$ by $I_{1,2}/g_0$ and the Poisson brackets by Dirac brackets. 

Thus, the relations \ref{eq:477} and \ref{eq:478} are equivalent to
those of \ref{eq:472}! 

\subsection{Emergence of the symplectic group $Sp(4,\mathbb{R})$}
We have seen that the functions $g_1,g_2$ and $g_3$ generate the Lie
algebra  $\mathfrak{so}(3)$ - Eqs.\ \ref{eq:436} - and
 the functions
$g_4,g_5$ and $g_0$ the Lie algebra  $\mathfrak{so}(1,2)$ - see
Eqs.\ \ref{eq:463}. These two groups are subgroups of the symplectic
group $Sp(4,\mathbb{R})$, the real symplectic group in 4 dimensions
which is an invariance group of the symplectic form \ref{eq:417}.
The group is 10-dimensional.

 The remaining 4 independent generators
of its Lie algebra may be obtained by the following 4 Poisson brackets:
\begin{eqnarray}  
\label{eq:1641}\{g_1,g_4\}&=& g_6 \equiv -\frac{1}{2}(q_1p_1+q_2p_2)\,
 \label{ge6}\\
\label{eq:1402}\{g_2,g_4\} &=& -g_8 \equiv 
-\frac{1}{4}(q_1^2-p_1^2 -q_2^2 +p_2^2)\,, \label{ge8}\\
\label{eq:1403}\{g_1,g_5\} &=& g_9 \equiv \frac{1}{4}(q_1^2-p_1^2 +q_2^2
 -p_2^2)\,,
\label{ge9} \\
\label{ge7}\{g_2,g_5\} &=& -g_7 \equiv -\frac{1}{2}(q_1p_1-q_2p_2)\,.  
\end{eqnarray}
Properties of the group $Sp(4, \mathbb{R})$ have been discussed
 extensively in the
mathematical and physical literature (see Appendix C for more details).

As the 10 functions $g_j,\,j=0,\ldots,9$, form a complete set of observables
for a group theoretical analysis of the phase space \ref{eq:416} in the 
framework of the present approach, let me list the remaining six of
 them here, too:
\begin{alignat}{2}
g_0 & = \frac{1}{4}(q_1^2+p_1^2 +q_2^2+p_2^2)\,,& ~~~ g_1& = 
\frac{1}{2}(q_1q_2 +p_1p_2)\,, \label{ge10} \\
g_2& = \frac{1}{2}(q_1p_2-q_2p_1)\,,& ~~~g_3&= \frac{1}{4}(q_1^2+p_1^2-q_2^2-
p_2^2)\,, \label{ge23} \\
g_4&=\frac{1}{2}(q_1q_2-p_1p_2)\,,&~~~
g_5& = \frac{1}{2}(q_1p_2+q_2p_1)\,,\label{ge45} \end{alignat}

It is also instructive to express the ten functions in terms of the
variables $I_1,I_2,\vp_1$ and $\vp_2$:
\begin{eqnarray}
  \label{eq:706}
g_0 &=& \frac{1}{2}(I_1+I_2)\,, \\
  g_1 &=& \sqrt{I_1I_2}\cos(\vp_1-\vp_2)\,,\label{eq:707} \\
 g_2 &=& \sqrt{I_1I_2}\sin(\vp_1-\vp_2)\,, \label{eq:708} \\
g_3 &=& \frac{1}{2}(I_1-I_2)\,,\label{eq:709} \\
 g_4 &=& \sqrt{I_1I_2}\cos(\vp_1+\vp_2)\,,\label{eq:710} \\
 g_5 &=& \sqrt{I_1I_2}\sin(\vp_1+\vp_2)\,, \label{eq:711} \\
g_6 &=& \frac{1}{2} (I_1\sin 2\vp_1 + I_2 \sin 2\vp_2)\,, \label{eq:712} \\
g_7 &=& \frac{1}{2} (-I_1\sin 2\vp_1 + I_2 \sin 2\vp_2)\,, \label{eq:713} \\
g_8 &=& \frac{1}{2} (I_1\cos 2\vp_1 - I_2 \cos 2\vp_2)\,, \label{eq:714} \\
g_9 &=& \frac{1}{2} (I_1\cos 2\vp_1 + I_2 \cos 2\vp_2)\,, \label{eq:715}
\end{eqnarray}
 
The 4 functions $g_0,g_j,j=1,2,3,$ are the generators of the (maximal) compact
subgroup $U(2) \simeq U(1)\times SU(2)$ of the symplectic group
$Sp(4,\mathbb{R})$. (See Sec.\ 8.2.3 and Appendix C!)
\subsection{The $Z_2$ ``gauge'' symmetry}
The ``observables'' $g_j,\,j=0,\ldots,9$, are all invariant under the
 $Z_2$ type
``gauge'' symmetry \ref{eq:705}. In case of the explicit form 
\ref{eq:706}-\ref{eq:715} this means invariance under
\begin{equation}
  \label{eq:716}
  I_j \to I_j\,,~~\vp_j \to \vp_j \pm \pi\,,\,j=1,2\,.
\end{equation}

We have here the same but more general situation we encountered in case
of the group $Sp(2,\mathbb{R})$,\, 
see the discussions
in Secs.\ 1.4, 2.3, 6.3 and in Appendix A.3:

Passing from the phase space
\begin{equation}
  \label{eq:800}
  {\cal S}^4_{q,p} =\{(q_1,p_1,q_2,p_2) \in \mathbb{R}^4 \}
\end{equation}
 to the phase space
 \begin{equation}
   \label{eq:801}
 {\cal S}^4_{\vp,I} = \{\vp_j \in \mathbb{R}\bmod{2\pi},~~I_j >0,\,j=1,2\}  
 \end{equation}
 means to factor out a $Z_2$ gauge symmetry, i.e.\ passing to the orbifold
 \begin{equation}
   \label{eq:802}
    {\cal S}^4_{q,p}/Z_2\,.
 \end{equation}
{\em Actually one can use the $Z_2$ invariance for a definition
 of ``observables'' on the space \ref{eq:800}. 
This implies that the original canonical variables $q_j$ and $p_j$
 are {\em not}
``observables'' in such a framework\,}! 

This is especially so for the
corresponding quantum theory (see below). The deeper reason is that the
deletion of the origin of the phase space \ref{eq:416} no longer allows
for arbitrary translations within that space. But the functions
$q_j$ and $p_j$, or the corresponding operators $Q_j$ and $P_j$ would
generate such translations. For that reason they have to be discarded!
But like in the 2-dimensional case one can define $Z_2$ invariant  
``composite'' coordinates 
$\tilde{q}_j,\tilde{p}_j$ and operators $\tilde{Q}_j, \tilde{P}_j$!

As to the group theoretical side this means the following: The reflection
 \ref{eq:705}
constitutes the non-trivial center element $-E_4$ ($E_4$: unit matrix in
$\mathbb{R}^4$) of the center
\begin{equation}
  \label{eq:717}
  Z_2 = \{E_4,-E_4\} \subset Sp(4,\mathbb{R})\,.
\end{equation}
As (see Appendix C.3)
\begin{equation}
  \label{eq:718}
  Sp(4,\mathbb{R})/Z_2 \cong SO^{\uparrow}(2,3)\,,
\end{equation}
we see that the actual structure group is not $Sp(4,\mathbb{R})$, but the
pseudo-orthogonal group $SO^{\uparrow}(2,3)$,  like in the 2-dimensional case,
where the effective structure group is $SO^{\uparrow}(1,2)
 =Sp(2,\mathbb{R})/Z_2$ not $Sp(2,\mathbb{R})\,$! This will be important for
the selection of the appropriate irreducible unitary representations.
(Compare also the well-known example of $SU(2)$ and $SU(2)/Z_2 \cong SO(3)$!)
\subsection{Two important remarks}
Let me conclude this section with two important remarks: 
\begin{enumerate} \item
 During the whole discussions above I have ignored any time- and
space-dependence of the quantities involved. In classical interference
phenomena, however, the intensity \ref{eq:410} and the amplitudes \ref{eq:411}
in general will vary from point to point (e.g.\ on a screen) 
according to the interferences
involved. The same may hold for the phase $\alpha$ in the transformations
\ref{eq:420} or \ref{eq:465}  which can depend on the time $t$, too.
The same applies to the $Z_2$ symmetry \ref{eq:705}.

 Thus, we are
actually dealing with field theories and genuine gauge transformations
(of the 2nd kind!). Taking this into account will not change the core of 
the above results in an essential way. 
\item
 The same mathematical structures which emerge from the interference
quantities \ref{eq:413}-\ref{eq:1375} also occur in the description of
  polarization
properties of electromagnetic waves in terms of  Stokes parameters
\cite{born2}: If
\begin{eqnarray}
\label{eq:1642}E_x &=& a_1\cos(\tau +\delta_1) =\Re(a_1e^{-i(\tau
 +\delta_1)})\,, \\
\label{eq:1405}E_y &=&  a_2\cos(\tau +\delta_2) =\Re(a_2e^{-i(\tau
 +\delta_2)})\,, \\
\label{eq:1406}&& a_1,a_2 >0\,,~\tau = \omega t-\vec{k}\cdot\vec{x}~,
\end{eqnarray}
then the observable properties of the wave may be characterized
by the parameters \begin{eqnarray}
\label{eq:1643}s_1 &=& 2a_1a_2 \cos\delta\,,~\delta =\delta_2-\delta_1~,
 \label{es1} \\
\label{eq:1407}s_2 &=& 2a_1a_2\sin\delta\,, \label{es2}\\
\label{eq:1408}s_3 &=& a_1^2-a_2^2 = I_1 -I_2~, \label{es3}\\
\label{eq:1409}s_0 &=& a_1^2 + a_2^2 = I_1 + I_2 ~, \label{es0} \end{eqnarray}
which obviously correspond exactly to the quantities $g_1$ (or $ h_1$),
\,$ g_2$ (or $-h_2$),\,$ 2g_3$ and $2g_0$ from above. 

All the formal results derived previously
may be applied to the  physical observables \ref{es1}-\ref{es0} as well.
\end{enumerate}
\section{Quantum theory}
Quantizing the  classical system of the last section  requires the group
 theoretical
approach: This follows from the structure of the phase spaces \ref{eq:416}
and \ref{eq:418} both of which have their origin deleted. Let us start with
the space \ref{eq:418}  which has been at the center of the
present paper! 
\subsection{$SO^{\uparrow}(1,2)$ quantization of interference patterns}
 According to the Eqs.\ \ref{ha1} and \ref{ha2} the functions
\begin{equation}
  \label{eq:479}
  h_1(\vp,I_{1,2})=I_{1,2}\cos\vp\,,~h_2(\vp,I_{1,2})=-I_{1,2}\sin\vp\,,
~I_{1,2}=\sqrt{I_1 I_2}~, 
\end{equation} characterize the classical ``observable'' interference
pattern. (The additional sum $I_1+I_2$ is a background quantity which does
not contribute to the dominant structure of the interference patterns.)

 Eqs.\ \ref{eq:441} show that the quantities \ref{eq:479} obey the Lie algebra 
 of the group $SU(1,1)$ and the Eqs.\ \ref{eq:472} mean that the same
 relations
hold for the functions $\tilde{h}_j(\tilde{\vp},I_{1,2})$ of the interference
pattern \ref{ha4} and \ref{ha5}! 

 In order to quantize the (reduced)
phase spaces \ref{eq:418} and \ref{eq:480} we again have to use the
unitary irreducible representation of the positive discrete series
of the group $SU(1,1)$ etc.

 As the ``observable'' $I_{1,2}$ corresponds to the
operator $K_0$, its eigenvalues in the quantum theory 
are $n+k, n=0,1,\ldots$ where $k$
characterizes the representation (see Ch.\ 2 and Appendix B for details).

 It is, of course, tempting - but certainly not necessary - to use
 the tensor product 
 $ \mbox{$\cal
H$}^{osc}_1\otimes \mbox{$\cal H$}^{osc}_2$ of  two harmonic
oscillator Hilbert spaces as the carrier space of the unitary representations
to be employed here (see the discussion in Sec.\ 6.4):
The SU(1,1) generators then have the form (see \ref{eq:266})\begin{equation}
  \label{eq:481}
 K_+=a_1^+a_2^+~,
~K_-=a_1a_2~,~K_0=\frac{1}{2}(a_1^+a_1+a_2^+a_2+1)=\frac{1}{2}(H_1+H_2)\,.
\end{equation}

  The product $ \mbox{$\cal
H$}^{osc}_1\otimes \mbox{$\cal H$}^{osc}_2$ contains all
 the irreducible unitary
representations of the group $SU(1,1)$ (for which
$k=1/2,1,3/2,\ldots$) as follows: 

 Let
$|n_i\rangle_j,~n_j =0,1,\ldots\,, \, j=1,2,$ be the eigenstates
of the number operators $N_j$, generated by $a^+_j$ from the
oscillator ground states $|n_j=0\rangle_j,j=1,2$.
 
 Each of those two subspaces of
 $\mbox{$\cal H$}^{osc}_1\otimes \mbox{$\cal H$}^{osc}_2= \mbox{ spanned by }
\{|n_1 \rangle_1\otimes
 |n_2\rangle_2\}$ with fixed $|n_1-n_2|\neq 0$  contains an irreducible
  representation
 with
\begin{equation}
  \label{eq:482}
 k=1/2+|n_1-n_2|/2=1,3/2,2,\ldots~.
\end{equation}
The Casimir operator $L = K_+K_-+K_0(1-K_0)$ now has the form
\begin{equation}
  \label{eq:483}
  L= \frac{1}{4}-(H_1-H_2)^2 = \frac{1}{4}-(N_1-N_2)^2~.
\end{equation}
For a given index $k$ according to \ref{eq:482} the number
  $n=0,1,\ldots $ in the eigenvalue $n+k$ of the above $K_0$ is given by
\begin{equation}
  \label{eq:484}
  n=\min\{n_1,n_2\}~.
\end{equation}
For the ``diagonal'' case $n_2=n_1$ we have the unitary
   representation with $k=1/2$. 

 For $n_1 > n_2$ and $n_1-n_2 =2k-1$
fixed - i.e.\ for a subspace of
 $\mbox{$\cal H$}^{osc}_1\otimes \mbox{$\cal H$}^{osc}_2$ with the basis
$\{|n_1 = n_2+2k-1 \rangle_1 \otimes |n_2
 \rangle_2\}$ which carries
 an irreducible unitary representation
with index $k$ - we have $n=n_2$! 

 A word of caution may be appropriate
here: According to \ref{eq:440} the quantity $I_{1,2}$  classically is
given by $\sqrt{I_1I_2}\,,\,I_j = (q_j^2+p_j^2)/2\,,\,j=1,2$.
When quantizing naively one expects that $I_1I_2 \to H_1H_2$ so that
$I_{1,2} \to \sqrt{H_1H_2}$. However, this argument is on the level of the
phase space \ref{eq:416}, not on the level of the reduced phase space
\ref{eq:418}! Only for the subspace with $n_1=n_2$ i.e. $k=1/2$ according
to Eq.\ \ref{eq:482}  the naive interpretation just mentioned appears to
be possible. 

In the other cases the use of the tensor product 
$\mbox{$\cal H$}^{osc}_1\otimes \mbox{$\cal H$}^{osc}_2$ is just a convenient
way to implement an
appropriate irreducible unitary
 representations of the group $SU(1,1)$ for an analysis of the quantum 
version of the interference pattern \ref{ha1} and \ref{ha2}. It 
serves the purpose to exploit the group theoretical structures \ref{eq:441}
or \ref{eq:472} of the reduced phase space of ``observable'' quantities
which characterize the interference pattern. Other Hilbert spaces for the
irreducible unitary representations in question might be more appropriate.
\subsection{Experimental aspects}
The question is, of course, how to determine the expectation values
of the operators $K_j = \hat{h}_j, j=1,2,\, K_0 = \hat{I}_{1,2},$
 and their statistical distributions experimentally. 

 Let $|\centerdot \rangle$
be an appropriate state which belongs to the domains of definition of  the
operators $K_j,j=1,2,$ and $K_0$, e.g.\ a 2-mode state of the type discussed
above or a 2-mode generated Barut-Girardello or Perelomov or 
Schr\"odinger-Glauber
coherent state associated
with a unitary representation indexed by $k$, i.e.\ eigenstates of the 
operators $K_- $ or $(K_0+k)^{-1}K_-$ or $(K_0+k)^{-1/2}K_-$ 
as discussed in Ch.\ 3. 

The classical relations \ref{eq:415} and \ref{we56} suggests the following
quantum relations
\begin{eqnarray}
\label{eq:1644}\bar{n}_3 - \bar{n}_4 &\equiv& \langle \centerdot |N_3| 
 \centerdot \rangle -
\langle \centerdot |N_4|  \centerdot \rangle = 4\,\langle \centerdot |K_1|
  \centerdot \rangle~,
\label{qwe34} \\
\label{eq:1410}\bar{n}_5 - \bar{n}_6 &\equiv& \langle \centerdot |N_5| 
 \centerdot \rangle -
\langle \centerdot |N_6|  \centerdot \rangle = 4\,\langle \centerdot |K_2| 
 \centerdot \rangle~,
\label{qwe56} \end{eqnarray}
where $N_3,N_4,N_5$ and $N_6$ are the number operators corresponding to
the classical intensities.
Here I have assumed that the quantum versions of the intensities $w_3$ and
$w_4$ as well as those of $w_5$ and $w_6$ have the same vacuum contributions so
that these cancel in the corresponding differences \ref{qwe34} and \ref{qwe56}.

 Similarly we have for the expectation values of $K_1^2$ and $K_2^2$:
\begin{equation}
  \label{eq:485}
  \langle \centerdot |K_1^2|\centerdot \rangle  = \frac{1}{16} \langle
 \centerdot |
(N_3-N_4)^2 |\centerdot \rangle\,,~
 \langle \centerdot |K_2^2|\centerdot \rangle  = \frac{1}{16} \langle
 \centerdot |
(N_5-N_6)^2 |\centerdot \rangle\,.~
\end{equation}

The corresponding determination of the expectation values of
 $ K_0 = \hat{I}_{1,2}$ and $K_0^2$
is not so obvious. Let us begin with $K_0^2$: 

Here one has to take into account that the classical
 Py\-tha\-gorean relation $ I_{1,2}^2=h_1^2 + h_2^2$ in general is
 no longer valid on the quantum level. For an irreducible representation
with index $k$ we have instead
\begin{equation}
  \label{eq:486}
  K_0^2 = K_1^2 + K_2^2 + k(k-1)~.
\end{equation}
So, if we know $k$ we may use that relation in order determine
 $\langle \centerdot | K_0^2 |\centerdot
 \rangle$
with the help of the relations \ref{eq:485}. However, in general one will
not have a definite value of $k$ in a given experimental situation. 

Additional information about
 $\langle \centerdot | K_0^2 |\centerdot
 \rangle$
 may come by exploiting the classical relations
\begin{equation}
  \label{eq:487}
  2\,I_{1,2}^2 = (I_1+I_2)^2 - I_1^2 -I_2^2\,,~I_1+I_2 =w_3+w_4 = w_5 +w_6~.
\end{equation}
On the quantum level we have to take vacuum contributions to $\hat{I}_j, 
j=1,2,$ and $\hat{w}_l, l=3,4,5,6, $ into account which here do not
cancel. A well-known experimental approach employs multi-port
 homodyning to be discussed below. 

A new problem is posed as to the determination of
\begin{equation}
  \label{eq:719}
 \langle \centerdot |K_0| \centerdot \rangle\,. 
\end{equation}
  Here the relation
  \begin{equation}
    \label{eq:722}
 K_0 =i\,[K_1,K_2] = [K_-,K_+]/2   
  \end{equation}
may be useful: Together with Eq.\ \ref{eq:486} it relates properties of
$K_1$ and $K_2$ to those of $K_0$.
 This can be seen for the special states $|\centerdot\rangle =
|k,z\rangle $ or $=|k, \lambda \rangle$ from section 3.1 and 3.2:

The  Eqs.\ \ref{delK} and \ref{eq:1078} or \ref{eq:686}, \ref{eq:1123} and
\ref{eq:163}  show how 
the expectation values $\langle k,z|K_0| k,z \rangle$ etc.\ may be obtained
from $\langle K_j^2 \rangle$ and the mean-square fluctuations 
$(\Delta K_j)^2_{k,z}\,,j=1,2$ etc.\ 

 How those coherent
states can be generated was indicated in section 6.5 above. 

Additional information about the expectation value \ref{eq:719} may come
from the relation \ref{eq:405}.

In general, however, one probably needs some new ideas for
 measuring the expectation values \ref{eq:719} of $K_0$ directly! 

The quantities $h_j(\vp,I_{1,2}),j=1,2,$ and $I_1+I_2 =2g_0$
 occuring in the relations
\ref{ha1}, \ref{ha2} and \ref{eq:423} (or the corresponding quantities
\ref{es1}, \ref{es2} and \ref{es0}) may be determined experimentally
by ``balanced'' multi-port homodyning \cite{walk1,mand,schlei1,leon2};
 reviews:
 \cite{loud,wel,opat}; textbooks \cite{homod}, especially Ref.\ \cite{schlei3}!

  (``Balanced''  means that the
employed beam splitters are of the $50:50$ type). Using the relations
\ref{qwe34} and \ref{qwe56} such a device should be
 suitable for measuring the quantized versions of $h_j \to K_j, j=1,2,$
simultaneously by making clever use of certain vacuum contributions \cite{ar}.
The squared fluctuations of these operators are then to be deduced from
Eqs.\ \ref{eq:485}.

Mandel and coworkers \cite{Noh1} have used an eight-port
 homodyning
 experimental setup in order to promote and analyze 
an ``operational'' approach to the concept of a ``quantized phase''.

 It is
a  central assumption  in their analysis of the experimental data that the
 Pythagorean trigonometric relation \ref{eq:438} is valid in the
 quantum regime, too! We have seen that this assumption is not justified
in general, especially for small numbers of the quanta involved. 

As to examples of other experiments using similar homodyning techniques see,
e.g.\ \cite{smith,hrad}.

In the ``balanced'' homodyning scheme one does not determine the difference
 $I_1-I_2 = 2g_3$ or the Stokes parameter $s_3$ (Eq.\ \ref{es3}).
Several proposals have been made \cite{luis1} to pass to an ``unbalanced''
 setup in order
to determine the quantized counterpart of the quantity $I_1-I_2 =2g_3 =s_3 $
 simultaneously with those of the 3 others measured
in the balanced scheme ($h_1,h_2$ and $I_1+I_2$ or $s_1,s_2$ and $s_0$). 

If one knows the 2 quantities $I_1+I_2$ and $I_1-I_2$ - or their quantum
counterparts - then one may calculate $I_1$ and $I_2$ and one may be able
to measure $4 I_{1,2}^2 = 4 I_1I_2 =(I_1 +I_2)^2 -(I_1-I_2)^2$ or
 the corresponding
quantum expectation value.

 But one still needs an idea in order to determine
the quantum version of $I_{1,2} = \sqrt{I_1I_2}$! 
 \subsection{Relations to unitary representations of $Sp(4,\mathbb{R})$}
In the quantum theory of the interference patterns we are entering
 again the realm of the symplectic group $Sp(4,\mathbb{R})$
already mentioned in section 8.1.2 . 

As explained in appendix C, an
irreducible unitary representation of the positive discrete series
of $Sp(4,\mathbb{R})$ may be characterized by a pair $(\epsilon_0,j_0)\,,
\epsilon_0 > j_0\,,$ of numbers, where $\epsilon_0$ is the lowest eigenvalue
of the operator $K_0$ (now embedded in the 10-dimensional Lie algebra
of $Sp(4,\mathbb{R})$) which generates the commuting $U(1)$ subgroup
 of the maximal
compact subgroup $U(2) \simeq U(1) \times SU(2)$. The number $\epsilon_0$
corresponds to the number $k$ which characterizes the irreducible unitary
representations of $SU(1,1)$ etc. The other compact
 subgroup $SU(2) \subset U(2)$ has the
usual finite dimensional unitary representations characterized by $j=0,1/2,1,
3/2,\ldots$. For a given eigenvalue  $\epsilon_0$ the associated eigenstates
carry an irreducible unitary representation of $SU(2)$ characterized by the
number $j_0$ which may take the values $j_0 = 0,1/2,1,\ldots$ but has to be
smaller than $\epsilon_0$! (see Appendix C!) 

 In the case of the tensor product
 $\mbox{$\cal H$}^{osc}_1\otimes \mbox{$\cal H$}^{osc}_2$ we can
construct two $Sp(4,\mathbb{R})$ irreducible unitary representations of
 this type: 

Employing creation and annihilation operators we have the correspondence
 \cite{barg3}
\begin{eqnarray}
\label{eq:1645}g_1 &\to & J_1 = \frac{1}{2}(a_1a_2^+ +a_2a_1^+)~,
 \label{J1} \\
g_2 &\to& J_2 = \frac{i}{2}(a_1a_2^+ -a_2a_1^+)~, \label{J2} \\
g_3 & \to & J_3 = \frac{1}{2}(a_1^+a_1 -a_2^+a_2)~,\label{J3} \\
g_0 &\to & K_0 = \frac{1}{2}(a_1^+a_1 +a_2^+a_2 +1) =H/2~. \label{K0}
\end{eqnarray}
The space 
 $\mbox{$\cal H$}^{osc}_1\otimes \mbox{$\cal H$}^{osc}_2$ may be decomposed
into 2 subspaces $\mbox{$\cal H$}_+$ and $\mbox{$\cal H$}_-$: one
 subspace is spanned by the basis 
\begin{equation}
  \label{eq:488}
|n_1,n_2 \rangle_+
\equiv \{|n_1\rangle_1
\otimes |n_2 \rangle_2; n_1 +n_2 \mbox{ even }\} \in \mbox{$\cal H$}_+~,  
\end{equation}
 the other by
 \begin{equation}
   \label{eq:489}
 |n_1,n_2 \rangle_-
\equiv \{|n_1\rangle_1
\otimes |n_2 \rangle_2; n_1 +n_2 \mbox{ odd }\}\in \mbox{$\cal H$}_-~.  
 \end{equation}
In the first case we have for the ground state ($n_1 = n_2=0$):
\begin{eqnarray} \label{eq:1646}K_0|0,0\rangle_+& =& \frac{1}{2}|0,0\rangle_+~, \\
\label{eq:1411}J_3|0,0\rangle_+ &=& 0~,
\end{eqnarray} i.e.\ we have a representation with $\epsilon_0 =1/2,\, j_0=0$.
It may be shown (see Appendix C) that it is irreducible and unitary. 

In the second case the ground state is degenerate because we  have the two
possibilities
$n_1=1,n_2=0$ and $n_1=0,n_1=1$. These ground states have the properties
\begin{eqnarray}
\label{eq:1647}K_0|1,0\rangle_-&=&|1,0\rangle_-~, \\
\label{eq:1412}K_0|0,1\rangle_-&=&|0,1\rangle_-~,\\
\label{eq:1413}J_3|1,0\rangle_-&=&\frac{1}{2}|1,0\rangle_-~, \\
\label{eq:1414}J_3|0,1\rangle_-&=&-\frac{1}{2}|0,1\rangle_-~,
\end{eqnarray}
i.e.\ here we have $\epsilon_0=1,\,j_0=1/2$. 

As to the higher levels in both representations we have the following
situation: 

If $n_1+n_2 =2n,n=0,1,2, \ldots$ we have the eigenvalues $\epsilon_n =
n+1/2$ for $K_0$ with a ($2n+1$)-fold degeneracy. 
The associated $(2n+1)$-dimensional
subspace carries an irreducible representation of the group $SU(2)$ with
$j=n$. 
 If $n_1+n_2=2n+1,n=0,1,2,\ldots$ we get the eigenvalues 
$\epsilon_n = n+1$ with a ($2n+2$)-fold degeneracy. The associated subspace
carries an irreducible $SU(2)$-representation with $j=n+1/2$. 

In the present context it is of special interest in which way the irreducible
unitary representations of the subgroup $SU(1,1) \cong SL(2,\mathbb{R}) =
Sp(2,\mathbb{R})$, as given by the
relations \ref{eq:481}-\ref{eq:484}, are contained in the two irreducible
unitary representations $(1/2,0)$ and $(1,1/2)$ of $Sp(2,\mathbb{R})$
just described.

 One sees
immediately that the $SU(1,1)$ representations with $k=1/2,$ $3/2,\ldots$ are
contained in $(1/2,0)$ and the representations with $k=1,2,\ldots$
 in $(1,1/2)$. (Recall that we have the correspondences $g_0 \to
K_0,\, g_4 \to K_1,\,g_5 \to -K_2\,$!)

It is important to realize that the use of irreducible unitary representations
(of the positive  discrete series) of the group $Sp(4,\mathbb{R})$
 is mandatory if one wants to quantize the phase space \ref{eq:416} with
its origin deleted! And it might be necessary to employ other irreducible
unitary representations of $Sp(4,\mathbb{R})$ than just the 2 discussed
above (see Appendix C)! 

Accordingly one has to use the unitary representations of the group
$SO^{\uparrow}(2,3)$ for quantizing the orbifold ${\cal S}^4_{q,p}/Z_2\, $!

There is another topic left which is to be discussed: In the previous
section when discussing the unquantized classical properties of interference
pattern we encountered the typical situation of  gauge invariances: The
interference observables \ref{ha1} and \ref{ha2} are invariant under 
phase transformations generated by $g_0$ and the  observables
\ref{ha4} and \ref{ha5} are invariant under phase transformations generated
by $g_3$. This lead to the constraints $\phi_0=0$ of  Eq.\ \ref{eq:433} or
 $\tilde{\phi}_3=0$ of Eq.\ \ref{eq:469},
 respectively.

 According to the ideas of Dirac \cite{dirac} one has
 two possibilities
in order to quantize such systems: 
\begin{enumerate} \item
  One quantizes the ``reduced'' system, i.e.\ the subspace of
gauge invariant quantities by associating a canonically conjugate 
gauge fixing function $\chi$ with the originally first class constraint
function $\phi$ (see the discussion around Eq.\ \ref{eq:449})
 and by eliminating 
 the unphysical gauge degrees of freedom by the conditions $\phi=0$ and
$\chi=0$. We have seen how
we arrived in this way at the reduced phase spaces \ref{eq:418} and
\ref{eq:480} and how these may be quantized in terms of irreducible
unitary representations of the group $SU(1,1)$ or $SO^{\uparrow}(1,2)\,$.
 \item
 The second approach of quantizing such a system with gauge constraints
consists in quantizing the original phase space (here \ref{eq:416})  first
(now in terms of irreducible unitary representations of the group $Sp(4,
\mathbb{R})$) and constructing the physical Hilbert space by requiring
the quantized version $\hat{\phi}\to 0$ of the classical first class 
constraint $\phi=0$ to be implemented by the condition that the constraint
operator $\hat{\phi}$ annihilates the physical states.

 Let us see how this works in our case: 

The quantized version of the classical constraint function \ref{eq:433}
is 
\begin{equation}
  \label{eq:490}
  \phi_0 \to \hat{\phi}_0 = K_0-\epsilon\,,~\epsilon=E/2\,.
\end{equation}
When we apply this $\hat{\phi}_0$ to a state $|n_1,n_2\rangle_+$ of the
representation $(1/2,0)$ we see that
\begin{equation}
  \label{eq:491}
  \hat{\phi}_0|n_1,n_2\rangle_+ = 0 \mbox{ iff } \epsilon =\epsilon_n
=n+1/2\,,\,n=(n_1+n_2)/2~.
\end{equation}
We know already that the corresponding subspace is (2n+1)-dimensio\-nal
and  carries an irreducible unitary representation of $SU(2)$ with
$j=n$. 

 Analogous results hold for the representation $(1,1/2)$.
The constraint operator $\hat{\phi}_0$ merely implements the conservation
of energy and decomposes the Hilbert spaces $\mbox{$\cal H$}_+$ and
$\mbox{$\cal H$}_-$ into energy eigenstate subspaces with unitary
representations of $SU(2)$\,. 

In the case of the constraint \ref{eq:469} we get the operator 
\begin{equation}
  \label{eq:492}
  \hat{\phi}_3 =J_3-\tilde{\epsilon}~.
\end{equation}
The physical state condition for $|n_1,n_2\rangle_+$  here is
\begin{equation}
  \label{eq:493}
  \hat{\phi}_3|n_1,n_2\rangle_+= [\frac{1}{2}(n_1-n_2)-\tilde{\epsilon}]
\,|n_1,n_2\rangle_+ =0
\end{equation}
which means that the number $\tilde{\epsilon}$ has to be quantized:
\begin{equation}
  \label{eq:494}
  \tilde{\epsilon} =\frac{1}{2}(n_1-n_2)~.
\end{equation}
In view of the relation \ref{eq:482} this means that
\begin{equation}
  \label{eq:495}
  |\tilde{\epsilon}|=k-1/2~,
\end{equation}
i.e.\ $\hat{\phi}$ essentially projects onto  irreducible unitary 
representations of $SU(1,1)$, here (in the case of the $Sp(4,\mathbb{R})$
representation $(1/2,0)$) onto
 representations with $k=1/2,3/2,\ldots$. For the $Sp(4,\mathbb{R})$ 
representation
$(1,1/2)$ the operator $\hat{\phi}_3$ projects onto
 $SU(1,1)$ or $SO^{\uparrow}(1,2)$ representations
with $k=1,2,\ldots$. 

For related discussions of first class constraints of the type $\hat{\phi}_0$
and $\hat{\phi}_3$ see the Refs.\ \cite{Klaud4}.
\end{enumerate}

\chapter*{Acknowledgments}
 \addcontentsline{toc}{chapter}{\protect 
\numberline{}
Acknowledgments}
During the more than two years I have been working on the present article
I enjoyed the support, help and encouragements of many people and
 institutions:

During the academic year 2000/2001 I was an invited guest of the CERN
Theory Division at Geneva. I am still very grateful for that
 invitation and I deeply thank
the Division for its friendly and supporting hospitality.

Since the fall of 2001 I have the privilege of being a permanent guest 
of the DESY Theory Group in Hamburg. I thank the Theory Group and the
DESY Directorate for the very friendly invitation to come to that
 eminent scientific
institution and for their very generous hospitality and kind support.

In 2002 I was invited by H.\ Nicolai to spend 3 months at the
 Albert-Einstein-Institut f\"ur Gravitationsphysik of the Max-Planck-Society
in Golm near Potsdam. I had a very stimulating and fruitful stay at that
institute and I am very grateful for that invitation, too.

The first 2 months of this year I enjoyed the very pleasant and highly
stimulating hospitality of the Quantum Field Theory Group in the Physics
Institute of the Humboldt-University in Berlin. I am very grateful to
D.\ L\"ust for his very kind invitation to be a guest of that group.

During all those stays I had the very competent and friendly support from
secretaries and librarians of the institutions mentioned. I thank
all of them and my son David for his excellent help with the Linux-version
 of Latex.

I am grateful to W.\ Schleich, H.\ Walther and H.\ Paul for
 their kind in\-vi\-tations
to give seminars on the subject of this paper in Ulm, Munich and Berlin.
I thank them and A.\ W\"unsche, Berlin, for very stimulating dis\-cus\-sions.

Finally there is one single person I owe unlimited thanks for her permanent
and loving support, encouragements and her unparalled patience
 with my relentless
pursuit of the problems of this paper: my wife Dorothea!

\begin{appendix}

 \chapter{Basic properties of group theoretical quantizations}
\section{Generalities}
In the following I shall sketch the main ideas of the group theoretical
approach to quantizing classical phase spaces (symplectic manifolds). 
For far  more thorough and more detailed discussions I refer to the two main
expositions of the subject:  Refs.\ \cite{is1,gui}. I shall also borrow
heavily from Ref.\ \cite{bo}. 

I have already indicated in the Introduction how the conventional quantization
procedure may be interpreted as a group theoretical quantization in terms
of translations in coordinate and momentum space. That interpretation appears
complicated by the fact that one hat to extend the abelian group of
 translations on $\mathbb{R}^{2n}$ by the abelian additive group of the 
real numbers $\mathbb{R}$ (see the group law \ref{eq:26} for $n=1$).
 Those complications do not occur for simple
groups (i.e.\ non-abelian Lie groups $G$ the Lie algebra $\mathfrak{g}$
 of which does
not contain any ideal except $\{0\}$ and $\mathfrak{g}$ itself). The Lie
groups we are interested in, namely $SU(1,1) \cong Sp(2,\mathbb{R})\,$,
$SO^{\uparrow}(1,2)$ and $Sp(4,\mathbb{R})$ are all
simple. 

 Group theoretical quantization is a genuine generalization of the
conventional quantization procedure to phase spaces (symplectic spaces)
$ {\cal S}^{2n}$ which {\em globally} are not diffeomorphic to
 $\mathbb{R}^{2n}$!

  On  a $2n$--dimensional symplectic space\footnote{In this section I use
the ``covariant'' notation: upper indices for the coordinates, $q^j$, and
lower ones, $p_j$, for the momenta.}
 (manifold) 
 \begin{equation}
   \label{eq:322}
{\cal S}^{2n} = \{s =(q^1,\ldots,q^n,p_1,\ldots,p_1)\}    
 \end{equation}
 with a non-degenerate (local) symplectic form
 \begin{equation}
   \label{eq:323}
   \omega =dq^1\wedge dp_1 + \cdots + dq^n \wedge dp_n~,~ dp_1\wedge dq^1 =
-dq^1 \wedge dp_1\,,\,\ldots\,,
 \end{equation}
 one has several Lie algebra structures: 
\begin{enumerate} \item  Smooth functions $f_j(s),j=1,2,\ldots, $ on
 ${\cal S}^{2n}$ form a Lie algebra by means of their Poisson brackets:
 \begin{equation}
   \label{eq:324}
   \{f_1,f_2\}_{q,p} = \sum_{j=1}^n \partial_{q^j}f_1\,\partial_{p_j}f_2 -
\partial_{p_j}f_1\,\partial_{q^j}f_2\,.
 \end{equation}
\item The Lie algebra of smooth (tangent) vector
 fields $X(s)$ on ${\cal S}^{2n}$: 

Let $C_{\tau_1 \to \tau_2} =\{ s(\tau)\,,\tau \in [\tau_1,\tau_2]\,\} \subset
{\cal S}^{2n} $ be a smooth curve and $f(s)$ a smooth function. Then
\begin{equation}
  \label{eq:325}
  \frac{d}{d\tau}f[s(\tau)] = \sum_{j=1}^n (\partial_{q^j}f(s))\dot{q}^j(\tau)
+ (\partial_{p_j}f(s))\dot{p}_j(\tau)\,,~\dot{q}^j \equiv \frac{dq^j}{d\tau}\,,
\end{equation}
where $(\dot{q}^1,\ldots,\dot{p}_n)$ determines the tangent vector of $C_{
\tau_1 \to \tau_2}$ at $s(\tau)$. 

The relation \ref{eq:325} may be interpreted as follows: The $2n$
 partial derivatives
$\partial_{q^j},\,\partial_{p_j}$ form a basis of the tangent space
at $s$. A general tangent vector $X(s)$ at $s$ has the form
\begin{equation}
  \label{eq:326}
  X(s) = \sum_{j=1}^n A^j(s)\partial_{q^j}+B_j(s)\partial_{p_j}\,.
\end{equation}
If $A^j(s),B_j(s),j=1,\ldots,n,$ are smooth functions, then the relation
\ref{eq:326} defines a smooth vector field on ${\cal S}^{2n}$. 

Comparing with the relation \ref{eq:325} shows that any such vector field
defines a family of curves as solutions of the ordinary differential
equations
\begin{equation}
  \label{eq:327}
  \dot{q}^j = A^j[s(\tau)]\,,~\dot{p}_j=B_j[s(\tau)]\,,\,j=1,\ldots,n\,.
\end{equation}
Let $\phi_{\tau}(s_0)$ be a solution with the initial condition $\phi_{\tau =0}
(s_0) = s_0$. Such a solution may be interpreted as a 1-parameter
 transformation group on ${\cal S}^{2n}$ with the property
 \begin{equation}
   \label{eq:328}
   s_0 \to s(\tau)= \phi_{\tau}(s_0)\,,~ \phi_{\tau_2}[\phi_{\tau_1}(s_0)]
= \phi_{\tau_1+\tau_2}(s_0)\,.
 \end{equation}
If
\begin{equation}
  \label{eq:329}
  X_{\alpha} = \sum_{j=1}^n A^j(s;\alpha)\partial_{q^j}+B_j(s;\alpha)
\partial_{p_j}\,,~\alpha =1,2\,,
\end{equation}
are two vector fields then their commutator is again a vector field:
\begin{eqnarray}
\label{eq:1648}(X_2\,X_1-X_1\,X_2)f &=& X_3f~, \label{comX} \\
 \label{eq:1415}A^j(s;3) &=& \sum_{k=1}^n [A^k(2)\partial_{q^k}B_j(1)
+B_k(2)\partial_{p_k}
B_j(1)- \nonumber \\ \label{eq:1416}&&- A^k(1)\partial_{q^k}B_j(2)
-B_k(1)\partial_{p_k}
B_j(2)]~, \nonumber \\
\label{eq:1417}B_j(s;3) &=& \sum_{k=1}^n [A^k(2)\partial_{q^k}A^j(2)
+B_k(2)\partial_{p_k}
A^j(1)- \nonumber\\ \label{eq:1418}&&- A^k(1)\partial_{q^k}A^j(1)
-B_k(1)\partial_{p_k}
A^j(2)]~. \nonumber
\end{eqnarray}
In this way we get a Lie algebra of vector fields on ${\cal S}^{2n}$. 

Recall that the differential 1-forms
\begin{equation}
  \label{eq:330}
  \rho = \sum_{j=1}^n a_j(s)dq^j +b^j(s)dp_j 
\end{equation}
are dual to the vector fields \ref{eq:326}, i.e.\ we have $dq^j(\partial_{q^k})
= \delta^j_k$ etc. 
 
I also briefly recall three important operations on differential
forms: exterior differentiation $d$, interior multiplication $i_X$ by a vector
field $X$ and Lie derivation $L_X$ (for a more sytematic introduction into
these concepts see, e.g.\ Refs.\ \cite{cho,mar1,mar2}): 

 {\em Exterior differentiation} of a
function $f(s)$ means
\begin{equation}
  \label{eq:331}
  df(s)= \sum_{j=1}^n \partial_{q^j}f(s)dq^j + \partial_{p_j}f(s)dp_j~,
\end{equation}
supplemented by the important property $d^2=0$. This implies for the
differential 1-form \ref{eq:330}
\begin{equation}
  \label{eq:332}
  d\rho = \sum_{j=1}^n da_j(s)\wedge dq^j +db^j(s) \wedge dp_j~.
\end{equation}
In this way we get the symplectic form \ref{eq:323} from $\theta = \sum_{j=1}^n
p_jdq^j$ as $\omega = -d\theta$. We also have $d\omega =0$. 

Exterior differentiation converts a function (a 0-form) into a 1-form, a 1-form
into a 2-form etc. 

 {\em Interior multiplication by a vector field  $X$} on the other
 hand converts
a 2-form 
\begin{eqnarray}
  \label{eq:1649}\label{eq:333}
  \eta(\cdot,\cdot) &=& \sum_{j,k=1}^n\frac{1}{2} a_{j k}dq^j \wedge dq^k + 
\frac{1}{2} b^{j k}dp_j \wedge
dp_k + c_j^k dq^j \wedge dp_k\,,~~~  \\
\label{eq:1419}&& a_{kj}=-a_{jk}\,,\,b_{kj} = -b_{jk}\,, \nonumber
\end{eqnarray}
into the 1-form
\begin{equation}
  \label{eq:334}
  i_X\eta = \eta(X,\cdot) = \sum_{j,k=1}^n (A^j\,a_{jk}-B_j\,c^j_k)\,dq^k +
(B_j\,b^{jk}+A^j\,c^k_j)\,dp_k \,.
\end{equation}
For $\eta = \omega$ (Eq.\ \ref{eq:323}) this simplifies to
\begin{equation}
  \label{eq:335}
  i_X\omega = \sum_{j=1}^n -B_j\,dq^j + A^j\,dp_j~.
\end{equation}
For functions $f(s)$ one has $i_X(f)=0$. 

The notion of {\em Lie derivative $L_X$ with respect to a vector field $X$}
is important in connection with invariance properties of differential
forms: The concept is closely related to the 1-parameter transformation group
\ref{eq:328}. It suffices to define $L_X$ for functions $f(s)$ and their
differentials $df(s)$ because all differential forms of higher degree may
built from them. The definitions are
\begin{eqnarray}
  \label{eq:1650}\label{LX}
  L_Xf(s_0) &=&Xf(s_0)= \lim_{\tau \to 0}\frac{1}{\tau}[f(\phi_{\tau}(s_0))
 -f(s_0)]~, \\ \label{eq:1420}L_Xdf(s_0) &=& d(Xf)(s_0) = d[\lim_{\tau
 \to 0}\frac{1}{\tau}
[f(\phi_{\tau}(s_0)) -f(s_0)]\,. \label{dLX}
\end{eqnarray}
Of considerable practical importance is the  identity
\begin{equation}
  \label{eq:337}
  L_X= d\,i_X + i_X\,d \,.
\end{equation}
If a p-form $\rho^{\,p},\,p=0,1,2,\ldots, $ is invariant under a
 transformation \ref{eq:328} this can be expressed as
\begin{equation}
  \label{eq:336}
  L_X\rho^{\,p} = 0\,.
\end{equation}
\item Of special interest are the so-called ``Hamiltonian'' vector fields:

For a Hamiltonian system with Hamilton function $H$ we have the Eqs. of motion
\begin{equation}
  \label{eq:338}
  \dot{q}^j = \frac{\partial H}{\partial p_j}\,,~\dot{p}_j = 
-\frac{\partial H}{\partial q^j}\,,~j=1,\ldots,n\,,~\tau = t\mbox{ (time)}.
\end{equation}
Comparing with Eqs.\ \ref{eq:327} we see that the Hamilton function $H(s)$
generates the vector field
\begin{equation}
  \label{eq:339}
  X_H = \sum_{j=1}^n(\partial_{p_j}H)\,\partial_{q^j} - (\partial_{q^j}H)\,
\partial_{p_j}~.
\end{equation}
We have (see Eq.\ \ref{eq:335})
\begin{equation}
  \label{eq:340}
  i_{X_H}\omega = \sum_{j=1}^n -(-\partial_{q^j}H)\,dq^j + (\partial_{p_j}H)\,
dp_j = dH~,
\end{equation}
i.e.\ we get  from the identity \ref{eq:337} 
that (recall $d\omega=0$)
\begin{equation}
  \label{eq:341}
  L_{X_H}\omega =d(i_{X_H}\omega)=d(dH) = 0\,.
\end{equation}
The last equation expresses the important property that the 1-parameter
transformations $s_0 = s(t=0) \to s(t)= \phi_t^{(H)}(s_0)$ generated by
 the Hamilton
function $H$ and its associated vector field $X_H$ are ``canonical'', i.e.\ 
they leave the symplectic form $\omega$ invariant:
\begin{equation}
  \label{eq:342}
  \omega_{s(t)} = \omega_{s_0}\,,~s(t) = \phi_t^{(H)}(s_0)\,.
\end{equation}
Important is the generalization:

 Any smooth function $f(s)$ generates a Hamilton-type
vector field
\begin{equation}
  \label{eq:343}
  X_f = \sum_{j=1}^n (\partial_{p_j}f)\,\partial_{q^j} -(\partial_{q^j}f)\,
\partial_{p_j}\,,
\end{equation}
with the properties
\begin{equation}
  \label{eq:344}
  i_{X_f}\omega = df\,,~L_{X_f}\omega =0\,.
\end{equation}
If $X_{f_1}$ and $X_{f_2}$ are two such  vector fields then one
may define the Poisson brackets \ref{eq:324} of $f_1$ and $f_2$ as
\begin{equation}
  \label{eq:345}
  \{f_1,f_2\} =\omega(X_{f_1},X_{f_2}) = -X_{f_1}(f_2)= X_{f_2}(f_1)\,.
\end{equation}
Essential  is the following relationship between the Lie algebra structure
induced by the Poisson brackets \ref{eq:324} and the Lie algebra structure
\ref{comX} of Hamiltonian vector fields:
\begin{equation}
  \label{eq:348}
  [X_{f_1},X_{f_2}] = -X_{\{f_1,f_2\}}\,.
\end{equation}

These remarks on Hamiltonian vector fields show that we have a homomorphism
\begin{equation}
  \label{eq:346}
  f \mapsto -X_f
\end{equation}
of the Lie algebra of smooth functions $f$ on ${\cal S}^{2n}$
 onto the Lie algebra of
 smooth Hamiltonian vector fields $X_f$ on ${\cal S}^{2n}$. 

It is also important
to notice that this mapping has a non-trivial kernel, namely the constant
functions:
\begin{equation}
  \label{eq:347}
  f_0 = \mbox{ const. } \mapsto -X_{f_0} = 0~.
\end{equation} \end{enumerate}
 After all these preliminaries we now come closer to the gist of
the group theoretical quantization approach: 

Let us assume that there is
a $r$-dimensional Lie transformation group $\mbox{G}=\{g\}$ acting 
on
\begin{equation}
  \label{eq:803}
  {\cal S}^{2n}= \{s\}\,:~ s \to g\cdot s, 
\end{equation}
 and which has the
 following properties
\begin{enumerate}
\item The transformations $s \to g\cdot s$ leave the 
 symplectic form $\omega$ invariant:
 \begin{equation}
   \label{eq:349}
   \omega_{g\cdot s} = \omega_s~.
 \end{equation}
\item Let $g(t) \subset G$ be 1-parameter subgroup  generated by an
element $A \in \mathfrak{g}$ of the Lie algebra $\mathfrak{g}$ of $G$:
\begin{equation}
  \label{eq:350}
  g(t) =e^{-A\,t}~,~A \in \mathfrak{g}~.
\end{equation}
The action of such a subgroup in turn generates a vector field $\tilde{A}(s)$ 
on ${\cal S}^{2n}$:
\begin{equation}
  \label{eq:351}
  [\tilde{A}f](s)= \lim_{t \to 0}\frac{1}{t}[f(e^{-A\,t}\cdot s)-f(s)]~.
\end{equation}
We shall discuss some properties of these $G$-induced vector fields
below.
\item For the implementation of the intended quantization procedure
one wants to have an isomorphismen between the $r$-dimensional Lie algebra
$\mathfrak{g}$ and the corresponding Lie algebra $\tilde{\mathfrak{g}}$
of the induced vector fields $\tilde{A}(s)$. This is the case if the
 action of the group is effective, 
 i.e.\ if $ g\cdot s = s~ \forall s$, then $g=e$ (unit group
 element). 

 The latter condition may be relaxed to almost effective actions, i.e.\ if
  $ g \cdot s = s~ \forall s$, then $g$ is an element of a discrete center
  subgroup. This generalization is possible because the existence of such
a discrete center subgroup does not affect the structure of the Lie algebra
which is the same for a group, all of its covering groups and all groups
which may be obtained by factoring out a discrete center group.
\item $G$ should act transitively on ${\cal S}^{2n}$, i.e.\ if $s_1$ and
$s_2$ are any two points of ${\cal S}^{2n}$ then there exist a group element
$g_{1\to 2} \in G$ such that $s_2 = g_{1\to 2}\cdot s_1\,.$ 

Transitivity of
the group action means that the group $G$ can map any given ``state'' to any
other ``state'' of the symplectic space, i.e.\ it takes the global structure
of the ``phase space'' into account! 

The transitivity requirement of the action in general will imply that
 the dimension $r$ of the group $G$ is larger than the dimension $2n$
of the space ${\cal S}^{2n}$. This is certainly so if the latter may be
described as a homogenous space $G/H$, where $H \subset G$ is an appropriate
subgroup of $G$.
\item 
 As the transformations \ref{eq:350}
  leave the symplectic form $\omega$ invariant its Lie-derivatives
 $L_{\tilde{A}}$ have the
  property 
  \begin{equation}
    \label{eq:352}
    L_{\tilde{A}}\omega =  0\,,
  \end{equation}
 which -- together with $d\omega=0$ and according to the relation \ref{eq:337}
 -- implies 
  \begin{equation}
    \label{eq:353}
d\,(i_{\tilde{A}}\omega)=0~.
   \end{equation}
The last
  relation means that $i_{\tilde{A}}\omega$ is a closed 1-form on
 ${\cal S}^{2n}$.
  The corresponding vector fields $\tilde{A}$ are called
 ``locally Hamiltonian''. 

  According to Poincar\'{e}'s famous lemma one has locally
  \begin{equation}
    \label{eq:354}
    i_{\tilde{A}}\omega(s) = dh_A(s)\,,
  \end{equation}
  where $h_A(s)$ is some function. 

Under certain conditions (the first cohomology group
  $H^1(\mbox{${\cal S}^{2n}$})$ has to vanishes)  $i_{\tilde{A}}\omega$ is
  even  exact and we have a
  globally defined Hamiltonian vector field, i.e.\ we have
  \begin{equation}
    \label{eq:355}
    \tilde{A}(s) = -X_{h_A}(s)~ \forall s \in {\cal S}^{2n}~.
  \end{equation}

  If the Lie algebra element  $A$ can be written as the commutator of two other
  ones, $A=[A_1,A_2]$, then, because of $i_{[\tilde{A}_1, \tilde{A}_2]}\omega =
  d(i_{\tilde{A}_1}i_{\tilde{A}_2}\omega)$, $\tilde{A}$ is globally
 Hamiltonian. This is so
   for semisimple transformation groups $G$, a case
we are mainly interested in in this paper. \end{enumerate}
We now come to the central part of the group theoretical 
quantization program: 

Up to now we have established \begin{enumerate} \item
an isomorphism between the 
$r$-dimensional Lie algebra $\mathfrak{g}$ and a corresponding 
$r$-dimensional Lie subalgebra $\tilde{\mathfrak{g}}$ of Hamiltonian
 vector fields on
${\cal S}^{2n}$ and \item a homomorphism \ref{eq:346} of functions $f(s)$
 on ${\cal S}^{2n}$
into the Lie algebra of Hamiltonian vector fields, with the constant
functions as kernel. \end{enumerate}  What we are aiming at
is the following: In general one will select some special functions $h_j(s),
\,j=1,2,\ldots$, as basic ``observables'' associated with the given symplectic
space ${\cal S}^{2n}$, in such a way that these function form a Lie algebra
with respect to the Poisson brackets \ref{eq:324}. In the conventional
case these special functions are the canonical variables $q^j,p_j,\,j=1,\ldots,
n$ and the number $1 \in \mathbb{R}$. In the well-known quantization procedure
described in Sec.\ 1.3.1 these special functions become 
self-adjoint operators representing the 
generating Lie
algebra of the Weyl-Heisenberg group. 

 We are interested in the
following generalizations: \begin{enumerate} \item We want to introduce
an appropriate set of basic functions $h_{A_{\rho}},\rho=1,\ldots,r$, 
forming a Poisson Lie algebra which is isomorphic to the Lie algebra
$\mathfrak{g}$ with basis $\{A_{\rho},\rho=1,\ldots,r\}$:
\begin{eqnarray}
  \label{eq:1651}\label{isoAh1}
  A_{\rho} \mapsto \tilde{A}_{\rho}& = &-X_{h_{A_{\rho}}} \leftrightarrow
 h_{A_{\rho}}\,,\\ \label{isoAh2}
\{h_{A_{\rho}},h_{A_{\sigma}}\} &=& h_{[A_{\rho},A_{\sigma}]}\,,
~\rho,\sigma =1,\ldots,r\,.
\end{eqnarray}
 The relations \ref{isoAh1} and \ref{isoAh2} are by no means trivial
and cannot always be satisfied. The deeper reason is that the functions
$h_A$ and $h_A + const.$ generate the same Hamiltonian vector field 
$X_{h_A}$. This
implies that the relation \ref{isoAh2} may acquire an additional constant
$c(A_{\rho},A_{\sigma})$ on the r.h.\ side which cannot be made to vanish! 
Such complications occur for the group theoretical quantization approach 
to the conventional quantization procedure 
(for details see Refs.\ \cite{is1,gui}) I shall not discuss these important
 features here, because they do not occur for the (simple)
 groups we are
dealing with in the present paper: $SU(1,1) \cong Sp(2,\mathbb{R})$ and
$Sp(4,\mathbb{R})$. 

In modern symplectic differential geometry the existence of the isomorphism
\ref{isoAh1} and \ref{isoAh2} is closely related to properties of the
so-called ``momentum map'' \cite{sour,guil2,mar2a,cush4}; for a recent
historical review  see \cite{mar3}).

If the relations \ref{isoAh1} and \ref{isoAh2} do hold, then one calls
the $r$-dimensional group $G$ the ``canonical group'' of the symplectic
space ${\cal S}^{2n}$.
 \item Having established the above isomorphism between the Lie algebra
  $\mathfrak{g}$
 and a corresponding Poisson Lie algebra of a system $\{h_A\}$ of preferred
 observables on ${\cal S}^{2n} $, one then can quantize the classical system
 by using the irreducible unitary representations of the transformation
 group $G$ where the self-adjoint generators $K_{\rho}(A_{\rho})$ of the
  unitary 1-parameter subgroups
  \begin{equation}
    \label{eq:357}
    U[g_{\rho}(t)=\exp(-A_{\rho}t)]=\exp[-iK_{\rho}(A_{\rho})t]\,,~
\rho=1,\ldots,r,
  \end{equation}
 represent the corresponding original classical observables $h_{A_{\rho}}$.
 \item As there may be different groups with symplectic, transitive
 and effective action on ${\cal S}^{2n}$,
 one has to make a choice which one to use.

 Here physical considerations come into play: One wants a group such that
 the corresponding observables $h_{A_{\rho}}(s)$
 constitute basic functions on $ {\cal
 S}^{2n}$, so that all  physically interesting observables can be expressed
 by them. For additional discussions of these problems see Refs.\
 \cite{is1,lo,bo2} \end{enumerate} 
\section{The canonical group $SO^{\uparrow}(1,2)$ of the symplectic
space ${\cal S}^2_{\vp,I} =\{ \vp \in \mathbb{R} \bmod{2\pi}\,,I >0\}$}
I now want to apply the general remarks of the last section to the concrete
phase space
\begin{equation}
  \label{eq:356}
 {\cal S}^2_{\vp,I} =\{ \vp \in \mathbb{R} \bmod{2\pi}\,,I >0\}\,, 
\end{equation}
with the Poisson brackets 
\begin{equation}
  \label{eq:585}
  \{k_1,k_2\}= \partial_{\vp}k_1\partial_I k_2 -\partial_I k_1 
\partial_{\vp}k_2\,.
\end{equation}
As the definition of the space \ref{eq:356} means that the origin $\{0\}$
 of the underlying plane $\mathbb{R}^2$ is deleted, one
cannot use the $2$-dimensional translations as the canonical quantizing
group because it cannot avoid the origin as the result of special elements
of the group! One therefore has to find another appropriate group which has
all the desired properties listed in the last section. 

The appropriate canonical group for the phase space \ref{eq:356} is
 the proper orthochronous
 Lorentz group $SO^{\uparrow}(1,2)$  which leaves the
quadratic form 
\begin{equation}
  \label{eq:358}
 (x^0)^2-(x^1)^2 -(x^2)^2\;,~ x^0>0\,, 
\end{equation}
 invariant, has determinant $+1$ and also leaves the time direction unchanged
 \cite{bo,lo}. 

The first reason for the choice of the group $SO^{\uparrow}(1,2)$ is that the
  cone 
  \begin{equation}
    \label{eq:359}
   (x^0)^2-(x^1)^2 -(x^2)^2 =0\;,~ x^0>0\,,  
  \end{equation}
is homeomorphic (and diffeomorphic) to the space \ref{eq:356}.
 In order to see this put 
\begin{equation}
  \label{eq:766}
  x^0=I>0\,,
~x^1=I\cos\varphi\,,~ x^2= -I\sin\varphi\,, 
\end{equation}
which provides a smooth parametrization of that space: any given triple
 $(x^0,x^1,x^2)$ of Eqs.\ \ref{eq:766} uniquely determines $I >0$ and
$\vp \in (-\pi,\pi]$!

 In the following it is
advantageous to employ the twofold covering group $SU(1,1)$ of
$SO^{\uparrow}(1,2)$ (see  Appendix B) the elements $g_0$ of
 which are given by
 \begin{equation}
   \label{eq:360}
 g_0= \left( \begin{array}{ll} \alpha & \beta \\
 \bar{\beta} & \bar{\alpha}
\end{array} \right)\,,~~ \det{g_0} =|\alpha|^2-|\beta|^2=1\,.  
 \end{equation}
 If we define the
matrix 
\begin{equation}
  \label{eq:361}
   X =
    \left( \begin{array}{cc} x^0 & x^1-i\,x^2 \\ x^1+i\,x^2 & x^0
\end{array} \right)\,,~~ \mbox{det}X= (x^0)^2- (x^1)^2-(x^2)^2\,,
\end{equation}
the transformations $x^{\mu}\rightarrow \hat{x}^{\mu},~ \mu
=0,1,2,$ under $SO^{\uparrow}(1,2)$ are implemented by
\begin{equation}
  \label{eq:362}
  X \rightarrow \hat{X}= g_0\cdot X\cdot g_0^+\,,~~ \det\hat{X} =
\det X\,, 
\end{equation}
  where $g_0^+$ denotes the hermitian conjugate of
the matrix $g_0$.

 Applying a general  $g_0$ to the
matrix 
\begin{equation}
  \label{eq:363}
X= \left(
\begin{array}{cc} I &  I\,e^{\ds -i\varphi} \\  I\,e^{\ds
i\varphi}
 & I
\end{array} \right)~ 
\end{equation}
  yields the mapping:, 
 \begin{eqnarray}
 (I,\varphi)
&\rightarrow& (\hat{I}, \hat{\varphi})\,: \label{eq:804} \\
 \hat{I}&=& |\alpha +e^{\ds i\varphi}\, \beta|^2 \; I~, \label{g0I_{12}}  \\
 \label{eq:1424}e^{\ds i\hat{\varphi}}&=& \frac{\bar{\alpha}\,e^{\ds i\varphi} 
 + \bar{\beta}}{\alpha+e^{\ds i\varphi}\;\beta}~. \label{g0vp}\end{eqnarray}
As 
\begin{equation}
  \label{eq:364}
 \frac{\partial \hat{\varphi}}{\partial \varphi} = |\alpha +
e^{\ds i \varphi} \beta |^{-2}  ~, 
\end{equation}
 we have the equality
 \begin{equation}
   \label{eq:365}
  d\hat{\varphi}\wedge d\hat{I} = d\varphi \wedge dI~, 
 \end{equation}
  that
is, the transformations \ref{g0I_{12}} and \ref{g0vp} are symplectic.

 One sees
immediately that $g_0$ and $-g_0$ lead to the same transformations
of $I$ and $\varphi$. Thus, the group $SU(1,1)$ acts on on the space
\ref{eq:356}
only almost effectively with the kernel $Z_2$ representing the
center of the twofold covering group $SU(1,1)$ of $SO^{\uparrow}(1,2)$. It
is well-known that the latter group acts effectively and
transitively on the forward light cone and thus on \ref{eq:356} (see
also the remarks below after Eq.\ \ref{Nvps}).

 For later we need the actions of the
1-parametric subgroups $\mbox{R}_0,\mbox{A}_0$ and $\mbox{N}_0$
which form the Iwasawa decomposition $SU(1,1)=\mbox{R}_0\cdot
\mbox{A}_0 \cdot \mbox{N}_0~$ (see the Eqs.\ \ref{ka0}-\ref{n0}),
 with the general element
\begin{eqnarray}  \label{eq:1653}r_0\cdot a_0 \cdot n_0& =& \left(
\begin{array}{cc} e^{ i\theta/2} & 0 \\ 0 & e^{ -i\theta/2}
\end{array} \right)
\cdot
\left( \begin{array}{cc} \cosh(t/2) &  i\, \sinh(t/2) \\ -i\, \sinh(t/2)
& \cosh(t/2)
\end{array} \right)\nonumber
  \\  \label{eq:1427}&& \cdot \left( \begin{array}{cc} 1+i\xi/2 &  \xi/2 \\
\xi/2 & 1-i\,\xi/2
\end{array} \right)~,  \end{eqnarray}
where $\theta \in (-2\pi, +2\pi];~ t,\xi \in \mathbb{R}$.
According to \ref{eq:362}
 the actions
of the 1-parameter subgroups $\mbox{R}_0,\mbox{A}_0,\mbox{N}_0$, respectively,
are:
 \begin{eqnarray} \label{eq:1654}\mbox{R}_0:~&& \hat{I}=I~,\label{RI_{12}}\\
&&  e^{ i\hat{\varphi}}= e^{ i(\varphi-
\theta)}~.  \label{Rvp} \\
   \label{eq:1430}\mbox{A}_0:~ && \hat{I}= \rho(t,\varphi)\,I~,~~
\rho(t,\varphi)=\cosh t-\sinh t\,\sin\varphi~, \label{AI_{12}} \\
&&\cos\hat{\varphi}=
\cos\varphi/\rho(t,\varphi)~, \label{Avpc}\\ \label{eq:1432}
&&\sin\hat{\varphi}=(\cosh
t\sin\varphi - \sinh t) /\rho(t,\varphi)~.\label{Avps} \\
\label{eq:1433}\mbox{N}_0:~ && \hat{I}=
\rho(\xi,\varphi)\,I~,~\rho(\xi,\varphi)=1+\xi \cos\varphi +
\xi^2(1+\sin\varphi)/2~,~~~~ \label{NI_{12}} \\ \label{eq:1434}
&&\cos\hat{\varphi}=
[\cos\varphi+\xi(1+\sin\varphi)]/\rho(\xi,\varphi)~, \label{Nvpc} \\
\label{eq:1435}&&\sin\hat{\varphi}= [\sin\varphi-\xi \cos\varphi -
\xi^2(1+\sin\varphi)/2]/\rho(\xi,\varphi)~. \label{Nvps}
\end{eqnarray}

 Transitivity of the $SU(1,1)$ group action on \ref{eq:356}
can be seen as follows: 

 Any point $s_1=(\varphi_1,I_1)$ may be transformed into any
other point $s_2=(\varphi_2, I_2)$: first
transform $(\varphi_1,I_1)$ into $(0,I_1)$ by
$r_0(\theta=\varphi_1)$, then map this point into $(\varphi_0
= -\arctan(\sinh t_0\,),I_2)$ by $a_0(t_0;\,
\cosh t_0=I_2/I_1)$ and finally transform $(\varphi_0,I_2)$
by $r_0(\theta = \varphi_0-\vp_2)$ into
$s_2=(\varphi_2,I_2)$. 

 These transitivity properties reflect the
fact that any element $g_0$ of $SU(1,1)$ may be written as
$r_0(\theta_2)\cdot a_0(t) \cdot r_0(\theta_1)$ (see \ref{eq:497}).

The transformation formulae \ref{Nvpc} and \ref{Nvps} show that
 the group $\mbox{N}_0$
leaves the half-line $\varphi = -\pi/2,\, I>0,$ invariant, that is,
$\mbox{N}_0$ is the stability group of these points. This means
that the symplectic space \ref{eq:356} is diffeomorphic to the coset space
$ SU(1,1)/(Z_2 \times \mbox{N}_0) \simeq
SO^{\uparrow}(1,2)/\mbox{N}_0$. Notice that $\mbox{N}_0$, and
$\mbox{A}_0$ as well, does not contain the second center element
$-e$ of $SU(1,1)$. The center $Z_2$ is a subgroup of
$\mbox{R}_0$. 

We also give the action of the group $B_0$,(see \ref{be0}), on the points
$s=(\vp,I)$: \begin{eqnarray} \label{eq:1655}B_0:~&&\hat{I}=\rho(s,\vp)
\,I\,,~ \rho(s,\vp)
= \cosh t + \sinh t\,\cos \vp\,,\label{BI_{12}} \\ \label{eq:1439}&&
 \cos\hat{\vp} =
 (\cosh s \cos \vp +\sinh s)/\rho(s,\vp)\,, \label{Bvpc}\\
\label{eq:1440}&& \sin\hat{\vp} = \sin\vp/\rho(s,\vp)\,. \label{Bspc}
 \end{eqnarray}

  We next determine the Hamiltonian vector fields
  induced  on \ref{eq:356} by the above $SU(1,1)$ transformations and 
-- most important -- the
  corresponding classical observables $h_R,h_A$ and $h_B$. 

 For infinitesimal values of the
  parameters $\theta,\, t\,$ and $ s $ the transformations
 \ref{RI_{12}}--\ref{Avps}
and \ref{BI_{12}}--\ref{Bspc}
 take the
  form
 \begin{eqnarray}\label{eq:1656}\mbox{R}~:&& \delta \varphi =
 -\theta,~|\theta|\ll 1,~~~ \delta I
 =0~,\\ \label{eq:1441}\mbox{A}~:&& \delta \varphi =-(\cos\varphi)\,t,
~~\delta I =
 -I\,(\sin\varphi)\,t,~~
 |t|\ll 1~, \\ \label{eq:1442}\mbox{B}~:&& \delta\varphi=-(\sin\varphi)
\,s,~~\delta I =
 I\,(\cos\varphi)\,s,~~|s|\ll 1~~. \end{eqnarray} According to 
\ref{eq:351} they induce on 
\ref{eq:356}
 the vector fields \begin{eqnarray} \label{eq:1657}\tilde{A}_R &=&
 \partial_{\varphi}\,,
\label{tAR}\\
 \label{eq:1443}\tilde{A}_A &=&
 \cos\varphi\,\partial_{\varphi}+I\,\sin\varphi\,\partial_I\,,
\label{tAA} \\
 \label{eq:1444}\tilde{A}_B& =& \sin\varphi\,
 \partial_{\varphi}-I\,\cos\varphi\,\partial_I\,. \label{tAB}\end{eqnarray}

  It is easy to check that their Lie algebra is isomorphic to the
 Lie algebra of $SO^{\uparrow}(1,2)$ (see Sec.\ B.2 of Appendix B),
 and all its covering groups, of course. 

According to the general relations \ref{eq:343}, \ref{eq:346} and \ref{eq:355}
we get the following  relations (recall that $X_f= \partial_If\,
\partial_{\vp} - \partial_{\vp}f\,\partial_I$)
\begin{eqnarray} \label{eq:1658}\tilde{A}_R &=& -X_{f_R}\,,~f_R(\vp,I)
 = -I\,, \label{efR}\\
\label{eq:1445}\tilde{A}_A &=& -X_{f_A}\,,~f_A(\vp,I) = -I\,\cos\vp
\,,\label{efA}\\
\label{eq:1446}\tilde{A}_B &=& -X_{f_B}\,,~f_B(\vp,I) = -I\,\sin\vp\,.
 \label{efB}
\end{eqnarray}

 The functions $f_R,f_A$ and $f_B$ obey the Lie algebra
$\mathfrak{so}(1,2)$ with respect to the Poisson brackets \ref{eq:585}:
\begin{equation}
  \label{eq:604}
 \{f_R,f_A\} =-f_B\,,~\{f_R,f_B\} = f_A\,,~\{f_A,f_B\}=f_R\,. 
\end{equation}

In order to avoid two of the minus signs we finally define
as our three  basic classical observables the functions
\begin{equation}
  \label{eq:367}
 h_0(\vp,I) \equiv -f_R =I\,,~ h_1(\vp,I) \equiv -f_A=
I\,\cos\varphi\,,~h_2(\vp,I) \equiv f_B = - I\,\sin\varphi\,.  
\end{equation}
Their Poisson brackets 
\begin{equation}
  \label{eq:586}
  \{h_0,h_1\}_{\vp,I} =-h_2\,,~ \{h_0,h_2\}_{\vp,I} =h_1\,,~
 \{h_1,h_2\}_{\vp,I} =h_0\,,
\end{equation} again form the Lie algebra $\mathfrak{so}(1,2)$. 

 The
Eqs.\ \ref{eq:367} constitute one of our principal results: 

 The canonical
group $SO^{\uparrow}(1,2)$ of the symplectic space \ref{eq:356} {\em 
determines}
the basic ``observables'' \ref{eq:367} of that classical space. 

 The functions \ref{eq:367} are indeed
suitable in order to fulfill the desired purposes: 

  Any smooth
function $f(\varphi,I)$ periodic in $\varphi$ with period $2\pi$
can, under quite general  conditions, be
 expanded in a Fourier series and as $\sin(n\varphi)$ and
$\cos(n\varphi)$ can be expressed as polynomials of $n$-th order
in $\sin\varphi =-h_2/I$ and $\cos\varphi=h_1/I$, the observables
 \ref{eq:367} are
indeed sufficient. 

 Actually the functions \ref{eq:367} are just the cone
 coordinates \ref{eq:363}   we started  from! We merely have
to identify
\begin{equation}
  \label{eq:756}
 I=h_0\,,~~ I\,e^{-i\vp} =h_1 +i\,h_2\,.
\end{equation}
The transformations \ref{eq:362} imply that $(h_0,h_1,h_2)$ transforms
as a 3-vector with repect to the group $SO^{\uparrow}(1,2)$. The explicit
transformation formulae for the three subgroups \ref{ka0}, \ref{a0} and
\ref{be0} are
\begin{eqnarray}
  \label{eq:757}
  R_0:~~~  h_0 &\to& \hat{h}_0 = h_0\,, \\
h_1 &\to& \hat{h}_1 = \cos\theta\,h_1 - \sin\theta\,h_2\,,\label{eq:758} \\
h_2 &\to& \hat{h}_2 = \sin\theta\,h_1 + \cos\theta\,h_2\,,\label{eq:759} \\
&& \nonumber \\
A_0:~~~ h_0 &\to & \hat{h}_0 = \cosh t\,h_0 + \sinh t\,h_2\,, \label{eq:760} \\
h_1 &\to& \hat{h}_1 =h_1\,, \label{eq:761} \\
 h_2 &\to & \hat{h}_2 = \sinh t\,h_0 + \cosh t\,h_2\,, \label{eq:762} \\
&& \nonumber \\
B_0:~~~  h_0 &\to & \hat{h}_0 = \cosh s\,h_0 + \sinh s\,h_1\,,\label{eq:763} \\
 h_1 &\to & \hat{h}_1 = \sinh s\,h_0 + \cosh s\,h_1\,, \label{eq:764} \\
h_2 &\to& \hat{h}_2 =h_2\,. \label{eq:765}
\end{eqnarray}
So we have  rotations in the $h_1-h_2$ plane and two Lorentz ``boosts'',
one in the $h_0-h_2$ plane and the other in the $h_0-h_1$ plane!
All transformations leave the form $ h_0^2-h_1^2-h_2^2 $ invariant.

In addition we know from the general result \ref{eq:365} that
 these transformations
are symplectic transformations of the phase space \ref{eq:356}.

After all the efforts the quantization of the phase space \ref{eq:356}
 is now straightforward:

The irreducible unitary representations of the group $SO^{\uparrow}(1,2)$ and
its covering groups are well-known (see Appendix B). Their  
1-parameter unitary subgroups are generated by self-adjoint generators
$K_j,\,j=0,1,2,$ corresponding to the three observables \ref{eq:367}:
\begin{equation}
  \label{eq:368}
  h_0 \to K_0~,~h_1 \to K_1~,~h_2 \to K_2~.
\end{equation}

Because $h_0 \equiv I > 0$ the quantized theory has to use the positive
discrete series of the irreducible unitary representations (Appendix B).
\section{The symplectic space $ {\cal S}^2_{\vp,I}$ and the \protect\\ orbifold
$\mathbb{R}^2/Z_2$} We now come to a very interesting relationship between the
symplectic space \ref{eq:356} and the original phase
space 
\begin{equation}
  \label{eq:587}
  {\cal S}^2_{q,p} = \{ x=\begin{pmatrix}q \\p \end{pmatrix} \in \mathbb{R}^2
\}\,,
\end{equation} on which the symplectic group $Sp(2,\mathbb{R})$, \ref{eq:375},
 acts as
\begin{equation}
  \label{eq:588}
  x \to \hat{x} = g_1 \cdot x\,,~g_1 \in Sp(2,\mathbb{R})\,,
\end{equation}
with the property 
\begin{equation}
  \label{eq:589}
  d\hat{q} \wedge d\hat{p} = dq \wedge dp \,.
\end{equation} The group action \ref{eq:588} has some intriguing other
properties: 

The whole group transforms the point $x=0$ into itself and acts
 transitively
on the complement 
\begin{equation}
  \label{eq:591}
 {\cal S}^2_{q,p;0} \equiv {\cal S}^2_{q,p} -\{x=0\}.
\end{equation}
 It also acts effectively on
the latter because the second element $-e$ of its  center $Z_2 = \{e,-e\}$,
where 
\begin{equation}
  \label{eq:590}
  e = E_2 \equiv \begin{pmatrix} 1&0\\0&1 \end{pmatrix}\,,
\end{equation}
acts non-trivially on  ${\cal S}^2_{q,p;0}$:
\begin{equation}
  \label{eq:592}
  (-e)\cdot x =-x \neq x\,.
\end{equation} This is in obvious contrast to the action \ref{eq:362} 
of the group
$SU(1,1) \cong Sp(2,\mathbb{R})$ on the space \ref{eq:356} for
 the points $s$ of which
one has 
\begin{equation}
  \label{eq:593}
  (-e)\cdot s= s\,,~e =E_2 \in SU(1,1)\,,
\end{equation}
as the Eqs.\ \ref{g0I_{12}} and \ref{g0vp} show. 

 How can this be reconciled
especially in view of the fact that {\em locally}
\begin{equation}
  \label{eq:594}
  d\vp \wedge dI =dq \wedge dp\,?
\end{equation}
Recall also that the space \ref{eq:356} is diffeomorphic to a cone with the
tip (vertex) deleted, but that the space \ref{eq:587}
 is {\em globally} different! 

The neat reconciliation of this apparent difficulty is the following: 

The mapping \ref{eq:588} has the same property \ref{eq:593} of the mapping
\ref{g0I_{12}},\ref{g0vp} if we identify the points $-x$ and $x$ of the
space \ref{eq:587}, i.e.\
if we pass from the space \ref{eq:587} to the quotient space
\begin{equation}
  \label{eq:595}
  {\cal S}^2_{q,p}/Z_2 \equiv \check{ {\cal S}}^2_{q,p} =\{\check{s} = 
\pm x\,,\,,x \in \mathbb{R}^2 \} \cong 
\mathbb{R}^2/Z_2\,.
\end{equation} 
Such a space is called an ``orbifold'' \cite{cush3}.
 
An orbifold may be generated
from a manifold $\mathtt{M}$ by identifying points which are connected by
a finite discontinuous group $D_n$ of $n$ elements so that the orbifold
is given by the quotient space $\mathtt{M}/D_n$.

 An orbifold generally has
additional singularities as compared to the mani\-fold from which it is
constructed, as we shall see now: 

In our case the orbifold \ref{eq:595} is a cone: Take the lower half
of the $(q,p)$-plane and rotate it around the $q$-axis till it coincides
with the upper half of the plane such that the negative $p$-axis lies
on the positive one. Then rotate the left half of the upper half plane
around the positive $p$-axis till the negative $q$-axis coincides with
the positive one. Finally glue the two $q$-half-axis together. The
resulting space is a cone with its ``tip'' (vertex) at $x=0$.
 (See, e.g.\ Fig. 1 in Ref.\ \cite{sha1}.)

 We thus arrive at the cone structure for the symplectic
space \ref{eq:356} by a differerent route and the quantization of that
space appears to be equivalent to the quantization of the orbifold
\ref{eq:595}, with the vertex deleted\,! 

Next let us see which vector fields are induced on \ref{eq:587}
by the groups \ref{ka1}, \ref{a1} and \ref{be1} and which
are the associated Hamiltonian functions. 

The same procedure as in the
previous section yields
\begin{eqnarray} \label{eq:1659}\tilde{A}_{R_1} &=&\frac{1}{2}(q\,
\partial_p-p\,\partial_q)\,,
\label{vk1}\\
\label{eq:1447}\tilde{A}_{A_1} &=& -\frac{1}{2}(q\,\partial_q +p\,
\partial_p)\,,
 \label{va1} \\
\label{eq:1448}\tilde{A}_{B_1} &=& -\frac{1}{2}(p\,\partial_q +q\,
\partial_p)\,, \label{vb1}
\end{eqnarray}
and the corresponding Hamiltonian functions (\ref{eq:343} and \ref{eq:344}) are
\begin{eqnarray} \label{eq:1660}\check{g}_0(q,p) &=&
 \frac{1}{4}(q^2+p^2)\,,\label{ch0} \\
\label{eq:1449}\check{g}_1(q,p) &=& -\frac{1}{2}\,q\,p\,, \label{ch1} \\
\label{eq:1450}\check{g}_2(q,p)&=& \frac{1}{4}(q^2-p^2)\,. \label{ch2}
\end{eqnarray} Their Poisson brackets again obey the Lie algebra $\mathfrak{
so}(1,2)$:
\begin{equation}
  \label{eq:596}
  \{\check{g}_0,\check{g}_1 \}_{q,p} = -\check{g}_2\,,~
 \{\check{g}_0,\check{g}_2 \}_{q,p} = \check{g}_1\,,~
 \{\check{g}_1,\check{g}_2 \}_{q,p} = \check{g}_0\,.
\end{equation} Notice that
\begin{equation}
  \label{eq:599}
  \check{g}_0^2-\check{g}_1^2 -\check{g}_2^2 =0\,.
\end{equation}
Inserting
\begin{equation}
  \label{eq:598}
  q = \sqrt{2I}\cos\vp\,,~p =-\sqrt{2I} \sin\vp\,,
\end{equation} into the expressions \ref{ch0}-\ref{ch2}
we get another set of functions $\check{h}_j(\vp,I)\,,j=0,1,2,$ which
obey the Lie algebra $\mathfrak{so}(1,2)$ with respect to the Poisson
brackets $\{\cdot,\cdot\}_{\vp,I}$:
\begin{eqnarray} \label{eq:1661}\check{h}_0(\vp,I) &=& \frac{1}{2}I\,,
\label{ch01} \\
\label{eq:1451}\check{h}_1(\vp,I) &=& \frac{1}{2}I \sin(2\vp)\,, 
\label{ch11} \\
\label{eq:1452}\check{h}_2(\vp,I)&=& \frac{1}{2}I\cos(2\vp)\,.
 \label{ch21}
\end{eqnarray}

We observe that the vector fields \ref{vk1}-\ref{vb1} are non-trivial
 only for $(q,p) \neq (0,0)$
and that the origin (0,0) here has to be excluded, too! 

As the vector fields \ref{vk1}-\ref{vb1} and the Hamiltonian functions
\ref{ch0}-\ref{ch2} are invariant against the substitution $(q,p) \to
-(q,p)$, they are defined on \ref{eq:587} and on the orbifold \ref{eq:595}
as well. 

 The identification of the points $(q,p)$ and $(-q,-p)$ implies
the identification of $\vp$ and $\vp \pm \pi$ for
 the angle $\vp$ in \ref{eq:598}.
It leaves the functions \ref{ch01}-\ref{ch21} invariant! 

The last point may also be disscused in terms of the complex amplitude
$a = \sqrt{I}\exp(-i\vp)$ from \ref{eq:578}: Identification of $a$ and $-a$
means identification of $a$ and $\exp(\pm i \pi)\,a$ and passing to the 
functions \ref{ch01}-\ref{ch21} is equivalent to
 passing to $I$ and $a^2$ or $\bar{a}^2$.

  The functions $\check{g}_j(q,p)\,,j=0,1,2,$ and $\check{h}_j(\vp,I)\,,
j=0,1,2,$ provide another parametrization of the cone (without its tip)
representing  the symplectic space \ref{eq:356} (with $\vp$ now $ \in (-\pi/2,
\pi/2]$). This parametrization is equivalent to that by
the functions \ref{eq:367} from above! 

 The functions $h_j(\vp,I)\,,\,
\check{g}_j(q,p)$ and $\check{h}_j(\vp,I)$, respectively, transform
 as 3-vectors
with respect to the group $SO^{\uparrow}(1,2)=Sp(2,\mathbb{R})/Z_2
 =SU(1,1)/Z_2$. This follows immediately from
the fact that they obey the Lie algebra $\mathfrak{so}(1,2)$, i.e.\ they
 transform according to the adjoint representation. These transformations
may be induced, e.g. by the action \ref{eq:588} of the group \ref{eq:375}.

Take the subgroup \ref{ka1} as an example: For the coordinates $x$, 
\ref{eq:587}, we have
\begin{equation}
  \label{eq:600}
  x \to \hat{x} = \begin{pmatrix}\hat{q}\\\hat{p} \end{pmatrix} =
\begin{pmatrix}\cos(\theta/2)\,q + \sin(\theta/2)\,p \\
-\sin(\theta/2)\,q + \cos(\theta/2)\,p \end{pmatrix}\,.
\end{equation}
These transformations induce the following mappings (rotations!) of the
functions \ref{ch0}-\ref{ch2}:
\begin{eqnarray}
  \label{eq:1662}\check{g}_0(q,p)&=& \frac{1}{2}I(q,p)  \to\frac{1}{2}
 I(q,p)\,, \\ 
  \label{eq:1454}\check{g}_1(q,p)& =&-\frac{1}{2}q\,p \to \cos\theta\,
\check{g}_1(q,p)
 +\sin\theta\,
\check{g}_2(q,p)\,, \\ \label{eq:1455}\check{g}_2(q,p)&=&\frac{1}{4}(q^2-p^2)
 \to -\sin\theta\,
\check{g}_1(q,p) + \cos\theta\,\check{g}_2(q,p) \,.
\end{eqnarray}

The transformations \ref{eq:600} and \ref{eq:1454}-\ref{eq:1455} illustrate
 the central message of this section very clearly: the canonical pair $x$
transforms as a ``spinor'' (see also
Sec.\ 6.3), whereas the pair $\check{g}_1, \check{g}_2$ transforms as a
vector as to the group  $SO^{\uparrow}(1,2)$!

If one chooses $\theta = 2\pi$, then we have the identity transformation
for \ref{eq:1454} and \ref{eq:1455}, but $x$ is replaced by $-x$ in 
\ref{eq:600}!
 \chapter{The group $SO^{\uparrow}(1,2)$, some
of its covering groups and their \protect  ir\-re\-ducible unitary
represen\-ta\-tions of the positive discrete series}
 In the present appendix I summarize
 properties of the group $SO^{\uparrow}(1,2)$, 
some of its covering groups, their Lie
algebra and their irreducible unitary representations, especially those of
the positive discrete series. These properties are important for the
group theoretical quantization procedures of the phase space \ref{eq:356}.

The following material is essentially  taken from  Appendix A and Ch.\ 
V of Ref.\ \cite{bo}. Practically all of it is contained
in a wealth of literature about the group $SO^{\uparrow}(1,2)$ 
which is the most elementary
of noncompact semisimple Lie groups. The readers of the present paper
will probably find it convenient to have the required properties assembled 
in an appendix here, too. 

The  essential classical paper on the group $SO^{\uparrow}(1,2)$
 and its irreducible unitary
 repesentations is (still!) that of Bargmann \cite{bar1}. In the
 meantime there are a number of monographs (and reviews) which deal with the
 group $SO^{\uparrow}(1,2)$, its covering groups and their representations
 \cite{gel0,sa1,wa,lan,hel2,kna,sug,vil,how,muk0}. As these
 textbooks contain many
 references to the original literature we mention only the most essential
 ones for our purposes. \section{The group and some covering groups}
\subsection{The groups $SU(1,1) \protect\cong SL(2,\protect\mathbb{R}) =
 Sp(2,\protect\mathbb{R})$}
In order to see the homomorphism between $SO^{\uparrow}(1,2)$ and
its isomorphic twofold covering groups $SU(1,1),SL(2,\mathbb{R})$ and
the symplectic group $Sp(2,\mathbb{R})$ in 2 dimensions, it is
convenient to start from the action of the group $SL(2,\mathbb{C})$
-- the twofold covering group of
 the proper orthochronous
Lorentz group $SO^{\uparrow}(1,3)$ --  on Minkow\-ski space $M^4$
 with the scalar product
$x\cdot x=(x^0)^2-(x^1)^2-(x^2)^2-(x^3)^2$:

 Define the hermitean
matrix 
\begin{equation}
  \label{eq:369}
  X= \left( \begin{array}{cc} x^0+x^3 & x^1-ix^2 \\
x^1+ix^2& x^0-x^3 \end{array} \right)= \sum_{j=0}^3 x^j\sigma_j\,,
\end{equation} where
\begin{equation}
  \label{eq:601}
  \sigma_0=\begin{pmatrix}1&0\\0&1\end{pmatrix}\,,\;
 \sigma_1=\begin{pmatrix}0&1\\1&0\end{pmatrix}\,,\;
 \sigma_2=\begin{pmatrix}0&-i\\i&0\end{pmatrix}\,,\;
 \sigma_3=\begin{pmatrix}1&0\\0&-1\end{pmatrix}\,,
\end{equation}
are the Pauli matrices. 

 As
\begin{equation}
  \label{eq:602}
 \mbox{det}X =
(x^0)^2-(x^1)^2-(x^2)^2-(x^3)^2\,,  
\end{equation}
the proper orthochronous Lorentz transformations 
\begin{equation}
  \label{eq:603}
  x^j \to \hat{x}^j= \sum_{k=0}^3\Lambda^j_{\;k}x^k\,,~\det (\Lambda^j_{\;k})
=1\,,~\Lambda^0_{\;0} > 0\,,
\end{equation} 
may be implemented as follows: 

 If $A \in SL(2,\mathbb{C}) \subset GL(2,\mathbb{C}),~
\det A=1,$ then  
\begin{equation}
  \label{eq:370}
 X \rightarrow \hat{X}=A\cdot X\cdot
A^+=\sum_{j=0}^3\hat{x}^j\,\sigma_j\,,~~ \mbox{det}\hat{X}= \mbox{det}X\,.  
\end{equation}
  Here $A^+$ means
the hermitean conjugate of the matrix $A$ and $\hat{X}$ the matrix
\ref{eq:369} with $x^j$ replaced by $\hat{x}^j$. 

 The well-known
properties of the Pauli matrices allow to express the parameters
$\Lambda^j_{\;k}$ in terms of the matrices $A$ as follows
\begin{equation}
  \label{eq:597}
  \Lambda^j_{\;k}(A) = \frac{1}{2}\,\text{tr}(\sigma_j\cdot A\cdot \sigma_k
\cdot A^+)\,.
\end{equation} 

 Those subgroups
$SO^{\uparrow}(1,2)$ which interest us here 
may be obtained by looking for the
transformations \ref{eq:370} which leave one of the coordinates
$x^j,~j=1,2$ or $3$ fixed: 

 The transformations with the
property 
\begin{equation}
  \label{eq:506}
  A\cdot \left( \begin{array}{cc} 1 & 0 \\ 0& -1
\end{array} \right)\cdot A^+= \left( \begin{array}{cc} 1 & 0 \\ 0&
-1 \end{array} \right) 
\end{equation}
 leave the coordinates $x^3$ invariant
and represent the subgroup $SU(1,1)=\{g_0\} \subset
SL(2,\mathbb{C})$: 
\begin{equation}
  \label{eq:371}
  g_0= \left( \begin{array}{cc} \alpha & \beta
\\ \bar{\beta}& \bar{\alpha} \end{array} \right)~,~ \det g_0 =
|\alpha|^2-|\beta|^2=1~. 
\end{equation}
  $\bar{\alpha}$: complex conjugate of
$\alpha$. The group elements $g_0$ act on a 2-dimensional complex vector
space $\mathbb{C}^2$ as 
\begin{equation}
  \label{eq:372}
  g_0 \cdot \left(
\begin{array}{c} z_1 \\ z_2 \end{array} \right)= \left(
\begin{array}{c} \hat{z}_1 \\ \hat{z}_2 \end{array}
\right)\,,~~\text{with}~|\hat{z}_1|^2 -|\hat{z}_2|^2 = |z_1|^2-|z_2|^2~.
\end{equation}
  If
 $|z_2|>|z_1|$ and $z=z_1/z_2$ then $SU(1,1)$ maps the interior
 \begin{equation}
   \label{eq:366}
 \mathbb{D}_1 =\{z;|z|<1\}  
 \end{equation}
 of the unit disc in the complex
$z$-plane (transitively) onto itself: 
\begin{equation}
  \label{eq:373}
 z \in \mathbb{D}_1 \rightarrow
\hat{z}=\frac{\alpha z + \beta}{ \bar{\beta} z +
\bar{\alpha}} \in \mathbb{D}_1~. 
\end{equation}
This property is important for the construction of the irreducible
unitary representations of $SU(1,1)$ (see Sec.\ B.3 below). 

 Notice that the group elments $g_0$ and $-g_0$ yield the same transformation
\ref{eq:373}. Thus, the transformation group is actually $SO^{\uparrow}(1,2)
=SU(1,1)/Z_2$ where $Z_2 =\{e,-e\}\,,\,e:$ unit group element, is the
(discrete) center of $SU(1,1)$. 

   The subgroup of
$SL(2,\mathbb{C})$ with the property 
\begin{equation}
  \label{eq:374}
  A\cdot \left(
\begin{array}{cc} 0 & -i \\ i & 0 \end{array} \right)\cdot A^+=
\left(
\begin{array}{cc} 0 & -i \\ i & 0 \end{array} \right) 
\end{equation}
  leaves
the coordinates $x^2$ invariant. It constitutes the group
$SL(2,\mathbb{R})=Sp(2,\mathbb{R})$: 
\begin{equation}
  \label{eq:375}
  C=g_1= \left( \begin{array}{cc} a_{11} & a_{12}
\\ a_{21} & a_{22} \end{array} \right)\,,~ a_{jk}\in
\mathbb{R}\,,~\det g_1=1\,. 
\end{equation}

  As 
  \begin{equation}
    \label{eq:376}
  g_1^T\cdot \left(
\begin{array}{cc} 0 & 1 \\ -1 & 0 \end{array} \right)\cdot
g_1= \left(
\begin{array}{cc} 0 & 1 \\ -1 & 0 \end{array} \right)\,,   
  \end{equation}
 the
group $SL(2,\mathbb{R})$ is identical with the {\em real symplectic group}
$Sp(2,\mathbb{R}) $ in 2 dimensions. \\ 
(I follow the  convention to denote the real symplectic
group of a $2n$-dimensio\-nal vector space by $Sp(2n,\mathbb{R})$. 
 In many  papers the
convention $Sp(n,\mathbb{R})$ is being used instead.
The number $n$
coincides with the rank of the group.) 

The unitary matrix 
\begin{equation}
  \label{eq:377}
 C_0 =
\frac{1}{\sqrt{2}}\left( \begin{array}{cc} 1 & -i \\ -i & 1
\end{array} \right)\,,~\mbox{det}C_0=1\,,~C_0^{-1}=
\frac{1}{\sqrt{2}}\left( \begin{array}{cc} 1 & i \\ i & 1
\end{array} \right)=C_0^+~, 
\end{equation}
   has the property 
   \begin{equation}
     \label{eq:378}
   C_0\cdot
\left( \begin{array}{cc} 0 & -i \\ i & 0 \end{array} \right)\cdot
C_0^{-1}=  \left( \begin{array}{cc} 1 & 0 \\ 0 & -1 \end{array}
\right)   
   \end{equation}
  and therefore implements an isomorphism between
$SU(1,1)$ and $SL(2,\mathbb{R})$: 
\begin{equation}
  \label{eq:379}
  C_0\cdot g_1\cdot C_0^{-1}=g_0~. 
\end{equation}

 It is obvious from \ref{eq:370} that the 
isomorphic groups $SU(1,1),SL(2,\mathbb{R})$
and $Sp(2,\mathbb{R})$ are twofold covering groups of
$SO^{\uparrow}(1,2)$, because $A$ and $-A$ induce the same
 Lorentz transformation!  

 The group $SL(2,\mathbb{R})=Sp(2,\mathbb{R})$
 maps the (``Siegel'') complex
upper half plane
\begin{equation}
  \label{eq:605}
  \mathbb{S}_1=\{z=x+iy,~y>0\}
\end{equation}
 transi\-tive\-ly
onto itself: 
\begin{equation}
  \label{eq:496}
  z \in \mathbb{S}_1 \rightarrow \hat{z}=\frac{a_{11}z
+a_{12}}{a_{21}z+a_{22}}~,~~
\Im(\hat{z})=\frac{y}{(a_{22}+a_{21}x)^2+ a_{21}^2y^2}\,>0. 
\end{equation}
This feature corresponds to the property \ref{eq:373} of the group
$SU(1,1)$.

 Again, the group elements $g_1$ and $-g_1$ give the same 
transformations \ref{eq:496}. They, too, are important for the explicit
 construction of irreducible
unitary representations (see   Sec.\ B.3). 

  Convenient for 
several of our purposes is the  {\em Iwasawa}
decomposition \cite{hel1,sug}
 of the groups
$G_1\equiv SL(2,\mathbb{R})=Sp(2,\mathbb{R})$ and $G_0\equiv SU(1,1)$: $
G_1\equiv
 R_1\cdot A_1\cdot N_1$,
$ G_0=R_0\cdot A_0\cdot N_0$, where R
is the maximal compact subgroup, A a maximally abelian noncompact
subgroup and N a nilpotent group.

 For $G_1$ this
decomposition is
\begin{eqnarray}\label{eq:1663}R_1: && r_1= \left(
\begin{array}{cc} \cos(\theta/2) & \sin(\theta/2) 
 \\ -\sin(\theta/2) &
\cos(\theta/2) \end{array} \right)\,,~ \theta \in (-2\pi,+2\pi]~, \label{ka1}
 \\ A_1: && a_1 = \left(
\begin{array}{cc} e^{\ds t/2} & 0 \\ 0 &
e^{\ds -t/2} \end{array} \right)\,,~t \in \mathbb{R}~, \label{a1} \\ N_1: &&
n_1 = \left(
\begin{array}{cc} 1 & \xi \\ 0 &
1 \end{array} \right)\,,~ \xi \in \mathbb{R}~~. \label{en1} \end{eqnarray} Each
element $g_1$ has a unique decomposition $g_1=k_1\cdot a_1 \cdot
n_1$. 

The isomorphism \ref{eq:379}  gives the corresponding decomposition
of $G_0$:
 \begin{eqnarray}\label{eq:1664}R_0: && r_0= \left(
\begin{array}{cc}  e^{\ds i\theta/2} & 0\\  0 &
e^{\ds -i\theta/2} \end{array} \right)\,,~ \theta \in
(-2\pi,+2\pi]~, \label{ka0}\\ A_0: && a_0 = \left(
\begin{array}{cc} \cosh(t/2) & i\sinh(t/2) \\ -i\sinh(t/2) &
\cosh(t/2) \end{array} \right)\,,~t \in \mathbb{R}~, \label{a0}\\ N_0: &&
n_0 = \left(
\begin{array}{cc} 1+i\xi/2 & \xi/2 \\ \xi/2 &
1-i\xi/2 \end{array} \right)\,,~ \xi \in \mathbb{R}~~. \label{n0}
\end{eqnarray} In addition to the above subgroups the following
two ones are of interest to us:
\begin{eqnarray} \label{eq:1665}\mbox{B}_1: && b_1=
\left(
\begin{array}{cc} \cosh(s/2) & \sinh(s/2) \\ \sinh(s/2) &
\cosh(s/2) \end{array} \right)\,,~s \in \mathbb{R}~, \label{be1}
 \\ \mbox{B}_0: &&
b_0 =C_0\cdot b_1 \cdot C_0^{-1} = \left(
\begin{array}{cc} \cosh(s/2) & \sinh(s/2) \\ \sinh(s/2) &
\cosh(s/2)
 \end{array} \right)\,, \label{be0}\\ \bar{N}_1: && \bar{n}_1 =
\left(
\begin{array}{cc} 1 & 0 \\ \zeta &
1 \end{array} \right)\,,~ \zeta \in \mathbb{R}~, \label{baren1}
 \\ \bar{N}_0: &&
\bar{n}_0= \left(
\begin{array}{cc} 1-i\zeta/2 & \zeta/2 \\ \zeta/2 &
1+i\zeta/2 \end{array} \right)\,. \end{eqnarray} 

Two more
decompositions of $SL(2,\mathbb{R})$ or $SU(1,1)$ are important for the
construction of their unitary representations:

 {\em Cartan} (or ``polar'') decomposition \cite{hel1,sug}: 

 Each
element of $SL(2,\mathbb{R})=Sp(2,\mathbb{R})$ can be written as 
\begin{equation}
  \label{eq:497}
  g_1=  r_1(\theta_2)
\cdot a_1(t)\cdot r_1(\theta_1)~, 
\end{equation}
  where $a_1(t)$ is determined
uniquely and $r_1(\theta_1),r_1(\theta_2)$ up to a relative sign, that
is, up to the center $Z_2$ of $SL(2,\mathbb{R})$. 

 {\em Bruhat}
decomposition \cite{hel1,wa1,lan1}: 

 From \begin{gather}
  r_1(\theta)\cdot a_1(t)\cdot r_1(-\theta)= 
 \\  \label{eq:1462}= \left( \begin{array}{cc}
\cos^2(\theta/2)e^{t/2}+\sin^2(\theta/2)e^{-t/2} &
\sin(\theta/2)\cos(\theta/2)(e^{-t/2}-e^{t/2}) \\
\sin(\theta/2)\cos(\theta/2)(e^{-t/2}-e^{t/2}) &
\cos^2(\theta/2)e^{-t/2}+\sin^2(\theta/2)e^{t/2} \end{array} \right)
\nonumber  \end{gather}
one sees that \begin{eqnarray}
 \label{eq:1666}r_1(\theta)\cdot a_1(t)\cdot r_1(-\theta)&=& a_1(t)~ 
\mbox{for}~
\theta=0, 2\pi~, \\ \label{eq:1463}r_1(\theta)\cdot a_1(t)\cdot r_1(-\theta)& 
\subset
& A_1 ~\mbox{for}~ \theta=0,\pm \pi, 2\pi~, \end{eqnarray}
 which means that the centralizer
$C_{R_1}(A_1)$ and normalizer
$N_{R_1}(A_1)$ of $A_1$ in $R_1$ are
given by 
\begin{equation}
  \label{eq:498}
  C_{R_1}(A_1)=\left\{ \pm \left(
\begin{array}{cc} 1 & 0 \\ 0 & 1 \end{array} \right)
\right\}=\mathsf{M}~, 
\end{equation}
\begin{equation}
  \label{eq:499}
  N_{R_1}(A_1) =
\left\{ \pm \left(
\begin{array}{cc} 1 & 0 \\ 0 & 1
\end{array} \right),~
 \pm \left( \begin{array}{cc}0  & 1 \\ -1 & 0 \end{array}
\right) \right\}= \mathsf{M}^*~. 
\end{equation}
 The quotient group 
 \begin{equation}
   \label{eq:500}
   W=\mathsf{M}^*/\mathsf{M} \cong Z_2
 \end{equation}
   is called the {\em Weyl} group of $SL(2,\mathbb{R})$. The associated Bruhat
  decomposition of $SL(2,\mathbb{R})$ is \cite{war3}
  \begin{equation}
    \label{eq:501}
    G_1=\mathsf{M}\cdot A_1 \cdot N_1 \cup N_1\cdot
   w \cdot \mathsf{M} \cdot A_1
   \cdot N_1~,~ w=
 \left( \begin{array}{cc} 0  & -1 \\ 1 & 0 \end{array}
\right)~. 
  \end{equation}
    Here $\mathsf{M}\cdot A_1$ is the group 
    \begin{equation}
      \label{eq:502}
    D_1=\mathsf{M}\cdot
  A_1 = \left\{ \left(
  \begin{array}{cc} c &0 \\ 0 & c^{-1} \end{array} \right)~,~ c
 \in \mathbb{R} -\{0\}
  \right\}~.  
    \end{equation}
  The relation \ref{eq:501} means that each element of $SL(2,\mathbb{R})$
  is either  an element of the ``parabolic''
  subgroup $P_1 =D_1\cdot N_1$
   or 
  an element of $N_1\cdot w \cdot P_1$. 

The Bruhat decomposition of $SL(2,\mathbb{R})$ plays a central role in
Sally's construction \cite{sa1} of the irreducible unitary
representations of the universal covering group
$\widetilde{SL(2,\mathbb{R})}$. 
\subsection{The universal covering group of $SO^{\uparrow}(1,2)$}
 As the compact subgroups
$SO(2) \simeq S^1 \subset SO^{\uparrow}(1,2)$ and
$R_1$
 or $R_0\; \simeq S^1$ are  infinitely connected, the groups $SO^{\uparrow}
(1,2)\,,\,
 SL(2,\mathbb{R})$ and
  $SU(1,1)$
 have an infinitely sheeted universal covering group which, according to
 Bargmann \cite{bar1}, may be parametrized as follows: 

Starting from $SU(1,1)$ one defines
 \begin{eqnarray} \label{eq:1667}\gamma& =& \beta/\alpha~,~|\alpha|^2-
|\beta|^2=1~(
 \Rightarrow |\gamma|<1);~
  ~\omega = \arg(\alpha)~;
 \\ \label{eq:1464}\alpha & =& e^{\ds i\omega}(1-|\gamma|^2)^{-1/2}~,~
 |\gamma|<1~,~\beta =
e^{\ds i\omega}\gamma(1-|\gamma|^2)^{-1/2}~. \end{eqnarray} Then
\begin{eqnarray} \label{eq:1668}SU(1,1)& =& \{g_0=(\omega,\gamma),~\omega \in
 (-\pi, \pi],~|\gamma|<1 \}~,~ \\ \label{eq:1465}\tilde{G}\equiv 
\widetilde{SU(1,1)}
 &=&
 \widetilde{Sp(2,\mathbb{R})}=  \{\tilde{g}=(\omega,\gamma),~\omega
 \in \mathbb{R},~
 |\gamma|<1 \}~. \end{eqnarray} The group composition law for $\tilde{g}_3 =
 \tilde{g}_2 \cdot \tilde{g}_1$ is given by \begin{eqnarray} \label{eq:1669}
\gamma_3 & = &
 (\gamma_1 +\gamma_2 e^{\ds
 -2i\omega_1})(1+\bar{\gamma}_1\gamma_2e^{\ds -2i\omega_1})^{-1}~~, \\
 \label{eq:1466}\omega_3&=& \omega_1 + \omega_2 +\frac{1}{2i}\ln[(
1+\bar{\gamma}_1\gamma_2e^{\ds - 2i\omega_1})(
1+\gamma_1\bar{\gamma}_2e^{\ds 2i\omega_1})^{-1}]~. \end{eqnarray}

 For the universal covering
group  the transformations
 \ref{g0I_{12}} and \ref{g0vp} take the form
\begin{eqnarray}
 \label{eq:1670}\hat{I}&=& \rho(\tilde{g}, \varphi)\; I\,,~\rho
 (\tilde{g}, \varphi)=
 |1 +e^{\ds i\varphi}\, \gamma|^2\, (1-|\gamma|^2)^{-1}~,  \\
 \label{eq:1467}e^{\ds i\hat{\varphi}}&=& e^{\ds -2i\omega}\;
 \frac{e^{\ds i\varphi} +
  \bar{\gamma}}{1+e^{\ds i\varphi}\gamma}~. \end{eqnarray}
  As $\partial\hat{\varphi}/\partial\varphi= 1/\rho(\tilde{g},\varphi)$, the
  equality \ref{eq:365} holds again.

  With the elements of the group $SU(1,1)$  given by the restriction
  $ -\pi < \omega \leq +\pi,\; \alpha = \exp(i\omega)(1-|\gamma|^2)^{-1/2},\;
  \beta = \gamma\, \alpha~,$ the homomorphisms \begin{eqnarray}
   \label{eq:1671}h^{\#} &:&~~ \widetilde{SU(1,1)} \rightarrow SU(1,1)\cong
 Sp(2,\mathbb{R})\,, \\
   \label{eq:1468}h^0&:&~~ SU(1,1) \rightarrow SO^{\uparrow}(1,2)\,,
 \end{eqnarray}
  have the kernels
  $\mbox{ker}(h^{\#})=2\pi\,\mathbb{Z}\,,~\mbox{ker}(h^0)=Z_2$,
  respectively,
   and the composite
  homomorphism $h^0 \circ h^{\#}$ has the kernel $\pi \mathbb{Z}$.

  As the space ${\cal S}^{2}_{\vp,I}$, Eq.\ \ref{eq:356}, 
is homeomorphic to $\mathbb{R}^2-\{0\}
  =\mathbb{C}-\{0\}$,
  its universal covering space is given by $\varphi \in \mathbb{R},~ I \in
  \mathbb{R}^+$, which is
  the infinitely sheeted Riemann surface of the logarithm.

 The transformations
  \ref{eq:1670} and \ref{eq:1467} may be interpreted as
 acting transitively and effectively
  on that universal covering space.
\subsection{The group $Sp_c(2) \protect\cong Sp(2,\mathbb{R}) 
\protect\cong SU(1,1)$}
For the interpretation of the crucial role the groups $SO^{\uparrow}(1,2)$
etc. play for our approach to the quantization of the symplectic space
\ref{eq:356} the following isomorphic version of $Sp(2,\mathbb{R})$ is of
interest: 

 Consider  $x=(q,p)^T \in \mathbb{R}^2$. Then $g_1 \in Sp(2,
\mathbb{R})$ transforms $x$ as
\begin{equation}
  \label{eq:576}
  x \to \hat{x} =\begin{pmatrix}\hat{q}\\\hat{p} \end{pmatrix} = g_1\cdot x =
g_1\cdot \begin{pmatrix} q \\ p \end{pmatrix}\,,
\end{equation} with the property
\begin{equation}
  \label{eq:580}
  d\hat{q} \wedge d\hat{p} = d q \wedge d p \,.
\end{equation}
If we define 
\begin{equation}
  \label{eq:578}
  b = \begin{pmatrix}a = \frac{1}{\sqrt{2}}(q+ip)= |a|e^{-i\vp} \\
\bar{a} = \frac{1}{\sqrt{2}}(q-ip)= |a|e^{i\vp} \end{pmatrix}\,, 
\end{equation}
then
  \begin{equation}
    \label{eq:579}
     b = C_1\cdot x\,,~C_1 =\frac{1}{\sqrt{2}}\begin{pmatrix}1&i \\1&-i
\end{pmatrix}\,,~~C_1^{-1} =\frac{1}{\sqrt{2}}\begin{pmatrix}1&1 \\-i&i
\end{pmatrix}=C_1^+\,,
  \end{equation} with
  \begin{equation}
    \label{eq:582}
     dq \wedge dp = i\,da \wedge d\bar{a}\,.
  \end{equation}
The matrix $C_1$ is unitary with $\det C_1 = -i$. \\
The transformations \ref{eq:576} of $x$ imply for those of $b$:
\begin{equation}
  \label{eq:581}
  b \to \hat{b} =g_c \cdot b\,,~~g_c =C_1\cdot g_1\cdot C_1^{-1}
 \in Sp_c(2)\,.
\end{equation}
The group $Sp_c(2)$ is obviously isomorphic to the groups $Sp(2,\mathbb{R})$
and $SU(1,1)$. With respect to the latter we have
\begin{equation}
  \label{eq:583}
  g_c = C_2\cdot g_0\cdot C_2^{-1}\,,~C_2 = C_1\cdot C_0^{-1} = \begin{pmatrix}
0&i\\1&0 \end{pmatrix}\,.
\end{equation} The transformations \ref{eq:581} have the 
property
\begin{equation}
  \label{eq:584}
  d\hat{a} \wedge d\bar{\hat{a}} =da \wedge d \bar{a} \,.
\end{equation}
For the group $Sp_c(2)$ the subgroups \ref{ka1}-\ref{en1} and \ref{be1}
 have the form
 \begin{eqnarray}\label{eq:1672}R_c: && r_c= \left(
\begin{array}{cc}  e^{\ds -i\theta/2} & 0\\  0 &
e^{\ds i\theta/2} \end{array} \right)~,~ \theta \in
(-2\pi,+2\pi]~, \label{kac}\\ A_c: && a_c = \left(
\begin{array}{cc} \cosh(t/2) & \sinh(t/2) \\ \sinh(t/2) &
\cosh(t/2) \end{array} \right)\,,~t \in \mathbb{R}~, \label{ac}\\ N_c: &&
n_c = \left(
\begin{array}{cc} 1-i\xi/2 & i\xi/2 \\ \xi/2 &
1+i\xi/2 \end{array} \right)~,~ \xi \in \mathbb{R}~,\\ \label{nc}
 B_c: && b_c = \begin{pmatrix} \cosh(s/2)& i\,\sinh(s/2)\\
-i\,\sinh(s/2)& \cosh(s/2) \end{pmatrix}\,,~s \in \mathbb{R}\,. \label{bc}
\end{eqnarray} 
\section{Lie algebra} As the structure of the 3-dimensional
Lie algebra $\mathfrak{so}(1,2)= \{l\}$ of
$SO^{\uparrow}(1,2)$ is the same as that of all its covering
groups we may calculate it by using any of them.

 For $SL(2,\mathbb{R})$
we get from \ref{ka1}-\ref{en1}, \ref{be1} and \ref{baren1}: 
\begin{equation}
  \label{eq:503}
 l_{R_1}=
\frac{1}{2}\left( \begin{array}{cc} 0& 1\\-1&0 \end{array}
\right),~ l_{A_1}= \frac{1}{2}\left(
\begin{array}{cc} 1& 0\\0&-1 \end{array} \right),~ l_{B_1}=
\frac{1}{2}\left(
\begin{array}{cc} 0& 1\\1&0 \end{array} \right)~, 
\end{equation}
\begin{equation}
  \label{eq:504}
  l_{N_1}=
\left( \begin{array}{cc}0 & 1\\0&0 \end{array} \right)~,~~
l_{\bar{N}_1}= \left( \begin{array}{cc}0 & 0\\1&0
\end{array} \right)~, 
\end{equation}
  which are not independent
(in the following I skip the indices ``1'', because the
structure relations are independent of them): 
\begin{equation}
  \label{eq:505}
 l_{N} +
l_{\bar{N}}= 2\; l_{B}~,~~~ l_{N} - l_{\bar{N}}= 2\; l_{R}~. 
\end{equation}
We have the commutation relations
\begin{eqnarray} \label{eq:1673}[l_{R},l_{A}]&=&-l_{B}~,~~[l_{R},l_{B}]=l_{A}~,~~[l_{A},l_{B}]
=l_{R}~, \label{kab}\\ &&
[l_{R},l_{N}]=l_{A}~,~~[l_{R},l_{\bar{N}}]=l_{A}~,\\
\label{eq:1469}&&[l_{A},l_{N}]=l_{N}~,~~ [l_{A},l_{\bar{N}}]=-l_{\bar{N}}~,
\label{aen}\\ \label{eq:1470}&&[l_{B},l_{N}]=-l_{A}~,~~
[l_{B},l_{\bar{N}}]=l_{A}~,\\ \label{eq:1471}&& [l_{N},l_{\bar{N}}]=2\; l_{A}~.
\end{eqnarray} The relations \ref{kab} show that the algebra is
simple, \ref{aen} that $A$ and $N$ combined form a
2-dimensional subgroup, as do $A$ and $\bar{N}$.
\section{Irreducible unitary representations \\
of the positive discrete series} As the group $SO^{\uparrow}(1,2)$
is noncompact, its irreducible unitary representations are
infinite-dimensional. Their structure can be seen already from its
Lie algebra: In unitary representations the elements
$-il_{R},-il_{A},-il_{B}$ of the Lie algebra correspond to
self-adjoint operators $K_0,K_1,K_2$ which obey die commutation
relations 
\begin{equation}
  \label{eq:507}
 [K_0,K_1]=iK_2~,~~[K_0,K_2]=-iK_1~,~~[K_1,K_2]=-iK_0~, 
\end{equation}
 or,
with the definitions 
\begin{equation}
  \label{eq:508}
   K_+ =K_1+iK_2\,,~~K_-=K_1-iK_2\,,
\end{equation}
\begin{equation}
  \label{eq:509}
 [K_0,K_+]=K_+~,~~[K_0,K_-]=-K_-~,~~[K_+,K_-]=-2K_0~. 
\end{equation}
  The
relations \ref{eq:507} are invariant under the replacement $K_1
\rightarrow -K_1, K_2 \rightarrow -K_2$ and the relations \ref{eq:509}
invariant under $K_+ \rightarrow \omega K_+, K_- \rightarrow
\bar{\omega} K_-, |\omega|=1$. These relations are in addition
invariant under the transformations $ K_+ \leftrightarrow K_-, K_0
\rightarrow -K_0$.

 In irreducible unitary representations with a scalar product $(f_1,f_2)$ the
operator $K_-$ is the adjoint operator of $K_+:\;
(f_1,K_+f_2)=(K_-f_1,f_2)$,
 and vice versa, where it is assumed that $f_1,f_2$ belong to the domains
 of definition of $K_+$ and $K_-$.

The Casimir operator $L$ of a representation is defined by 
\begin{equation}
  \label{eq:510}
 L=K_1^2+K_2^2-K_0^2 
\end{equation}
  and we have the relations 
  \begin{equation}
    \label{eq:511}
  K_+K_-=L+K_0(K_0-1)~,~~K_-K_+=L+K_0(K_0+1)~.  
  \end{equation}
  All unitary
representations make use of the fact that $K_0$ is the generator
of a compact group and that its eigenfunctions $g_m$ are
normalizable elements of the associated Hilbert space $\cal H$.

The relations \ref{eq:509} imply \begin{eqnarray}
 \label{eq:1674}K_0\,g_m&=&m\,g_m~,
 \label{k0} \\
\label{eq:1472}K_0\,K_+g_m&=&(m+1)K_+g_m~, \label{k+}\\ 
\label{eq:1473}K_0\,K_-g_m &=& (m-1)K_-g_m~, \label{k-}
\end{eqnarray} which, combined with \ref{eq:511}, lead to
\begin{eqnarray} \label{eq:1675}(g_m,K_+K_-g_m)&=&(K_-g_m,K_-g_m)=l+m(m-1) \ge
0~,\label{k+k-} \\ \label{eq:1474}(g_m, K_-K_+g_m)&=&l+m(m+1)\ge 0 ~,
 ~l=(g_m,Lg_m),
\label{k-k+}
\end{eqnarray} implying 
\begin{equation}
  \label{eq:512}
  (K_+g_m,K_+g_m)= 2m+(K_-g_m,K_-g_m)
\ge 0. 
\end{equation}
  In the following we assume that we have an irreducible
representation for which the functions $g_m$ are eigenfunctions of
the Casimir operator $L$, too: $Lg_m=l\,g_m$.

 The relations
\ref{k0}-\ref{k-}  show that the eigenvalues $m$ of $K_0$ in principle can
be any real number, where, however, different eigenvalues
 differ by an integer.  

 For the ``principle'' and the ``complementary''
 series the spectrum of $K_0$ is unbounded from below and above \cite{prin}.
As $K_0$ corresponds to the classical positive definite quantity $I$,
these unitary representations are of no interest here. 

 Important for us is the
positive discrete series $D_+$ of irreducible  unitary
representations. It is
 characterized by the property that there exists a lowest eigenvalue
 $m=k$ such that 
 \begin{equation}
   \label{eq:513}
    K_-\,g_k=0\,,~~K_0\,g_k =k\,g_k\,.
 \end{equation}
  Then the relations \ref{k+k-}-\ref{eq:512} imply
  \begin{equation}
    \label{eq:514}
    l=k(1-k)~,~~~ k>0~,~~m=k+n,~ n=0,1,2,\ldots . 
  \end{equation}
  The relations
 \ref{k0}-\ref{k-} now take the form \begin{eqnarray}
 \label{eq:1676}K_0g_{k,n}&=&(k+n)\,g_{k,n}~, \label{k0+}\\
 K_+g_{k,n}&=&\omega_n\,[(2k+n)(n+1)]^{1/2}\,g_{k,n+1}~,~~|\omega_n|=1~,
\label{k++} \\
 K_-g_{k,n}&=&\frac{1}{\omega_{n-1}}[(2k+n-1)n]^{1/2}\,g_{k,n-1}~.
\label{k-+}\end{eqnarray}
 The phases $\omega_n$ guarantee that $(f_1,K_+f_2)=(K_-f_1,f_2)$. In
most cases $\omega_n$ is independent of $n$. Then one can absorb it
into the definition of $K_{\pm}$ and forget the phases $\omega_n$!

 Up to now we have allowed for any value of $k>0$. It turns out
 \cite{bar1,puk,sa1} that this
 is so for the irreducible representations of the universal covering
 group $\widetilde{SO^{\uparrow}(1,2)}$. These representations may
 be realized for $k \ge 1/2$ in the Hilbert space of holomorphic
 functions on the unit
 disc $\mathbb{D}_1=\{z,|z|<1\}$ with the scalar product 
 \begin{equation}
   \label{eq:515}
  (f,g)_k=\frac{2k-1}{\pi}\int_{\mathbb{D}_1}\bar{f}(z)g(z)
(1-|z|^2)^{2k-2}dxdy~. 
 \end{equation}
  
 The unitary operators representing the universal covering group
are given by \begin{eqnarray}
 \label{eq:1677}[U(\tilde{g},k)f](z)&=&e^{\ds 2ik\omega}(1-|\gamma|^2)^{\ds k}
 (1+\bar{\gamma}z)^{\ds -2k}
 f\left( \frac{\alpha z+\beta}{\bar{\beta}z+\bar{\alpha}}\right)\;,
 \label{Ug0} \\
  \tilde{g}&=&(\omega,\gamma)\,,~~ \left( \begin{array}{cc} \alpha & \beta \\
 \bar{\beta}& \bar{\alpha} \end{array} \right)~=~h^{\#}(\tilde{g})~
 \in SU(1,1)\,. \end{eqnarray}
Because $|\gamma\,z|<1$, the function $(1+\bar{\gamma}z)^{-2k}$ is, for $k>0$,
 defined
in terms of a series expansion.

 For $SU(1,1)$ we have $\omega \in
\mathbb{R} \bmod 2\pi$. Uniqueness of the phase factor then requires
$k=1/2,1,3/2,\cdots$.

 For $SO^{\uparrow}(1,2)$ itself we
have $\omega \in \mathbb{R} \bmod \pi$ which implies $k=1,2,\cdots$.
\subsection{Hilbert space of holomorphic functions \\ inside the unit
disc $\mathbb{D}_1$}
   One of the more important Hilbert spaces for the explicit construction
of irreducible unitary representations of the group $SU(1,1)$ is
  the (Bargmann) Hilbert space $ {\cal H}_{\mathbb{D}_1,\, k}$
   of holomorphic functions in the unit
  disc $\mathbb{D}_1=\{z=x+iy,\, |z| <1 \}$, with the scalar product
  \begin{equation}
    \label{eq:516}
   (f,g)_{\mathbb{D}_1,\, k} =\frac{2k-1}{\pi}\int_{\mathbb{D}_1}
 \bar{f}(z)g(z)\,(1-|z|^2)^{2k-2}
  dxdy~.  
  \end{equation}
  It can be used for any real $k > 1/2$ and also in the
  limiting case $k\rightarrow 1/2$. As
  \begin{equation}
    \label{eq:517}
    (z^{n_1},z^{n_2})_{\mathbb{D}_1,\, k}
   = \frac{n_1!}{(2k)_{n_1}}
 \; \delta_{n_1n_2}~,\, (2k)_n =\frac{\Gamma(2k+n)}{\Gamma(2k)}\,,
  \end{equation}
   and since any holomorphic
 function in $\mathbb{D}_1$ can be
  expanded in powers of $z$, the functions
  \begin{equation}
    \label{eq:518}
  e_{k,n}(z)= \sqrt{\frac{(2k)_n}{n!}}\,
  z^n~~,~~ n \in \mathbb{N}_{ 0}=\{n =0,1,2,\ldots\}~,   
  \end{equation}
  form an orthonormal basis of
  $\mbox{$ \cal H$}_{\mathbb{D}_1,\, k}$.

 It follows from the the unitary transformation \ref{Ug0} that the 
operators $K_j,j=0,1,2,$ here have the explicit forms
 \begin{eqnarray}  \label{eq:1678}K_0 &=&k+z\frac{d}{dz}\,,
  \\ \label{eq:1475}K_+&\equiv &K_1+i\,K_2 =2kz+z^2\frac{d}{dz}\,, \\
 \label{eq:1476}K_-&\equiv &K_1-iK_2=\frac{d}{dz}\,. \end{eqnarray} 
The basis functions 
\ref{eq:518}
 are the eigenfunctions of $K_0$ with eigenvalues $k+n$, the operators $K_+$
 and $K_-$ being raising and lowering operators:
 \begin{eqnarray} \label{eq:1679}K_0\,e_{k,n}&=&(k+n)\,e_{k,n}~~, \\
 \label{eq:1477}K_+\,e_{k,n}&=&[(2k+n)(n+1)]^{1/2}\,e_{k,n+1}~~, \\
 \label{eq:1478}K_-\, e_{k,n}&=&[(2k+n-1)n]^{1/2}\, e_{k,n-1}~~. \end{eqnarray}
 If we have on $ \mathbb{D}_1$ the
holomorphic functions 
\begin{equation}
  \label{eq:519}
 f(z)=\sum_{n=0}^{\infty}a_n\,z^n~,~~g(z)=
\sum_{n=0}^{\infty}b_n\,z^n~, 
\end{equation}
  then, according to Eq.\ \ref{eq:517},
 their scalar product $(f,g)_{\mathbb{D}_1,\, k}$ is given by
 \begin{equation}
   \label{eq:520}
   (f,g)_{\mathbb{D}_1,\, k}=\sum_{n=0}^{\infty}
\frac{n!}{(2k)_n}
 \bar{a}_n \,b_n~. 
 \end{equation}
This series can be used as a scalar product to extend the
definition of the Hilbert spaces ${\cal H}_{\mathbb{
D},\, k}$ to all real $k>0$ \cite{sa2,vil}.
 \subsection{Unitary representations in the Hilbert
space \\ of  holomorphic functions on the upper half plane} The
unit disc $\mathbb{D}_1$ and its associated Hilbert space with
 the scalar product \ref{eq:515}
is especially suited for the construction of unitary
representations of $SU(1,1)$ because that group acts transitively
on $\mathbb{D}_1$. Similarly, the group
$SL(2,\mathbb{R})=Sp(2,\mathbb{R})$,
 isomorphic to $SU(1,1)$,
acts transitively on the upper complex half plane
$\mathbb{S}_1=\{w=u+iv,\; v>0\}$. The mapping \begin{eqnarray}
\label{eq:1680}w&=&\frac{1-iz}{z-i}=
\frac{2x+(1-x^2-y^2)\,i}{x^2+(y-1)^2}~,\\\label{eq:1479}
z&=&\frac{iw+1}{w+i}\,,~~|z|^2
=\frac{u^2 +(v-1)^2}{u^2+(v+1)^2}\,, \end{eqnarray} provides a
holomorphic diffeomeorphism from $\mathbb{D}_1$ onto
$\mathbb{S}_1$ and back.

 Because of 
\begin{equation}
  \label{eq:526}
   \frac{dudv}{4v^2}
=\frac{dxdy}{(1-|z|^2)^2}\,,~~1-|z|^2 =
\frac{2^2\,v}{(w+i)(\bar{w}-i)}\,,
\end{equation}
 we have for $k=1/2,1,3/2,2,
\ldots $ the following isomorphism: 
\begin{equation}
  \label{eq:527}
(f,g)_{\mathbb{D}_1,\,
k}=(\tilde{f},\tilde{g})_{\mathbb{S}_1,\,
k}=\frac{1}{\Gamma(2k-1)}
 \int_{\mathbb{S}_1}\overline{\tilde{f}}~\tilde{g}\, v^{2k-2}dudv\,,  
\end{equation} 
 where \begin{eqnarray} \label{eq:1681}E_k:&& \tilde{f}(w)=
\sqrt{\frac{\Gamma(2k)}{\pi}}
\, 2^{2k-1}(w+i)^{-2k} f\left(z=\frac{1+iw}{i+w}\right)\,,\\
\label{eq:1480}E^{-1}_k:&& f(z)= 2\sqrt{\frac{\pi}{\Gamma(2k)}}\,
(z-i)^{-2k}\tilde{f}\left(w=
\frac{1-iz}{z-i}\right)\,. \end{eqnarray}
The  (unitary) transformation $E_k$ maps the basis \ref{eq:517} of
 $ {\cal H}_{\mathbb{D}_1,\, k}$ onto the basis
 \begin{equation}
   \label{eq:528}
  \tilde{e}_{k,n}(w)= \sqrt{\frac{\Gamma(2k+n)}{\pi\,n!}}
\;2^{2k-1}\;i^n\;(w-i)^n\;(w+i)^{-2k-n}~,n \in \mathbb{N}_{0}~,  
 \end{equation}
 of $ {\cal H}_{\mathbb{S}_1,\, k}$. One can, of
course, discard the phase factor $i^n$. 

 On this Hilbert space
 the irreducible unitary
representations $T_k^+$ of the positive discrete series of
$SL(2,\mathbb{R})$ are given by \begin{eqnarray}
\label{eq:1682}[T^+(g_1,k)\tilde{f}](w)& =&
(a_{12}w+a_{22})^{-2k}\,\tilde{f}\left(\frac{a_{11}w+a_{21}}{a_{12}
w+a_{22}}\right)\,, \label{tk+} \\ \label{eq:1481}g_1& =& \left(
 \begin{array}{cc} a_{11}&
a_{12}\\ \label{eq:1482}a_{21}&a_{22} \end{array}\right) \in SL(2,
\mathbb{R})\,,
\end{eqnarray} which is defined for $k=1/2,1,3/2,2, \ldots $ only.

The subgroups
\begin{eqnarray} \label{eq:1683}\mbox{R}_1: && r_1 = \left(
 \begin{array}{cc} \cos(\theta/2)&
\sin(\theta/2)\\ \label{eq:1483}-\sin(\theta/2)& \cos(\theta/2) \end{array}
\right)\,,\\ \label{eq:1484}\mbox{A}_1:&& a_1 = \left(
 \begin{array}{cc} e^{\ds t/2}&
  0\\ \label{eq:1485}0 & e^{\ds -t/2} \end{array} \right)\,, \\ \mbox{B}_1:&&
  b_1 =
\left( \begin{array}{cc} \cosh(s/2)&
\sinh(s/2)\\ \sinh(s/2)& \cosh(s/2) \end{array} \right)\,,
\end{eqnarray}
are associated with the following generators of their unitary
representations (the sign of $\tilde{K_0}$ is chosen such that its
spectrum is positive): \begin{eqnarray} \label{eq:1684}\tilde{K}_0&=
&\frac{1}{i}( k\,w+\frac{1}{2}(w^2+1)\frac{d}{dw})\,, \\
\label{eq:1486}\tilde{K}_{\pm}&=& \pm k\,(w\mp i)\pm \frac{1}{2}(w\mp
i)^2\frac{d}{dw}\,.
\end{eqnarray} Their action on the basis \ref{eq:528} is given by
 \begin{eqnarray} \label{eq:1685}\tilde{K}_0\,\tilde{e}_{k,n}&=&(k+n)\,
\tilde{e}_{k,n}~~, \\
\label{eq:1487}\tilde{K}_+\,\tilde{e}_{k,n}&=& i[(2k+n)(n+1)]^{1/2}\,
\tilde{e}_{k,n+1}\,, \\
\label{eq:1488}\tilde{K}_-\tilde{e}_{k,n}&=& \frac{1}{i}[(2k+n-1)n)]^{1/2}
\tilde{e}_{k,n-1}\,.
\end{eqnarray}
For the limiting case $k \rightarrow 1/2$ the Hilbert space with
the scalar product \ref{eq:527} now can be replaced by the ``Hardy space
$H^2(\mathbb{R},\,du)$ of the upper half plane'' \cite{shiop,hard2}, 
 the elements
of which are the functions $\tilde{f}(u)$, limits for
$\Im(w)=v \rightarrow 0$ of the
 previous holomorphic functions $\tilde{f}(w)$
  on the upper half plane and the Hilbert
 space of which  has the scalar product 
 \begin{equation}
   \label{eq:529}
  (\tilde{f_1},\tilde{f_2})
 =\int_{-\infty}^{\infty}du \overline{\tilde{f}}_1(u)\,\tilde{f}_2(u)~. 
 \end{equation}
Applications of this space are discussed in Subsec.\ 4.4, where further
details can be found.
\chapter{The symplectic group $Sp(4,\mathbb{R})$ and the positive discrete
series of its irreducible unitary representations}
A reader might wonder why I include a rather long appendix on the symplectic
group in four dimensions. But we have seen in Ch.\ 8 that the canonical
group  for the 4-dimensional phase space \ref{eq:416} of interference phenomena
is the group $Sp(4,\mathbb{R})$. The crucial point  here is  again that
the necessary deletion of the origin of that phase space requires a 
quantization procedure which is different from the conventional one!
Therefore one needs the appropriate ``observables'' associated with the 
group $Sp(4, \mathbb{R})$ and with the quotient group $SO^{\uparrow}(2,3) =
 Sp(4, \mathbb{R})/Z_2$ and one needs the irreducible unitary representations
 of the positive discrete series of these  groups. 

For that reason the present appendix collects - incompletely - some
 properties of 
$Sp(4,\mathbb{R})$ and $SO^{\uparrow}(2,3)$ and tries to point out
 some of the essential References.
\section{Properties of $Sp(4,\mathbb{R})$}
The main References for this and the next Section
 are \cite{sie,car1,barg3,foll}.
  
 Let
\begin{equation}
  \label{eq:521}
  \langle y,x \rangle = y^T\Omega\, x = y_1x_3-y_3x_1 +y_2x_4  -y_4x_2~,
\end{equation}
where \begin{eqnarray} \label{eq:1686}\Omega &=& \begin{pmatrix} 0_2 & E_2 \\ -E_2 & 0_2
 \end{pmatrix}\,,
~E_2 = \begin{pmatrix} 1 &0 \\ 0 &1 \end{pmatrix}\,,~0_2 = \begin{pmatrix}
0 & 0 \\ 0 & 0 \end{pmatrix}\,, \\
&& x= \begin{pmatrix} x_1 \\ x_2 \\ x_3 \\ x_4 \end{pmatrix}\,,~
y^T = (y_1,y_2,y_3,y_4)\,, \nonumber \end{eqnarray}
be a skew symmetric bilinear form on $\mathbb{R}^4$. 

The matrix $\Omega$
has the property
\begin{equation}
  \label{eq:522}
  \Omega^2= -1\,,~\Omega^{-1} = -\Omega = \Omega^T~.
\end{equation}
If $e_j,j=1,...,4,$ is the cartesian basis of $\mathbb{R}^4$, then
\begin{equation}
  \label{eq:523}
  \langle e_1,e_3 \rangle = -\langle e_3,e_1 \rangle = 1\,,\,
 \langle e_2,e_4 \rangle = -\langle e_4,e_2 \rangle = 1\,,\, 
\, \langle e_j,e_k\rangle = 0~ \text{else}\,.
\end{equation} In the following we in general will identify the vector $x$ as
\begin{equation}
  \label{eq:547}
  x= \begin{pmatrix} q_1 \\ q_2 \\ p_1 \\ p_2 \end{pmatrix}
\end{equation}

The group $Sp(4,\mathbb{R})$ consists of all $4 \times 4$ real
matrices (mappings) $w$ which leave the bilinear form \ref{eq:521} invariant:
\begin{equation}
  \label{eq:524}
 \langle w\cdot y,w\cdot x\rangle = \langle y, x \rangle~. 
\end{equation}
This implies for the matrices $w$
\begin{equation}
  \label{eq:525}
  w^T\Omega\, w = \Omega~,~(w^{-1})^T\Omega w^{-1}=
 \Omega~,~\text{or}~w^{-1} = 
\Omega w^T \Omega^{-1}~.
\end{equation}
The relations \ref{eq:523} show that the group $Sp(4,\mathbb{R})$ 
may also be defined as the transformations which leave the sum of exterior
products \cite{gre,mar2}
\begin{equation}
  \label{eq:575}
  e_1 \wedge e_3 + e_2 \wedge e_4~~(\text{or}~~dq_1\wedge dp_1 + dq_2 \wedge
dp_2\,)
\end{equation}
invariant. \\
If we write $w$ in block form in terms of $2\times 2$-matrices,
\begin{equation}
  \label{eq:530}
  w = \begin{pmatrix}A_{11} & A_{12} \\ A_{21} & A_{22} \end{pmatrix}~,
\end{equation}
then the conditions \ref{eq:525} imply for the submatrices:
\begin{eqnarray}
  \label{eq:1687}\label{eq:531}
  A_{11}^TA_{22}-A_{21}^TA_{12} &=& E_2~,\\ A_{11}^TA_{21} =
 A_{21}^TA_{11}\,, &~&
A_{12}^TA_{22} = A_{22}^TA_{12}~,
\label{eq:531a} \\ w^{-1}& =& \begin{pmatrix} A_{22}^T& -A_{12}^T \\
-A_{21} T& A_{11}^T \end{pmatrix}~. \label{eq:531b}
\end{eqnarray} It follows from the last of the relations \ref{eq:525}
 that $w^T \in Sp(4,\mathbb{R})$ if $w \in
Sp(4,\mathbb{R})$ and one can show that $\det w = 1$. 

Important is the identification of the maximal compact subgroup $K=U(2)$
of $Sp(4,\mathbb{R})$: Any real non-singular $4\times4$ matrix $A$ may
be uniquely written as $A= O\cdot P$, where $P=+(A^TA)^{1/2}$ is a positive
definite matrix and $O=A\cdot P^{-1}$ an orthogonal one, $O^{-1} = O^T$.
If $A \in Sp(4,\mathbb{R})$, then $P,O \in Sp(4,\mathbb{R})$. As the orthogonal
group $O(4)$ is compact, the maximal compact subgroup $K$ of $Sp(4,\mathbb{R})$
 is given by $Sp(4,\mathbb{R}) \cap O(4)$. By identifying $\mathbb{R}^4$
with $\mathbb{C}^2$, one can see that $K=U(2)$: If $u \in U(2)$, then the
corresponding element in $Sp(4,\mathbb{R})$ is given by
\begin{equation}
  \label{eq:532} w(K) = \begin{pmatrix} \Re( u)& -\Im (u) \\ \Im (u) & \Re (u)
\end{pmatrix}~,  
\end{equation}
where the unitarity relations $u^+u = u\, u^+ = E_2$, when rewritten for
$ \Re (u)$ and $ \Im (u)$, are just the relations \ref{eq:531} and
 \ref{eq:531a}!
Notice that $\Omega\, w(K) = w(K)\, \Omega$. 
 
Any element $u \in U(2)\cong U(1)\times SU(2) $ may be parametrized as
\begin{eqnarray}
  \label{eq:1688}\label{eq:534}
  u&=& e^{i\beta/2}\,\check{u}\,,~\beta \in (-2\pi,2\pi]\,,
 \\ \check{u}(\alpha,\vec{n})
 &=& \cos\frac{\alpha}{2}\,E_2 -
i\sin\frac{\alpha}{2} \sum_{j=1}^3 n_j\sigma_j\,,~\vec{n}^2 =1\,,~
\alpha \in (-2\pi,2\pi]\,,~~~~ \\
&& \sigma_j,j=1,2,3:  \text{Pauli's matrices}\,; \nonumber \\
\det \check{u} &=& 1\,,~~~ \\
\det u &=& e^{i\beta}~. 
\end{eqnarray}
For the abelian subgroup $U(1) = \{e^{i\beta/2}\}$ we have
\begin{equation}
  \label{eq:535}
  w(e^{i\beta/2}) = \begin{pmatrix} \cos\frac{\beta}{2}&0&\sin\frac{\beta}{2}
&0 \\ 0& \cos\frac{\beta}{2}&0&\sin\frac{\beta}{2} \\
-\sin\frac{\beta}{2}&0&\cos\frac{\beta}{2}&0 \\
0&-\sin\frac{\beta}{2}&0&\cos\frac{\beta}{2} \end{pmatrix}\,,
\end{equation}
and for the 1-parameter subgroup $\{\check{u}(\alpha, \vec{n}=(0,0,1))\}$
of $SU(2)$ we get
\begin{equation}
  \label{eq:536}
  w[\check{u}(\alpha, \vec{n} = (0,0,1))] = \begin{pmatrix}
 \cos\frac{\alpha}{2}&0&\sin\frac{\alpha}{2}
&0 \\ 0& \cos\frac{\alpha}{2}&0&-\sin\frac{\alpha}{2} \\
-\sin\frac{\alpha}{2}&0&\cos\frac{\alpha}{2}&0 \\
0&\sin\frac{\alpha}{2}&0&\cos\frac{\alpha}{2} \end{pmatrix}\,, 
\end{equation}
Another important subgroup is the 2-parameter abelian subgroup
\begin{align}
  A =& \{a_1(t_1),a_2(t_2)\}\,,\nonumber 
 \\  w(A)=& w(a_1(t_1))\cdot w(a_2(t_2))=
 \begin{pmatrix} e^{t_1/2}&0&0&0 \\
0&e^{t_2/2} &0&0 \\0&0&e^{-t_1/2} &0 \\0&0&0&e^{-t_2/2} \end{pmatrix}\,,
\label{GA}\\
  &~ t_j \in \mathbb{R}\,,\,j=1,2\,. \nonumber
\end{align}
This group constitutes the maximal abelian non-compact subgroup
of an Iwasawa decomposition
\begin{equation}
  \label{eq:616}
  K\cdot A\cdot N 
\end{equation}
 of the group $Sp(4,\mathbb{R})$, the group
$K$ being the maximal compact subgroup. The remaining nilpotent subgroup
\begin{equation}
  \label{eq:537}
   N = \{n_1(\xi_1),n_2(\xi_2),n_3(\xi_3),n_4(\xi_4)\} 
\end{equation}
 of that decomposition is generated by  four
 1-parameter subgroups which have the following elements
\begin{alignat}{2}
  \label{GN}
  n_1(\xi_1)=&
 \begin{pmatrix} 1&0&\xi_1&0 \\
0&1&0&0 \\0&0&1 &0 \\0&0&0&1 \end{pmatrix}\,,&~~
n_2(\xi_2)&= \begin{pmatrix}1&0&0&0 \\0& 1&0&\xi_2 \\
0&0&1&0 \\0&0&0&1 \end{pmatrix}\,, \\ n_3(\xi_3)=&
\begin{pmatrix} 1&0&0&\xi_3 \\
0&1&\xi_3&0 \\0&0&1 &0 \\0&0&0&1 \end{pmatrix}\,,&~~
n_4(\xi_4)&= \begin{pmatrix}1&\xi_4&0&0 \\0& 1&0&0 \\
0&0&1&0 \\0&0&-\xi_4&1 \end{pmatrix}\,, \\& \xi_j \in \mathbb{R}\,,\,
j=1,2,3,4\,. \nonumber
\end{alignat}
The proof for this will be indicated below in connection with the Lie
algebra of the group. 

 Notice that $N_3 = \{n_1(\xi_1),n_2(\xi_2),n_3(
\xi_3)\}$ forms a 3-parameter abelian subgroup of $N$. 

Another subgroup of interest is the (commuting) product $Sp(2,\mathbb{R})_1 
\otimes Sp(2,\mathbb{R})_2$, where the first factor acts on the $(q_1,p_1)$-
subspace as described in appendix A and the second factor on the $(q_2,p_2)$-
subspace correspondingly. The matrices related to the first factor are 
(see the formulae \ref{ka1}, \ref{a1} and \ref{en1}):
\begin{equation}
  \label{eq:548}
  \begin{pmatrix} \cos(\frac{\theta_1}{2})& 0&\sin(\frac{\theta_1}{2})&0
 \\0&1&0&0 \\
-\sin(\frac{\theta_1}{2})& 0& \cos(\frac{\theta_1}{2})&0
 \\0&0&0&1 \end{pmatrix}\,,\,
\begin{pmatrix} e^{t_1/2}&0&0&0 \\0&1&0&0 \\0&0&e^{-t_1/2}&0 \\
0&0&0&1 \end{pmatrix}\,,\, \begin{pmatrix}1&0&\xi_1&0 \\0&1&0&0 \\
0&0&1&0 \\0&0&0&1 \end{pmatrix}\,,
\end{equation} and those for the second factor 
\begin{equation}
  \label{eq:549}
  \begin{pmatrix}1&0&0&0 \\ 0&
 \cos(\frac{\theta_2}{2})& 0&\sin(\frac{\theta_2}{2}) \\0&0&1&0 \\
0&-\sin(\frac{\theta_2}{2})& 0& \cos(\frac{\theta_2}{2}) \end{pmatrix}\,,\,
\begin{pmatrix}1&0&0&0 \\0& e^{t_2/2} &0&0 \\0&0&1&0 \\
0&0&0&e^{-t_2/2} \end{pmatrix}\,,\, \begin{pmatrix}1&0&0&0 \\0&1&0&\xi_2 \\
0&0&1&0 \\0&0&0&1 \end{pmatrix}\,.
\end{equation}

In Appendix B.1 we have seen that the group
 $Sp(2,\mathbb{R})=SL(2,\mathbb{R})$
is isomorphic to the complex group $SU(1,1)$. A corresponding property holds
for the general case $Sp(n,\mathbb{R})$: This group is isomorphic to a 
subgroup $SU(n,n)_{sp}$ of the complex group $SU(n,n)$ acting
 on $\mathbb{C}^{2n}$, the elements of which
leave the quadratic form
\begin{equation}
  \label{eq:533}
  \bar{z}_1z_1 + \cdots +\bar{z}_nz_n -\bar{z}_{n+1}z_{n+1}-\cdots -
\bar{z}_{2n}z_{2n}
\end{equation}
invariant and have  determinant $=1$. 

 The subgroup $SU(n,n)_{sp}$ is defined as
\begin{equation}
  \label{eq:538}
  SU(n,n)_{sp}= \{w_c\}: ~w_c \in SU(n,n)~
 \text{and}~ w_c^T\Omega\, w_c= \Omega=
\begin{pmatrix} 0_n& E_n \\ -E_n& 0_n \end{pmatrix}\,.
\end{equation}
Next I specialize to $n=2$: We identify the vector $x$
in \ref{eq:521} accor\-ding to \ref{eq:547},
  define the complex numbers
\begin{equation}
  \label{eq:540}
  a_j =\frac{1}{\sqrt{2}}(q_j+ip_j)= |a_j|\,e^{-i\vp_j}\,,\,j=1,2\,,
\end{equation}
and the complex vectors $\in \mathbb{C}^4$ 
\begin{equation}
  \label{eq:541}
  a = \begin{pmatrix} a_1 \\a_2 \\\bar{a}_1\\ \bar{a}_2 \end{pmatrix}\,. 
\end{equation}
The column vectors $x$ and $a$ are related by a unitary transformation:
\begin{equation}
  \label{eq:542}
  a=C\cdot x\,,~C= \frac{1}{\sqrt{2}}\begin{pmatrix} E_2 & iE_2\\
E_2& -iE_2\end{pmatrix}\,,~C^{-1} =C^+ =\frac{1}{\sqrt{2}}
\begin{pmatrix} E_2& E_2 \\ -iE_2&iE_2 \end{pmatrix}\,.
\end{equation}
The elements  $w_c \in SU(2,2)_{sp} \equiv Sp_c(4)$ are then given by
\begin{equation}
  \label{eq:543}
  w_c = C\cdot w \cdot C^{-1}\,.
 \end{equation}
 The first of the conditions \ref{eq:525} now becomes
 \begin{equation}
   \label{eq:545}
   w_c^+\Delta\, w_c = \Delta\,,~\Delta = \begin{pmatrix} E_2& 0_2 \\
0_2 & -E_2 \end{pmatrix}\,.  
\end{equation} This says that $w_c$ is an element of $SU(2,2)$. \\
A general element $w_c$ has the form
\begin{equation}
  \label{eq:614}
  w_c =\begin{pmatrix}B_1&B_2\\ \bar{B}_2& \bar{B}_1 \end{pmatrix}\,,\,
B_1^+ \cdot B_1-B_2^T\cdot \bar{B}_2 =E_2\,,\,B_2^+\cdot B_1=B_1^T\cdot 
\bar{B}_2\,.
\end{equation}
For the compact subgroup \ref{eq:532} we get
\begin{equation}
  \label{eq:544}
  w_c(K) = \begin{pmatrix} u& 0_2 \\ 0_2 & \bar{u} \end{pmatrix}\,,
\end{equation}
where $u$ is the unitary matrix \ref{eq:534}. 

 The maximal abelian non-compact subgroup \ref{GA} now takes the form
 \begin{equation}
   \label{eq:546}
   w_c(A) = \begin{pmatrix}\cosh(t_1/2)&0&\sinh(t_1/2)&0 \\0&\cosh(t_2/2)&0&
\sinh(t_2/2) \\ \sinh(t_1/2)&0& \cosh(t_1/2)&0 \\
0& \sinh(t_2/2)&0&\cosh(t_2/2) \end{pmatrix}\,,
 \end{equation}
and for the nilpotent group \ref{eq:537} one gets
\begin{equation}
  \label{eq:550}
  w_c(N_3) = \begin{pmatrix} E_2 -i\,\Xi_3/2 & i\,\Xi_3/2 \\
-i\,\Xi_3/2& E_2+i\,\Xi_3/2 \end{pmatrix}\,,\,\Xi_3 =
\begin{pmatrix} \xi_1&\xi_3 \\\xi_3&\xi_2 \end{pmatrix}\,,
\end{equation}
and 
\begin{equation}
  \label{eq:551}
  w_c[n_4(\xi_4)] = \begin{pmatrix}1&\xi_4/2&0&\xi_4/2 \\-\xi_4/2&1&\xi_4/2&
0\\ 0& \xi_4/2&1&\xi_4/2 \\ \xi_4/2&0&-\xi_4/2&1 \end{pmatrix}\,.
\end{equation}
\section{The Lie algebra $\mathfrak{sp}(4,\mathbb{R})$}
In the neighbourhoods of the unit element $E_4$ of $Sp(4,\mathbb{R})$
the group elements $w$ may be approximated as
\begin{equation}
  \label{eq:552}
  w = E_4 + \hat{w}\,\epsilon\,,\, |\epsilon| \ll 1\,,
\end{equation}
where the matrix $\hat{w}$ is an element of the Lie algebra $\mathfrak{
sp}(4,\mathbb{R})$ of the group $Sp(4,\mathbb{R})$. It follows from the first
of the relations \ref{eq:525} that $\hat{w}$ has to obey the condition
\begin{equation}
  \label{eq:539}
  \hat{w}^T\Omega + \Omega\,\hat{w} = 0\,.
\end{equation}
 For the  ansatz 
\begin{equation}
  \label{eq:553}
  \hat{w} = \begin{pmatrix} \hat{B}_{11}& \hat{B}_{12}\\
\hat{B}_{21}&\hat{B}_{22} \end{pmatrix}
\end{equation}
the condition \ref{eq:539} implies
\begin{equation}
  \label{eq:554}
  \hat{B}_{12}^T = \hat{B}_{12}\,,\,\hat{B}_{21}^T=\hat{B}_{21}\,,\,
\hat{B}_{22} = -\hat{B}_{11}^T\,.
\end{equation}
If $g(t)$ is a 1-parameter group we shall denote the generating Lie algebra
element by $\hat{g}, g(t) = \exp(\hat{g}\,t)$, in the following. 

 A basis of
the Lie algebra $\mathfrak{sp}(4,\mathbb{R})$ is easily obtained from the
subgroups \ref{eq:534}, \ref{GA} and \ref{eq:537}. For the Lie algebra
$\mathfrak{k}$ of the compact subgroup \ref{eq:532} we get from \ref{eq:534}
the 4 basis elements:
\begin{equation}
  \label{eq:555}
  \hat{u}_0 = \frac{1}{2}\begin{pmatrix} 0&0&1&0\\0&0&0&1\\-1&0&0&0\\
0&-1&0&0 \end{pmatrix}
\end{equation}
for the $U(1)$ subgroup \ref{eq:535} and
\begin{align}
  \label{hu3}
  \hat{u}_1& = \frac{1}{2}\begin{pmatrix} 0&0&0&1\\0&0&1&0\\0&-1&0&0\\
-1&0&0&0 \end{pmatrix}\,,~~ \hat{u}_2 =
 \frac{1}{2}\begin{pmatrix} 0&-1&0&0\\1&0&0&0\\0&0&0&-1\\
0&0&1&0 \end{pmatrix}\,,\,\\ \hat{u}_3& =
\frac{1}{2}\begin{pmatrix} 0&0&1&0\\0&0&0&-1\\-1&0&0&0\\
0&1&0&0 \end{pmatrix}\,, \label{u3}
\end{align}
for the group $SU(2)$. The latter obey the usual commutation relations
\begin{equation}
  \label{eq:557}
  [\hat{u}_j,\hat{u}_k] = \epsilon_{jkl}\,\hat{u}_l\,.
\end{equation}

For the 2 generators of the abelian subgroup \ref{GA} we get
\begin{equation}
  \label{eq:556}
  \hat{a}_1 =\frac{1}{2}\begin{pmatrix} 1&0&0&0\\0&0&0&0\\0&0&-1&0\\
0&0&0&0 \end{pmatrix}\,,\,
\hat{a}_2 =\frac{1}{2}\begin{pmatrix} 0&0&0&0\\0&1&0&0\\0&0&0&0\\
0&0&0&-1 \end{pmatrix}\,.
\end{equation} The algebra $\mathfrak{a}$ is maximal abelian, non-compact in
both dimensions and constitutes one of the 4 Cartan subalgebras of 
$\mathfrak{sp}(4,\mathbb{R})$ \cite{kna2}. Its general elements have the form
\begin{equation}
  \label{eq:558}
  \hat{a} = 2\omega_1\, \hat{a}_1 + 2\omega_2\, \hat{a}_2\,,\,\omega_j
 \in \mathbb{R}\,.
\end{equation} (The factor $2$ is convention. As to the following see, e.g.\
the Refs.\ \cite{helg2,war2,ros2}) 

As all the elements of $\mathfrak{a}$ commute with each other, they
have a common set of eigenvectors in the adjoint representation 
which here are to be understood as  $4 \times 4$
matrices $E_{\lambda}$ satisfying
\begin{equation}
  \label{eq:559}
  [\hat{a}, E_{\lambda}] = \lambda E_{\lambda}\,.
\end{equation}
The eigenvalues $\lambda$ are called ``roots''. 
In our case there are 8 different eigenvectors with 8 different eigenvalues
\begin{equation}
  \label{eq:560}
 \pm 2\omega_1\,,\,\pm 2 \omega_2\,,\, \pm(\omega_1+\omega_2)\,,\,
\pm(\omega_1-\omega_2)\,.
\end{equation}
Choosing $\omega_1 > \omega_2 > 0$ we have 4 positive and 4 negative 
roots. In the following it is convenient to introduce the matrices
\begin{equation}
  \label{Ejk}
  E_{jk}\,=\, \begin{pmatrix}\text{number $1$ at
 the crossing of the $j$-th row} \\
 \text{with the $k$-th column, all other elements $=0$} \end{pmatrix}\,.
\end{equation}
We then have the eigenvectors 
\begin{equation}
  \label{eq:562}
  E_{2\omega_1} = E_{13}\,,\,E_{2\omega_2} =E_{24}\,,\,E_{\omega_1+\omega_2} =
E_{14}+E_{23}\,,\,E_{\omega_1-\omega_2} = E_{12}-E_{43}\,,
\end{equation}
 of the four positive roots and the eigenvectors of the corresponding
negative roots are
\begin{equation}
  \label{eq:561}
    E_{-2\omega_1} = E_{31}\,,\,E_{-2\omega_2} =E_{42}\,,
\,E_{-\omega_1-\omega_2} =
E_{32}+E_{41}\,,\,E_{-\omega_1+\omega_2} = E_{21}-E_{34}\,.
\end{equation}
The eigenvectors \ref{eq:562} form a basis of the Lie algebra $\mathfrak{n}$
of the nilpotent group \ref{eq:537}, i.e.\ we have the relations
\begin{equation}
  \label{eq:563}
  E_{2\omega_1} = \hat{n}_1\,,\,E_{2\omega_2} = \hat{n}_2\,,\,
E_{\omega_1+\omega_2} =
\hat{n}_3 \,,\,E_{\omega_1-\omega_2} = \hat{n}_4\,, 
\end{equation}
which are exactly the generators of the four 1-parameter groups
\ref{GN}. \\
 The associated commutation relations are
\begin{alignat}{3} [\hat{n}_1,\hat{n}_2] &=0\,,& ~~ \label{n41}
 [\hat{n}_1,\hat{n}_3] &=0\,,&~~ [\hat{n}_1,\hat{n}_4] &=0\,, \\ \label{n42}
 [\hat{n}_2,\hat{n}_3] &= 0\,,&~~ [\hat{n}_2,\hat{n}_4] &= -\hat{n}_3\,,&~~
 [\hat{n}_3,\hat{n}_4] &= -2 \hat{n}_1 \,. \end{alignat}
They prove that $\mathfrak{n}$ is indeed a (nilpotent) subalgebra and its
construction shows that 
$\mathfrak{n}$ and $\mathfrak{a}$ combined form a subalgebra, too. 

The eigenvectors \ref{eq:561} of the negative roots constitute another
4-dimen\-sio\-nal nilpotent algebra $\bar{\mathfrak{n}}$ of a 4-dimensional
nilpotent group $\bar{N}$ with obvious 1-parameter subgroups. 

The  Lie algebra of the  subgroup \ref{eq:548} is generated by
\begin{equation}
  \label{eq:564}
  l_0^{(1)} = \frac{1}{2}(\hat{u}_0 +\hat{u}_3)\,,\, l_A^{(1)} = \hat{a}_1\,,\,
l_N^{(1)} = \hat{n}_1\,,
\end{equation} and that of the subgroup \ref{eq:549} by
\begin{equation}
  \label{eq:565}
  l_0^{(2)} = \frac{1}{2}(\hat{u}_0 -\hat{u}_3)\,,\, l_A^{(2)} = \hat{a}_2\,,\,
l_N^{(2)} = \hat{n}_2\,,
\end{equation}
where the relations \ref{eq:555}, \ref{u3}, \ref{eq:556} and \ref{eq:563}
have been used. 

 In Ch.\ 8 we encountered the ten  Hamiltonian functions
$g_0,\ldots,g_9$ on the phase space \ref{eq:416} which generate the Lie
algebra $\mathfrak{sp}(4,\mathbb{R})$ via their Poisson brackets. 
If we denote the corresponding $4\times 4$ matrices by $\hat{g}_0,\ldots,
\hat{g}_9$ we get the following relations the validity of which will
be demonstrated below:
 \begin{gather} \label{ghat} \hat{g}_0 = \hat{u}_0\,,~\hat{g}_1 =
 \hat{u}_1\,,~\hat{g}_2
=\hat{u}_2\,,~\hat{g}_3 =\hat{u}_3\,,~\hat{g}_4= 
\hat{n}_3-\hat{u}_1\,,~
\hat{g}_5= \hat{n}_4-\hat{u}_2\,, \\ \hat{g}_6 =\hat{a}_1 +\hat{a}_2\,,~
\hat{g}_7= \hat{n}_1+\hat{n}_2-\hat{u}_0\,,~\hat{g}_8= \hat{n}_1-\hat{n}_2
-\hat{u}_3\,,~\hat{g}_9 = \hat{a}_2 -\hat{a}_1\,.~~~~~~ \notag  \end{gather}  
\section{On the isomorphism between the groups $Sp(4,\mathbb{R})$ 
\protect \\ and
  $\mbox{Spin}[SO^{\uparrow}(2,3)]$}
(Here again I closely follow Bargmann's paper \cite{barg3} to a large
extent.) 

There is a very interesting and helpful isomorphism between the group
$Sp(4,\mathbb{R})$ and the double covering - i.e.\ the spinor variant -
of the identity component $SO^{\uparrow}(2,3)$ of the group which leaves
the quadratic form
\begin{equation}
  \label{eq:566}
  u^T\cdot \eta\cdot u  = u^j\eta_{jk}u^k =-(u^1)^2 -(u^2)^2 
-(u^3)^2 +(u^4)^2+(u^5)^5\,,\, u= \begin{pmatrix} u^1\\ \vdots \\u^5 
\end{pmatrix}\,, 
\end{equation}
invariant. The matrix
\begin{equation}
  \label{eq:568}
  \eta = \eta^{-1} =  (\eta_{jk}) = (\eta^{jk}) = \begin{pmatrix}
-1&0&0&0&0 \\0&-1&0&0&0 \\0&0&-1&0&0\\0&0&0&1&0\\
0&0&0 &0&1 \end{pmatrix} 
\end{equation}
 defines the metric. 
It may be used for lowering and raising indices, e.g.\ $u_j=\eta_{jk}
u^k$ or $v^j = \eta^{jk}v_k$. Here and in the following  Einstein's summation
convention is assumed: summation over equal indices. 

A transformation
\begin{equation}
  \label{eq:567}
  u \to v=B\cdot u\,,\, v^j = b^j_{\;k}u^k\,,\,B=(b^j_{\;k})\,,
\end{equation}
leaves the quadratic form \ref{eq:566} invariant iff
\begin{equation}
  \label{eq:569}
  B^T\cdot\eta\cdot B = \eta~~ \text{or}~~ B\cdot \eta\cdot B^T = \eta\,.
\end{equation}
The transformations \ref{eq:567} form a group $O(2,3)$ with 4 components
 (unconnected pieces). The identity component $SO^{\uparrow}(2,3)$ 
 (i.e.\ the component which contains the unit element) 
is given by the conditions
\begin{equation}
  \label{eq:570}
  \det B =1\,,\,b^4_{\;4}\,b^5_{\;5}-b^4_{\;5}\,b^5_{\;4} >0\,.
\end{equation}
In the early days the group has been called the ``2+3 de Sitter group''
  because it is the  group of motions for the de Sitter space
 \begin{equation}
   \label{eq:571}
  -(u^1)^2 -(u^2)^2 
-(u^3)^2 +(u^4)^2+(u^5)^2 =-a^2\,. 
 \end{equation}
There is another, different de Sitter group, the ``1+4 de Sitter group''
 $O(1,4)$,  which leaves a
corresponding quadratic form of a ``1+4 de Sitter space'' invariant
 \cite{wig3} 

As the global geometries of the two de Sitter spaces are quite different it has
become customary in gravity, field and string theory \cite{hawk,pol}
 to call the
4-dimensional space \ref{eq:571}  ``anti--de Sitter space'', $AdS_4$,
and the transformation group $O(2,3)$ the $AdS_4$ group! 

The isomorphismen between the Lie algebras $\mathfrak{sp}(4,\mathbb{R})$
and $\mathfrak{so}(2,3)$ was first shown by E.\ Cartan  \cite{car2}.
As both groups are ``infinitely'' connected (they contain the subgroups
$U(1)$ or $O(2)$ respectively) the situation is more subtle on the group
level: The group $Sp(4,\mathbb{R})$ is actually isomorphic to a double
covering (i.e.\ a spinor version) of the group $SO^{\uparrow}(2,3)$. 

Siegel in his famous paper on symplectic geometry \cite{sie2}
 gave 2 independent
proofs of this isomorphism. One by using a complex
version of the quadratic form \ref{eq:566} and its transformations and by using
 properties of a so-called ``Siegel domain''  (see Sec.\ C.4 below).
 The second one
by exploiting the fact that the Clifford algebra associated with the group
$SO(2,3)$ can be realized by purely real $4\times 4$ matrices which in
turn ``support'' an associated spinor group isomorphic
 to $Sp(4,\mathbb{R})\,$. 

Dirac, at the end of his paper \cite{dir3}  on two ``remarkable''
 repesentations
of the $2+3$ de Sitter group, points out the general 
relationship between the two groups after having learnt about it
from Res Jost. Dirac's arguments are in the spirit of Siegel's
 first proof, but
he was not aware of Siegel's work and he calls the two groups ``equivalent'',
without detailing the isomorphism. 

Bargmann in his paper \cite{barg3} gives a detailed proof of the isomorphism
which is completely equivalent to Siegel's second proof. He also was not
aware of Siegel's paper. 

 I shall indicate the idea of the proof for the
isomorphism mentioned above without
giving any details. These may be found in Bargmann's paper (or Siegel's).  

The procedure is well-known from the use of Dirac matrices for the construction
of the spinor version $SL(2,\mathbb{C})$ of the homogeneous Lorentz group
as required in the context of the Dirac equation (see Appendix B.1). 

The Clifford algebra associated with the quadratic form \ref{eq:566} is
defined by
\begin{equation}
  \label{eq:572}
  \{\gamma_j,\gamma_k\} =-2\,\eta_{jk}\,E_4\,,~j,k=1,\ldots,5\,.
\end{equation}
These relations may be satisfied by 5 {\em real} $4 \times 4 $ matrices
which have the property
\begin{equation}
  \label{eq:573}
  \gamma_j^T = \Omega \gamma_j \Omega^{-1}\,,
\end{equation}
where $\Omega$ is the matrix from \ref{eq:521}. \\ Using the irreducibility
of a given set of matrices $\gamma_j$ satisfying the relations \ref{eq:572},
 observing that any other set $\alpha_j$ obeying them has to be related to the
given one by an equivalence transformation and that $\alpha_j =
 b_j^{\;k}\gamma_k\,,\, b_j^{\;k}=\eta_{jl}\,\eta^{km}\, b^{l}_{\,m}\,$, obeys
\ref{eq:572},  gives
\begin{equation}
  \label{eq:574}
  w\cdot\gamma_j\cdot w^{-1} = b^k_{\;j}\gamma_k\,,~~w^T\cdot\Omega\cdot w =
\Omega\,,~(b^k_{\;j}) \in SO^{\uparrow}(2,3)\,.
\end{equation}
The  relations \ref{eq:574} show that $w \in Sp(4,\mathbb{R})$ and that $w$ and
$-w$ correspond to the same $B \in SO^{\uparrow}(2,3)$. Thus, $Sp(4,
\mathbb{R})$ is isomorphic to the spinor group (double covering) of
  $ SO^{\uparrow}(2,3)$. 

Using the properties \ref{eq:572} of the $\gamma$-matrices and the
relation \ref{eq:574} one may express
the coefficients $b^j_{\;k}$ in terms of $w$ as
\begin{equation}
  \label{eq:617}
  b^j_{\;k} = -\frac{1}{4}\mbox{tr}(\gamma^j\cdot w\cdot
 \gamma_k\cdot w^{-1})\,.
\end{equation}
There is a close relationship between the Lie algebra generators $\hat{g}_0,
\ldots,\hat{g}_9$ of Eqs.\ \ref{ghat} and the Lie algebra generators $
\hat{m}_{jk} =-\hat{m}_{kj} \in \mathfrak{so}(2,3)$ of the rotations or
 special Lorentz transformations in the $u^j-u^k$--planes which will be
demonstrated below. Here we just give the result: 
\begin{alignat}{5} \label{img}
 \hat{m}_{12}&=\hat{g}_3\,,&~~\hat{m}_{23}&=\hat{g}_1\,,&~~
\hat{m}_{31}&=\hat{g}_2\,,&~~\hat{m}_{41}&=\hat{g}_7\,,&~~ 
\hat{m}_{42}&=\hat{g}_9\,,\\ \hat{m}_{43}&=-\hat{g}_5\,,&~~ 
\hat{m}_{51}&=\hat{g}_8\,,&~~\hat{m}_{52}&=\hat{g}_6\,,&~~
\hat{m}_{53}&=-\hat{g}_4\,,&~~\hat{m}_{54}&=\hat{g}_0\,. \notag
\end{alignat}
These relations make the isomorphism of the Lie algebras  obvious.

 One
can also verify the $\mathfrak{so}(2,3)$ commutation relations
\begin{equation}
  \label{eq:577}
  [\hat{m}_{jk},\hat{m}_{lm}] = \eta_{kl}\,\hat{m}_{jm} + \eta_{jm}\,
\hat{m}_{kl}
-\eta_{jl}\,\hat{m}_{km} -\eta_{km}\,\hat{m}_{jl}
\end{equation}
by means of the commutation relations of the $\hat{g}_0,\ldots,\hat{g}_9$.
\section{Transformations induced by $Sp(4,\mathbb{R})$ on certain spaces}
In the following I briefly discuss the actions of the group $Sp(4,\mathbb{R})$
on the spaces
\begin{equation}
  \label{eq:606}
  {\cal S}^4_{q,p} = \{x=\begin{pmatrix}q_1\\q_2\\p_1\\p_2 \end{pmatrix}
\in \mathbb{R}^4\}\,,
\end{equation}
\begin{equation}
  \label{eq:607}
  {\cal S}^4_{q,p;0}={\cal S}^4_{q,p}-\{x=0\}\,, 
\end{equation}
\begin{equation}
  \label{eq:608}
  \check{{\cal S}}^4_{q,p}={\cal S}^4_{q,p}/Z_2\,,
\end{equation} (which are
generalizations of the 2-dimensional spaces \ref{eq:587}, \ref{eq:591} and
\ref{eq:595}), on
\begin{equation}
  \label{eq:609}
  {\cal S}^4_{\vp,I} = \{\vp_1 \in \mathbb{R}\bmod{2\pi}, I_1 >0,
\vp_2 \in \mathbb{R}\bmod{2\pi}, I_2 >0\}\,,
\end{equation}
a generalization of \ref{eq:356},
and on the Siegel domains $\mathbb{S}_2$ and $\mathbb{D}_2$, generalizations
of the spaces \ref{eq:605} and \ref{eq:366}. 

A discussion of the group action on the spaces \ref{eq:606}-\ref{eq:608}
 and \ref{eq:609}
essentially amounts to a generalizing repetition of the arguments in 
section A.3. So I shall be very brief here and stress only a few
 important  points: 

The orbifold \ref{eq:608} is not just the topological product of the
orbifold \ref{eq:595} with itself. The non-trivial center elements
 of the subgroups \ref{eq:548} and \ref{eq:549} are
 \begin{equation}
   \label{eq:618}
   \begin{pmatrix}-1&0&0&0 \\0&1&0&0 \\0&0&-1&0\\0&0&0&1 \end{pmatrix}\,,~~
 \begin{pmatrix}1&0&0&0 \\0&-1&0&0 \\0&0&1&0\\0&0&0&-1 \end{pmatrix}\,.
 \end{equation}
They do {\em not} belong to the center of $Sp(4,\mathbb{R})$! According to
\ref{eq:617} they correspond to the element
\begin{equation}
  \label{eq:619}
   \begin{pmatrix}-1&0&0&0& \\0&-1&0&0&0 \\0&0&1&0&0\\0&0&0&-1&0
\\ 0&0&0&0&-1 \end{pmatrix}\,
\end{equation} of $SO^{\uparrow}(2,3)$. 

 Eq.\ \ref{eq:608} means that
the canonical 2-dimensional subspaces are ``conified'' simultaneously.
The conic space \ref{eq:609} now may be represented by the orbifold
 \ref{eq:608}
(with the $Z_2$ fix point $x=0$ deleted) on which the group $Sp(4,\mathbb{R})$
 acts in the same way as the group
$Sp(2,\mathbb{R})$ on the orbifold \ref{eq:595}. The effective transformation
group  here is $SO^{\uparrow}(2,3) \cong Sp(4,\mathbb{R})/Z_2$. 

For the explicit construction of the irreducible unitary representations
of the positive discrete series of $Sp(4,\mathbb{R})$ the following 
generalizations of the ``Siegel upper half plane'' \ref{eq:605} and of the
``Siegel unit disc'' \ref{eq:366} which got their name from Siegel's
paper \cite{sie} but where already introduced by E.\ Cartan \cite{car3}.
(See Appendix B.3 for the important role $\mathbb{S}_1$ and $\mathbb{D}_1$
play for the discrete series of $Sp(2,\mathbb{R})$ and $SU(1,1)$.) 

In our present case the generalized Siegel upper plane is given by 
symmetric complex $2\times 2$ matrices with positive definite imaginary
part:
\begin{eqnarray} \label{eq:1689}W&=& U+iV =\begin{pmatrix}u_{11}&u_{12}
\\u_{12}&u_{22}
\end{pmatrix}+i\,\begin{pmatrix}v_{11}&v_{12}\\v_{12}&v_{22}
\end{pmatrix}\,, \label{W} \\ && W^T=W\,,\,u_{jk},v_{jk} \in
 \mathbb{R}\,,  \\
V&>& 0 \Leftrightarrow v_{11} >0\,,\,v_{11}v_{22} -(v_{12})^2 >0\,.\label{V}
\end{eqnarray} $W$ has $3$ complex and $6$ real dimensions. 

The  group $Sp(4,\mathbb{R})$ with its elements $w$ (see \ref{eq:530}) acts on 
$W$ as follows
\begin{equation}
  \label{eq:610}
  W \to \hat{W} = (A_{11}\cdot W +A_{12})\cdot (A_{12}\cdot W +A_{22})^{-1}\;.
\end{equation}
The group action is transitive, but only almost effecive because $w$ and
$-w$ give the same transformation on $\mathbb{S}_2$, i.e.\ here again the
effective group is $SO^{\uparrow}(2,3)$! 

The stability group of the points of $\mathbb{S}_2$ is the compact
subgroup $K$ (see \ref{eq:534}). Thus, $\mathbb{S}_2$ is the homogeneous
space 
\begin{equation}
  \label{eq:620}
  \mathbb{S}_2 = Sp(4,\mathbb{R})/K\,.
\end{equation}
The invariant positive definite metric on $\mathbb{S}_2$ is
\begin{equation}
  \label{eq:611}
  \tr V^{-1}\cdot dW\cdot V^{-1}\cdot d\bar{W})\, 
\end{equation}and the associated volume element is
\begin{eqnarray}
  \label{eq:1690}\label{deo}
  d\omega &=& 2\, (\det V)^{-3}\,d^3u\,d^3v = 2\, d^3u\,d^3t\,, \\
&& d^3u =du_{11}du_{12}du_{22}\,, \nonumber \\
&& V^{-1} =\begin{pmatrix}t_{11}&t_{12}\\ t_{12}&t_{22}
\end{pmatrix}\,,\,d^3t =dt_{11}dt_{12}dt_{22}\,.
\end{eqnarray}
The Cayley transform
\begin{equation}
  \label{eq:615}
  W \to Z =(E_2 +iW)\cdot(E_2-iW)^{-1} = Z^T =\begin{pmatrix} z_{11}&z_{12} \\
z_{12}&z_{22} \end{pmatrix}\,,\,z_{jk} \in \mathbb{C}\,,
\end{equation}
maps $\mathbb{S}_2$ one-to-one onto the generalized Siegel disc
\begin{equation}
  \label{eq:612}
  \mathbb{D}_2 = \{Z:\, Z=Z^T\,, E_2- Z\cdot \bar{Z} >0\}\,.
\end{equation}
The group $Sp_c(4)$ with its elements \ref{eq:614} acts transitively
on $\mathbb{D}_2$:
\begin{equation}
  \label{eq:613}
  Z \to \hat{Z} = (B_1\cdot Z + B_2)\cdot(\bar{B}_2\cdot Z+ \bar{B}_1)^{-1}\,.
\end{equation} Like $\mathbb{S}_2$ the disc $\mathbb{D}_2$ is a homogeneous
space, namely 
\begin{equation}
  \label{eq:621}
  \mathbb{D}_2 = Sp_c(4)/K_c\,.
\end{equation}
For more details as to this section see the Refs. \cite{sie,car1,foll}.
\section{Irreducible unitary repesentations of the positive discrete series
for $Sp(4,\mathbb{R})$}
Like in the case of the group $SL(2,\mathbb{R})$ we are mainly interested in
the positive discrete series of the irreducible unitary representations
of the group $Sp(4,\mathbb{R})$ or $SO^{\uparrow}(2,3)$, because 
in general one wants the Hamilton operator $H$ of a physical system to be
bounded from below!  In Ch.\  8 we saw  that in the case of 2
classical oscillators the Hamilton function is given by $2g_0$ where
$g_0$ is the function which corresponds to the generator of the abelian
subgroup $U(1)$ of the maximal compact subgroup $U(2)$ of $Sp(4,\mathbb{R})$
(see the relations \ref{eq:423} and \ref{eq:424} and on the quantum level
\ref{K0}; see also the relation $\hat{g}_0 = \hat{m}_{45}$: generator of
the $O(2)$ subgroup of $SO^{\uparrow}(2,3)$), Eq.\ \ref{eq:577}.

 If denote by $G_0$ the 
self-adjoint operator which represents $\hat{g}_0$ in an irreducible unitary
representation then $G_0$ has only positive (discrete) eigenvalues
for the positive discrete series! Its relation to the Hamilton operator $H$
is
\begin{equation}
  \label{eq:622}
  H= 2\,G_0\,,
\end{equation}
i.e.\ the energy levels are {\em twice} the eigenvalues of $G_0$!! 

 Before I discuss more details let
me make some historical remarks:
\subsection{A few historical remarks} If one tries to trace the 
work on  irreducible unitary representations of the group $Sp(4,\mathbb{R})$
or (and) $SO^{\uparrow}(2,3)$ in the literature  one makes the surprising
 discovery that most authors are not aware of important previous work 
pertaining to their own. As there is no comprehensive monograph or review
article on the subject - at least to my knowledge - one has to collect
bits and pieces from many (more or less) original papers!

 I shall try
to give a brief guide to the more important papers on the subject - as far
as I found and judged them! I also certainly missed essential ones. My 
selection is primarily motivated by the interests of a physicist, not by
those of a mathematician. 

Let me begin with an exceptional case: There is a beautiful and comprehensive
paper by Bargmann \cite{barg3} on those unitary representations of
 $Sp(2n;\mathbb{R})$
the Lie algebra of which is generated by the self-adjoint 
operator versions $G_j,j=0,\ldots,9,$
 of the
functions $g_j,j=0,\ldots,9$ encountered in Ch.\ 8. The associate 
representations of the group are actually those of a double covering
of $Sp(4,\mathbb{R})$, named ``metaplectic'' bei the mathematician Weil
\cite{wei} and denoted by $M_p(2n,\mathbb{R})$ for general $Sp(2n,\mathbb{R})$.

Bargmann's paper is hardly quoted by authors of later papers on the subject.
It is preceded by an important similar paper by Itzykson \cite{itz}.

Bargmann's results were extended by Folland in Ch.\ 4 of his very
recommendable book \cite{foll} which is also not well-known. 

Let me now turn to earlier times: This is characterized by the feature that
the isomorphism between $Sp(4,\mathbb{R})$ and the spin version of
$SO^{\uparrow}(2,3)$ is not taken into account. 

The story of the irreducible
unitary representations of the latter appears to begin with the 1954 Princeton
thesis by Ehrman (supervised by Bargmann), excerpts of which are published
in Ref.\ \cite{ehr}. Ehrman's aim was to determine all irreducible unitary
 representations  of the universal covering group. Following Harish-Chandra
he used the Iwasawa decomposition \ref{eq:616} and the Cartan subgroup
of the compact group $K$ generated by $\hat{m}_{12}$ and $\hat{m}_{45}$
(see Eqs.\  \ref{img})  as well as the Casimir operators
\begin{eqnarray} \label{eq:1691}L_2 &=& \frac{1}{2}\hat{m}_{jk}
\hat{m}^{jk}\,, \label{cas} \\
L_4 &=& \hat{w}_j\hat{w}^j\,,\,\hat{w}_j= \frac{1}{8}\epsilon_{jklmn}
\hat{m}^{kl}\hat{m}^{mn}\,,
\\ && \epsilon_{jklmn}\,:~\text{completely antisymmetric in}~j,k,\ldots\,, 
\nonumber\\ 
&& \epsilon_{12345} =1\,, \nonumber
\end{eqnarray}
for classifying the irreducible representations. 

 Ehrman dealt only briefly
at the end of his paper with the positive discrete series, where he coined
the term ``singleton'' which has been used frequently afterwards:

 We
have already seen at the end of Ch.\ 8 and will again discuss it below
that the irreducible unitary representations of the discrete series may
be characterized by a pair $(\epsilon_0,j_0)$, where $\epsilon_0$ denotes 
the lowest
eigenvalue of the self-adjoint operator $J_{45} =G_0$ which represents
$\hat{m}_{45} = \hat{g}_0$. 

 $j_0$ is the largest eigenvalue of the angular
 momentum operator $J_{12}$ associated with that irreducible 
representation of $SU(2)$ which has the smallest dimension of all the
$SU(2)$ representations contained in the infinite dimensional representation
 of the full group $SO^{\uparrow}(2,3)$. If $j_0 > 0$ this means the ground
state is degenerate with a degeneracy $2j_0+1$. Higher levels with 
$\epsilon_n =\epsilon_0 +n\,,\,n=1,2,\ldots,$ also carry angular momenta $j 
 >j_0$. There may be several of them. If there is just one, Ehrman called
the corresponding representation of the full group a ``singleton''! 

In 1963 Dirac discovered and discussed \cite{dir3} the two
 unitary representations
of the positive discrete series which may be generated by the operator
version of the functions $g_0,\ldots,g_9$, of Ch.\  8. They are the
``singletons'' $(1/2,0)$ and $(1,1/2)$. 

 In 1965 Fronsdal gave a brief
characterization of irreducible unitary representations
 of $SO^{\uparrow}(2,3)$
- especially those of the discrete series - in the appendix of his paper
\cite{fro} on the possible role of the anti-de Sitter space in particle
physics. 

 In 1966 an important paper by Evans appeared \cite{eva} in
which he gave a complete clear classification of all irreducible unitary
representations of the universal covering group of $SO^{\uparrow}(2,3)$
in terms of the pairs $(\epsilon_0,l_0)$. I shall come back to this paper
below. It was almost never quoted by later papers on the subject. 
Evans did not give {\em explicit} constructions of the representations he
classified. 

 The explicit constructions of the positive discrete series of
 $Sp(2n, \mathbb{R})$, inspired by the work
 of Bargmann for the
group $SU(1,1)$,  had already been given by the mathematician
Godement in 1958 \cite{god}, but they remained unnoticed for a long time. 

 In 1966 non-compact ``dynamical groups'' became popular in particle
physics including the use of Majorana-type infinite component wave 
equations \cite{frad}. In this context irreducible unitary representations
of  $SO^{\uparrow}(2,3)$ were employed by several research groups
 \cite{baru1,dir4,suda,fro2}, mainly using Dirac's (singleton)
 representations from 1963. 

Later Howe discussed some of the related mathematics \cite{howe2}.

 In 1968 Goshen and Lipkin extended their previous work 
on  applications of $Sp(2,\mathbb{R})$ \cite{lip} to questions of
 nuclear structure
to the group $Sp(4,\mathbb{R})$ \cite{lip2}. 

In 1971 Moshinsky and Quesne discussed \cite{mosh} unitary representations of 
symplectic groups (without knowning the results of Itzykson and Bargmann). \\
In 1972 Kirillov in his textbook \cite{kir} briefly discussed unitary spinor
representation of the symplectic group. 

In 1973 Gelbart published a paper \cite{gelb} on the explicit
 construction of the
discrete series for $Sp(2n,\mathbb{R})$ in terms of Weil's representation
 \cite{wei},
 using Godement's previous work. 

In 1977 Rosensteel and Rowe used Godement's results as to the positive discrete
series \cite{god} for their description
of nuclear models in terms of unitary representations of
 symplectic groups \cite{ros1}.

 In 1983 a (not easy to read) paper by Angelopoulos with a complete (?) list
of the irreducible unitary representations of $SO^{\uparrow}(2,3)$
appeared \cite{ang}. 

 In 1984 the first paper on possible applications
of the group $Sp(4,\mathbb{R})$ in quantum optics was published,
by Milburn \cite{mil}. Later several papers on that subject followed
 \cite{syqo}. 

 From 1985 on  Mukunda, Simon, Sudarshan et al.\ published
a series \cite{muku2} of interesting papers on the structure and
 possible physical (quantum optical)
applications of the groups $Sp(4,\mathbb{R})$ and $SO^{\uparrow}(2,3)$. 

There are also a number of papers on the isomorphism between the Lie algebras
of these groups and a number of their subalgebras \cite{kim2}.

On the more mathematical side there is, of course, the important monograph
by Guillemin and Sternberg \cite{gui}. 

In the textbook by Knapp \cite{kna3} the group $Sp(4,\mathbb{R})$ serves
as an example in many chapters. 

In an important memoire from 1989  on a certain (Howe) duality 
between the groups $O(2,2)$ and $Sp(4,\mathbb{R})$ Przebinda \cite{prz}
also gives a complete list of the irreducible unitary representations
of $Sp(4,\mathbb{R})$, including those of the discrete series. The article
is not introductory, but a research report with many cross-references
as to technical details. It mainly appeals to the experts and is not easy
to read for somebody who is not! 

What is urgently needed is an introductory monograph which combines
the results of Godement, Evans and Przebinda, written in the style
of Barg\-mann!
\subsection{About the positive discrete series of $Sp(4,\mathbb{R})$}
I shall start with the two representations of the Lie algebra 
$\mathfrak{so}(2,3)$ found by
Dirac \cite{dir3} and thoroughly analysed by Bargmann on the group level 
\cite{barg3}. Then I shall mention some elements of the classification scheme 
used by Evans \cite{eva}, with a few remarks as to the results of Przebinda
\cite{prz}. And finally I shall briefly describe the principle of the explicit
construction of the positive discrete series due to Godement \cite{god}.
I want to stress that all this is very fragmentary and merely intended to
give a superficial  overview as to the problems involved and to be dealt
with more systematically in the future,  with the aim to apply the results
to quantum optical and other physical phenomena! 
\subsubsection{The two Dirac-Bargmann representations}
These representations have already been mentioned at the end of Ch.\ 8.2
and in Sec.\ C.5.1 above. Their Lie algebra corresponds to the classical
Hamiltonian functions $g_j,j=0,\ldots,9,$ discussed in Ch.\ 8 in connection
with interference patterns. As these functions constitute the ``observables''
associated with the canonical group $Sp(4,\mathbb{R})$ of the phase space
\ref{eq:607} and it appears to be worthwhile to collect them here once more:
\begin{alignat}{2}
g_0 & = \frac{1}{4}(q_1^2+p_1^2 +q_2^2+p_2^2)\,,& ~~~ g_1& = 
\frac{1}{2}(q_1q_2 +p_1p_2)\,, \label{ge01} \\
g_2& = \frac{1}{2}(q_1p_2-q_2p_1)\,,& ~~~g_3&= \frac{1}{4}(q_1^2+p_1^2-q_2^2-
p_2^2)\,, \nonumber \\
g_4&=\frac{1}{2}(q_1q_2-p_1p_2)\,,&~~~
g_5& = \frac{1}{2}(q_1p_2+q_2p_1)\,,\nonumber \\ g_6& =
 -\frac{1}{2}(q_1p_2+q_2p_1)\,,
& ~~~ g_7&=\frac{1}{2}(q_1p_1 -
q_2p_2)\,, \nonumber \\
 g_8 &=
 \frac{1}{4}(q_1^2-p_1^2 -q_2^2 + p_2^2)\,,& ~~~g_9 &=
 \frac{1}{4}(q_1^2-p_1^2+q_2^2-p_2^2)\,. \nonumber
 \nonumber \end{alignat} We have seen that these functions fulfill
the Lie algebra $ \mathfrak{so}(2,3)$ (or $\mathfrak{sp}(4,\mathbb{R})$) in
terms of their Poisson brackets. 

Dirac discussed \cite{dir3} the operator version of \ref{ge01} and two 
related representations, Bargmann \cite{barg3} the irreducible
 unitary representations of the associated (metaplectic) group. 

The self-adjoint operators $G_j, j=0,\ldots,9,$ or $J_{jk}=-J_{kj}\,$,
corresponding to the functions \ref{ge01} or the $\hat{m}_{jk}$ of
\ref{img} may conveniently expressed by the associated creation and
annihilation operators:
\begin{alignat}{2} G_0 &=J_{54} = \frac{1}{2}(a_1^+a_1 + a_2^+a_2+1)\,,&
G_1 &= J_{23} =\frac{1}{2}(a_1^+a_2 +a_1a_2^+)\,, \nonumber \\
G_2 &= J_{31} =\frac{i}{2}(a_1a_2^+ -a_1^+a_2)\,,&
G_3 &= J_{12} =\frac{1}{2}(a_1^+a_1 -a_2^+a_2)\,, \nonumber \\
G_4 &= -J_{53} = \frac{1}{2}(a_1^+a_2^++a_1a_2)\,,&
G_5 &=-J_{43} = \frac{i}{2}(a_1^+a_2^+-a_1a_2)\,, \nonumber \\
G_6 &= J_{52} = \frac{i}{4}( a_1^2 -(a_1^+)^2 + a_2^2 -(a_2^+)^2)\,,&~~~& 
 \label{Jik}\\
G_7 & = J_{41}= \frac{i}{4}( (a_1^+)^2 -a_1^2 + a_2^2 -(a_2^+)^2)\,.&~~~&
\nonumber \\
G_8 &= J_{51} =\frac{1}{4}( a_1^2 +(a_1^+)^2 - a_2^2 -(a_2^+)^2)\,,&~~~&
\nonumber \\
G_9 &= J_{42} =\frac{1}{4}( a_1^2 +(a_1^+)^2 + a_2^2 +(a_2^+)^2)\,.&~~~& 
\nonumber 
 \end{alignat}
These operators act in a Hilbert space $ {\cal H}_1^{osc}\otimes {\cal H}_2^{
osc}= {\cal H}_+ \oplus {\cal H}_-$, the tensor product of two
 harmonic oscillator Hilbert spaces, with
the even and odd states
\begin{eqnarray}\label{eq:1692}{\cal H}_+ &\ni& |n_1,n_2 \rangle_+
 \equiv |n_1\rangle_1
 \otimes |n_2 
\rangle_2\,,\,n_1+n_2~\text{even}\,, \label{H+} \\
{\cal H}_- &\ni& |n_1,n_2 \rangle_+ \equiv |n_1\rangle_1
 \otimes |n_2 
\rangle_2\,,\,n_1+n_2~\text{odd}\,. \label{H-}
\end{eqnarray}
For the ``even'' ground state we have
\begin{eqnarray} \label{eq:1693}G_0|0,0\rangle_+ &=&\frac{1}{2}|0,0\rangle_+\,,
 \label{G00+} \\
J_3 |0,0\rangle_+&=&0\,,~~J_3 \equiv J_{12}\,, \label{J30+}
\end{eqnarray} and for a state with $n_1+n_2=2n$
\begin{align}
  \label{G0n+}
  G_0|n_1,n_2;n_1+n_2=2n\rangle_+ &= \epsilon_n^{(+)}
 |n_1,n_2;n_1+n_2=2n\rangle_+\,, \\
\epsilon_n^{(+)} &= n+\frac{1}{2}\,,\,n=0,1,\ldots.
\end{align}
The state \ref{G0n+} is $(2n+1)$-fold degenerate. The corresponding subspace
carries an irreducible unitary representation of the group $SU(2)$ with angular
momentum $j=n$. The 3 generators of that represenation are $G_1,G_2$ and
$G_3=J_3$. The eigenstate of $J_3$ with eigenvalue $j=n$ is given by
$n_1=2n,n_2=0$. 

 The ground state of ${\cal H}_-$ is degenerate  with
respect to $G_0=J_{54}$, too:
\begin{eqnarray} \label{eq:1694}G_0|1,0\rangle_-&=&|1,0\rangle_-\,,
 \label{G00-1} \\
 G_0|0,1\rangle_-&=&|0,1\rangle_-\,, \label{G00-2}\,.
\end{eqnarray} If $n_1+n_2 =2n+1$ then
\begin{align}
  \label{G0n-}
 G_0 |n_1,n_2;n_1+n_2=2n+1\rangle_-& =\epsilon_n^{(-)}
|n_1,n_2;n_1+n_2=2n+1\rangle_-\,, \\ \epsilon_n^{(-)}&= n+1\,,
\,n=0,1,\ldots\,. 
\end{align}
Here the degeneracy of the eigen-subspace is $(2n+2)$-fold. It carries
an irreducible unitary $SU(2)$ representation with $j=n+1/2$. 

Notice that in the ``even'' case \ref{H+} the eigenvalues of $G_0$ are
half-integer, whereas the values $j$ of the angular momentum are integer.
For the ``odd'' case \ref{H-} it is the other way round. This is what
Dirac found ``remarkable'' \cite{dir3}. But notice also (see \ref{Jik})
    that $G_0$ is
just half the Hamiltonian $H$ of the 2-dimensional harmonic oscillator! 
\subsubsection{Sketch of the classification scheme}
The nucleus of the classification scheme for the irreducible unitary 
representations of the positive discrete series of $Sp(4,\mathbb{R})$
is already visible in the case of the Dirac-Bargmann representations. 

I mainly follow Evans \cite{eva} in this brief sketch: 

Crucial is the existence of the compact Cartan subalgebra spanned by
the two commuting Lie algebra elements $\hat{g}_0 =\hat{m}_{54}$
 and $\hat{g}_3 =
\hat{m}_{12}$. They are two of the four generators of the maximal compact
group $U(2)$ of $Sp(4,\mathbb{R})$. 
It is then  possible to characterize
the representations in question uniquely by the lowest eigenvalue $\epsilon_0$
of $G_0=J_{54}$ and the angular momentum  value $j_0$ of the 
lowest-dimensional unitary representation of $SU(2)$ contained in the
representation of $Sp(4,\mathbb{R})$. In general this $SU(2)$ representation
 will characterize the
degeneracy of the ground state. 

Thus, any irreducible unitary representation
of the positive discrete series is characterized by a pair $(\epsilon_0,j_0)$.
The $2j_0+1$ associated eigenvalues $m_0$  of $G_3 = J_{12}$ are
 $m_0=j_0,j_0-1,\ldots,-j_0+1,-j_0$. 

One may also use the eigenvalues of the two Casimir operators \ref{cas},
for a classification, but that appears here to be much more
 cumbersome than using the pair $(\epsilon_0,j_0)$ (see below). 

 Notice that
$\epsilon_0$ might be interpreted as half the energy of the ground state, at
least this is so for the two Dirac-Bargmann representations. 

 The operator
$G_0=J_{54}$ is the generator of the representations of the $U(1)$ subgroup
of $U(2)$. It commutes with the $SU(2)$ generators $G_1,G_2$ and $G_3$ of the 
representations. 

 Like in the case of the representations of $Sp(2,\mathbb{R})
\cong SU(1,1)$
 (see Appendix B) where the lowest eigenvalue $k$ of the generator $K_0$
for the corresponding subgroup $U(1)$ is determined by the choice of the
covering group of $U(1)$ (there are infinitely many of them!), so does the
value of $\epsilon_0$ depend on that covering group.

 In the case of the Dirac-Bargmann representations
we have met the values $\epsilon_0 = 1/2$ and $=1$. For the universal
covering group  of $Sp(4,\mathbb{R})$ (see Ref.\ \cite{barg3}) $\epsilon_0$
may take any value in the interval $(0,1]$. 

Given the lowest eigenvalue $\epsilon_0$ the higher eigenvalues of $G_0$ in
a given representation are
\begin{align}
  \label{epsn}
  G_0|n;(\epsilon_0,j_0)\rangle &= \epsilon_n |n;(\epsilon_0,j_0)\rangle\,, \\
\epsilon_n &= n+\epsilon_0\,,\,n=0,1,\ldots\,,
\end{align}
The eigenstates \ref{epsn} are degenerate. The carry at least one irreducible
representation of $SU(2)$ with angular momentum $j,\, j=0,1/2,1,3/2,\ldots,$
where $G_3$ has the eigenvalues $m=j,j-1,\ldots,-j+1,-j\,$.
The whole infinite dimensional Hilbert space may be built up from such
subspaces. 

 Important restrictions on $j$ and $m$ for a given $\epsilon_n$ 
follow from the following observation: According to the Eqs.\ \ref{eq:564}
and \ref{eq:565} the operators $(G_0+G_3)/2$ and $(G_0-G_3)/2$ are the
 generators $K_0^{(1)}$ and $K_0^{(2)}$ of two  positive discrete series
sub-representations of two independent $SL(2,\mathbb{R})=Sp(2,\mathbb{R})$
subgroups of $Sp(4,\mathbb{R})$. As both $K_0^{(j)},j=1,2,$ are positive
definite we have the important inequlities
\begin{equation}
  \label{eq:624}
  \frac{1}{2}(\epsilon_n+m) >0\,,~~\frac{1}{2}(\epsilon_n-m) >0\,,\,m=j,j-1,
\ldots,-j+1,-j\,.
\end{equation} This implies
\begin{equation}
  \label{eq:623}
  |m| < \epsilon_n\,,~~\text{especially}~j < \epsilon_n\,.
\end{equation}
These inequalities applied to the ground state(s) give
\begin{equation}
  \label{eq:625}
  |m_0| < \epsilon_0\,,~~j_0 < \epsilon_0\,.
\end{equation}
The last conditions are obviously fulfilled for the Dirac-Bargmann
representations. 

 The above remarks in connection with the two subalgebras
\ref{eq:564} and \ref{eq:565} lead to another important consequence: If we
denote the Bargmann indices of the corresponding irreducible
 unitary representations
of the discrete series by $k_1$ and $k_2$, then a decomposition of
 the irreducible
representation $(\epsilon_0,j_0)$ of $Sp(4,\mathbb{R})$ with respect  
to  sub-representations $k_1$ and $k_2$ of $Sp(2,\mathbb{R})$ leads to the
$2j_0+1$ possible values
\begin{equation}
  \label{eq:626}
  k_1 =\frac{1}{2}(\epsilon_0+m_0)\,,~~k_2 = \frac{1}{2}(\epsilon_0-m_0)\,.
\end{equation}
For $(\epsilon_0,j_0)=(1/2,0)$ we have $k_j=1/4,j=1,2,$
and for $(1,1/2)$ we get $k_j=1/4,3/4,$ which are just the metaplectic
representations we encountered in Ch. 6.2. 

As to the list of possible irreducible unitary representations of the
positive discrete series I refer to Evan's paper \cite{eva}. Notice that
his list applies to the universal covering group of $Sp(4,\mathbb{R})$,
not to $Sp(4, \mathbb{R})$. 

For illustration I give Evan's expressions for the eigenvalues $l_2$ and
$l_4$ of the Casimir operators
\ref{cas}:
\begin{equation}
  \label{eq:627}
  l_2 =-[\epsilon_0(\epsilon_0-3)+j_0(j_0+1)]\,,~l_4 = -j_0(j_0+1)
(\epsilon_0-1)(\epsilon_0-2)\,.
\end{equation}
The list
of Przebinda \cite{prz1} is more special because he is only interested in true
representations of $Sp(4,\mathbb{R})$ itself. He parametrizes the 
representations by integers $m=\epsilon_0+j_0$ and $n=\epsilon_0-j_0$ which
excludes the Dirac-Bargmann representations.
\subsubsection{Sketch of Godement's construction}
Godement's explicit construction of the positive discrete series of the
irreducible unitary representations of the groups $Sp(2n,\mathbb{R})$
is a generalization of the corresponding construction for $Sp(2,\mathbb{R})=
SL(2,\mathbb{R})$ on the Siegel upper half plane as discussed in
 Appendix B.3.2 above, Eq.\ \ref{tk+}. 

In order to briefly describe Godement's construction for $Sp(4,\mathbb{R})$
we need some facts from the theory of certain finite dimensional
 representations of the groups $GL_+(2,\mathbb{C \text{ or }R})$ and $U(2)$ on
 complex vector spaces  with an hermitian scalar product. \\
  These vector
spaces  are constructed \cite{wey2,waer,tdie} from a 2-dimenional one 
\begin{equation}
  \label{eq:628}
  V^2 =\{ b=\begin{pmatrix}z_1\\z_2\end{pmatrix},z_j \in
 \mathbb{C}\,,\,j=1,2\}\,,\,
\langle b,b \rangle = b^+\cdot b\,.
\end{equation} The homogeneous polynomials of degree $n$,
\begin{equation}
  \label{eq:630}
  P_k^{(n)}(b)=z_1^kz_2^{n-k}\,,\,k=0,1,\ldots,n\,,
\end{equation}
span a $(n+1)$-dimenional complex vector space $V^{n+1}$. \\
The group elements
\begin{equation}
  \label{eq:631}
  g =\begin{pmatrix}c_{11}&c_{12} \\c_{21}&c_{22} \end{pmatrix}
 \in GL(2,\mathbb{C})
\end{equation} act on the basis \ref{eq:630} as
\begin{equation}
  \label{eq:632}
  P_k^{(n)}(b) \to (g\cdot P_k^{(n)})(b) = P_k^{(n)}(b^T\cdot g) = \sum_{l=0}^n
\tilde{D}_{kl}(g)\,P_k^{(n)}(b)\,,
\end{equation} and thus induce a $(n+1)$ dimensional representation of $GL(2,
\mathbb{C})$ on $V^{n+1}$ in terms of the matrices $\tilde{D}(g) = 
(\tilde{D}_{kl}(g))$. 

For $GL(2,\mathbb{C})= U(2)$ the representations are equivalent to unitary
ones: If one introduces the normalized  basis 
\begin{equation}
  \label{eq:633}
  e_k(b)=\frac{P_k^{(n)}(b)}{\sqrt{k!(n-k)!}}\,,~k=0,1,\ldots,n,
\end{equation}
and replaces the basis $P_k^{(n)}(b)$ in \ref{eq:632} by $e_k(b)$ then one
gets representation matrices
\begin{equation}
  \label{eq:634}
 D(g)= (D_{kl}(g))\,,~k,l =0,1,\ldots,n,
\end{equation} instead of $\tilde{D}(g)$.
If
$g=g_u \in U(2)$, then the matrices $S(g_u)\equiv D(g_u)$ 
are unitary. This procedure is well-known for $SU(2)$ where $n=2j$,
j: angular momentum. 

If we put
\begin{equation}
  \label{eq:635}
 v_j = \sum_{k=0}^n c_k(j)\,e_k(b)\,,\,j=1,2\,, 
\end{equation}
then we have the hermitian scalar product
\begin{equation}
  \label{eq:636}
 \langle v_2,v_1\rangle = \sum_{k=0}^n
\bar{c}_k(2)c_k(1)\,. 
\end{equation} The representations \ref{eq:632} are irreducible and the
matrices $D(g)$ are polynomials in the matrix elements $c_{jk}$
of the $2 \times 2$ matrices \ref{eq:631}. Of special interest are the
following cases: Let $g$ be the positive definite diagonal matrix
\begin{equation}
  \label{eq:637}
  g=g_a =\begin{pmatrix}a_1&0\\0&a_2 \end{pmatrix}\,,~a_j >0\,,\,j=1,2\,,
\end{equation} then $D(g_a)$ is diagonal and positive definite, too:
\begin{equation}
  \label{eq:638}
  D(g_a)=\,\text{diag}\,(a_2^n,a_1a_2^{n-1},\ldots,a_1^ka_2^{n-k},\ldots,
a_1^n)\,,
\end{equation} with
\begin{equation}
  \label{eq:629}
  \det D(g_a) = (\det g_a)^{n(n+1)/2}\,.
\end{equation}
Relations like \ref{eq:638} and \ref{eq:629} hold for arbitrary diagonal
matrices $g_d =\, \text{diag}\,(\lambda_1,\lambda_2)$. 

It  follows from \ref{eq:638} that
\begin{equation}
  \label{eq:639}
  D(g_a^s)=D^s(g_a)\,,~s \in \mathbb{R}\,.
\end{equation}
 One can also show that $D(g_h)$ is hermitian if $g_h$ is hermitian.
In our context the following special case of \ref{eq:639} is of interest:
\begin{equation}
  \label{eq:640}
  g_a = y= g^T\cdot g \,,\,g \in GL(2,\mathbb{R})\,,\, \det g > 0\,.
\end{equation} We then have from the hermiticity of $D(y)$ and \ref{eq:639}
that
\begin{equation}
  \label{eq:641}
  \langle D(y^{1/2})\,v,D(y^{1/2}v \rangle = \langle v,D(y = g^T\cdot g)
\, v \rangle = \langle D(g)v,D(g)v \rangle \,.
\end{equation}
I am now ready to sketch Godement's construction (see also Ref. \cite{ros1}):

 In Sec.\ C.4  the symmetric $2 \times 2$ matrices of
 the Siegel upper
half plane $\mathbb{S}_2$ were denoted by $W=U+iV,\,V>0,$ and
 those of the Siegel unit
disc $\mathbb{D}_2$ by $Z=X+iY$. As we shall not discuss the disc here (see,
however, Ref.\ \cite{god2}), I switch the notation and use 
$Z=X+iY \in \mathbb{S}_2\,,\,,Y >0$ instead of $W$. 
The notation for the invariant 
volume element
\ref{deo} has to be changed accordingly. 

  Let $f(Z)$ be a matrix-valued
function on $\mathbb{S}_2$ with values in the complex vector space $V^{n+1}$
from above. Then a Hilbert space $ {\cal H}(D)$ may be defined in terms of
the scalar product
\begin{equation}
  \label{eq:642}
  (f,f) \equiv \int_{\mathbb{S}_2} d^3x\,d^3y \,(\det Y)^{-3} \langle
D(Y^{1/2})\cdot f(Z),D(Y^{1/2})\cdot f(Z)\rangle < \infty \,.
\end{equation} On this space the group elements (see
 \ref{eq:530}-\ref{eq:531b})
\begin{equation}
  \label{eq:643}
  w = \begin{pmatrix}A_{11} & A_{12} \\ A_{21} & A_{22} \end{pmatrix} 
\end{equation} act as
\begin{equation}
  \label{eq:644}
  f(Z) \to D(A_{21}\cdot Z+A_{22})^{-1} \cdot f[(A_{11}\cdot Z + A_{12})\cdot
(A_{21}\cdot Z + A_{22})^{-1}]\,.
\end{equation} Godement shows that these transformations are irreducible
and unitary. 

 In general the representation matrix $D(g)$ will be of the
type
\begin{equation}
  \label{eq:645}
  D(g) = (\det g)^{\alpha_2}\,\hat{D}(g)\,,
\end{equation}
where $\hat{D}(g)$ is again a polynomial in the matrix elements $c_{jk}$
of \ref{eq:631}. If $\alpha_2$ is not an integer, then one needs additional
discussions \cite{god}. 
\chapter{Estimates and asymptotic expansions of some functions}
\section{Proof that $I_{\nu+1}(x)/I_{\nu}(x) < 1$ } In Sec.\ 3.1.2 of
Ch.\ 3 the ratio \ref{eq:1062},
\begin{equation}
  \label{eq:646}
\rho_k(|z|) = I_{2k}(2|z|)/I_{2k-1}(2|z|)\,,  
\end{equation} plays a major role. I shall prove now that this ratio
is smaller than $1$ for all finite $|z|$. 

It follows from the relation \cite{wat1}
\begin{equation}
  \label{eq:647}
  x\, \frac{dI_{\nu}}{dx}(x) = \nu\,I_{\nu}(x) + x\,I_{\nu +1}
\end{equation}
 that
 \begin{equation}
   \label{eq:648}
  I_{\nu +1}(x)/I_{\nu}(x) = \frac{d}{dx} \ln (I_{\nu}(x)/x^{\nu}) \,.
 \end{equation}
As \cite{wat2}
\begin{equation}
  \label{eq:649}
  I_{\nu}(x) = \frac{x^{\nu}}{2^{\nu}\,\sqrt{\pi}\,\Gamma (\nu + 1/2)}\,
\int_0^{\pi} d\theta\, e^{x\, \cos \theta}\, \sin^{2 \nu} \theta\,,
\end{equation}
we get for \ref{eq:648}
\begin{equation}
  \label{eq:650}
  I_{\nu +1}(x)/I_{\nu}(x) =\frac{\int_0^{\pi} d\theta\,(\cos \theta)\, e^{x\, 
\cos \theta}\, \sin^{2 \nu} \theta}{
\int_0^{\pi} d\theta\, e^{x\, \cos \theta}\, \sin^{2 \nu} \theta} < 1\,.
\end{equation} This proves the assertion\footnote{The inequality \eqref{eq:650} is valid only for $k \geq 1/4$. In the 
interval $k \in (0,1/4)$ the ratio \eqref{eq:647} can be larger than $1\,$!}!
\section{Asymptotic expansion of an integral}
Next I want to prove the asymptotic expansion of the integral in Eq.\
\ref{eq:1085},
\begin{equation}
  \label{eq:651}
  l_k(|z|^2) = \frac{1}{\sqrt{\pi}}\int_0^{\infty}dt\,t^{-1/2}e^{-2kt}
g_k(|z|^2e^{-t})\,,
\end{equation} for large $|z|$ 
 which leads to the expansion \ref{eq:1087}.

From Eq.\ \ref{eq:1519} we have
\begin{equation}
  \label{eq:652}
  g_k(|z|^2) \asymp \frac{\Gamma(2k)}{2 \sqrt{\pi}}|z|^{1/2 -2k}\, 
e^{2|z|}\,[1-a_{-1}/|z| + O(|z|^{-2})]~\text{for}~|z| \to \infty\,.
\end{equation} Inserting this expression into the integral \ref{eq:651}
and making the change of variables
\begin{equation}
  \label{eq:653}
  1-u = e^{-t/2} 
\end{equation} leads to
\begin{align}
  \label{eq:654}
 l_k(|z|^2) \asymp \frac{\Gamma(2k)}{\sqrt{2} \pi}|z|^{1/2 -2k} 
e^{2|z|}   \int_0^1&du\,u^{-1/2}[-\frac{1}{u}\ln(1-u)]^{-1/2}
(1-u)^{2k-1/2} \times \nonumber
\\& \times e^{-2|z|u}\, [1-\frac{a_{-1}}{|z|}(1-u)^{-1} + O(|z|^{-2})]\,. 
\end{align} For a large $|z|$ expansion the value $u=0$ is the critical
point under the integral! 

Expanding
\begin{equation}
  \label{eq:655}
  [-\frac{1}{u}\ln(1-u)]^{-1/2} = 1-\frac{1}{4}\,u + O(u^2)\,,~
(1-u)^{2k-1/2} =1 -(2k-1/2)\,u +O(u^2)\,,
\end{equation}
and introducing $v= 2|z|u$ gives, up to next-to-leading order,
\begin{align}
  \label{eq:656}
  l_k(|z|^2)& \asymp \frac{\Gamma(2k)}{2 \pi}|z|^{1/2 -2k} 
e^{2|z|}(1-a_{-1}/|z|)\times \\ &\times |z|^{-1/2} \int_0^{2|z|}dv\,e^{-v}
v^{-1/2}[1-(2k-1/4)\frac{v}{2
|z|}]\,.  
\end{align}
Letting the upper limit $2|z|$ of the integral go to $\infty$ and
recalling that \cite{gam}
\begin{equation}
  \label{eq:657}
  \int_0^{\infty}dv\,e^{-v}\,v^{\mu-1}= \Gamma(\mu)\,,~\Gamma(\frac{1}{2}) =
\sqrt{\pi}\,,~\Gamma(\frac{3}{2}) = \Gamma(\frac{1}{2}+1)=
\sqrt{\pi}/2\,,
\end{equation} we obtain
\begin{equation}
  \label{eq:658}
  l_k(|z|^2) \asymp \frac{\Gamma(2k)}{2 \sqrt{\pi}}|z|^{1/2 -2k} 
e^{2|z|}(1-a_{-1}/|z|)\, |z|^{-1/2}\,[1-(2k-1/4)\frac{1}{4|z|}]\,. 
\end{equation}
Combining this with the relation \ref{eq:652} finally gives the
expansion \ref{eq:1087}
\section{Asymptotic expansions of certain series}
In Sec.\ 3.3.2 we encountered (Eqs.\ \ref{eq:173} and \ref{eq:1157})
 the functions
\begin{eqnarray}
  \label{eq:659}
  h_1^{(k)}(|\alpha|) &=& e^{-|\alpha|^2} \sum_{n=0}^{\infty} \sqrt{n+2k}
\frac{|\alpha|^{2n}}{n!}\,, \\
\label{eq:660} 
 h_2^{(k)}(|\alpha|) &=& e^{-|\alpha|^2} \sum_{n=0}^{\infty} \sqrt{(2k+n)(
2k+n+1)}
\frac{|\alpha|^{2n}}{n!}\,,
\end{eqnarray} and the need for their asymptotic expansions if $|\alpha|$
becomes large.  I used those expansions in Sec.\ 3.3.2 
without justifying them. This will be done now. 

The asymptotic expensions of the functions \ref{eq:659} and \ref{eq:660}
may be reduced to that of the series
\begin{equation}
  \label{eq:661}
  F_{a,s}(x) = \sum_{n=0}^{\infty} \frac{x^n}{n!\,(n+a)^s}\,,~x,a,s \in
 \mathbb{R} \,,\,>0\,.
\end{equation} How that reduction is to be done will be indicated below.

The asymptotic expansion of the series \ref{eq:661} was derived about
simultaneously by Hardy \cite{hard} and Barnes \cite{barn1} around 1905.
Both discussed the series for complex values of $x,a$ and $s$, too,
but we do not need the more general case here. Their nice result is
\begin{equation}
  \label{eq:662}
  F_{a,s}(x) \asymp \frac{x^{-s}\,e^x}{\Gamma(s)}[\sum_{n=0}^N c_n\,
\frac{\Gamma(s+n)}{x^n} + O(x^{-N-1})]\,,
\end{equation}
where  $c_n$ is the coefficient of $u^n$ in the Taylor expansion at
$u=0$ of the function
\begin{equation}
  \label{eq:663}
  f(u) = (1-u)^{a-1}[-\frac{1}{u}\ln(1-u)]^{s-1}=
 \sum_{n=0}^{\infty}c_n\,u^n\,.
\end{equation} For the first three $c_n$ one gets
\begin{eqnarray}
  \label{eq:664}
  c_0&=&1\,, \\ c_1&=& \frac{1}{2}(s+1) -a\,, \label{eq:665} \\
c_2 &=& \frac{1}{8}(s-1)(s-4a+14/3)+ \frac{1}{2}(a-1)(a-2)\,. \label{eq:666}
\end{eqnarray} For higher terms in the expansion of $f(u)$ it
 is helpful to know that
\begin{equation}
  \label{eq:667}
  [-\frac{1}{u}\ln(1-u)]^{\alpha} = 1+\alpha \sum_{n=0}^{\infty}\psi_n
(\alpha + n)\,u^{n+1}\,,
\end{equation} where the $\psi_n(y)$ are Stirling's polynomials \cite{niel}.
The first three of them are \cite{niel1}
\begin{eqnarray}
  \label{eq:668}
  \psi_0(y)&=& \frac{1}{2}\,,\\
\psi_1(y)&=& \frac{1}{4!}(3y+2)\,, \label{eq:669} \\
\psi_2(y)&=& \frac{y(y+1)}{4!\,2} \label{eq:670}\,.
\end{eqnarray}  (See also Ref.\ \cite{hans}.)

For the derivation of \ref{eq:664}-\ref{eq:666} one needs only
$\psi_0$ and $\psi_1$. 

Let us denote the sum in Eq.\ \ref{eq:659} by $\hat{h}_1^{(k)}(|\alpha|^2)$.
It may be rewritten as
\begin{equation}
  \label{eq:672}
  \hat{h}_1^{(k)}(|\alpha|^2)= \sum_{n=0}^{\infty}(2k+n)\frac{|\alpha|^{2n}}{
n!\,\sqrt{2k+n}}\,.
\end{equation} Expressed in terms of the function \ref{eq:661} this means
\begin{equation}
  \label{eq:673}
   \hat{h}_1^{(k)}(|\alpha|^2)= 2k\, F_{2k,1/2}(|\alpha|^2) + |\alpha|^2\,
F_{2k+1,1/2}(|\alpha|^2)\,.
\end{equation} The asymptotic expansion \ref{eq:1557} of $h_1^{(k)}(|\alpha|)$
 follows from the relations \ref{eq:673} and \ref{eq:662}.

If we denote the sum in Eq.\ \ref{eq:660} by $\hat{h}_2^{(k)}(|\alpha|^2)$, it
can be written as
\begin{equation}
  \label{eq:674}
  \hat{h}_2^{(k)}(|\alpha|^2) = \sum_{n=0}^{\infty} \frac{|\alpha|^{2n}}{n!}
\frac{(2k+n)(2k+n+1)}{\sqrt{2k+n}\sqrt{2k+n+1}}\,.
\end{equation} The asymptotic expansion \ref{eq:1171} of the series
 \ref{eq:674} can be
derived by using a generalization of the above procedure which is also
due to Barnes \cite{barn2}: 

For $n$ large enough we have
\begin{align}
  \label{eq:671}
  (1+2k+n)^{-1/2} &= (2k+n)^{-1/2}\,(1+\frac{1}{2k+n})^{-1/2} \,, \\
(1+\frac{1}{2k+n})^{-1/2} & \approx 1-\frac{1}{2}\frac{1}{2k+n} + \frac{3}{8}
\frac{1}{(2k+n)^2} + O[(2k+n)^{-3}]\,. \label{eq:675} 
\end{align} Inserting this into the series \ref{eq:674} yields
\begin{equation}
  \label{eq:676}
  \hat{h}_2^{(k)}(|\alpha|^2) \approx e^{|\alpha|^2}(|\alpha|^2 +2k +1/2)
-\frac{1}{8}\sum_{n=0}^{\infty}\frac{|\alpha|^{2n}}{n!}[\frac{1}{2k+n} +
O(1/(2k+n)^2)]\,. 
\end{equation} The asymptotic behaviour of the sum can  again be deduced
with the help of \ref{eq:662}, so that we finally obtain
\begin{equation}
  \label{eq:677}
   \hat{h}_2^{(k)}(|\alpha|^2) \asymp e^{|\alpha|^2}|\alpha|^2 [1 +(2k+1/2)
|\alpha|^{-2} -\frac{1}{8}|\alpha|^{-4} + O(|\alpha|^{-6}]\,.
\end{equation}
\end{appendix}

\begin{thebibliography}{300}
\addcontentsline{toc}{chapter}{\protect\numberline{}{Bibliography}}
\bibitem{rev1} P.\ Carruthers and M.M.\ Nieto, Rev.\ Mod.\ Phys. {\bf
 40}, 411 (1968)
\bibitem{pau1} H.\ Paul, Fortschr.\ Physik {\bf 22}, 657 (1974)
 \bibitem{rev2} Physica Scripta  {\bf  T48} (1993): Special issue on
 {\em Quantum Phase and Phase Dependent Measurements}, ed.\ by
 W.P.\ Schleich and S.M.\ Barnett
 \bibitem{rev2a} A.\ Luk\v{s} and V.\ Pe\v{r}inov\'{a}, Quantum
 Opt.\ {\bf 6}, 125 (1994)
 \bibitem{rev2b} U.\ Leonhardt and H.\ Paul, Prog.\ Quant.\
 Electr.\ {\bf 19}, 89 (1995)
 \bibitem{rev3} R.\ Lynch, Phys.\ Reports {\bf 256}, 367 (1995)
 \bibitem{rev4}
 M.\ Heni, M.\ Freyberger and W.P.\ Schleich, in {\em Coherence and
 Quantum Optics VII}, ed.\ by J.H.\ Eberly, L.\ Mandel and E.\ Wolf (Plenum
Press, New York and London, 1996), p.\ 239
\bibitem{rev5} D.A.\ Dubin, M.A.\ Hennings and T.B.\ Smith, Intern.\ Journ.\
Mod.\ Phys.\ B {\bf 9}, 2597 (1995)
\bibitem{rev6} D.T.\ Pegg and
S.M.\ Barnett, Journ.\ Mod.\ Optics {\bf 44}, 225 (1997)
\bibitem{rev8} V.V.\ Dodonov, J.\ Opt.\ B: Quantum Semiclass.\
Opt.\ {\bf 4}, R1 (2002); this review contains a wealth of
References!
\bibitem{texb1}
D.F.\ Walls and G.J.\ Milburn, {\em Quantum Optics}
(Springer-Verlag, Heidelberg etc., 1994)
 \bibitem{texb2} L.\
Mandel and E.\ Wolf, {\em Optical Coherence and Quantum Optics}
(Cambridge University Press, Cambridge etc., 1995)
\bibitem{texb3} U.\ Leonhardt, {\em Measuring the Quantum State of
Light} (Cambridge University Press, Cambridge etc., 1997)
\bibitem{texb4} P.\ Meystre and M.\ Sargent III, {\em Elements of Quantum
Optics}, 3rd ed.\ (Springer, Berlin etc., 1998)
\bibitem{texb5}
 V.\ Pe\v{r}inov\'{a}, A.\ Luk\v{s} and J.\ Pe\v{r}ina, {\em
Phase in Optics} (World Scientific Publ.\ Co., Singapore, 1998)
\bibitem{texb6} D.A.\ Dubin, M.A.\ Hennings and T.B. Smith, {\em Mathematical
Aspects of Weyl Quantization and Phase} (World Scientific Publ.\
Co.\,, Singapore, 2000)
 \bibitem{texb7} W.\ Vogel, D.-G.\ Welsch
and S.\ Wallentowitz, {\em Quantum Optics. An Introduction}, 2nd
ed.\ (Wiley-VHC Verlag, Weinheim, 2001)
\bibitem{texb8} W.P.\
Schleich, {\em Quantum Optics in Phase Space} (Wiley-VHC Verlag,
Weinheim, 2001)
\bibitem{texb9} R.R.\ Puri, {\em Mathematical Methods of Quantum Optics}
(Springer Series in Optical Sciences 79, Springer, Berlin etc., 2001)
\bibitem{is1}C.J.\ Isham  in {\em Relativity, Groups and Topology II}
 (Les Houches
Session XL, 1983), ed.\ by B.S.\ Dewitt and R.\ Stora (North-Holland,
Amsterdam etc., 1984), p.\ 1059
\bibitem{gui} V.\ Guillemin and S.\
Sternberg, {\em Symplectic techniques in physics} (Cambridge University
 Press, Cambridge etc., 1984; paperback edition: 1990)
\bibitem{ahar} Y.\ Aharonov and D.\ Bohm, Phys.\ Rev.\ {\bf 115}, 485 (1959);
\\ reviews are: \\
M.\ Peshkin and A.\ Tonomura, {\em The Aharonov-Bohm effect} (Lecture Notes
in Physics 340, Springer-Verlag, Berlin and Heidelberg, 1989); \\
J.\ Hamilton, {\em Aharonov-Bohm and other Cyclic Phenomena} (Springer Tracts
in Modern Physics 139, Springer-Verlag, Berlin etc., 1997)
\bibitem{dir1} P.A.M.\ Dirac, Proceed.\ Royal Soc.\ London, Ser.\
A, {\bf 114}, 243 (1927)
\bibitem{hei} W.\ Heitler, {\em The Quantum Theory of Radiation,
3rd Ed.} (Oxford Univ. Press, London, 1954; reprinted by Dover
Publications, New York, 1984), \S \,7.4
\bibitem{lon1} F.\ London, Zeitschr.\ f.\ Physik {\bf 37}, 915 (1926)
\bibitem{lon2} F.\ London, Zeitschr.\ f. Physik {\bf 40}, 193 (1927)
\bibitem{dir2} P.A.M.\ Dirac, Proceed.\ Royal Soc.\ London, Ser.\
A, {\bf 109}, 642 (1925); {\bf 110}, 561 (1926); in the first of these
papers Dirac, still a ``Senior Research Student'' at Cambridge, invented
the formalism of matrix mechanics all by himself, after he saw
Heisenberg's fundamental paper (Zeitschr.\ f.\ Physik {\bf 33}, 879 (1925));
he realized the close  relationship of the algebraic structure of 
the new formalism to
that of the Poisson brackets and called the commutators of the new
redefined canonical quantities ``quantum Poisson brackets''.
\bibitem{reed1} M.\ Reed and B.\ Simon, {\em Methods of Modern
Mathematical Physics II: Fourier Analysis, Self-Adjointness}
(Academic Press, New York, 1975), Ch. X;\\ as to the impossibility to
promote the classical phase to a self-adjoint operator see also
G.W.\ Mackey, {\em Mathematical Foundations of Quantum Mechanics}
(W.A.\ Benjamin, Inc., New York and Amsterdam, 1963), p.\ 103
\bibitem{suss} L.\ Susskind and J.\ Glogower, Physics {\bf 1},
49 (1964)
\bibitem{carr2} P.\ Carruthers and M.M.\ Nieto, Phys.\ Rev.\
Lett.\ {\bf 14}, 387 (1965)
\bibitem{shiop}
 K.\ Hoffmann, {\em Banach Spaces of Analytic Functions}
 (Prentice-Hall, Inc., Englewood Cliffs, N.J., 1962); \\
 B.\ Sz-Nagy and C.\ Foia\c{s}, {\em Harmonic analysis of
operators on Hilbert space} (North-Holland Publ.\ Co., Amsterdam,
1970); \\
 W.\ Rudin, {\em Real and Complex Analysis}, 3rd ed.\ 
(McGraw-Hill Book Co., N.Y.\ etc., 1987), Ch.\ 17; \\
 J.B.\ Conway, {\em The Theory of Subnormal Operators}
(Mathematical Surveys and Monographs 36, Amer.\ Math.\ Soc.,
Providence, R.\ I., 1991)
\bibitem{niet0} Ref.\ \cite{rev2}, p.\ 5
\bibitem{hel} S.\ Helgason, {\em Differential Geometry, Lie Groups
and Symmetric Spaces} (Academic Press, New York etc., 1978), Ch.\
III
\bibitem{mes} See, e.g. A.\ Messiah, {\em Quantum Mechanics, vol.\
I} (North-Holland Publ.\ Co., Amsterdam, 1961), p.\ 442; \\ R.M.\
Wilcox, Journ.\ Math.\ Phys.\ {\bf 8}, 962 (1967); \\ Formula \ref{eq:25}
is a very special case of the general relation $\exp(A)\circ \exp(B)= \exp[
C(A,B)]$, where the operators $A$ and $B$ are given. \\ The most elaborate 
discussion of calculating the exponent $C$ in terms of $A$ and $B$ is
due to Dynkin: \\
E.B.\ Dynkin, Uspekhi Matem.\ Nauk. (N.S.) {\bf 5}, 135 (1950); Engl.\ transl.:
{\em Normed Lie Algebras and Analytic Groups} (Amer. Mathem.\
Soc.\ Translations No.\ 97, Amer.\ Mathem.\ Soc.\, Providence, R.I., 1953); \\
 Dynkin's early 2 papers (in Russian) on the subject are: 
Doklady Akad.\ Nauk SSSR (N.S.) {\bf 57}, 323 (1947); Mat.\ Sbornik (N.S.)
{\bf 25}, 155 (1949).
 \\ Similarly elaborate is 
J.-P.\ Serre, {\em Lie Algebras and Lie Groups, 1964 Lectures Given at
Harvard University} (W.A.\ Benjamin, Inc., New York and Amsterdam, 1965),
Ch.\ IV, \S 7 \\ The inverse (or ``dual'') formula, namely $\exp(A+B) =
\exp(A)\circ \exp(B)\circ \exp(C_2)\circ \exp(C_3)\circ \cdots \exp(C_n) \cdots
$, where the $C_n$ are to be expressed in terms of $A$ und $B$, was
 investigated by Zassenhaus (unpublished); the ``Zassenhaus formula''
 has been discussed by 
W.\ Magnus, Comm.\ Pure and Appl.\ Mathem.\ {\bf VII}, 649 (1954) and \\
Wilcox, l.c.
\bibitem{ish3} V.\ Bargmann, Ann.\ Mathematics {\bf 59}, 1 (1954); \\
 C.M.\ Isham, Ref.\ \cite{is1}; \\
 P.\ Goddard and D.\ Olive, Intern.\ J.\ of Mod.\
Physics A {\bf 1}, 303 (1986)
\bibitem{wey} H.\ Weyl, Zeitschr.\ f.\ Phys.\ {\bf 46}, 1 (1927); \\
idem, {\em Gruppentheorie und Quantenmechanik}, 2.\ Aufl.\
(S.\ Hirzel Verlag, Leipzig, 1931); Engl.\ translation: {\em The
Theory of Groups and Quantum Mechanics} (Dover Publications, Inc.,
New York, 1950), Ch.\ IV, \S 14
\bibitem{sto} M.H.\ Stone, Proc.\
Nat.\ Acad.\ Sci.\ US {\bf 15}, 198 and 423 (1929); {\bf 16},
 172 (1930); (the last of these three communications contains the
celebrated proof); Stone's more general results are discussed
in his monograph {\em
Linear Transformations in Hilbert Space and Their Applications to
Analysis} (Am.\ Math.\ Soc.\ Coll.\ Publ.\ XV, Providence, R.I., 1932); \\
J.\ von Neumann, Math.\ Annalen {\bf 104}, 570 (1931);\\ the subject is
discussed in detail by C.R.\
Putnam, {\em Commutation Properties of Hilbert Operators and
Related Topics} (Springer-Verlag, Berlin, Heidelberg, New York,
1967); \\
see also the illuminating comments on that important theorem by
G.W.\ Mackey in {\em Functional Analysis and Related Fields}, ed.\
by F.E.\ Browder, (Springer-Verlag, Berlin etc., 1970), p.\ 132. \\
As to the harmonic analysis of the Weyl-Heisenberg group see: \\
 R.\ Howe, Bull.\
Amer.\ Math.\ Soc.\ (New Ser.) {\bf 3}, 821 (1980); \\
G.B.\ Folland, {\em Harmonic Analysis in Phase Space} (Ann.\ Mathem.\
Studies, no.\ 122, Princeton
University Press, Princeton, N.J., 1989); \\
E.M.\ Stein, {\em Harmonic Analysis: Real-Variable Methods, Orthogonality,
and Oscillatory Integrals} (Princeton University Press, Princeton, N.J.,
1993, 2nd printing 1995), Chs.\ XII and XIII
\bibitem{hela} \cite{hel}, Ch.\ II
\bibitem{lo} R.\ Loll, Phys.\ Rev.\ D {\bf 41}, 3785 (1990)
\bibitem{bo} M.\ Bojowald, H.A.\ Kastrup, F.\ Schramm and T.\ Strobl,
  Phys.\ Rev.\ D {\bf 62}, 044026 (2000) = e-print gr-qc/9906105
\bibitem{ka2} H.A.\ Kastrup, Ann.\ Physik (Leipzig)
{\bf 9}, 503 (2000) = e-print gr-qc/9906104
 \bibitem{ka1} H.A.\ Kastrup, e-print quant-ph/0005033
\bibitem{ka3} H.A.\ Kastrup, report CERN-TH/2001-238 = e-print quant-ph/0109013
\bibitem{ka4} H.A.\ Kastrup, in {\em Group 24: Physical and Mathematical Aspects of Symmetries},
Proceed.\ 24th Intern.\ Coll.\ Group Theor.\ Meth.\ Phys., Paris, July 2002, 
Inst.\ Phys.\ Conf.\ Ser.\ 173, ed.\ by J.-P.\ Gazeau et al.\ (Inst.\ Phys.\ Publishing, Bristol
and Philadelphia, 2003)\,p.\ 661
\bibitem{lou} W.H.\ Louisell, Phys.\ Lett.\ {\bf 7}, 60 (1963); \\ in the same 
year G.W.\ Mackey suggested the use of $\cos\vp$ and $\sin\vp$ instead of
$\vp$ itself for mathematical reasons: see Mackey, Ref.\ \cite{reed1}, p.\ 103
 \bibitem{is2} See Isham, Ref.\ \cite{is1}, Ch.\ 4.6
\bibitem{suga} Ref.\ \cite{sug}, Ch.\ IV
\bibitem{orbi} Orbifolds were introduced as ``V-manifolds'' by \\
I.\ Satake, Proceed.\ Nat.\ Acad.\ Science USA {\bf 42}, 359 (1956);
Journ.\ Math.\ Soc.\ Japan {\bf 9}, 464 (1957); \\
the term {\em ``orbifold''} was coined in Thurston's 
1976/77 Princeton lectures. The lecture notes are available under 
http://www.msri.org/publications/books/gt3m/, where or\-bi\-folds
 are discussed
in Ch.\ 13.  \\  In the first  volume of the printed version on
the subject of those lectures the term orbifold does not appear,
 but the concept itself is
discussed in a fascinating variety of examples: \\
W.P.\ Thurston, ed.\ by S.\ Levy, {\em Three-dimensional Geometry and
Topology, vol.\ 1} (Princeton Univ.\ Press, Princeton, N.J., 1997); \\ 
for a brief modern introduction to orbifolds see \\
R.H.\ Cushman and L.M.\ Bates, {\em Global Aspects of Classical Integrable
Systems} (Birkh\"auser Verlag, Basel, Boston and Berlin, 1997), Appendix B,
Sec.\ 2.3; \\
orbifolds play a larger role in string theories: \\
J.\ Polchinski, {\em String Theory I} (Cambridge Univ.\ Press,
 Cambridge, 1998), Sec.\ 8.5; \\ as to the state of the art see \\
{\em Orbifolds in mathematics and physics}, Proceed.\ Conf.\ Mathem.\ Aspects
of Orbifold String Theory, Univ.\ Wisconsin May 2001, ed.\ by A.\ Adem, 
J.\ Morava and Y.\ Ruan (Contemp.\ Mathem.\ 310, Amer.\ Mathem.\ Soc.,
Providence, R.I., 2002)
\bibitem{prok1} L.V.\ Prokhorov, Sov.\ Journ.\ Nucl.\ Phys.\ {\bf 35}, 129 
(1982); \\
S.V.\ Shabanov, Theor.\ Mathem.\ Phys.\ {\bf 78}, 292 (1989); \\
L.V.\ Prokhorov and S.V.\ Shabanov, Phys.\ Lett.\ B {\bf 216}, 341 (1989);\\
S.V.\ Shabanov, Intern.\ Journ.\ Mod.\ Phys.\ A {\bf 6}, 845 (1991);\\
L.V.\ Prokhorov and S.V.\ Shabanov, Sov.\ Phys.\ Uspekhi {\bf 34}, 108 (1991)
\bibitem{sha1} S.V.\ Shabanov, Phys.\ Reports {\bf 326}, 1 (2000)  
\bibitem{mlo} The realization \ref{eq:45} was apparently first discussed
by L.D.\ Mlodinow and N.\ Papanicoulaou, Ann.\ Phys.\ (N.Y.) {\bf
128}, 314 (1980); see also \\ C.C.\ Gerry, J.\ Phys.\ A: Math.\
Gen.\ {\bf 16}, L1 (1983); \\ J.\ Katriel, A.I.\ Solomon, G.\
D'Ariano and M.\ Rasetti, Phys.\ Rev.\ D {\bf 34}, 2332 (1986); \\
H.\ Bacry, Journ.\ Mathem.\ Phys.\ {\bf 31}, 2061 (1990); 
\\ C.C.\ Gerry and R.\ Grobe, Quantum Semiclass.\ Opt.\ {\bf 9},
59 (1997);
\\ A.\ W\"{u}nsche, Acta physica slov.\ {\bf 49}, 771 (1999); Journ.\
Opt.\ B: Quantum Semiclass.\ Opt.\ {\bf 5}, S429 (2003)
\bibitem{ho} T.\ Holstein and H.\ Primakoff, Phys.\ Rev.\ {\bf
58}, 1098 (1940)
\bibitem{ba1} A.O.\ Barut and L.\ Girardello, Commun.\ math.\ Phys.\
{\bf 21}, 41 (1971)
\bibitem{per0} A.M.\ Perelomov, Commun.\ Math.\ Phys.\ {\bf 26},
 222 (1972); see also the Refs.\ \cite{per1,per}
\bibitem{schro1} E.\ Schr\"odinger, Die Naturwissenschaften {\bf 14}, 664
(1926)
\bibitem{glau1} R.J.\ Glauber, Phys.\ Rev.\ Lett.\ {\bf 10}, 84 (1963);
Phys.\ Rev.\ {\bf 130}, 2529 (1963) and {\bf 131}, 2766 (1963); \\
Mathematical and field theoretical properties of what is now called
``coherent'' states where extensively discussed in \\
K.O.\ Friedrichs, {\em Mathematical Aspects of the Quantum Theory of
Fields} (Interscience Publ., Inc., New York, 1953);
\\ Coherent states in their special form \ref{eq:1527} were 
 introduced  by J.R.\ Klauder
in connection with problems in quantum field theory (Ann.\ Phys.\ (N.Y.)
 {\bf 11}, 123 (1960)). Their importance
for quantum optical problems was first and thoroughly discussed by Glauber.
\bibitem{phst} See, e.g.\ the  following textbooks and reviews with many
References: Ref.\ \cite{texb3}, Ch.\ 6.3; Ref.\
\cite{texb5}, Ch.\ 4.7; Ref.\ \cite{texb6}, Ch.\ 10; Ref.\
\cite{texb7}, Ch.\ 3.5; Ref.\ \cite{texb8}, Ch.\ 8.5
\bibitem{mil} G.J.\ Milburn, J.\ Phys.\ A: Math.\ Gen.\ {\bf 17},
737 (1984)
\bibitem{fish} R.A.\ Fisher, M.M.\ Nieto and V.D.\ Sandberg,
Phys.\ Rev.\ D {\bf 29}, 1107 (1984)
 \bibitem{wod} K.\ W\'{o}dkiewicz and J.H.\ Eberly,
Journ.\ Opt.\ Soc.\ Am.\ B {\bf 2}, 458 (1985);
K.\ W\'odkiewicz, Journ.\ Mod.\ Opt.\ {\bf 34}, 941 (1987)
 \bibitem{schu} B.L.\ Schumaker and C.M.\ Caves, Phys.\ Rev.\ A {\bf
31}, 3093 (1985); \\ B.L.\ Schumaker, Phys.\ Reports {\bf 135}, 317
(1986)
\bibitem{ger1} C.C.\ Gerry, Phys.\ Rev.\ A {\bf 31}, 2721 (1985);
{\bf 35}, 2146 (1987); {\bf 38}, 1734 (1988)
 \bibitem{bis} R.F.\ Bishop and A.\ Vourdas,
Journ.\ Phys.\ A: Math.\ Gen.\ {\bf 19}, 2525 (1986) and {\bf 20},
3727 (1987)
\bibitem{squ} E.H.\ Kennard, Zeitschr.\ Physik {\bf 44}, 326 (1927); \\
H.\ Takahashi, in {\em Advances in Communication Systems, Theory and
Applications, vol.\ 1}, ed.\ by A.V.\ Balakrishnan (Academic Press, New York
and London, 1965), p.\ 227; \\
 D.\ Stoler, Phys.\ Rev.\ D {\bf 1}, 3217 (1970); D {\bf 4}, 1925 (1971); \\
E.Y.C.\ Lu, Lett.\ Nuovo Cim.\ {\bf 2}, 1241 (1971); {\bf 3}, 585 (1972); \\
H.P.\ Yuen, Phys.\ Rev.\ A {\bf 13}, 2226 (1976); \\
 J.N.\ Hollenhorst, Phys.\ Rev.\ D {\bf 19}, 1669 (1979); \\
 For more see  the textbooks \cite{texb1}-\cite{texb9} and
the vast literature on squeezed states quoted by Dodonov in
Ref.\ \cite{rev8}; see also \\
Y.S.\ Kim and M.E.\ Noz, {\em Phase Space Picture of Quantum Mechanics}
(Lecture Notes in Physics Series, vol.\ 40, World Scientific, Singapore,
1991); \\
early reviews on squeezed states are \\
D.F.\ Walls, Nature {\bf 306}, 141 (1983); \\ R.\ Loudon and P.L.\
Knight, Journ.\ Mod.\ Optics {\bf 34}, 709 (1987); \\ R.W.\ Henry
and S.C.\ Glotzer, Am.\ J.\ Phys.\ {56}, 318 (1988)
\bibitem{coh1} R.J.\ Glauber, in {\em Quantum Optics and Electronics}
(XIV.\ Les Houches Summer School 1964), ed.\ by C.\ DeWitt, A.\
Blandin and C.\ Cohen-Tannoudji (Gordon and Breach, New York, 1965), 
p.\ 63; \\ idem, in {\em Fundamental Problems in Statistical Mechanics II}
(Proceed.\ 2nd NUFFIC Intern.\ Summer Course at Noordwijk, The Netherlands,
20 June - 8 July 1967), ed.\ by E.G.D.\ Cohen (North-Holland Publ.\ Co., 
Amsterdam, 1968), p.\ 140
\bibitem{coh1a} J.R.\ Klauder and E.C.G.\ Sudarshan, {\em
Fundamentals of Quantum Optics} (W.A.\ Benjamin, Inc., New York,
Amsterdam, 1968)
\bibitem{coh1b} J.R.\ Klauder and B.-S.\ Skagerstam, {\em
Coherent States -- Applications in Physics and Mathematical
Physics} (World Scientific Publ.\ Co., Singapore, 1985)
\bibitem{coh2} W.-M.\ Zhang,
D.H.\ Feng and R.\ Gilmore, Rev.\ Mod.\ Phys.\ {\bf 62}, 867
(1990)
\bibitem{coh3a} S.T.\ Ali, J.-P.\ Antoine, J.-P.\ Gazeau
and U.A.\ Mueller, Rev.\ Mathem.\ Phys.\ {\bf 7}, 1013 (1995)
\bibitem{coh3} D.A.\ Trifonov,
J.\ Opt.\ Soc.\ Amer. A {\bf 17}, 2486 (2000) =
quant-ph/0012072
\bibitem{coh4} B.C.\ Hall, Contemp.\ Mathem.\ {\bf 260}, 1 (2000) =
quant-ph/9912054
\bibitem{ar} E.\ Arthurs and J.L.\ Kelly, Jr., Bell System
Technical J.\ {\bf 44}, 725 (1965); \\ H.P.\ Yuen, Phys.\ Lett.\
 {\bf 91 A}, 101 (1982); \\ Y.\ Yamamoto
and H.A.\ Haus, Rev.\ Mod.\ Phys.\ {\bf 58}, 1001 (1986);  \\
S.L.\ Braunstein, C.M.\ Caves and G.J.\ Milburn,
Phys.\ Rev.\ A {\bf 43}, 1153 (1991); \\
S.\ Stenholm, Ann.\ Physics (NY) {\bf 218}, 233 (1992);\\
U.\ Leonhardt and H.\ Paul, Journ.\ Mod.\ Optics {\bf 40}, 1745
(1993); \\ M.G.\ Raymer, Amer.\ J.\ Phys.\ {\bf 68}, 986 (1994);
\\ M.\ Freiberger, M.\ Heni and
W.P.\ Schleich, Quantum Semiclass.\ Opt.\ {\bf 7}, 187 (1995); \\
U.\ Leonhardt, B.\ B\H{o}hmer and H.\ Paul, Opt.\ Commun.\ {\bf
119}, 296 (1995) 
\bibitem{er1} A.\ Erd\'{e}lyi et al.\ (Eds.), {\em Higher Transcendental
 Functions I} (McGraw-Hill Book Co., Inc., New York etc., 1953), here p.\ 52
\bibitem{span} J.\ Spanier and K.B.\ Oldham, {\em An Atlas of Functions}
(Hemisphere Publ.\ Corpor.\ - Taylor \& Francis Group - , New York etc., 1987),
Ch.\ 18
 \bibitem{er2} A.\ Erd\'{e}lyi et al.\ (Eds.), {\em Higher Transcendental
 Functions II} (McGraw-Hill Book Co., Inc., New York etc., 1953), Ch.\ VII
 \bibitem{er1c} Ref.\ \cite{er1}, Ch.\ IV
 \bibitem{hil} E.\ Hille, {\em Analytic Function Theory II}, 2nd Ed.\, (Chelsea
 Publ.\ Co., New York, 1987), Ch.\ 14
\bibitem{GR} Ref.\ \cite{er2}, here p.\ 51, Eq.\ (27)
\bibitem{aro} N.\ Aronszajn, Transact.\ Amer.\ Math.\ Soc.\
 {\bf 68}, 337 (1950)
 \bibitem{barg2} V.\ Bargmann, Commun.\ Pure and Appl.\
Math.\ {\bf 14}, 187 (1961); {\bf 20}, 1 (1967)
 \bibitem{neeb} K.-H.\ Neeb, {\em Holomorphy and Convexity in Lie
 Theory} (de Gruyter Expos.\ in Math.\ 28; Walter de Gruyter, Berlin, 1999),
here Ch.\ I; see also \\
H.\ Hedenmalm, B.\ Korenblum and K.\ Zhu, {\em Theory of Bergman Spaces} 
(Graduate Texts in Math.\ 199; Springer-Verlag Inc., New York, 2000), here
Ch.\ 9
\bibitem{er4} Ref.\ \cite{er1}, here p.\ 251, Eq.\ (13)
\bibitem{er5} Ref.\ \cite{er1}, Ch.\ VI
 \bibitem{seg} I.\ Segal, {\em Mathematical Problems of
 Relativistic Physics} (Lectures in Applied Mathematics, Proceed.\ 
Summer Seminar, Boulder,
 Colorado, 1960, ed.\ by M.\ Kac) vol.\ II, (Amer.\ Math.\ Soc.,
 Providence, R.I., 1963); Illinois Journ.\
 Math.\ {\bf 6}, 500 (1962)
\bibitem{pau2} H.\ Paul, {\em Photonen}, 2., durchges.\ Aufl.\ (B.G.\ Teubner,
Stuttgart und Leipzig, 1999), p.\ 179
 \bibitem{mand1} See Ref.\ \cite{texb2}\,, p.\ 627
 \bibitem{agar1} G.S.\ Agarwal, J.\ Opt.\ Soc.\ Amer.\ B {\bf 5},
 1940 (1988)
 \bibitem{buz1} V.\ Bu\v{z}ek, Journ.\ Mod.\ Opt.\ {\bf 37}, 303
 (1990)
 \bibitem{agar2} A.\ Joshi and R.R.\ Puri, Phys.\ Rev.\ A {\bf
 42}, 4336 (1990)
 \bibitem{gerr1} C.C.\ Gerry and R.F.\ Welch, J.\ Opt.\ Soc.\
 Amer.\ B {\bf 8}, 868 (1991)
 \bibitem{ban0} M.\ Ban, Journ.\ Math.\ Phys.\ {\bf 33}, 3213 (1992); 
Journ.\ Opt.\ Soc.\ Amer.\ B {\bf 10}, 1347 (1993); 
 Phys.\ Rev.\ A {\bf 47}, 5093 (1993) (the exponent in
Eq.\ (2.6) has to be $-2$ instead of $2$)
 \bibitem{trif1} D.A.\ Trifonov, Journ.\ Math.\ Phys.\ {\bf 35},
 2297 (1994)
 \bibitem{agar3} G.S.\ Prakash and G.S.\ Agarwal, Phys.\ Rev.\ A
 {\bf 50}, 4258 (1994)
 \bibitem{ban1}M.\ Ban, Phys.\ Lett.\ A {\bf 193}, 121 (1994)
 \bibitem{brif1} C.\ Brif and Y.\ Ben-Aryeh, Quantum Opt.\ {\bf
 6}, 391 (1994); 
  J.\ Phys.\ A: Math.\
 Gen.\ {\bf 27}, 8185 (1994)
 \bibitem{agar4} B.A.\ Bambah and G.S.\ Agarwal, Phys.\ Rev.\ A {\bf 51}, 4918
 (1995)
 \bibitem{gerr2a} C.C.\ Gerry and R.\ Grobe, Phys.\ Rev.\ A {\bf
 51}, 1698 (1995)
 \bibitem{gerr2} C.C.\ Gerry and R.\ Grobe, Phys.\ Rev.\ A {\bf
 51}, 4123 (1995)
 \bibitem{brif3} C.\ Brif, Quantum Semiclass.\ Opt.\ {\bf 7}, 803
 (1995);  Ann.\ Phys.\ (N.Y.) {\bf 251}, 180 (1996); \\
 Intern.\ Journ.\ Theor.\ Phys.\ {\bf
 36}, 1651 (1997)
 \bibitem{brif4} C.\ Brif and A.\ Mann, Phys.\ Lett.\ A {\bf 219},
 257 (1996); \\
 Quantum Semiclass.\ Opt.\
 {\bf 9}, 899 (1997)
 \bibitem{brif5} C.\ Brif, A.\ Vourdas and A.\ Mann, J.\ Phys.\ A:
 Math.\ Gen.\ {\bf 29}, 5873 and 5587 (1996)
 \bibitem{trif2} D.A.\ Trifonov, J.\ Phys.\ A: Math.\ Gen.\ {\bf
 30}, 5941 (1997);  {\bf 31}, 5673 (1998)
 \bibitem{pera1} V.\ Pe\v{r}inov\'{a}, A.\ Luk\v{s} and J.\
 K\v{r}epelka, J.\ Opt. B: Semiclass.\ Opt.\ {\bf 2}, 81 (2000)
 \bibitem{wan1} X.\ Wang, B.\ Sanders and S.-h.\ Pan,
J.\ Phys.\ A: Math.\ Gen.\ {\bf 33}, 7451 (2000)
\bibitem{peri1} M.S.\ Abdalla, F.A.A.\ El-Orany and J.\
Pe\v{r}ina, Acta Phys.\ Slov.\ {\bf 50}, 613 (2000)
\bibitem{er3} Ref.\ \cite{er1}, p.\ 52, formulae (5) and (6)
\bibitem{span1} Ref.\ \cite{span}, p.\ 45, formula 6:5:1 and p.\  150,
formula 18:3:3
\bibitem{hel0} S.\ Helgason, {\em Differential Geometry and Symmetric
Spaces} (Academic Press,  New York and London, 1962), p.\ 316, Lemma 7.7; 
idem, Ref.\ \cite{hel}, p.\ 387, Lemma 7.7
 \bibitem{gil} F.T.\ Arecchi, E.\ Courtens, R.\ Gilmore and H.\
 Thomas, Phys.\ Rev.\ A {\bf 6}, 2211 (1972); \\
  R.\ Gilmore, Revista Mexic.\ Fis.\ {\bf 23}, 143 (1974);
  Journ.\ Math.\ Phys.\ {\bf 15}, 2090 (1974); \\
idem, {\em Lie Groups, Lie Algebras, and Some of Their Applications}
(John Wiley \& Sons, New York etc., 1974), Ch.\ 5, Sec.\ VI
 \bibitem{per1} A.M.\ Perelomov, Sov.\ Phys.\ Usp.\ {\bf 20}, 703
 (1978)
 \bibitem{wue1} A.\ W\"unsche, Ref.\ \cite{mlo}
\bibitem{wan} X.-G.\ Wang, Intern.\ Journ.\ Mod.\ Phys.\ B {\bf
 14}, 1093 (2000)
 \bibitem{sud} V.I.\ Man'ko, G.\ Marmo, E.C.G.\ Sudarshan and F.\
 Zaccaria, in {\em Proceed.\ of the IV.\ Wigner Symposium,
 Guadalajara 1995)\,}, ed.\ by N.\ Atakishiyev, T.\ Seligman and K.B.\ Wolf
  (World Scientific, Singapore, 1996), p.\ 421;
 Physica Scr.\ {\bf 55}, 528 (1997)
 \bibitem{vog} R.L.\ de Matos Filho and W.\ Vogel, Phys.\ Rev.\ A {\bf 54},
 4560  (1996)
 \bibitem{lern2} E.C.\ Lerner, H.W.\ Huang and G.E.\ Walters, J.\
 Math.\ Phys.\ {\bf 11}, 1679 (1970) \\
 Y.\ Aharonov, H.W.\ Huang, J.M.\ Knight and E.C.\
 Lerner, Lett.\ Nuovo Cim.\ {\bf 2}, 1317 (1971); \\ Y.\ Aharonov,
 E.C.\ Lerner, H.W.\ Huang and J.M.\ Knight,  J.\ Math.\ Phys.\
 {\bf 14}, 746 (1973)
 \bibitem{lebl} J.-M.\ L\'{e}vy-Leblond, Ann.\ Phys.\ (N.Y.) {\bf
 101}, 319 (1976)
 \bibitem{sha} J.H.\ Shapiro and S.R.\ Shepard, Phys.\ Rev.\ A
 {\bf 43}, 3795 (1991); \\ G.S.\ Agarwal, Phys.\ Rev.\ A {\bf 44},
 8398 (1991)
 \bibitem{vour1} A.\ Vourdas, Phys.\ Rev.\ A {\bf 45}, 1943
 (1992);
 \\ G.S.\ Agarwal, Opt.\ Commun.\ {\bf 100}, 479 (1993)
\bibitem{agar4a} G.S.\ Agarwal, Phys.\ Rev.\ A {\bf 45}, 1787
(1992)
 \bibitem{er1a} See Ref.\ \cite{er1}, here p.\ 9
 \bibitem{basu} D.\ Basu, Proc.\ Roy.\ Soc. London A {\bf 455}, 975
 (1999); see also the Refs.\ therein
\bibitem{span2} Ref.\ \cite{span}, p.\ 151, formula 18:5:6
\bibitem{er1b} See Ref.\ \cite{er1}, here p.\ 27-31
\bibitem{lew}  L.\ Lewin, {\em Polylogarithms and Associated Functions}
(North Holland,  New York, Oxford, 1981); see also \\
http://functions.wolfram.com/ZetafunctionsandPolylogarithms
\bibitem{er3a} See Ref.\ \cite{er1}, p.\ 115
\bibitem{barn3} E.W.\ Barnes, Proceed.\ London Mathem.\ Soc.\ (Ser.\ 2)
{\bf 4}, 284 (1906)
\bibitem{hard2} See Ref.\ \cite{shiop} and \\
H.\ Dym and H.P.\ McKean, {\em Fourier Series and Integrals} (Probability
and Mathematical Statistics 14, Academic Press, New York and London, 1972), 
Ch.\ 3; \\
J.B.\ Garnett, {\em Bounded Analytic Functions} (Academic Press, New
 York etc., 1981); \\
P.\ Koosis, {\em Introduction to $H_p$ Spaces}, 2nd ed.\ (Cambridge Tracts
in Mathematics 115, Cambridge University Press, Cambridge, 1998), Ch.\ VI
\bibitem{mesa} See, e.g.\ Messiah, Ref.\ \cite{mes}\,, Ch.\ XII and Appendix
B III
\bibitem{garr} J.C.\ Garrison and J.\ Wong, Journ.\ Math.\ Phys.\
{\bf 11}, 2242 (1970); the imaginary part in the exponent should, however, be $-i|\alpha|^2(\vp+\beta)$
\bibitem{jost} R.\ Jost, {\em The General Theory of Quantized
Fields} (Amer.\ Math.\ Soc.\, Providence, Rhode Island, 1965), Ch.\
II.3
\bibitem{barn} D.T.\ Pegg and S.M.\ Barnett, Europhys.\ Lett.\
{\bf 6}, 483 (1988);\\ S.M.\ Barnett and D.T.\ Pegg, Journ.\ Mod.\
Opt.\ {\bf 36}, 7 (1989);\\ D.T.\ Pegg and S.M.\ Barnett, Phys.\
Rev.\ A {\bf 39}, 1615 (1989)
\bibitem{dub} Ref.\ \cite{rev5}, Ch.\ 15 and Ref.\ \cite{texb6},
Chs.\ 10 and 16;\\  see also K.\ Fujikawa, Phys.\ Rev.\ A {\bf 52}, 3299 (1995) and
\\ M.G.A.\ Paris, Fizika B {\bf 6}, 63 (1997)
\bibitem{grad} See, e.g.\ I.S.\ Gradsteyn and I.M.\ Ryzhik, A.\
Jeffrey (Editor), D.\ Zwillinger (Assoc.\ Ed.), {\em Table of
Integrals, Series, and Products}, 6th ed.\ (Academic Press, San
Diego etc., 2000), p.\ 62
\bibitem{cau1} N.L.\ Johnson and S.\ Kotz, {\em distributions in statistics,
continuous univariate distributions - 1} (Houghton Mifflin Co., Boston etc.,
1970), Ch.\ 16; \\
W.\ Feller, {\em An Introduction to Probability Theory and Its Applications,
vol.\ II}, 2nd ed.\ (John Wiley \& Sohns, Inc., New York etc., 1971); \\
M.\ Fisz, {\em Probability Theory and Mathematical Statistics}, 3rd ed.\
(Robert E.\ Krieger Publ.\ Co., Malabar, Florida, 1980), Sec.\ 5.10; \\
A.\ Papoulis and S.U.\ Pillai, {\em Probability, Random Variables,
 and Stochastic
Processes}, 4th ed.\ (McGraw-Hill Co., Boston etc., 2002)
\bibitem{vank} N.G.\ van Kampen, {\em Stochastic Processes in Physics and
Chemistry} (North-Holland Publ.\ Co., Amsterdam etc., 1981)
\bibitem{caup} See, e.g.\ Feller, Ref.\  \cite{cau1}; \\
C.W.\ Gardiner, {\em Handbook of Stochastic Methods for Physics, Chemistry
and the Natural Sciences}, 2nd ed.\ (Springer-Verlag, Berlin etc., 1985); \\
K.-I.\ Sato, {\em L\'{e}vy Processes and Infinitely Divisible Distributions}
(Cambridge Studies in Advanced Mathematics 68, Cambridge Univ.\ Press, 
Cambridge, 1999); \\
P.\ Garbaczewski and R.\ Olkiewicz, Journ.\ Math.\ Phys.\ {\bf 41}, 6843
 (2000); see also the Refs.\ in this paper
\bibitem{cau2} see, e.g.\ Johnson and Kotz, Ref.\ \cite{cau1} and
Fisz, Ref.\ \cite{cau1}, Sec.\ 3.5
\bibitem{pal} See, e.g.\ Rudin, Ref.\ \cite{shiop}, Ch.\ 19; Garnett, Ref.\
\cite{hard2}, Ch.\ II; Koosis, Ref.\ \cite{hard2}, Ch.\ VI
\bibitem{ertrI} A.\ Erd\'{e}lyi et al., {\em Tables of Integral Transforms,
vol.\ I} (McGraw-Hill Book Co., Inc., New York etc., 1954), p.\ 119, formula
(12) (one has to take the limit $\mu + \nu \to 1/2$); or ibidem, p.\ 174,
formula (25)
\bibitem{lag} Ref.\ \cite{er2}, Ch.\ 10; \\
G.\ Szeg\"{o}, {\em Orthogonal Polynomials}, 4th  ed.\ 
 (Amer.\ Math.\ Soc.\ Publ. XXIII, Amer.\ Math.\ Soc., Providence, R.I.,
 1975), Ch.\ V
\bibitem{confl} See Ref.\ \cite{er1}, Ch.\ 6
\bibitem{mo} See Ref.\ \cite{magn}, p.\ 278
\bibitem{h2l2} See Ref.\ \cite{hard2}; 
the most explicit expressions are contained in \\
E.C.\ Titchmarsh, {\em Introduction to the Theory of Fourier Integrals},
2nd ed.\ (Oxford University Press, Oxford, 1948), Ch.\ V (there appears to
be a sign misprint in the exponents inside the Fourier integrals following
Theorem 98 on p.\ 131)
\bibitem{fher} See, e.g.\ Titchmarsh, Ref.\ \cite{h2l2}, Sec.\ 3.5
\bibitem{laplp2} See, e.g.\ Ref.\ \cite{ertrI}, p.\ 146, formula (21)
\bibitem{errf} See, e.g.\ Ref.\ \cite{magn}, Sec.\ 9.2.3; Ref.\ \cite{span}, 
Ch.\ 40
\bibitem{errf1} Ref.\ \cite{span}, Ch.\ 40
\bibitem{erdht2} A.\ Erdelyi et al., {\em Tables of Integral Transforms,
vol.\ II} (McGraw-Hill Book Co., Inc., New York etc., 1954), p.\ 293, formula
(4)
\bibitem{boa} Compared to Ref.\ \cite{bo} the phases of
$K_{\pm}$ and $\chi_{k,n}(\vp)$ have been chosen differently.
\bibitem{boy} C.P.\ Boyer and K.B.\ Wolf, Journ.\ Math.\ Phys.\ {\bf 16},
1493 (1975)
\bibitem{lern3} E.C.\ Lerner, Nuovo Cim.\ {\bf 56 B}, 183 (1968);
Errat. Nuovo Cim.\ {\bf 57 B}, 251 (1968)
\bibitem{lyn1} R.\ Lynch, Journ.\ Opt.\ Soc.\ Amer.\ B {\bf 3}, 1006 (1986)
\bibitem{mu} R.\ Simon and N.\ Mukunda, Optics Commun.\ {\bf 95},
39 (1993)
\bibitem{mer} E.\ Merzbacher, {\em Quantum Mechanics}, 2nd ed.\
(J.\ Wiley \& Sons, Inc., New York etc., 1970), Ch.\ 8, \S 8
\bibitem{stol} D.\ Stoler, Ref.\ \cite{squ} 
\bibitem{lu}E.Y.C.\ Lu, Ref.\ \cite{squ}
\bibitem{yue}H.P.\ Yuen, Ref.\ \cite{squ}
\bibitem{hol} J.N.\ Hollenhorst, Ref.\ \cite{squ}
\bibitem{rob} H.\ P.\ Robertson, Phys.\ Rev.\ {\bf 34}, 163
(1929); Phys.\ Rev.\ {\bf 35}, 667 (1930)
\bibitem{schro2} E.\ Schr\"odinger, Sitz.\ Berichte Preuss.\ Akad.\
Wiss.\,(Berlin), Phys.-Math.\ Klasse, {\bf 1930}, p.\ 296
\bibitem{jack} R.\ Jackiw, Journ.\ Math.\ Phys.\ {\bf 9}, 339
(1968)
\bibitem{jor} T.F.\ Jordan, {\em Linear Operators for Quantum
Mechanics} (John Wiley \& Sons, Inc., New York etc., 1969), Ch.\ V
\bibitem{mer1} Ref.\ \cite{mer}, Ch.\ 8, \S 6
\bibitem{dod} V.V.\ Dodonov, E.V.\ Kurmyshev and V.I.\ Man'ko, 
Phys.\ Lett.\ {\bf 79 A}, 150 (1980); \\ R.R.\ Puri and G.S.\ Agarwal, Phys.\ Rev.\ A
{\bf 53}, 1786 (1996); Intern.\ Journ.\ Mod.\ Phys.\ B {\bf 10}, 1563 and 3893 (1996); \\
R.R.\ Puri, Pramana Journ.\ Phys.\ {\bf 48}, 787 (1997); \\
G.S.\ Agarwal, Fortschr.\ Physik {\bf 50}, 575 (2002) 
\bibitem{mann1} C.\ Brif and A.\ Mann, Quantum Semiclass.\ Opt.\
{\bf 9}, 899 (1997)
\bibitem{hew} see, e.g.\ E.\ Hewitt and K.\ Stromberg, {\em Real
and Abstract Analysis} (Springer-Verlag, Berlin etc., 1969), p.\
234
\bibitem{lip}S.\ Goshen (Goldstein) and H.J.\ Lipkin, Ann.\ Phys.\
(N.Y.) {\bf 6}, 301 (1959); \\  H.J.\ Lipkin, {\em Lie Groups for
Pedestrians} (North-Holland Publ.\ Co., Amsterdam, 1965), Ch.\ 5
\bibitem{barg3a} This case is a special one in Bargmann's analysis
\cite{barg3} of  constructing  Lie algebras $\mathfrak{sp}(2n,\mathbb{R})$
in terms of creation and anihilation operators.
\bibitem{nied} U.\ Niederer, Helv.\ Phys.\ Acta {\bf 46}, 191 (1973)
\bibitem{mosh3} M.\ Moshinsky, SIAM Journ.\ Appl.\ Math.\ {\bf 25}, 193 (1973)
\bibitem{cush1} R.H.\ Cushman and L.M.\ Bates, Ref.\ \cite{orbi}, 
 here p.\ 315, Example 4
\bibitem{pari} C.\ Brif, A.\ Mann and A.\ Vourdas, Journ.\ Phys.\ A:
Math.\ Gen.\ {\bf 29}, 2053 (1996)
\bibitem{hill1} M.\ Hillery, Opt.\ Commun.\ {\bf 62}, 135 (1987);
Phys.\ Rev.\ A {\bf 36}, 3796 (1987); Phys.\ Rev.\ A {\bf 40},
3147 (1989)
\bibitem{ger2} C.C.\ Gerry and E.R.\ Vrscay, Phys.\ Rev.\ {\bf
37}, 1779 (1988)
\bibitem{hill2} D.\ Yu and M.\ Hillery, Quantum Opt.\ {\bf 6}, 37
(1994)
\bibitem{bied}W.J.\ Holman, III and L.C.\ Biedenharn, Jr., Ann.\ Phys.\
(N.Y.) {\bf 39}, 1 (1966)
\bibitem{bar2} A.O.\ Barut in {\em Mathematical Methods in
Theoretical Physics} (Boulder Lectures in Theoretical Physics,
vol.\ IX A), ed. by W.E.\ Brittin, A.O.\ Barut and M.\ Guenin
(Gordon and Breach, Science Publishers, Inc.\, New York, 1967), p.\
125
\bibitem{sol} A.I.\ Solomon, Journ.\ Mathem.\ Phys.\ {\bf 12}, 390
(1971)
\bibitem{vour4} R.F.\ Bishop and A.\ Vourdas, Zeitschr.\ Physik B - 
Condensed Matter {\bf 71}, 527 (1988)
\bibitem{kat} J.\ Katriel, A.I.\ Solomon, G.\
D'Ariano and M.\ Rasetti, Phys.\ Rev.\ D {\bf 34}, 2332 (1986);
\\ J.\ Katriel, M.\ Rasetti and A.I.\ Solomon,  Phys.\ Rev.\ D {\bf
35}, 1248 (1987); \\ G.\ D'Ariano, S.\ Morosi, M.\ Rasetti, J.\
Katriel and A.I.\ Solomon, Phys.\ Rev.\ D {\bf 36}, 2399 (1987)
\bibitem{yur} B.\ Yurke, S.L.\ McCall and J.R.\ Klauder, Phys.\
Rev.\ A {\bf 33}, 4033 (1986); \\
U.\ Leonhardt, Phys.\ Rev.\ A {\bf 49}, 1231 (1994); \\
C.\ Brif and Y.\ Ben-Aryeh, Quantum Semiclass.\ Opt.\ {\bf 8}, 1 (1996); \\
C.\ Brif and A.\ Mann, Phys.\ Lett.\ A {\bf 219}, 257 (1996); Phys.\ Rev.\
A {\bf 54}, 4505 (1996); \\
V.\ Pe\v{r}inov\'a, A.\ Luk\v{s} and J.\ K\v{r}epelka, Journ.\ Opt.\ B:
Quantum Semiclass.\ Opt. {\bf 2}, 81 (2000)
\bibitem{nonl} See, e.g.\ Ref.\ \cite{texb1}, Ch.\ 5, Ref.\ 
\cite{texb2}, Ch.\ 22,  Ref.\ \cite{texb5}, Ch.\ 5.4  and
Ref.\ \cite{texb7}, Ch.\ 8
\bibitem{wig1} E.P.\ Wigner, Phys.\ Rev.\ {\bf 40}, 749 (1932);
in a footnote Wigner says that ``the expression \ref{eq:274} was found
by L.\ Szilard and the present author some years ago for another purpose''. \\
 Important
follow-up papers are: \\
J.E.\ Moyal, Proceed.\ Cambridge Philos.\ Soc.\ {\bf 45}, 99
(1949); \\
R.L.\ Stratonovich, Soviet Phys.\ JETP {\bf 4}, 891 (1957); {\bf 5}, 1206
 (1957)
\bibitem{tat} V.I.\ Tatarski\v{i}, Sov.\ Phys.\ Usp.\ {\bf 26}, 311 (1983)
\bibitem{wig2} M.\ Hillery, R.F.\ O'Connel, M.O.\ Scully
 and E.P.\ Wigner, Phys.\
Reports {\bf 106}, 121 (1984)
\bibitem{folla} G.B.\ Folland, Ref.\ \cite{sto}, Ch.\ 1 
\bibitem{kim3} Y.S.\ Kim and M.E.\ Noz, Ref.\ \cite{squ}
\bibitem{lee} H.-W. Lee, Physics Rep.\ {\bf 259}, 147 (1995)
\bibitem{gad} M.\ Gadella, Fortschr.\ Physik {\bf 43}, 229 (1995)
\bibitem{drag} D.\ Dragoman, Progr.\ in Optics {\bf 43}, 433 (2002)
\bibitem{uhl} G.E.\ Uhlenbeck, Journ.\ of Mathem.\ and Physics {\bf 14}, 
10 (1934)
\bibitem{hus} K.\ Husimi, Proceed.\ Physico-Mathematical Soc.\
Japan (3rd Ser.) {\bf 22}, 264 (1940); related papers by Husimi are: 
 Progr.\ Theor.\ Phys.\ {\bf 9}, 238
and 381 (1953)
\bibitem{kan}Y.\ Kano, Proceed.\ Phys.\ Soc.\ Japan {\bf 19}, 1555 (1964);
 Journ.\ Math.\ Phys.\ {\bf 6}, 1913 (1965)
\bibitem{su} E.C.G.\ Sudarshan, Phys.\ Rev.\ Lett.\ {\bf 10}, 277 (1963)
\bibitem{su1} E.C.G.\ Sudarshan, Intern.\ Journ.\ Theor.\ Phys.\
 {\bf 32}, 1069 (1993)
\bibitem{glau2} R.J.\ Glauber, Phys.\ Rev.\ {\bf 131}, 2766 (1963)
\bibitem{kla2} J.R.\ Klauder, J.\ McKenna and D.G.\ Currie, J.\ Math.\
Phys.\ {\bf 5}, 734 (1965); J.R.\ Klauder, Phys.\ Rev.\ Lett.\ {\bf 16},
 534 (1966)
\bibitem{su2} C.L.\ Mehta and E.C.G.\ Sudarshan, Phys.\ Rev.\ {\bf 138},
 B 274 (1965) 
\bibitem{kla3} J.R.\ Klauder and E.C.G.\ Sudarshan, Ref.\ \cite{coh1a},
Ch.\ 8-4
\bibitem{mill}  M.M.\ Miller and E.A.\ Mishkin, Phys.\ Rev.\ {\bf 164},
 1610 (1967); \\ M.M.\ Miller, Journ.\ Math.\ Phys.\ {\bf 9}, 1270 (1968)
\bibitem{gel} I.M.\ Gel'fand and G.E.\ Shilov, {\em Generalized Functions}
vol.\ I (Academic Press, New York and London, 1964), here Ch.\ II; vol.\ II 
(Academic Press, New York and London, 1968), here Ch.\ III
\bibitem{zem} A.H.\ Zemanian, {\em Distribution Theory and Transform Analysis}
 (McGraw-Hill Book Co., New York etc., 1965), here Secs.\ 7.6 - 7.10
\bibitem{jay} E.T.\ Jaynes, Phys.\ Rev.\ {\bf 106}, 620 (1957);
 {\bf 108}, 171 (1957); \\ idem, in {\em Statistical Physics}, vol.\ 3 of the
1962 Brandeis Summer Institute Lectures in Theor.\ Physics, ed.\ by K.W. Ford
(W.A.\ Benjamin, Inc., New York and Amsterdam, 1963), p.\ 181
\bibitem{wolf2} G.S.\ Agarwal and E.\ Wolf, Phys.\ Rev.\ D {\bf 2}, 2161, 2187,
2206 (1970)
\bibitem{fan} H.-yi Fan, Y.\ Fan and J.-h.\ Chen, Journ.\ Opt.\ B: Quantum
Semiclass.\ Opt.\ {\bf 4}, 336 (2002)
\bibitem{roc} F.\ Rocca, Le Journ.\ de Physique {\bf 28}, 113 (1967)
\bibitem{per4} J.\ Pe\v{r}ina and L.\ Mi\v{s}ta, Phys.\ Lett.\ {\bf 27A},
217 (1968); Ann.\ Physik (Leipzig), 7.\ Folge, {\bf 22}, 372 (1969);
see also J.\ Pe\v{r}ina, {\em Quantum Statistics of Linear and Nonlinear
Optical Phenomena}, 2nd completely rev.\ ed.\ (Kluwer Academic Publ., 
Dordrecht etc., 1991), Sec.\ 4.3
\bibitem{luks} A.\ Luk\v{s}, Czech.\ Journ.\ Phys.\ B {\bf 26}, 1095 (1976)
\bibitem{per5} A.\ Pe\v{r}ina, Czech.\ Journ.\ Phys.\ B {\bf 21}, 731 (1971)
\bibitem{wue3} A.\ W\"unsche, Acta physica slovaca {\bf 48}, 385 (1998);
Journ.\ Phys.\ A: Math.\ Gen.\ {\bf 32}, 3179 (1999)
\bibitem{rich} Th.\ Richter, Journ.\ Mod.\ Optics {\bf 48}, 1881 (2001)
\bibitem{meht2} C.L.\ Mehta and E.C.G.\ Sudarshan, Phys.\ Rev.\ {\bf 138},
B 274 (1965)
\bibitem{cahi} K.E.\ Cahill and R.J.\ Glauber, Phys.\ Rev.\ {\bf 177}, 1857
and 1882 (1969)
\bibitem{san} M.A.M.\ Santiago and A.N.\ Vaidya, Journ.\, Phys.\ A: Math.\
Gen. {\bf 9}, 897 (1976); in the numerator of Eq.\ (2.20) a factor $2$ is
missing
\bibitem{trua} D.R.\ Truax, Phys.\ Rev.\ D {\bf 31}, 1988 (1985)
\bibitem{magn} W.\ Magnus, F.\ Oberhettinger and R.P.\ Soni, {\em
Formulas and Theorems for the Special Functions of Mathematical Physics}
(Die Grundlehren der mathem. Wissenschaften vol.\ 52; Springer-Verlag,
Berlin etc., 1966), p.\ 70 
\bibitem{zem1} See, e.g.\ Ref.\ \cite{zem}, p.\ 55, problem 10
\bibitem{gel1} See Ref.\ \cite{gel}, vol.\ I, Appendix 1 of Ch.\ I;
Ref.\ \cite{zem}, Sec.\ 1.2
\bibitem{er3c} Ref.\ \cite{er1}, p.\ 187, Eq.\ (12)
\bibitem{er4a} Ref.\ \cite{er2}, p.\ 176, Eq.\ (20); see also Ref.\ \cite{er1},
Sec.\ 3.15
\bibitem{mann4} C.\ Brif and A.\ Mann, Journ.\ Phys.\ A: Math.\ Gen.\ {\bf 31},
L9 (1998); Phys.\ Rev.\ A {\bf 59}, 971 (1999)
\bibitem{helg} S.\ Helgason, {\em Topics in Harmonic Analysis on Homogeneous
Spaces} (Progress in Mathematics vol.\ 13, Birkh\"auser, Boston etc., 1981),
 \S 4; idem, Ref.\ \cite{hel2}, Introduction
\bibitem{er5a} See Ref.\ \cite{er1}, p.\ 175, Eq.\ (1)
\bibitem{meh} C.L.\ Mehta, Phys.\ Rev.\ Lett.\ {\bf 18}, 752 (1967);
 see also the first paper of Ref.\ \cite{wolf2}
\bibitem{born} M.\ Born and E.\ Wolf, {\em Principles of Optics}, 7th
 (expanded) ed. (Cambridge Univ.\ Press, Cambridge, 1999)
\bibitem{cho} Y.\ Choquet-Bruhat and C.\ DeWitt-Morette, {\em Analysis,
Manifolds and Physics}, rev.\ ed.\ (North-Holland Publ.\ Co., Amsterdam, 
New York and Oxford, 1982), Chs.\ III and IV
\bibitem{mar1} J.E.\ Marsden and T.S.\ Ratiu, {\em Introduction to Mechanics
and Symmetry}, 2nd ed.\ (Springer-Verlag, New York, Berlin and Heidelberg, 
1999), Ch.\ 4
\bibitem{dirac} P.A.M.\ Dirac, {\em Lectures on Quantum Mechanics} (Belfer
Graduate School of Science Monographs Series, no.\ 2, Yeshiva Univ., New York,
1964) \\
Later reviews and monographs on the subject are: \\
E.C.G.\ Sudarshan and N.\ Mukunda, {\em Classical Dynamics: A Modern
 Perspective} (John Wiley \& Sons, New York etc., 1974), Chs.\ 8 and 9; \\
A.\ Hanson, T.\ Regge and C.\  Teitelboim, {\em Constrained Hamiltonian Systems
} (Accademia Nazionale dei Lincei, Roma, 1976); \\
K.\ Sundermeyer, {\em Constrained Dynamics} (Lecture Notes in Physics 169,
Springer-Verlag, Berlin, Heidelberg and New York, 1982); \\
M.\ Henneaux and C.\ Teitelboim, {\em Quantization of Gauge Systems} (Princeton
University Press, Princeton, N.J., 1992); \\
Ref.\ \cite{mar1}, Ch.\ 8\,; \\
J.R.\ Klauder, in {\em Methods of Quantization, Lectures Held at the 39.\
Universit\"atswochen f\"ur Kern- und Teilchenphysik, Schladming, Austria},
ed.\ by H.\ Latal and W.\ Schweiger (Lecture notes in physics 572, 
Springer-Verlag, Berlin etc., 2001), p.\ 143 
\bibitem{ab1} R.\ Abraham and J.E.\ Marsden, {\em Foundations of Mechanics},
2nd ed.\ (The Benjamin/Cummings Publ.\ Co., Inc., Reading, Mass.\, etc.,
1978), pp.\ 301-302
\bibitem{ab2} Ref.\ \cite{ab1}, pp.\ 722-723
\bibitem{ste} N.\ Steenrod, {\em The Topology of Fiber Bundles} (Princeton
University Press, Princeton, N.J., 1951), p.\ 105
\bibitem{cush2} R.H.\ Cushman and L.M.\ Bates,  Ref.\ \cite{orbi}, Sec.\ I.4
\bibitem{thur1} W.P.\ Thurston, Ref.\ \cite{orbi}, book,  Sec.\ 2.7 
\bibitem{ler2} R.\ Cushman and R.\ Sjamaar, in {\em Symplectic Geometry and
Mathematical Physics, Actes du colloque en l'honneur de Jean-Marie Souriau},
ed.\ by P.\ Donato, C.\ Duval, J.\ Elhadad and G.M.\ Tuynman (Birkh\"auser,
Boston, Basel and Berlin, 1991), p.\ 114; \\ 
 E.\ Lerman, R.\ Montgomery, R.\ Sjamaar, in {\em Symplectic
Geometry}, ed.\ by D.\ Salamon (London Mathem.\ Soc.\ Lecture Note Series,
no.\ 192, Cambridge Univ.\ Press, Cambridge, 1993), p.\ 127
\bibitem{born2} Ref.\ \cite{born}, Secs.\ 6.2 and 6.3; \\
R.M.A.\ Azzam and N.M.\ Bashara, {\em Ellipsometry and Polarized Light},
 paperback ed.\ (North-Holland Personal Library, North-Holland, Amsterdam etc.,
1987), Ch.\ 1
\bibitem{walk1} N.G.\ Walker and J.E.\ Carroll, Electron.\ Lett.\ {\bf 20},
981 (1984); Opt.\ Quantum Electronics {\bf 18}, 355 (1986); \\
N.G.\ Walker, Journ.\ Mod.\ Optics {\bf 34}, 15 (1987)
\bibitem{mand} L.\ Mandel, in {\em Foundations of Quantum Mechanics, Santa Fe
Workshop, May 1991}, ed.\ by T.D.\ Black, M.M.\ Nieto, H.S.\ Pilloff, M.O.\
Scully and R.M.\ Sinclair (World Scientific, Singapore etc., 1992), p.\ 55
\bibitem{schlei1} M.\ Freyberger and W.\ Schleich, Phys.\ Rev.\ A {\bf 47},
R30 (1993); \\
M.\ Freyberger, K.\ Vogel and W.\ Schleich, Quantum Opt.\ {\bf 5}, 65 (1993);
Phys.\ Lett.\ A {\bf 176}, 41 (1993); \\
M.\ Freyberger, M.\ Heni and W.P.\ Schleich, Quantum Semiclass.\ Opt.\
{\bf 7}, 187 (1995); \\ M.T.\ Fontanelle, S.L.\ Braunstein, W.P.\ Schleich
and M.\ Hillery, Acta Phys.\ Slov.\ {\bf 46}, 373 (1996) = e-print
 quant-ph/9712032
\bibitem{leon2} U.\ Leonhardt and H.\ Paul, Phys.\ Rev.\ A {\bf 48}, 4598
 (1993); Phys.\ Rev.\ Lett.\ {\bf 72}, 4086 (1994); \\
U.\ Leonhardt, B.\ B\"ohmer and H.\ Paul, Optics Comm.\ {\bf 119}, 296 (1995);
\\ Th.\ Richter, Phys.\ Rev.\ A {\bf 56}, 3134 (1997)
\bibitem{loud} R.\ Loudon, in {\em Coherence, cooperation and fluctuations, 
Proceed.\ Sympos.\  60th Birthday Prof.\ R.J.\ Glauber, Oct.\ 1985},
ed.\ by F.\ Haake, L.M.\ Narducci and D.\ Walls (Cambridge Studies in Modern
Optics: 5, Cambridge Univ.\ Press, Cambridge etc., 1986), p.\ 240
\bibitem{wel} D.-G.\ Welsch, W.\ Vogel and T.\ Opatrn\'y, Progr.\ Optics
{\bf 39}, 63 (1999)
\bibitem{opat} T.\ Opatrn\'y, in {\em Coherence and Statistics of Photons
and Atoms}, ed.\ by J.\ Pe\v{r}ina (Wiley Series in Lasers and Applications 3,
John Wiley \& Sons, Inc., New York etc., 2001), p.\ 159 
\bibitem{homod} See, e.g. the textbooks Ref.\ \cite{texb5}, Ch.\ 4.12;
Ref.\ \cite{texb7}, Ch.\ 6.5
\bibitem{schlei3}  Ref.\ \cite{texb8}, ch.\ 13; see also Ref.\ \cite{schlei1}
\bibitem{Noh1} J.W.\ Noh, A.\ Foug\`{e}res and L.\ Mandel, Phys.\
Rev.\ Lett.\ {\bf 67}, 1426 (1991); \\ Phys.\ Rev.\ A {\bf 45}, 424
(1992);\\ Phys.\ Rev.\ A {\bf 46}, 2840 (1992);\\ Phys.\ Rev.\ A {\bf
47}, 4535; 4541 (1993);\\ Phys.\ Rev.\ Lett.\ {\bf 71}, 2579 (1993);\\
Phys.\ Rev.\ A {\bf 48}, 1719 (1993);\\ Physica Scripta {\bf T48},
29 (1993); \\
 A.\ Foug\`{e}res, J.W.\ Noh, T.P.\ Grayson and L.\
Mandel, Phys.\ Rev.\ A {\bf 49}, 530 (1994); \\
A.\ Foug\`{e}res, J.R.\ Torgerson and L.\ Mandel,
Optics Comm.\ {\bf 105}, 199 (1994); \\
J.R.\ Torgerson and L.\ Mandel, Phys.\ Rev.\ Lett.\
{\bf 76}, 3939 (1995);\\ Optics Comm.\ {\bf 133}, 153 (1997);
Physica Scripta {\bf T76}, 110 (1998). \\
As to controversies and further comments see \\
Z.\ Hradil, Phys.\ Rev.\ A {\bf 47}, 4532 (1993); reply by J.W.\ Noh,
A.\ Foug\`{e}res and L.\ Mandel, ibidem, p.\ 4535, \\
S.M.\ Barnett and D.T.\ Barnett, Phys.\ Rev.\ A {\bf 47}, 4537 (1993);
reply by J.W.\ Noh,A.\ Foug\`{e}res and L.\ Mandel, ibidem, p.\ 4541, \\
Z.\ Hradil and J.\ Bajer, Phys.\ Rev.\ A {\bf 48}, 1717 (1993);
reply by J.W.\ Noh,A.\ Foug\`{e}res and L.\ Mandel, ibidem, p.\ 1719, \\
P.\ Riegler and K.\ W\'{o}dkiewicz, Phys.\ Rev.\ A {\bf 49}, 1387 (1994); \\
B.-G.\ Englert and K.\ W\'{o}dkiewics, Phys.\ Rev.\ A {\bf 51}, R2661 (1995);
 \\ B.-G.\ Englert, K.\ W\'{o}dkiewics and P.\ Riegler, Phys.\ Rev.\ A
 {\bf 52}, 1704 (1995)
\bibitem{smith} D.T.\ Smithey, M.\ Beck, A.\ Faridani
 and M.G.\ Raymer, Phys.\ Rev.\
Lett.\ {\bf 70}, 1244 (1993); \\ M.\ Beck, D.T.\ Smithey and M.G.\ Raymer,
Phys.\ Rev.\ A {\bf 48}, R890 and 3159 (1993); \\ M.G.\ Raymer, J.\ Cooper,
H.J.\ Carmichael, M.\ Beck and D.T.\ Smithey, Journ.\ Optic.\ Soc.\ Amer.\
B {\bf 12}, 1801 (1995)
\bibitem{hrad}J.\ \v{R}eh\'a\v{c}ek, Z.\ Hradil, M.\ Zawisky,
 S.\ Pascazio, H.\ Rauch and
J.\ Pe\v{r}ina, Phys.\ Rev.\ A {\bf 60}, 473 (1999); \\
 J.\ \v{R}eh\'a\v{c}ek, Z.\ Hradil, M.\ Du\v{s}ek, O.\ Haderka
 and M.\ Hendrych,
Journ.\ Opt.\ B: Quantum Semiclass.\ Opt.\ {\bf 2}, 237 (2000)
\bibitem{luis1}  T.\ Hakio\u{g}lu, A.S.\ Shumovsky
 and O.\ Ayt\"ur, Phys.\ Lett.\
A {\bf 194}, 304 (1994); \\
A.\ Luis and J.\ Pe\v{r}ina, Quantum Semiclass.\ Opt.\ {\bf 8}, 873 and 887 
(1996); \\T.\ Hakio\u{g}lu, Phys.\ Rev.\ A {\bf 59}, 1586 (1999); \\
A.\ Cives-Esclop, A.\ Luis and L.L.\ S\'anchez-Soto, Optics Comm.\ {\bf 175},
153 (2000); J.\ Opt.\ B: Quantum Semiclass.\ Opt.\ {\bf 2}, 526 (2000)
\bibitem{Klaud4}J.R.\ Klauder, Ann.\ Physics (N.Y.) {\bf 254}, 419 (1997); \\
J.\ Govaerts and J.R.\ Klauder, Ann.\ Physics (N.Y.) {\bf 274}, 251 (1999); \\
J.R.\ Klauder, see Ref.\ \cite{dirac}
\bibitem{sour} J.-M.\ Souriau, {\em Structure of Dynamical Systems, A
 Symplectic View of Physics} (Progress in Mathematics 149, Birkh\"auser,
Boston, Basel and Berlin, 1997; translation of {\em Structure des syst\`emes
 dynamiques}, Dunod, Paris, 1970), \S 11
\bibitem{guil2} Ref.\ \cite{gui}, Ch.\ II
\bibitem{mar2a} Ref.\ \cite{mar1}, Ch.\ 11
\bibitem{cush4} R.H.\ Cushman and L.M.\ Bates, Ref.\ \cite{orbi}, Appendix B.3
\bibitem{mar3} J.E.\ Marsden and A.\ Weinstein, in {\em Quantization of 
Singular Symplectic Quotients}, ed.\ by N.P.\ Landsman, M.\ Pflaum and
M.\ Schlichenmaier (Progress in Mathematics 198, Birkh\"auser Verlag,
Basel, Boston and Berlin, 2001), p.\ 1
\bibitem{bo2} M.\ Bojowald and T.\ Strobl, Journ.\ Math.\ Physics {\bf 41},
2537 (2000); Intern.\ Journ.\ Mod.\ Physics D {\bf 12}, 713 (2003)
\bibitem{cush3} R.H.\ Cushman and L.M.\ Bates, Ref.\ \cite{orbi},
 p.\ 315, example 5
\bibitem{bar1} V.\ Bargmann, Ann.\ Math.\  {\bf 48}, 568 (1947)
\bibitem{gel0} I.M.\ Gel'fand, M.I.\ Graev and N.Ya.\ Vilenkin, {\em
Generalized Functions, vol.\ 5: Integral Geometry and
Representation Theory} (Academic Press, Inc., New York and London,
1966), Ch.\ 7
\bibitem{sa1} P.J.\ Sally, Jr., {\em Analytic continuation of the irreducible
unitary representations of the universal covering group of
$SL(2,\mathbb{R})$} (Memoirs of the Amer.\ Math.\ Soc.\ 69, Amer.\
Math.\ Soc., Providence, R.I., 1967)
\bibitem{wa} G.\ Warner, {\em Harmonic Analysis on Semi-Simple Lie Groups I}
 (Springer-Verlag, Berlin etc., 1972)
\bibitem{lan} S.\ Lang, $SL_2(\mathbb{R})$ (Graduate Texts in Mathematics 105;
Springer-Verlag, New York etc., 1975)
\bibitem{hel2} S.\ Helgason, {\em Groups and Geometric Analysis} (Academic
Press, Inc., Orlando etc., 1984; reprinted in the series Mathematical Surveys
and Monographs, no.\ 83, Amer.\ Mathem.\ Soc., Providence, R.I., 2000)
\bibitem{kna} A.W.\ Knapp, {\em Representation Theory of Semisimple
Groups, an Over\-view based on Examples} (Princeton University
Press, Princeton, N.J., 1986)
\bibitem{per} A.\ Perelomov, {\em
Generalized Coherent States and Their Applications}
(Springer-Verlag, Berlin etc., 1986), here Ch.\ 5
\bibitem{sug} M.\ Sugiura, {\em Unitary Representations and Harmonic
 Analysis,} 2nd ed.\ 
(North-Holland Mathem.\ Library 44, North-Holland,
 Amsterdam etc., 1990), here Ch.\ 5
\bibitem{vil}N.Ja.\ Vilenkin and A.U.\ Klimyk,
 {\em Representation of Lie Groups and Special Functions, vol.\ 1:
 Simplest Lie Groups, Special Functions and Integral Transforms}
 (Mathematics and Its Applications (Soviet Series) 72; Kluwer Academic
 Publishers, Dordrecht etc., 1991), here Ch.\ 7
\bibitem{how} R.\ Howe and E.C.\ Tan, {\em Non-Abelian Harmonic Analysis,
Applications of $SL(2,\mathbb{R})$} (Springer-Verlag, New York
etc., 1992)
\bibitem{muk0} R.\ Simon and N.\ Mukunda, in {\em Symmetries in Science VI, 
From the Rotation Group to Quantum Algebras}, ed.\ by B.\ Gruber
 (Plenum Press,
 New York and London, 1993), p.\ 659
\bibitem{hel1} Ref.\ \cite{hel}, Chs.\ VI and IX
\bibitem{wa1} Ref.\ \cite{wa}, Sec.\ 1.2
\bibitem{lan1} Ref.\ \cite{lan}, Chs.\ XI and XIII
\bibitem{war3} Ref.\ \cite{wa}, p.\ 58
\bibitem{prin} As to the other series of irreducible unitary representations
of the group $SU(1,1)$ see, e.g.\ the Refs.\ \cite{bar1,sa1,lan,sug}; \\
Kronecker (tensor) products of irreducible unitary representations are
 discussed by \\
L.\ Puk\'{a}nszky, Trans.\ Amer.\ Math.\ Soc.\ {\bf 100}, 116 (1961); \\
W.J.\ Holman, III and L.C.\ Biedenharn, Jr., Ref.\ \cite{bied} and Ann.\
 Physics (N.Y.) {\bf 47}, 205 (1968); \\
G.\ Rideau, Ann.\ Inst.\ Henri Poincar\'{e} A {\bf 4}, 67 (1966); \\
 S.S.\ Sannikov, Sov.\ Phys.- Doklady {\bf 11}, 1045 (1967); \\
I.\ Ferretti and M.\ Verde, Nuovo Cim.\ {\bf 55 A}, 110 (1968); \\
K.-H.\ Wang, Journ.\ Math.\ Phys.\ {\bf 11}, 2077 (1970); \\ N.\ Mukunda and
B.\ Radhakrishnan, Journ.\ Math.\ Phys.\ {\bf 15}, 1320, 1332, 1643, 1656
(1974); use is made of the authors paper Journ.\ Math.\ Phys.\ {\bf 14}, 254
(1973); \\ 
H.\ Neunh\"{o}ffer, Sitzungsber.\ Heidelberger Akad.\ Wissensch., 
Mathem.-naturw.\ Klasse, Jahrgang 1978, 3.\ Abhandl.\ (Springer-Verlag,
Berlin etc.\, 1978), p.\ 167; \\ Ref.\ \cite{vil}, Ch.\ 8.7 
\bibitem{puk} L.\ Puk\'anszky, Mathem.\ Annalen {\bf 156}, 96 (1964); \\
M.\ Montgomery and L.\ O'Raifeartaigh, Journ.\ Math.\ Phys.\ {\bf 15}, 
380 (1974); \\ P.J.\ Gambardella, Journ.\ Math.\ Phys.\ {\bf 16}, 1172 (1975)
\bibitem{sa2} P.J.\ Sally, Jr., Journ.\ Funct.\ Anal.\ {\bf 6}, 441 (1970)
\bibitem{sie} C.L.\ Siegel, Amer.\ Journ.\ Mathem.\ {\bf 65}, 1 (1943);
the article was reprinted as a book: C.L.\ Siegel, {\em Symplectic Geometry}
 (Academic Press, N.Y.\ and London, 1964)           
\bibitem{car1} H.\ Cartan, in {\em Fonctions Automorphes} (S\'eminaire Henri
 CARTAN, 10e ann\'ee: 1957/ 1958), \'ed.\  Ecole Normale Sup\'erieure
(Secr\'etariat math\'ematique, Paris, 1958), vol.\ 1, expos\'e 3
\bibitem{barg3} V.\ Bargmann, in {\em Analytical methods in mathematical
physics}, ed.\ by R.P.\ Gilbert and R.G.\ Newton (Based on the conference held
at Indiana University, Bloomington, Indiana, June 2-6, 1968; Gordon and Breach
Science Publ., N.Y., London and Paris, 1970), p.\  27
\bibitem{foll} G.B.\ Folland, Ref.\ \cite{sto}
\bibitem{gre} W.H.\ Greub, {\em Multilinear Algebra} (Springer-Verlag, Berlin,
Heidelberg and N.Y., 1967), Ch.\ V
\bibitem{mar2} R.\ Abraham, J.E.\ Marsden and T.\ Ratiu, {\em Manifolds,
Tensor Analysis, and Applications} (Addison-Wesley Publ.\ Comp., London etc.,
1983), Ch.\ 6
\bibitem{kna2} See Ref.\ \cite{kna}, p.\ 130
\bibitem{helg2} S.\ Helgason, Ref.\ \cite{hel}
\bibitem{war2} Ref.\ \cite{wa}, here especially p.\ 84
\bibitem{ros2} G.\ Rosensteel and D.J.\ Rowe, Intern.\ Journ.\ Theor.\
Phys.\ {\bf 15}, 453 (1976)
\bibitem{wig3} As to properties of this group, its unitary representations
and the related literature till 1968 see \\
T.O.\ Philips and E.P.\ Wigner, in {\em Group Theory and Its Applications,
 vol.\ I}, ed.\ by E.M.\ Loebl (Academic Press, New York and London, 1968),
 p.\ 631
\bibitem{hawk} S.W.\ Hawking and G.F.R.\ Ellis, {\em The large scale
structure of space-time} (Cambridge University Press, Cambridge, 1973),
p.\ 131
\bibitem{pol} J.\ Polchinski, {\em String Theory, vol.\ II} (Cambridge
University Press, Cambridge, 1998), p.\ 223
\bibitem{car2} E.\ Cartan, Ann.\ Sci.\ \'Ecole Norm.\ Sup.\ {\bf 31},
263 (1914)
\bibitem{sie2} Ref.\ \cite{sie}, here nos. 56-58; due to the use of
 ornamental Gothic capital letters there are several confusing 
mix-ups of the letters ``V'' and ``B''!
\bibitem{dir3} P.A.M.\ Dirac, Journ.\ Math.\ Phys.\ {\bf 4}, 901 (1963)
\bibitem{car3} E.\ Cartan, Abhandl.\ aus dem Mathem.\ Seminar der
Hansischen Universit\"at (Hamburg) {\bf 11}, 116 (1936)
\bibitem{wei} A.\ Weil, Acta mathematica {\bf 111}, 143 (1964)
\bibitem{itz} C.\ Itzykson, Commun.\ Mathem.\ Phys.\ {\bf 4}, 92 (1967); \\
see also D.\ Shale, Trans.\ Amer.\ Math.\ Soc.\ {\bf 103}, 149 (1962)
\bibitem{ehr} J.B.\ Ehrman, Proceed.\ Cambridge Philos.\ Soc.\ {\bf 53},
290 (1957)
\bibitem{fro} C.\ Fronsdal, Rev.\ Mod.\ Phys.\ {\bf 37}, 221 (1965)
\bibitem{eva} N.T.\ Evans, Journ.\ Mathem.\ Phys.\ {\bf 8}, 170 (1967)
\bibitem{god} R.\ Godement, in {\em Fonctions Automorphes, vol.\ 1},
see Ref.\ \cite{car1}, expos\'es 5, 6 et 10
\bibitem{frad} E.\ Majorana, Nuovo Cim.\ {\bf 9}, 335 (1932); as to the
early history of this type of equations see D.M.\ Fradkin, Amer.\ Journ.\
Phys.\ {\bf 34}, 314 (1966)
\bibitem{baru1} A.O.\ Barut and H.\ Kleinert, Phys.\ Rev.\ {\bf 156}, 1546
(1967); \\
A.\ B\"ohm, in {\em High Energy Physics and Fundamental Particles},
ed.\ by A.O.\ Barut and W.E.\ Brittin (Lectures in Theor.\ Physics X-B,
Gordon and Breach Science Publ., N.Y., London and Paris, 1968), p.\ 483; \\
A.\ B\"ohm, Phys.\ Rev.\ {\bf 175}, 1767 (1968); Phys.\ Rev.\ D {\bf 3},
367 (1971); \\
A.O.\ Barut and A.\ B\"ohm, Journ.\ Mathem.\ Phys.\ {\bf 11}, 2938 (1970); \\
L.\ Jaffe, Journ.\ Mathem.\ Phys.\ {\bf 12}, 882 (1971); idem,
 in {\em De Sitter
and Conformal Groups and Their Applications}, ed.\ by A.O.\ Barut and
W.E.\ Brittin (Lectures in Theor.\ Physics XIII, Colorado Assoc. Univ.\ Press,
Boulder, Colorado, 1971), p.\ 125, 
\bibitem{dir4} P.A.M.\ Dirac, Proceed.\ Roy.\ Soc.\ London A {\bf 322},
435 (1971); {\bf 328}, 1 (1972) 
\bibitem{suda} E.C.G.\ Sudarshan and N.\ Mukunda, Phys.\ Rev.\ D {\bf 1},
571 (1970);\\
N.\ Mukunda, H.\ van Dam and L.C.\ Biedenharn, {\em Relativistic Models
of Extended Hadrons Obeying a Mass-Spin Trajectory Constraint}, Lectures,
ed.\ by A.\ B\"ohm and J.D.\ Dollard (Lecture Notes in Physics 165,
Springer-Verlag, Berlin, Heidelberg and N.Y., 1982)
\bibitem{fro2} C.\ Fronsdal, Phys.\ Rev.\ D {\bf 10}, 589 (1974);\\
C.\ Fronsdal and R.B.\ Haugen, Phys.\ Rev.\ D {\bf 12}, 3810 (1975); \\
C.\ Fronsdal, Phys.\ Rev.\ D {\bf 12}, 3819 (1975); \\
J.\ Fang and C.\ Fronsdal, Lett.\ Mathem.\ Phys.\ {\bf 2}, 391 (1978);
 Phys.\ Rev.\ D {\bf 18}, 3630 (1978); Phys.\ Rev.\ D {\bf 22}, 1361 (1980); \\
M.\ Flato and C.\ Fronsdal, Lett.\ Mathem.\ Phys.\ {\bf 2}, 421 (1978); \\
W.\ Heidenreich, Journ.\ Mathem.\ Phys.\ {\bf 22}, 1566 (1981); \\
E.\ Angelopoulos, M.\ Flato, C.\ Fronsdal and D.\ Sternheimer, Phys.\ Rev.\
D {\bf 23}, 1278 (1981); \\
M.\ Flato and C.\ Fronsdal, Journ.\ Mathem.\ Phys.\ {\bf 22}, 1100 (1981);
 Phys.\ Lett.\ {\bf 97 B}, 236 (1980) 
\bibitem{howe2} R.\ Howe, in {\em Application of Group Theory in Physics
and Mathematical Physics}, ed.\ by M.\ Flato, P.\ Sally and G.\ Zuckerman
(Lectures in Appl.\ Mathem.\ 21, Amer.\ Math.\ Soc., Providence, R.I., 1985),
p.\ 179
\bibitem{lip2} S.\ Goshen and H.J.\ Lipkin, in {\em Spectroscopic and Group
Theoretical Methods in Physics, Racah Memorial Volume},  ed.\ by F.\ Bloch,
S.G.\ Cohen, A.\ De-Shalit, S.\ Sambursky and I.\ Talmi (North-Holland Publ.\
Co., Amsterdam, 1968), p.\ 245
\bibitem{mosh} M.\ Moshinsky and C.\ Quesne, Journ.\ Mathem.\ Phys.\
{\bf 12}, 1772 and 1780 (1971)
\bibitem{kir} A.A.\ Kirillov, {\em Elements of the Theory of Representations}
(Grundlehren der mathematischen Wissenschaften 220, Springer-Verlag, Berlin,
Heidelberg and N.Y., 1976; translation of the Russian edition from 1972), pp.\
287-290
\bibitem{gelb} S.\ Gelbart, Inventiones math.\ {\bf 19}, 49 (1973); closely
related papers of the same author are: Bull.\ Amer.\ Mathem.\ Soc.\ {\bf 78},
451 (1972); Trans.\ Amer.\ Mathem.\ Soc.\ {\bf 192}, 29 (1974)
\bibitem{ros1} G.\ Rosensteel and D.J.\ Howe, Intern.\ Journ.\ Theor.\
Phys.\ {\bf 16}, 63 (1977)
\bibitem{ang} E.\ Angelopoulos, in {\em Quantum Theory, Groups, Fields and
Particles}, ed.\ by A.O.\ Barut (D.\ Reidel Publ.\ Co., Dordrecht, Boston
and London, 1983), p.\ 101
\bibitem{syqo} R.F.\ Bishop and A.\ Vourdas, Zeitschr.\ Physik B - 
Condensed Matter {\bf 71}, 527 (1988);  \\
A.\ Luis and L.L.\ S\'anchez-Soto, Quantum Semiclass.\ Opt.\ {\bf 7}, 153
 (1995); \\
D.A.\ Trifonov, Journ.\ Phys.\ A: {\bf 31}, 5673 (1998); \\
J.\ Fiur\'a\v{s}ek and J.\ Pe\v{r}ina, Phys.\ Rev.\ A {\bf 62}, 033808-1
(2000) 
\bibitem{muku2} R.\ Simon, E.C.G.\ Sudarshan and N.\ Mukunda, Phys.\ Rev.\
A {\bf 31}, 2419 (1985); {\bf 36},  3868 (1987); {\bf 37}, 3028 (1988); \\
R.\ Simon, N.\ Mukunda and B.\ Dutta, Phys.\ Rev.\ A {\bf 49}, 1567 (1994);\\
Arvind, B.\ Dutta, N.\ Mukunda and R.\ Simon, Phys.\ Rev.\ A {\bf 52}, 1609
(1995); \\
B.\ Arvind and N.\ Mukunda, Journ.\ Phys.\ A: Math.\ Gen.\ {\bf 29}, 5855 
(1996); \\
Arvind, N.\ Mukunda and R.\ Simon, Journ.\ Phys.\ A: Math.\ Gen.\ {\bf 31},
565 (1998); \\
E.M.\ Rabei, Arvind, N.\ Mukunda and R.\ Simon, Phys.\ Rev.\ A {\bf 60},
3397 (1999)
\bibitem{kim2} D.\ Han, Y.S.\ Kim and M.E.\ Noz, Phys.\ Rev.\ A {\bf 41},
6233 (1990); \\
D.\ Han, Y.S.\ Kim, M.E.\ Noz and L.\ Yeh, Journ.\ Mathem.\ Phys.\ {\bf 34},
5493 (1993); \\
A.\ W\"unsche, Journ.\ Opt.\ B: Quantum Semiclass.\ Opt.\ {\bf 2}, 73 (2000);
{\bf 4}, 1 (2002)
\bibitem{kna3} Ref.\ \cite{kna}, pp.\ 130, 147, 152/53, 404-407, 498, 502-511,
606/7, 717/8
\bibitem{prz} T.\ Przebinda, {\em The oscillator duality correspondence
for the pair  $O(2,2),\,Sp(2, \mathbb{R})~~$} (Memoirs Amer.\ Mathem.\ Soc.\
vol.\ 79, no.\ 403, Amer.\   Mathem.\ Soc., Providence, R.I., 1989)
\bibitem{prz1} Ref.\ \cite{prz}, p.\ 36 
\bibitem{wey2} H.\ Weyl, {\em The Theory of Groups and Quantum Mechanics},
  Ref.\ \cite{wey}, p.\ 137
\bibitem{waer} B.L.\ van der Waerden, {\em Die gruppentheoretische Methode
in der Quantenmechanik} (Verlag Julius Springer, Berlin, 1932), \S 16
\bibitem{tdie} T.\ Br\"ocker and T.\ tom Dieck, {\em Representations of Compact
Lie Groups} (Graduate texts in mathematics 98, Springer-Verlag,
 New York etc., 1985), pp.\ 84-87
 \bibitem{god2} Ref.\ \cite{god}, expos\'e 10
\bibitem{wat1} G.N.\ Watson, {\em A Treatise on the Theory of Bessel Functions}
, 2nd ed.\ (Cambridge University Press, Cambridge, 1966), p.\ 79, formula 
(4)
\bibitem{wat2} Ref.\ \cite{wat1}, p.\ 79, formula (9)
\bibitem{gam} See, e.g.\ Ref. \cite{er1}, p.\ 1, formula (1) and p.\ 4, 
formula (10)  
\bibitem{hard} G.H.\ Hardy, Proceed.\ London Mathem.\ Soc.\ (Ser.\ 2)
{\bf 2}, 401 (1905)
\bibitem{barn1} E.W.\ Barnes, Philos.\ Transact.\ Roy.\ Soc.\ London
(Ser.\ A) {\bf 206}, 249 (1906); here Part III
\bibitem{niel} N.\ Nielsen, {\em Handbuch der Theorie der Gammafunktion}, Part
I in {\em Die Gammafunktion} (Chelsea Publ.\ Co.\, New York, 1965; originally
publ.\ by B.G.\ Teubner, Leipzig, 1906), p.\ 74, formula (14)
\bibitem{niel1} Ref.\ \cite{niel}, p.\ 76
\bibitem{hans} E.R.\ Hansen, {\em A Table of Series and Products} 
(Prentice-Hall, Inc., Englewood Cliffs, N.J., 1975), p.\ 149, formula (6.12.1)
\bibitem{barn2} Ref.\ \cite{barn1}, Part V
\end{thebibliography}
 \end{document}